\documentclass[final,12pt, 3p,times,onecolumn,compress]{elsarticle}

\usepackage[table, svgnames, dvipsnames]{xcolor}
\usepackage{amsmath, dsfont}
\usepackage{multirow}
\usepackage{amsfonts}
\usepackage{bbm}
\usepackage{mathdesign}
\usepackage{mathtools}
\usepackage{amsthm}
\usepackage{bm}
\usepackage{makecell}
\usepackage{graphicx}
\usepackage{amssymb}
\usepackage[colorlinks=true,citecolor=Blue,linkcolor=RubineRed,urlcolor=Blue]{hyperref}

\usepackage{float}
\usepackage{tikz}
\usepackage{ifthen}
\usepackage{stmaryrd}
\usetikzlibrary{tikzmark}
\usepackage{upgreek}
\usepackage{braket}
\usepackage{physics}
\usepackage{relsize}
\usepackage[export]{adjustbox}
\usepackage[scaled=1]{helvet}
\usepackage{accents}
\usepackage{yfonts}
\usepackage{braket, xcolor}
\usepackage{verbatim}
\usepackage{multirow}
\usepackage{scalerel}
\usepackage{soul}
\usepackage{makecell}
\usepackage{dcolumn}
\usepackage{amsfonts, mathtools, resizegather}
\usepackage{mathrsfs}
\DeclareMathOperator{\arcsinh}{arcsinh}
\newcommand{\id}{\mathbb{I}}
\newlength{\dhatheight}

\makeatletter
\newsavebox{\@brx}
\newcommand{\llangle}[1][]{\savebox{\@brx}{\(\m@th{#1\langle}\)}%
  \mathopen{\copy\@brx\kern-0.5\wd\@brx\usebox{\@brx}}}
\newcommand{\rrangle}[1][]{\savebox{\@brx}{\(\m@th{#1\rangle}\)}%
  \mathclose{\copy\@brx\kern-0.5\wd\@brx\usebox{\@brx}}}
\makeatother
\newcommand{\mc}[1]{\mathcal{#1}}
\newcommand{\mr}[1]{\mathrm{#1}}

\newcommand{\ha}[1]{\hat{#1}}

\journal{Physics Reports}

\begin{document}

\begin{frontmatter}

\title{Quantum Dynamics in Krylov Space: Methods and Applications}

%Pratik

\author[a,b]{Pratik Nandy
\href{https://orcid.org/0000-0001-5383-2458}
{\includegraphics[scale=0.05]{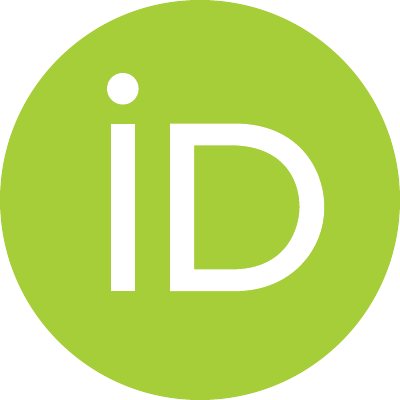}}}
\ead{pratik@yukawa.kyoto-u.ac.jp}
\affiliation[a]{organization={Center for Gravitational Physics and Quantum Information, Yukawa Institute for Theoretical Physics,\\ Kyoto University},
            addressline={Kitashirakawa Oiwakecho, Sakyo-ku}, 
           city={Kyoto},
            postcode={606-8502}, 
            country={Japan}}
\affiliation[b]{organization={RIKEN Center for Interdisciplinary Theoretical and Mathematical Sciences (iTHEMS)},
            addressline={\\2-1 Hirosawa}, 
           city={Wako},
            postcode={351-0198}, 
            state={Saitama},
            country={Japan}}
            
%Apollo
\author[c,d]{Apollonas S. Matsoukas-Roubeas \href{https://orcid.org/0000-0001-5517-0224}
{\includegraphics[scale=0.05]{orcidid.pdf}}}
\ead{sam310@cam.ac.uk}

\affiliation[c]{organization={Cavendish Laboratory, University of Cambridge}, addressline={J.J. Thomson Avenue, Cambridge, CB3 0HE}, 
country={UK}}

\affiliation[d]{organization={Department of Physics and Materials  Science,  University  of  Luxembourg}, addressline={L-1511}, 
country={Luxembourg}}

%Pablo
\author[d]{Pablo Martínez-Azcona \href{https://orcid.org/0000-0002-9553-2610}{\includegraphics[scale=0.05]{orcidid.pdf}}}
\ead{pablo.martinez@uni.lu}

%Anatoly
\author[e]{Anatoly Dymarsky \href{https://orcid.org/0000-0001-5762-6774}{\includegraphics[scale=0.05]{orcidid.pdf}}}
\ead{a.dymarsky@uky.edu} 
\affiliation[e]{organization={Department  of  Physics  and  Astronomy,  University  of  Kentucky}, addressline={Lexington, KY 40506}, country={USA}}

% Adolfo
\author[d,f]{Adolfo del Campo \href{https://orcid.org/0000-0003-2219-2851}{\includegraphics[scale=0.05]{orcidid.pdf}}\,}
\ead{adolfo.delcampo@uni.lu}

\affiliation[f]{organization={Donostia International Physics Center}, addressline={ E-20018, San Sebasti\'an}, country={Spain}}

%\email{qdks.review@gmail.com}

\begin{abstract}
The dynamics of quantum systems unfolds within a subspace of the state space or operator space, known as the Krylov space. This review presents the use of Krylov subspace methods to provide an efficient description of quantum evolution and quantum chaos, with emphasis on nonequilibrium phenomena of many-body systems with a large Hilbert space. It provides a comprehensive update of recent developments, focused on the quantum evolution of operators in the Heisenberg picture as well as pure and mixed states. It further explores the notion of Krylov complexity and associated metrics as tools for quantifying operator growth,  their bounds by generalized quantum speed limits,  the universal operator growth hypothesis, and its relation to quantum chaos, scrambling, and generalized coherent states. A comparison of several generalizations of the Krylov construction for open quantum systems is presented. A closing discussion addresses the application of Krylov subspace methods in quantum field theory, holography, integrability, quantum control, and quantum computing, as well as current open problems.

\end{abstract}
 
%%%%%%%%%%%%%%%%%%%%%%%%%%%%%

\begin{keyword}
%% keywords here, in the form: keyword \sep keyword, up to a maximum of 6 keywords
Krylov complexity \sep Lanczos algorithm \sep Quantum chaos \sep Operator growth



\end{keyword}

\end{frontmatter}

%~~~~~~~~~~~~~~~~~~~~~~~~~~~~~~RIKEN-iTHEMS-Report-24
%\maketitle

\tableofcontents

%%%%%%%%%%%%%%%%%%%%%%%%%%%  MAIN %%%%%%%%%%%%%%%%%%%%%%%%%%%%%%%%%%

%%%%%%%%%%%%%%%%%%%% INTRODUCTION %%%%%%%%%%%%%%%%%%%%%%%%%%%
\section{Introduction}
Krylov subspace methods constitute an essential toolbox in scientific computing \cite{viswanath1994recursion,parlett1998,liesenbook}. The core idea behind them is to project a high-dimensional problem onto a lower-dimensional Krylov subspace, thus making the problem more tractable \cite{krylov1931,lanczos1950}. The solution or approximation is then sought within this subspace, easing the computational resources required for solving the problem. As such, they are suited for tackling large-scale linear algebra problems, which are ubiquitous in science and engineering \cite{cullmbook}. Their use is common for solving linear systems, eigenvalue problems, and estimating spectral widths, among other applications. These methods leverage the properties of Krylov subspaces to approximate solutions. They are especially efficient for sparse or structured matrices where direct methods are computationally impractical \cite{Komzsikbook}.

Krylov subspace methods have become increasingly relevant in the study of classical and quantum many-body systems, where they are also known as the recursion method \cite{mori1965continued,cyrot1967electronic,haydock1980recursive,grigolini1983calculation,grosso1985memory,viswanath1994recursion}. For quantum systems, the time evolution is described by a trajectory of the quantum state in Hilbert space. Its dimension scales exponentially with the system size, motivating the quest for more efficient descriptions. This is apparent in the evolution of an isolated system generated by a Hamiltonian according to the Schr\"odinger equation. Krylov subspace methods offer a powerful approach by identifying the minimal subspace in which the dynamics unfolds, without the need to fully diagonalize the Hamiltonian and explicitly store the quantum state. Their formulation in the Heisenberg evolution is also frequent in this context and has long proved useful in the study of correlation functions, linear response theory, spectral functions, and other equilibrium properties.  

For any initial state or observable, the action of the Liouvillian as the generator of time evolution can be encoded in a tridiagonal matrix.
The powers of Liouvillian acting on an initial vector remain linearly independent up to a given order and span the subspace in which the vector evolves with time.
Any set of such linearly independent vectors can be transformed into an orthonormal basis, the \emph{Krylov basis} \cite{krylov1931,lanczos1950} (or the \textit{Lanczos basis}, as is known in the literature on numerical analysis \cite{liesenbook}). 
In this way, any unitary quantum evolution can be mapped to a one-dimensional hopping problem in the so-called Krylov lattice \cite{Mattis1981}.

Further developments have been spurred by the study of ergodic behavior and thermalization in nonequilibrium isolated quantum systems \cite{DAlessio16,Mori2018} and its relation to quantum chaos \cite{haake2019}. Leveraging the notion of Lyapunov exponents in classical chaos, early studies focused on the sensitivity of the quantum state evolution to external perturbations, as captured, e.g., by the Loschmidt echo and related fidelity measures \cite{Gorin06,PhysRevE.89.012923}. 
A bound on quantum chaos was introduced by shifting the emphasis to the time evolution of operators. A quantum analog of the Lyapunov exponent was introduced by analyzing the out-of-time-order correlators (OTOCs), and a universal bound to its value was proposed in  \cite{Maldacena2016,tsuji_bound_2018}.  However, its existence relies on the exponential behavior of OTOCs as a function of time, which is restricted to systems with a small parameter, such as large $N$ theories in the semiclassical regime.  Alternative approaches are thus needed for generic many-body systems beyond the semiclassical approximation.  
Progress to this end harness the notion of operator growth \cite{rakovszky_diffusive_18, keyserlingk_operator_18, nahum_operator_18, khemani_operator_18}.  In a many-body system, an operator is said to be local if its support is restricted to a small part of the system. Under unitary time evolution,  a simple operator becomes increasingly nonlocal and scrambled, acting nontrivially all over the system.  
The information needed to fully characterize the operator grows exponentially in time. This exponential growth prevents an exact description of the operator in a generic system. Yet, it allows for a hydrodynamic description, in terms of very few quantities, which emerges from the behavior of the exponentially large operator space as a heat bath. 

A landmark study by Parker \textit{et al.} \cite{parker2019} used Krylov subspace methods to characterize the operator growth.  
This led to the formulation of the operator growth hypothesis, relating the Krylov construction to the spectral features of the quantum system and its integrable, nonintegrable, and chaotic character. 
The work \cite{parker2019} also introduced the notion of Krylov complexity, which can be used to bound the quantum Lyapunov exponent and quantify operator growth. A surge of activity has ensued, advancing the understanding of operator growth and generalizing it to quantum field theory and open quantum systems with nonunitary dynamics. 
As a result, Krylov subspace methods in quantum dynamics have transcended the early scope as a computational method to provide a fundamental approach to understanding universal dynamical properties and complexity. These developments fulfill in the quantum domain the vision expressed in the correspondence between Lanczos and Einstein regarding the suitability of these methods to do ``justice to the inner nature of the problem'' \cite{liesenbook}. 

In parallel, recent years have witnessed the incipient use of Krylov subspace methods in quantum technologies.  Prominently, they are being harnessed in quantum computing for efficient quantum simulation of real and imaginary time evolution and for estimating ground-state, eigenstate, and thermal properties of a problem Hamiltonian \cite{Motta20, Stair2020, LiuLanczos2020, Cortes22}. Further applications of Krylov subspace methods have occurred in quantum control \cite{Larocca21, Takahashi24, Bhattacharjee23} and can be envisioned in other quantum technologies, such as parameter estimation and quantum metrology. The interplay of these ideas creates a fertile ground for further developments and motivates this contribution.  

This manuscript provides a comprehensive account of quantum dynamics in Krylov space and the role of Krylov complexity in describing quantum processes, with emphasis on methodology. In doing so,  we bridge the mathematical foundation of Krylov subspace methods with applications, highlighting challenges and opportunities in advancing quantum science and technology. This manuscript is structured as follows. Section \ref{secChaos} discusses the preliminaries of quantum chaos, including a brief background on spectral statistics such as level spacing ratio and spectral form factor in random matrix theory. It also introduces out-of-time-ordered correlations and other notions of quantum complexity in the context of quantum many-body systems and holography. Section \ref{secObservable} formally introduces the Krylov space for operators, including the notion of Krylov complexity, which is the main focus of this review. Section \ref{secLAMonic} introduces two versions of the Lanczos algorithms: the orthonormal and monic versions, followed by the introduction to the universal operator growth hypothesis and its subtleties in Section \ref{secOGH}. Section \ref{Lanc_analytic} Introduces two analytic methods: the method of moments through the lattice path and the Toda chain method to compute the Lanczos coefficients. It also outlines the relationship between the pole structure of the autocorrelation function and the spectral function, with explicit computations in the large $q$ SYK model. Section \ref{secKdyn} discusses the coherent state description and complexity algebra in Krylov space, and their connection with quantum speed limits. Section \ref{secOpSize} discusses the notion of ``operator size concentration'', a concept which is crucial for understanding the properties of Krylov basis elements. It provides a simple diagrammatic proof of the operator growth hypothesis in the large $q$ SYK model. Section \ref{secKtemp} formally introduces the concept of temperature in the Krylov space framework and discusses the generic bound of Krylov complexity at finite temperature, especially its relation to the Maldacena-Shenker-Stanford chaos bound. Section \ref{secStates} 
introduces the Krylov space framework in quantum states and the notion of spread complexity. It discusses its generic behavior in random matrix theory and its relationship with the spectral form factor. The section also covers spectral complexity and its similarity with spread complexity at early times. This is followed by Section \ref{secDensity} where Krylov construction is generalized to mixed states. Section \ref{secOpen} provides a formal introduction to open quantum systems and the Lindblad formalism in Markovian settings. It includes an extensive discussion of Krylov space construction and generic algorithms, such as Arnoldi iteration and the bi-Lanczos algorithm, in open quantum systems. An explicit example of the dissipative SYK model is provided from both numerical and analytical perspectives in Section \ref{Secopex}. Section \ref{QFTholo} briefly discusses the Krylov space aspects in quantum field theory and holography. Section \ref{secIntSys} and Section \ref{secKryapp} cover further aspects of the Krylov space framework, such as integrability with an example of mixed-field Ising model (MFIM), and applications in quantum control and quantum computing. Section \ref{secOP} discusses some open problems and a conclusion in Section \ref{secConc}, along with a list and acronyms used in the review.

\section{Elements of quantum chaos}\label{secChaos}
The study of quantum dynamics in Krylov space is largely motivated by progress in understanding quantum chaos. Much of the background has been reviewed in excellent references \cite{gutzwiller1991chaos,Guhr1998,Braun03,haake2019,Borgonovi2016}, and we provide only a succinct account with emphasis on recent developments, discussing a selection of measures to diagnose quantum chaos that are relevant to the study of quantum dynamics in Krylov space. 

\subsection{Spectral statistics and random matrix theory}
Random matrix theory (RMT) finds broad applications in science and engineering \cite{mehta1991random, Forrester2010,Eynard:2015aea, livan2018introduction, oxfordRMT}.
 A series of landmark works by Wigner \cite{Wigner1, Wigner2, Wigner3} and Dyson \cite{Dyson1962a,Dyson1962b,Dyson1962c,Dyson1962} introduced random matrix theory (RMT) in physics.
 This provided a way to describe the spectra of heavy nuclei and complex quantum systems with minimal information about the underlying Hamiltonian. In doing so, Dyson identified the role played by the symmetries of the system, introducing a classification known as the three-fold way \cite{Dyson1962}. The use of RMT in quantum physics was further spurred by the study of chaos across that quantum-to-classical transition.   
The Bohigas-Giannoni-Schmit (BGS) conjecture posits that the spectral statistics of quantum chaotic systems are described by RMT \cite{BGS, Wigner3}, while generic integrable systems show no correlations in the spectrum, as stated by the Berry-Tabor conjecture \cite{Berry_Tabor}. 
As a result,  random matrix Hamiltonians provide a reference framework in the study of quantum chaos \cite{Guhr1998, Atas2013distribution, haake2019}. Likewise, random matrix Lindbladians play a key role in generalizations of quantum chaos to dissipative quantum systems \cite{Xu2019,Can2019,Denisov2019,Can2019PRL,Sa2020,Wang2020,Lucas2020PRX,delCampo2020,Xu2021,HuiZhai2021,cornelius_spectral_2022,Roubeas2023,Lucas2023PRX,Kawabata2023,Xiao2023PRXQ,Roubeas2023PRA,ZhouZhouZhang23}.

A central focus of RMT is the characterization of an ensemble of $N \times N$ matrices, with probability measure defined as
\begin{align}
    p(H) = \frac{1}{Z_{\upbeta_D}} \exp\left(-\frac{\upbeta_D N}{4} \mathrm{Tr}\left( V(H) \right) \right) \,, \label{RMTdiffpot}
\end{align}
where $Z_{\upbeta_D}$ serves as the normalization constant and $V(H)$ represents the potential function of the Hamiltonian $H$.  
The choice of a quadratic potential $V(H) = H^2$ leads to the Gaussian ensembles: the Gaussian Orthogonal Ensemble (GOE) for real symmetric matrices ($\upbeta_D =1$), the Gaussian Unitary Ensemble (GUE) for Hermitian matrices ($\upbeta_D =2$), and the Gaussian Symplectic Ensemble (GSE) for Hermitian quaternionic matrices ($\upbeta_D =4$). These ensembles are named for their invariance under orthogonal, unitary, and symplectic transformations, respectively. In the Gaussian case, the Dyson index $\upbeta_D$ also specifies the nature of the matrix elements: real ($\upbeta_D = 1$), complex ($\upbeta_D = 2$), or quaternion ($\upbeta_D = 4$). 

Specifically, random matrices in the Gaussian ensembles have real-valued diagonal entries $a_{mm}\in \mathcal{N}(0,\sigma^{2})$, where $\sigma$ represents the standard deviation. The off-diagonal entries with $m\neq n$  are defined as $a_{mn}=e^{0}_{mn}$ for GOE, $b_{mn}=e^{0}_{mn}+ie^{1}_{mn}$ for GUE, and $e_{mn}=e^{0}_{mn}+ie^{1}_{mn}+je^{2}_{mn}+ke^{3}_{mn}$ for GSE, respectively. Here, $e^{l}_{mn}\in\mathcal{N}(0,\sigma^{2}/2)$ ($l=0,1,2,3$), and $i$, $j$, $k$ are the basis elements for a quaternion.

The joint probability distribution for the eigenvalues $\lambda_i$ within RMT is given by
\begin{align}
    p(\lambda_1 , \cdots \lambda_N) =  \frac{1}{Z_{\upbeta_D}} e^{-\frac{\upbeta_D N}{4} \sum_k V(\lambda_k)} \prod_{i<j} |\lambda_i - \lambda_j|^{\upbeta_D}\,.
\end{align}
In this expression, the exponential term suppresses configurations with widely separated eigenvalues, while the product term ensures that eigenvalues do not cluster too closely \cite{livan2018introduction}. Together, these factors critically influence the statistics of the eigenvalues, which lie at the heart of RMT's predictive power.

\subsection{Spectral Form Factor}\label{sec:ChaosSFF}

The focus on probing the spectral statistics of many-body systems motivated the introduction of the spectral form factor (SFF) as a diagnostic tool for quantum chaos. The SFF is defined in terms of the  Fourier transform of the energy spectrum \cite{
Leviandier1986,WilkieBrumer1991,Alhassid1993,Ma1995,Brezin1997}. For an isolated system described by a Hamiltonian with spectral decomposition $H=\sum_{n=1}^dE_n|E_n\rangle\langle E_n|$, the SFF reads 
\begin{eqnarray}
{\rm SFF}(t)=\sum_{n,m=1}^dG(E_n,E_m)e^{-it(E_n-E_m)}\,. \label{sffdef}
\end{eqnarray}
Here $G(E_n,E_m)$ represents a spectral filter \cite{Roubeas2023PRA}.  Its use is ubiquitous in numerical simulations \cite{Hammerich1989,Wall1995,Mandelshtam1997,prange_spectral_1997,Gharibyan2018}. When the filter function factorizes $G(E_n,E_m)=g(E_n)g(E_m)$, the function $g(E_n)$ acts as an eigenvalue filter, and $\mathrm{SFF}(t)=|\sum_ng(E_n)e^{-iE_nt}|^2$. A filter of the form $G(E_n,E_m)=G(E_n-E_m)$ acts as a frequency filter.
However, the original definition of the SFF involved no filtering. The use of a Boltzmann factor as an eigenvalue filter $g(E_n)=\exp(-\beta E_n)/Z(\beta)$ with $Z(\beta)=\sum_n\exp(-\beta E_n)$ being the partition function, allows to write the spectral form factor in terms of its analytic continuation at complex temperature as
 \begin{eqnarray}
     {\rm SFF}(t)=\left|\frac{Z(\beta+it)}{Z(\beta)}\right|^2\,.
 \end{eqnarray}
 This form
 has been extensively discussed in the context of black hole physics, RMT,  and conformal field theory \cite{Dyer2017, Cotler2017, Cotler2017b, del_campo_scrambling_2017}.

\begin{figure}[t]
   \centering
\includegraphics[width=0.5\textwidth]{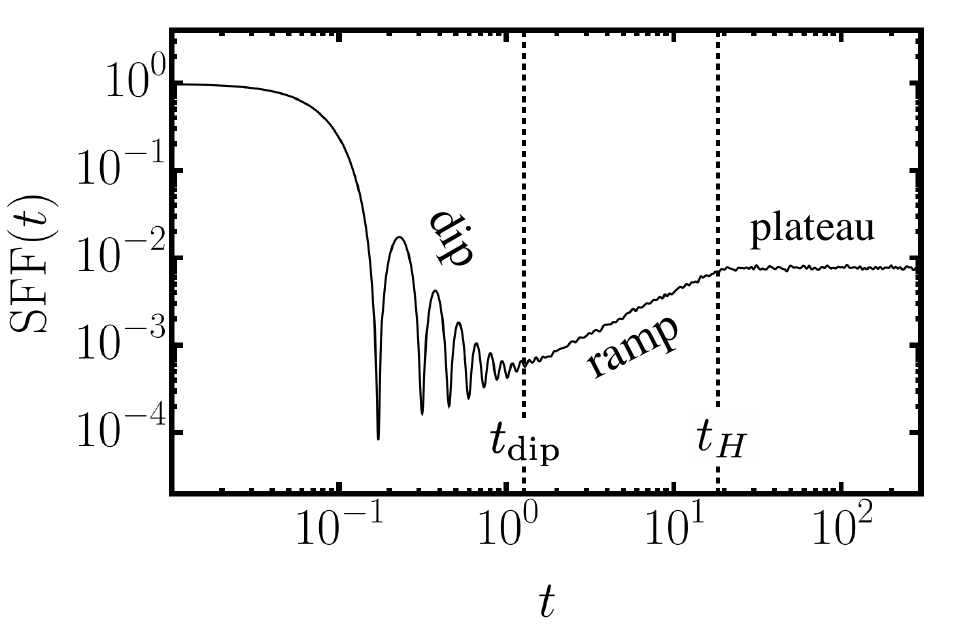}
\caption{Example of the dip-ramp-plateau structure of the SFF for characterizing quantum chaotic systems. The Hamiltonian average is taken over $1000$ $\mathrm{GUE}$ matrices with $\sigma=1$, $d=128$,  setting $\hbar=1$ and $\beta=0$.} \label{fig:sffgoe}
\end{figure}
 
The SFF is generally studied using a Hamiltonian ensemble, e.g., in the context of RMT or disordered systems \cite{TorresHerrera2018,Schiulaz2019}. It is sensitive to the average eigenvalue density $\langle \rho(E)\rangle_{\mathcal{E}}=\langle \sum_n\delta(E-E_n)\rangle_{\mathcal{E}}$ as well as the two-level correlation function of the energy spectrum $\langle \rho(E)\rho(E')\rangle_{\mathcal{E},c}=\langle \rho(E)\rho(E')\rangle_{\mathcal{E}}-\langle \rho(E)\rangle\langle\rho(E')\rangle_{\mathcal{E}}$, where $\langle \bullet\rangle_{\mathcal{E}}$ denotes the ensemble average. 
The SFF contains information about all $k$-th neighbor level spacings \cite{shir2024range}.  Probes sensitive to higher-order correlation functions of the energy eigenvalues can be built analogously and are related to the frame-potentials and unitary $t$-designs \cite{Cotler2017b,Chenu2019, RenZhangSFF}.

The characteristic behavior of the SFF in a quantum chaotic system is shown in Fig.\,\ref{fig:sffgoe}. The early decay from the initial unit value is governed by the energy fluctuations of the initial state and can be extended and approximated by a Gaussian function solely governed by the average density of the states. This is interrupted at the ``dip'' time, when the contribution of two-level correlations becomes significant, giving rise to a ramp that constitutes a dynamical manifestation of quantum chaos.  The ramp is followed by a plateau that approaches the ensemble average value of $Z(2\beta)/Z(\beta)^2$ in the absence of degeneracies.

%which extends for arbitrarily long times and is governed by the purity of the thermal state, associated with the complete dephasing of the initial coherent-Gibbs or thermofield double (TFD) state.  

The SFF admits an information-theoretic interpretation as the fidelity between a quantum state and its time evolution. Consider the
 coherent Gibbs state $|\psi(\beta)\rangle=\frac{1}{\sqrt{Z(\beta)}}\sum_ne^{-\beta E_n/2}|n\rangle$. Then
 \begin{eqnarray}
 \label{SFFdef}
 {\rm SFF}(t)=\left|\langle\psi(\beta)|e^{-it H}|\psi(\beta)\rangle\right|^2=\left|\frac{Z(\beta+it)}{Z(\beta)}\right|^2.
 \end{eqnarray}
This is also known as the survival probability (of the coherent Gibbs state) and has been the subject of extensive studies in the context of quantum dynamics \cite{Fonda1978,Facchi2008}. It is closely related to the Loschmidt echo \cite{Gorin06, peres_stability_1984}, and sometimes termed as such.
The same identity holds if the thermofield double state $|{\rm TFD(\beta)}\rangle=\frac{1}{\sqrt{Z(\beta)}}\sum_ne^{-\beta E_n/2}|n\rangle\otimes |n\rangle$ is considered as initial state and evolved under the Hamiltonian $H\otimes \id$, i.e., ${\rm SFF}(t)=\left|\langle{\rm TFD(\beta)}|e^{-it H\otimes \id}|{\rm TFD(\beta)}\rangle\right|^2$. The general eigenvalue-filtered SFF is obtained when the initial state is chosen to be $|\psi(\beta)\rangle=\sum_n\sqrt{g(E_n)}|n\rangle$. 

The fidelity-based interpretation of the SFF motivates its generalization to nonunitary dynamics and open quantum systems. When the time evolution is described by a quantum channel $\Phi_t(\cdot)$ (i.e., a completely positive and trace-preserving map),  the generalized SFF is then defined as  ${\rm SFF}(t)=\langle \psi(\beta)|\Phi_t[|\psi(\beta)\rangle\langle\psi(\beta)|]|\psi(\beta)\rangle$. Such generalization is particularly suited to probe how the dynamical manifestations of quantum chaos stemming from the Hamiltonian spectral statistics are modified by the nonunitary dynamics \cite{Xu2021,cornelius_spectral_2022,Roubeas2023,Roubeas2023PRA,ZhouZhouZhang23}. Alternative generalizations have been put forward focused on the spectral statistics of the generator of evolution in open quantum systems, e.g., using the Fourier transform of the corresponding complex spectrum \cite{LiProsen2021} or its singular values \cite{Xiao2023PRXQ, Roccati2024}.   There has been some recent interest in experimentally realizable protocols to measure directly the SFF \cite{joshi_probing_2022, vasilyev_monitoring_2020}, as well as closely related correlation functions \cite{das2024proposal}. The first successful direct experimental measurements of the SFF have recently been reported \cite{dong2024measuring}.

\subsection{Out-of-time order correlators (OTOCs) and scrambling}\label{sec:ChaosOTOCs}
\begin{figure}
    \centering
   \includegraphics[width=0.5\textwidth]{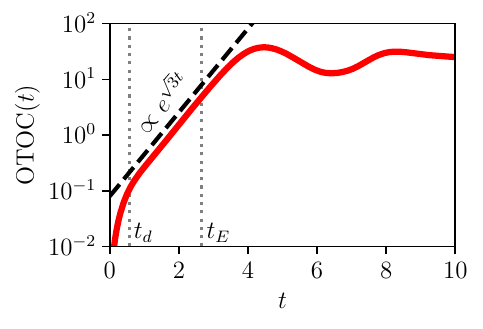}
    \caption{Infinite-temperature OTOC in the Lipkin-Meshkov-Glick (LMG) model with Hamiltonian $H = \hat x + 2 \hat z^2 $, see Sec.\,\ref{sec:linearGrowth_chaos?} for further details. The quantity shows a period of exponential growth between the dissipation $t_d$ and Ehrenfest $t_E$ times. The system shows saddle-dominated scrambling with the quantum Lyapunov exponent given by $\lambda_{\rm OTOC}=\sqrt{3}$. The operators are chosen as $V(0)=W(0)=\ha z$, the spin is $s = 100$ and the effective Planck's constant is $\hbar_{\rm eff} = 1/S$. The propagator $U(t)$ is constructed from the rescaled Hamiltonian $H/\hbar_{\rm eff}$. The figure is adapted from \cite{Xu:2019lhc} with different system parameters.}
    \label{fig:OTOC_LMG}
\end{figure}

Out-of-time-order correlators (OTOCs) were originally introduced in a quasiclassical theory of superconductivity \cite{Larkin_1969_OTOC} and have become a standard origin of a quantum analog of the Lyapunov exponent \cite{Kittu}, as well as a natural probe of quantum information scrambling \cite{hayden_black_2007, Sekino:2008he, Swingle2016}; see \cite{OTOCreviewSwingle} for a comprehensive review on OTOCs. Given two observables $ W(t)$ and $V(0)$ in the Heisenberg picture and in a $d$-dimensional Hilbert space, the infinite-temperature OTOC is defined as \cite{Larkin_1969_OTOC}
\begin{equation}
    %C(t) = \frac{1}{d}{\rm Tr}\left(W(t) V(0) W(t) V(0)\right). 
    \mathrm{OTOC}(t) = - \frac{1}{d}\Tr\left([W(t), V(0)]^2\right) \,.
\end{equation}
The exponential growth of $\mathrm{OTOC}(t) \sim \epsilon e^{\lambda_{\mathrm{OTOC}} t}$ for a certain time window $t_d \ll t \ll t_E$  is considered a direct probe of quantum information scrambling, where $\lambda_{\mathrm{OTOC}}$ is known as quantum Lyapunov exponent \cite{Maldacena2016}. In particular the time-scales correspond to the \textit{dissipation time} $t_d \sim 1/\lambda_{\mathrm{OTOC}}$, at which two-point functions $\Tr{V(0) V(t)}$ saturate, and the \textit{Ehrenfest time} or \textit{scrambling time} $t_E \sim \log(1/\epsilon)/\lambda_{\mathrm{OTOC}}$, at which the OTOC saturates. Figure \ref{fig:OTOC_LMG} shows the growth of the OTOC between the two timescales for the Lipkin-Meshkov-Glick (LMG) model \cite{Xu:2019lhc}. The existence of such exponential growth of the OTOC requires a parametrically large hierarchy between the two timescales and, thus, a very small parameter $\epsilon$. The latter is typically related to Planck's constant in systems with a well-defined semiclassical limit or to $1/N$ expansions in large $N$ systems, such as the Sachdev-Ye-Kitaev (SYK) model \cite{PhysRevLett.70.3339, Kittu}.

The exponential growth of OTOCs can be justified in the semiclassical limit \cite{Larkin_1969_OTOC, Maldacena2016} by taking the operators to be the canonically conjugated position and momentum $W(t) = x(t), \; V = p$. Therefore the classical limit of the OTOC is given by the Poisson bracket, which quantifies the classical sensitivity to initial conditions ${\rm OTOC}_{\rm cl}(t) = \hbar^2 \{q(t), p\}^2 = \hbar^2 \left( \frac{\partial q(t)}{\partial q(0)}\right)^2 \sim e^{2 \lambda_{\rm cl} t} $, where $\lambda_{\rm cl}$ is the classical Lyapunov exponent \cite{strogatz2019}. The introduction of the OTOC surpasses some previous definitions of the quantum Lyapunov exponent based on the Loschmidt echo \cite{jalabert_environment_2001, jacquod_golden_2001, cucchietti_decoherence_2003} in that one does not need to find a \textit{Lyapunov regime} in which the decay of the Loschmidt echo is environment-independent. The Loschmidt echo and OTOCs are related at the average level \cite{yan_information_2020} and through Fidelity OTOCs, in which one of the operators is chosen to be a projector over a state and the other has a ``kick'' form $V= \ket{\psi}\bra{\psi}, \; W = e^{i \delta G}$ with $\delta \ll 1$ \cite{lewis-swan_unifying_2019}. The presence of a positive Lyapunov exponent is a necessary but not sufficient condition for classical chaos, as saddle or unstable points can also give rise to a positive Lyapunov exponent. To have a sufficient condition for chaos, there are different approaches that either require aperiodicity of the trajectories at long times \cite{strogatz2019} or the mixing condition \cite{wimberger_nonlinear_2014}. In a similar spirit, the existence of a positive quantum Lyapunov exponent has been argued to be a signature of scrambling rather than quantum chaos \cite{Xu:2019lhc}. In particular, there exist integrable systems, such as the LMG model, see Fig.\,\ref{fig:OTOC_LMG}, showing a positive quantum Lyapunov exponent. This phenomenon has been termed \textit{saddle-dominated scrambling} and will be further discussed in the context of Krylov complexity and the operator growth hypothesis in Sec.\,\ref{sec:linearGrowth_chaos?}.

One key constraint on the quantum Lyapunov exponent is the Maldacena-Shenker-Stanford (MSS) bound \cite{Maldacena2016}, which states that the Lyapunov exponent $\lambda_{\mathrm{OTOC}}$ is universally upper bounded by the temperature $T$ of the system
\begin{align}
    \lambda_{\mathrm{OTOC}}(T) \leq 2\pi T\,, \label{mssbound}
\end{align}
where natural units $\hbar = k_{\rm B}=1$ are used for simplicity. The universal upper bound $2 \pi T$ is the Lyapunov exponent of a black hole at the same temperature, which is motivated by the conjecture that black holes are the fastest scramblers in nature \cite{Sekino:2008he}. An experimental proposal of verifying this bound using trapped ions in quantum optical systems is also proposed \cite{Tian2022MSSexp}. Indeed, when a system saturates the upper bound, e.g., the SYK model at low temperatures \cite{Maldacena:2016hyu, Kobrin_many_body_SYK}, it is holographically dual to a black hole; see \cite{Trunin_2021} for a review. For finite temperatures, the thermal OTOC, especially in quantum field theory, can present divergences. It is thus customary to introduce a regularization, e.g.,  splitting the thermal factor as $\mathrm{OTOC}_\beta(t) = -\Tr{[W(t), V(0)]e^{-\beta H/2}[W(t), V(0)]e^{-\beta H/2}}/Z_\beta$. The MSS bound can be rederived under different assumptions,  by introducing a one-parameter family of regularizations \cite{tsuji_bound_2018}, or from the fluctuation-dissipation theorem \cite{pappalardi_quantum_2022, tsuji_out--time-order_2018}. It can also be understood in the motion of particles in curved surfaces at low temperatures \cite{kurchan2018quantum, pappalardi_low_2022} and a similar bound constrains the early-time decay of the SFF \cite{del_campo_scrambling_2017,martinez-azcona_analyticity_2022, vikram_exact_2024, vikram2024proof}. However, at infinite temperature \eqref{mssbound} is trivially satisfied. A stricter bound in such cases will be discussed in Sec.\,\ref{secmssbound}.

The experimental measurement of OTOCs and scrambling is confronted with the requirement of time-inversion operations \cite{Swingle2016,GarciaAlvarez17}. Despite this difficulty, they have successfully been  
measured in several experimental platforms \cite{li_measuring_2017, garttner_measuring_2017, vermersch_probing_2019, joshi_QIScrambling_2020}, also at finite temperature \cite{green_experimentalOTOC_2022}. Even under experimental errors, an ideal OTOC may be extracted \cite{swingle_resilience_2018}. Furthermore, they have been extended to several nonunitary situations like finite open quantum systems \cite{syzranov_out--time-order_2018}, bipartite systems \cite{zanardi_information_2021}, dissipative spin chains \cite{zhang_information_2019}, random unitary circuits \cite{AltmanRUC23, schuster_operator_2022}, dissipative SYK model \cite{Zhang23Opgrowth, Bhattacharjee:2023uwx, Liu:2024stj}  and stochastic Hamiltonians \cite{martinez_azcona_SOV_2023}.

\subsection{Quantum complexity}
The notion of complexity in quantum systems arises in a wide variety of contexts, ranging from quantum computing to black hole physics \cite{Arorabook, susskind2020three}.
Measures of quantum complexity quantify the resources needed to perform a physical process (e.g., a computation) or prepare a quantum state. 
The complexity of a quantum unitary operation is defined in terms of the number of gates of the smallest circuit that implements it. This notion of quantum circuit complexity carries over to a quantum state in terms of the size of the shortest circuit that leads to its preparation from a reference state. In an $n$-qubit Hilbert space, preparing a typical pure state starting from a reference product state requires an exponentially long time when using physical local Hamiltonians. Most unitaries are maximally complex. This renders the preparation of typical pure states experimentally unfeasible \cite{Poulin2011}. 
Providing lower bounds for quantum complexity is challenging due to the possibility of shortcuts.
One approach relies on the geometric approach introduced by Nielsen \cite{Nielsen2005, Nielsen2006, Nielsen2006control} and its higher-order integrators of Suzuki-Trotter \cite{NandyPRL} in the complexity manifold \cite{Balasubramanianchaotic2020, Auzzi21, BrownSusskind19}, in accordance with AdS/CFT correspondence \cite{MaldacenaAdSCFT, WittenAdSCFT, GubserAdSCFT}.

The study of quantum state complexity
has been further advanced in the context of AdS/CFT by relating the quantum state complexity of the boundary theory with the volume of the bulk geometry \cite{Susskind:2014rva, Stanford2014,Brown16}.
Defining quantum state complexity in quantum field theory comes with additional difficulties stemming from the choice of the reference state, the set of generators for the corresponding elementary gates, the presence of UV divergences calling for a regularization procedure, and the formulation of a complexity measure \cite{Susskind:2014rva, Brown16, Brown16prd, Jefferson:2017sdb, Chapman18}. The continuous version of the Entanglement Renormalization tensor networks (cMERA) provides a framework to tackle these features \cite{Caputa2017, Caputa2017jhep, Czech2018, Molina-Vilaplana18, Takayanagi2018, CamargoQcurv2022}. See Ref.~\cite{Chapman:2021jbh} for a detailed exposure.

In finite-dimensional systems, random quantum circuits provide a natural framework to describe quantum chaotic dynamics, e.g., as characterized by OTOCs \cite{nahum_operator_18}. The Brown-Susskind conjecture posits that the quantum circuit complexity grows linearly with time for exponentially long times in the number of qubits in a random quantum circuit \cite{Brown2018}. 
The growth of quantum complexity has been rigorously established in connection to unitary $t$-resigns \cite{Brandao2021}, which are collections of unitaries that approximate a completely random unitary 
\cite{Gross2007,Dankert2009}. As a result,  the Brown-Susskind conjecture has been proved in local random quantum circuits \cite{Brandao2021,Haferkamp2022}.

%%%%%%%%%%%%%%%%   KRYLOV SPACE OF OBSERVABLE OPERATORS  %%%%%%%%%%%%%%%%%%%%
\section{Krylov Space of Observable Operators} \label{secObservable}

%\subsection{Preliminaries}

\subsection{Preliminaries: vectorization of operators}\label{SecVec}

The quantum evolution of a pure state is governed by the Schr\"odinger equation, in which the rate of change of the state vector equals the Hamiltonian acting linearly on the state vector itself. By contrast, when considering the unitary time evolution of  operators in the Heisenberg picture, or mixed states in the Schr\"odinger picture, the generator of the dynamics acts on a matrix to give another matrix, this is what is known as a linear \textit{superoperator}. In this case the superoperator that generates the evolution is the \textit{Liouvillian}, which appears in both the Heisenberg equation and the Liouville-von Neumann equation, respectively describing the evolution of operators and density matrices. Similarly, the dynamics in open quantum systems involves the Lindbladian superoperator, as the generator of evolution.  
In all these cases, the operator-vector correspondence known as \textit{vectorization} provides a convenient way for rendering the action of the superoperator on an operator as a matrix acting on a vector.  
In particular, vectorization expresses a $d\times d$ matrice as a $d^2$ dimensional column vector. 
For an operator $A$ with $d\times d$ matrix representation, we use the vectorization 
\cite{gyamfi2020}
\begin{equation}
A  = \sum_{i,j=0}^{d-1} a_{ij} \ketbra{i}{j} \; \rightarrow  \;  \mathrm{vec} \,A\,  = \sum_{i,j=0}^{d-1} a_{ij} \ket{i} \otimes \ket{j}^{*}. 
\end{equation}
where $\ket{j}^{*}$ is related to the state $\ket{j}$ by the complex conjugation operation \cite{wigner_antiunitary_1960}. This way of vectorization is known as row-wise or horizontal vectorization, where the rows of the matrices are stacked below. Such vectorization, i.e., expressing the operator in the doubled-Hilbert space $\mathscr H \otimes \mathscr H$ is commonly referred to as the \emph{operator-state correspondence}. A prototypical example is the identity operator expressed as maximally entangled EPR states. An analogous representation at finite temperature is provided by the thermofield double (TFD) states, which serve as the thermal counterpart to the EPR states.

The space of linear operators over a Hilbert space is itself a Hilbert space when endowed with the \textit{Hilbert–Schmidt} inner product, which has a particularly simple form when written in the vectorized notation since it is simply the standard euclidean inner product between vectors \cite{gyamfi2020}
\begin{align}
    \mathrm{Tr}(A^{\dagger} B) = (\mathrm{vec} \, A)^{\dagger} \,\mathrm{vec} \, B \,. \label{vecdef}
\end{align}
%The following identity is useful in reducing the action of superoperators to matrix multiplication 
A linear superoperator acts on an operator as a matrix multiplication from left and right. In the vectorized representation superoperators are $d^2 \times d^2$ matrices. The matrix expression of the superoperator can be found using the identity  \cite{gyamfi2020}
\begin{align}\label{EqAOB}
    \mathrm{vec} (A \, \mathcal{O} \, B ) =  (A \otimes B^{\intercal}) (\mathrm{vec} \, \mathcal{O})\,,
\end{align}
for any arbitrary operators $A$, $B$ and $\mathcal{O}$, where ``$\intercal$'' denotes the transpose operation. 

In quantum information theory, vectorization  arises 
in the realm of the Choi-Jamio\l kowski isomorphism or channel-state duality \cite{CHOI1975285, JAMIOLKOWSKI1972275,Watrous18}. In this context, vectorization is extensively used for the study of the spectral properties of quantum channels and open systems.  Operations within this framework are efficiently represented using tensor networks \cite{ArpanBhattacharyya2022, Mele2024introductiontohaar}. A different convention for vectorization, known as column-wise or vertical vectorization, is also common in the literature \cite{turkington2013}.  Column-wise and row-wise vectorizations are equivalent and give the same results when used consistently.

\subsection{Krylov basis and operators: Lanczos algorithm}

Consider an observable, described by a Hermitian operator $\mathcal{O}$, evolving in time under the action of a time-independent Hamiltonian $H$. 
In the Heisenberg picture, the dynamics of one such observable is governed by the \textit{Heisenberg equation}
\begin{equation}
    \partial_t \mc O(t) = i [H, \mc O(t)] =: i \mc L \mc O(t),
\end{equation}
where  $\mc L \,\bullet = [H, \bullet]$ denotes the \textit{Liouvillian} superoperator. The solution of the Heisenberg equation reads
\begin{align}
     \mathcal{O}(t) &= e^{i  H t}  \,\mathcal{O}\, e^{-i  H t}  =  \mathcal{O} +it[ H, \mathcal{O}] +\frac{(it)^2}{2}[  H,[  H, \mathcal{O}]] +\ldots=  \mathcal{O} +it\mathcal L\mathcal{O} +\frac{(it)^2}{2}\mc L^2 \mathcal{O} +\ldots = e^{i \mathcal{L} t} \mathcal{O}\,.
\label{Liou1} 
\end{align}
The terms in the expansion involve nested commutators with the Hamiltonian, or equivalently,   powers of the \textit{Liouvillian} superoperator acting on the initial observable, as  $\mathcal{L}^0\mathcal{O}=\mathcal{O}$, $\mathcal{L}\mathcal{O}=[ H, \mathcal{O}]$, $\mathcal{L}^2\mathcal{O}=[ H,[ H, \mathcal{O}]]$, and so on.   
The evolution of $\mathcal{O}(t)$ in general does not span the full operator space, instead it is contained in a subspace of the operator space, known as the \textit{Krylov space}, that is spanned by the set of all nested commutators, i.e., 
\begin{equation}
\mathrm{span}\{\mathcal{L}^n \mathcal{O}\}_{n=0}^\infty=\mathrm{span}\{ \mathcal{O}, \mathcal{L} \mathcal{O},\mathcal{L}^2 \mathcal{O},\dots \}.
\end{equation}
Describing the evolution in the Krylov space thus paves the way to ease computational resources in the study of many-body systems, and thus Krylov subspace methods have been used in condensed-matter problems\cite{viswanath1994recursion}. This is because the Krylov space under consideration is \textit{``adapted''} to only include the parts of the operator space that $O(t)$ visits.

The expansion (\ref{Liou1}) brings additional insights in the case of multipartite systems in which the Hilbert space $\mathscr H$ has a tensor-product structure $\mathscr H_1 \otimes \mathscr H_2 \otimes \dots \otimes \mathscr H_L$,  as in many-body spin systems and multi-qubit systems. To illustrate this, assume that the initial operator is local, i.e., acting on $p$-spins. If the Hamiltonian is $k$-local, i.e. having at most terms acting on $k$ spins, when the Liouvillian acts once $\mc L \mathcal O = [H, \mathcal O]$ will have support over $O(p+k)$ spins. After $n$ applications of the Liouvillian then one can expect the support or size of  $\mathcal{L}^n \mathcal{O}$ to be of order $O(nk+p)$. This reflects the phenomenon of \textit{operator growth}: as the time of evolution goes by,  the support of the operator,  known as the \textit{operator size}, increases. In addition, the number of terms needed to describe the operator in \eqref{Liou1} grows with time and leads to operator scrambling (cf. Sec. \ref{sec:ChaosOTOCs}). 
As we shall discuss below, these phenomena can be rigorously defined and quantified using different measures that have a natural representation in Krylov space.

The initial operator is restricted to be local and should not be a conserved quantity of the system; otherwise, the commutator $[H, \mathcal{O}]$ vanishes. 
To progress further, it is essential to introduce an inner product. 
While several options exist,  the ``infinite temperature'' inner product, which equals the Hilbert–Schmidt inner product divided by the dimension of state space \cite{speedlimitreview} is frequently employed. The finite-temperature inner-product will be discussed in Sec.\,\ref{sec:finiteT}. Specifically, for any two operators $A$ and $B$, the infinite-temperature inner product is defined as 
\begin{eqnarray} \label{IP}
(A|B) := \frac{\Tr( A^\dagger B)}{\Tr \id} = \frac{1}{\Tr \id} \, (\mathrm{vec} \, A)^{\dagger} \,\mathrm{vec} \, B \,,
\end{eqnarray}
where $\id$ denotes the identity matrix and $\Tr \id = \mathrm{dim}(\mathscr H) := d$ is the dimension of the Hilbert space.  The second equality in \eqref{IP} follows
from the vectorized inner product of the operators \eqref{vecdef} and should be used when the operators are vectorized. Consequently, $|A)$ is defined as 
\begin{align}
    |A) = \frac{1}{\sqrt{\mathrm{Tr}\,\mathbb{I}}} \mathrm{vec}\, A = \frac{1}{\sqrt{\mathrm{Tr}\,\mathbb{I}}}\, \sum_{i,j=0}^{d-1} a_{ij} \ket{i} \otimes \ket{j}^{*} \,.
\end{align}
This gives the norm of an operator as $\| A \| \equiv \sqrt{( A| A)}$, usually known as the \emph{Frobenius norm} \cite{golub1996,speedlimitreview,OTOCreviewSwingle}. The inner product 
%\footnote{We often denote $A := |A)$ and $A^{\dagger} := (A|$ interchangeably. This is aligned with the definition of the inner product \eqref{IP}.} 
in \eqref{IP} coincides with the conventional inner product used for vectors once the operators $A$ and $B$ are vectorized up to the normalization factor.  
Note that the Liouvillian is Hermitian with respect to the definition \eqref{IP}, i.e., $( A|\mathcal{L} B) = (\mathcal{L} A| B)$, which in the vectorized notation can be expressed simply as $\mathcal{L} = \mathcal{L}^{\dagger}$. 
We also take the initial operator $\mathcal{O}$ to be normalized, i.e., $\|\mathcal{O}\| = 1$.

With a choice of the inner product in hand, it is apparent that the basis elements formed by the nested commutators $\{\mathcal{L}^n |\mathcal{O})\}$ are in general neither normalized nor orthogonal with respect to \eqref{IP}. However, an orthonormal basis can be constructed recursively. We assume the initial operator $\mathcal{O}$ is  Hermitian so that it describes an observable, $\mathcal{O}= \mathcal{O}^\dagger$. An orthonormal basis is then constructed via the Gram-Schmidt-like \cite{gram1883,schmidt1908} procedure, also known as the \textit{Lanczos algorithm} \cite{lanczos1950}. The algorithm is as follows \cite{viswanath1994recursion, parker2019}:
\begin{enumerate}
    \item Define $|\mathcal{O}_{-1}) := 0$ and $b_{0} := 0$.
    \item $|\mathcal{O}_0) :=  |\mathcal{O}) $.
    \item $|\mathcal{A}_1)=\mathcal{L} |\mathcal{O}_0)$. If $\|\mathcal{A}_1\|=0$, stop. Otherwise define $b_1=\|\mathcal{A}_1\|$ and $|\mathcal{O}_1)= |\mathcal{A}_1)/b_1$.
    \item For $n>1$: $|\mathcal{A}_n) = \mathcal{L}|\mathcal{O}_{n-1}) - b_{n-1}  |\mathcal{O}_{n-2})$. If $\|\mathcal{A}_n\|=0$, stop. Otherwise define $b_n=\|\mathcal{A}_n\|$ and $|\mathcal{O}_n)= b_n^{-1}|\mathcal{A}_n)$.
\end{enumerate}

This process stops at $n=D_K$ where $D_K$ is the \textit{Krylov dimension}, which is also known as \textit{the grade of} $\mathcal{O}_0$ \textit{with respect to} $\mathcal{L}$ \cite{liesenbook}.  The output is a $D_K$-dimensional orthonormal \textit{ordered} basis, $\{|\mathcal{O}_n)\}_{n=0}^{D_K-1} = \{|\mathcal{O}_0), |\mathcal{O}_1), \dots,  |\mathcal{O}_{D_k-1})\}$ known as the \textit{Krylov basis} for the corresponding Krylov space. 
It follows that
\begin{eqnarray}
(\mathcal{O}_n|\mathcal{O}_m)=\delta_{nm},
\end{eqnarray}
while the identity in Krylov space reads
\begin{eqnarray}
\sum_{n=0}^{D_K-1}|\mathcal{O}_n)(\mathcal{O}_n|=\id.
\end{eqnarray}
While the elements of the Krylov basis are not Hermitian, all operators in the set $\{i^n\mathcal{O}_n\}_{n=0}^{D_K-1}$ are Hermitian. In addition, the Lanczos algorithm provides the set of non-negative coefficients, known as \textit{Lanczos coefficients}, $\{b_n\}_{n=1}^{D_K-1} = \{b_1,b_2,\dots,b_{D_K-1}\}$, which completely determine the dynamics of the operator in the Krylov basis, as discussed later. It is important to note that the Lanczos algorithm is numerically unstable due to the accumulation of rounding errors during the orthogonalization process. This limitation means that after a small number of numerically stable iterations, orthogonality is rapidly lost \cite{parlett1998, LanczosSO}. To ensure orthogonality, methods such as full orthogonalization (FO) or partial re-orthogonalization (PRO) are typically necessary \cite{LanczosPRO, Rabinovici:2020ryf}, ensuring adherence to the precision limits of the machine.

The Krylov space is a subspace in the operator space, with the latter having the dimension $d^2$. The dimension of Krylov space $D_K$ is equal to the number of {\it distinct} energy gaps $E_i-E_j$, for which (in case of degeneracies, at least one) corresponding matrix element is not zero.  This readily yields an upper bound on the Krylov space dimension \cite{Rabinovici:2020ryf},
\begin{align}
    1 \leq D_K\leq d^2-d+1\,. \label{Kbound}
\end{align}
Note that $D_K$ can be infinite if the Hilbert space dimension is infinite and depends on the initial probe operator. One might expect that integrable and non-integrable systems might follow the bound \eqref{Kbound} differently. While the upper bound is tight for systems like the SYK$_2$, which can be mapped to free fermions \cite{Rabinovici:2020ryf}, this is not always the case. Interacting integrable systems also tend to saturate the bound \cite{Rabinovici:2021qqt, Rabinovici:2022beu}. Hence, the Krylov dimension \eqref{Kbound} is not a proper diagnostic between integrability and chaos.

The crux of the Lanczos algorithm can be compactly denoted by the following identity \cite{parker2019}
\begin{align}
    \mathcal{L} |\mathcal{O}_{n}) = b_{n} |\mathcal{O}_{n-1}) + b_{n+1}  |\mathcal{O}_{n+1}) \,, \label{Lanc1}
\end{align}
for $n\geq1$, with $b_{0} = 0$. In other words, the Liouvillian operator in Krylov basis $\{|\mathcal{O}_n)\}_{n=0}^{D_K-1}$ reads
\begin{align}
    \mathcal{L} := \sum_{n=0}^{D_K-2} b_{n+1} \Big(|\mathcal{O}_{n})(\mathcal{O}_{n+1}|+ |\mathcal{O}_{n+1})(\mathcal{O}_{n}|  \Big)\,,
\end{align}
where we made a shift $n \rightarrow n+1$ in the first term of \eqref{Lanc1}, since $b_0 = 0$. The matrix representation of the Liouvillian, with elements $(\mathcal{O}_m|\mathcal{L}|\mathcal{O}_n)$, takes the following \emph{tridiagonal} form \cite{parker2019}
\begin{align}
\mathcal{L} = 
	\begin{pmatrix}
0 & b_1 &  &  &  \\
b_1 & 0 & b_2 &   &  \\
 & b_2 & 0 & & \\
 &  & &  \ddots  \\
 &  &  & & 0 &  b_{D_K-1}   \\
 &  &  &  &  b_{D_K-1}  & 0  
	\end{pmatrix}\,, \label{triLiouvillian}
	\end{align}
with the primary off-diagonal elements being the Lanczos coefficients $\{b_n\}_{n=1}^{D_K-1}$. The value $b_{D_K} = 0$ indicates the end of the Krylov space. 
This directly results from \eqref{Lanc1}, with the condition that the Krylov basis elements are orthonormal $(\mathcal{O}_n|\mathcal{O}_m) = \delta_{nm}$. The vanishing of the diagonal elements results from the initial operator being Hermitian, under the definition of the inner product \eqref{IP}. This renders the Liouvillian also Hermitian ($\mathcal{L} = \mathcal{L}^{\dagger}$) in the Krylov basis - a fact that can be traced back to the unitary evolution of the system. This property breaks down for open systems, where the evolution is non-unitary, as discussed in Sec.\,\ref{secOpen}.

\begin{figure}[t]
   \centering
\includegraphics[width=0.65\textwidth]{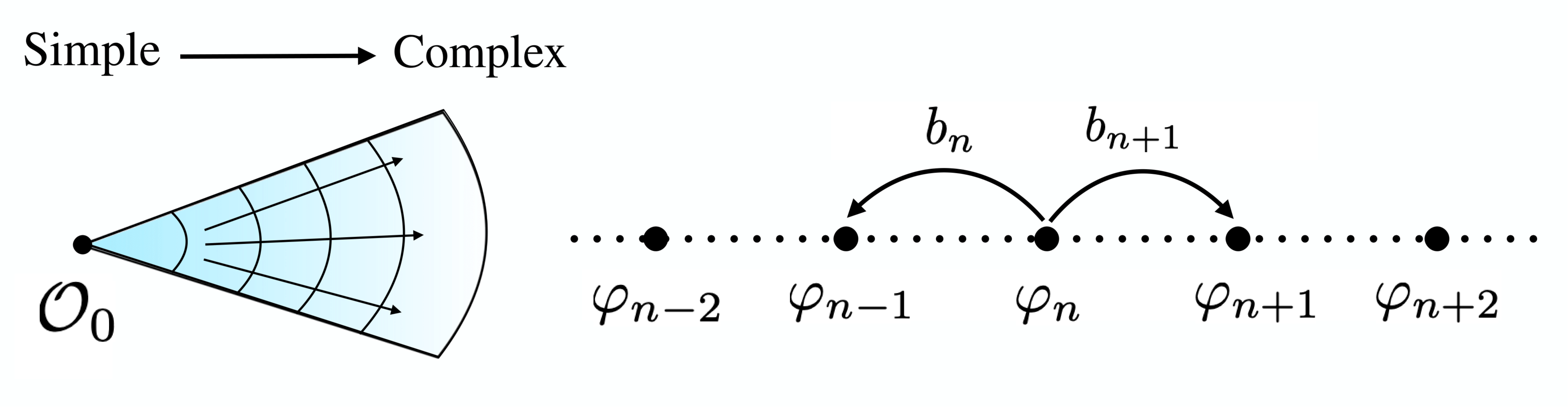}
\caption{The growth of an initial operator $\mathcal{O}_0 \equiv \mathcal{O}$ (left) is mapped to a single particle hopping problem in a one-dimensional Krylov chain (right). Here, $b_{n+1}$ and $b_n$ denote the hopping rates from the $n$-th site to the $(n+1)$-th and $(n-1)$-th sites, respectively. Adapted from \cite{parker2019}.} \label{fig:Hchain}
\end{figure}

The normalization of the initial operator  $|\mathcal{O})$ is preserved under unitary evolution, as $(\mathcal{O}(t)|\mathcal{O}(t)) = (\mathcal{O}|\mathcal{O})$. Thus, $\mathcal{O}(t)$ can be expanded in the (Hermitian) Krylov basis $\{i^n \mathcal{O}_n\}$ as \cite{parker2019, Rabinovici:2021qqt}
\begin{align}
     |\mathcal{O}(t)) = \sum_{n=0}^{D_K-1} i^n\varphi_n(t)  |\mathcal{O}_n)\,. \label{kexpansion}
\end{align}
%The evolved state has been expanded in the Hermitian basis $\{i^n \mathcal{O}_n\}$. 
The  real-valued functions $\varphi_n(t)$ are known as the \emph{Krylov-basis  wavefunctions} or \emph{operator wavefunctions} and denote the probability amplitude  (or weightage) 
of each Krylov basis element. From the Heisenberg equation, it is straightforward to see that they satisfy the difference-differential equation \cite{parker2019, Rabinovici:2021qqt}
\begin{align} \label{discschr}
    \dot{\varphi}_n(t) =  b_n \varphi_{n-1}(t)  - b_{n+1}\varphi_{n+1}(t)\,,
\end{align}
with initial conditions, $\varphi_{-1}(t)=0$ and $\varphi_n(0)=\delta_{0n}$. The ``dot'' indicates the time derivative. Equation \eqref{discschr} describes a single-particle hopping model, where the particle localized at site $n$ hops to site $(n-1)$ with rate $b_n$ and to the site $(n+1)$ with rate $b_{n+1}$, as illustrated in Fig.\,\ref{fig:Hchain}. The operator time evolution is thus equivalent to a single-particle hopping problem in the one-dimensional \emph{Krylov lattice} or \emph{Krylov chain}, with asymmetric nearest-neighbor transition rates. These equations show that the dynamics on the Krylov chain are fully determined by the Lanczos coefficients. 

The wavefunction $\varphi_0(t)$ associated with the initial state $|\mathcal{O}_0) = |\mathcal{O})$ is given by \cite{parker2019}
\begin{align}
    \mathcal{C}(t) := \varphi_0(t) =(\mathcal{O}|\mathcal{O} (t)) = \frac{\Tr(\mathcal{O}\, \mathcal{O}(t))}{\Tr \id}\,, \label{defauto}
\end{align}
and known as the \emph{autocorrelation
function}, being the correlation function between the initial operator and its evolution \cite{viswanath1994recursion}. This function encodes the full information of all the Lanczos coefficients and, thus, the entire information of operator growth in Krylov space.  Indeed, finding $\mathcal{C}(t)$ requires the exact time-evolved operator $\mathcal{O}(t)$.

Since the operator norm is preserved under unitary time evolution, 
\begin{align}
    \mathcal{Z} (t) := \sum_{n=0}^{D_K-1}  |\varphi_{n} (t)|^2  = 1 \,. \label{dsenorm} 
\end{align}
This can be viewed as the conservation of the probability density in the Krylov chain, normalized to unity. This feature no longer holds in the case of dissipative systems, where the effect of the environment tends to decrease the probability, $\mathcal{Z} (t)  < 1$ at $t>0$, leading to a generalization of \eqref{discschr} in the form of a non-Hermitian tight-binding model \cite{Bhattacharjee:2022lzy, Liu:2022god, Bhattacharjee:2022vlt}. We will return to this in Sec.\,\ref{secOpen} in greater detail.

\subsection{Krylov complexity and cumulants}
 
An important measure of operator growth is known as the \emph{Krylov complexity} and is defined as the average position in the Krylov chain  \cite{parker2019, Rabinovici:2021qqt}
\begin{align}
    K(t) = \sum_{n=0}^{D_K-1}  n |\varphi_{n} (t)|^2\,. \label{kcompcl}
\end{align} 
By definition, $K(t)\geq 0$ and it vanishes for the initial operator, $K(0)=0$. The Krylov complexity grows as the operator shifts away from the origin of the Krylov lattice. This measure reflects the fact that the basis elements $|\mathcal{O}_n)$ are more nonlocal as the lattice index is increased.
As a complexity measure, its growth with time indicates that the initial simple operator becomes complex over time.

There are several other conventional measures of quantum complexity, such as the computational complexity \cite{Susskind:2014rva}, circuit complexity \cite{Jefferson:2017sdb}, and the holographic complexity \cite{Susskind:2014rva,Brown16}. Some connections have been proposed between them and the Krylov complexity \cite{Lv:2023jbv, Craps:2023ivc, Chattopadhyay23}, with possible subtleties involved \cite{Aguilar-Gutierrez:2023nyk}.

One may ask whether the Krylov basis furnishes any advantages over the computational basis, an example being the $N$-qubit Pauli basis for the $N$ site system. Although both bases are time-independent, the computational basis is fixed, while the Krylov basis depends on the Hamiltonian and the initial operator. This fact is particularly leveraged by Eq. \eqref{discschr}, where the differential equation involving the coefficients is much simpler in the Krylov basis than in the computational one. Interestingly, the operator size can be defined over the computational basis in the same way the Krylov complexity is defined in \eqref{kcompcl} \cite{Roberts2015}.

Note that Krylov complexity, as the mean of a probability distribution, is insensitive to the actual spread of the operator in the Krylov chain. As such, it is more suited to capture operator growth than operator scrambling.
Complete information on operator growth in the Krylov lattice is provided by the normalized distribution \eqref{dsenorm}, which can be further characterized by its moments and cumulants. An example of the latter is the Krylov variance (second cumulant), which provides a complementary measure of the dynamics focused on the spread of the operator in the Krylov lattice. The \emph{Krylov variance} is defined as \cite{Bhattacharjee:2022lzy}
\begin{align}
    \Delta K (t)^2 :=
    \sum_{n=0}^{D_K-1} n^2 |\varphi_n (t)|^2 - \Big(\sum_{n=0}^{D_K-1} n |\varphi_n (t)|^2\Big)^2= \sum_{n=0}^{D_K-1} |\varphi_n (t)|^2 (n - K(t))^2 \,. \label{kvar0}
\end{align}
An alternative definition was considered in \cite{Caputa:2021ori, Bhattacharjee:2022ave}, which in our notation stands for $\Delta K (t)^2 /K(t)^2$. The higher moments of the distribution can be similarly defined \cite{Bhattacharjee:2022ave}. For this, it is interesting to consider the Krylov operator $\mathcal{K}$ such that \cite{barbon2019}
\begin{align}
    \mathcal{K} |\mathcal{O}_n) = n |\mathcal{O}_n)\,, ~~ \Leftrightarrow ~~ \mathcal{K} := \sum_{n=0}^{D_K-1}  n |\mathcal{O}_n)(\mathcal{O}_{n}| \,.  \label{kop}
\end{align}
In other words, the Krylov operator acts as a number operator on the Krylov basis, $\mathcal{K} = \mathrm{diag}(0, 1, 2 \cdots, D_K-1)$. In terms of the definition \eqref{kop}, the  Krylov complexity is associated with the expectation value of the Krylov operator $\mathcal{K}$ in the time-evolved operator $|\mathcal{O}(t))$, i.e.,
\cite{barbon2019, Bhattacharjee:2022ave}
\begin{align}
    K(t) = (\mathcal{O}(t)|\mathcal{K}|\mathcal{O}(t))\,.
\end{align}
Beyond the expectation value, the probability distribution in the Krylov lattice determines the eigenvalue statistics of the Krylov complexity operator
\begin{align}
P(n,t)=(\mathcal O(t)|\delta(\mathcal{K}-n)|\mathcal O(t))\,.
\end{align} 
The distribution in the Krylov lattice is associated with $D_K$ independent discrete random variables, in which the $n$-th variable takes the measurement outcome $n$ with probability $\varphi_n(t)^2$ and the value $0$ with probability $1-\varphi_n(t)^2$. 
This allows us to define the cumulant-generating function
\cite{Bhattacharjee:2022ave, Jian:2020qpp}
\begin{align}
    G(\theta,t) := \log(\mathcal{O}(t)|e^{\theta \mathcal{K}}|\mathcal{O}(t)) =   \log\Big(\sum_{n=0}^{D_K-1} e^{\theta n} \, |\varphi_n (t)|^2 \Big)\,,
\end{align}
that is continuous and differentiable with respect to  $\theta \in \mathbb{R}$.

The $n$-th cumulant of the distribution $\kappa_n$ is given by the $n$-th derivative of the generating functional, which reads \cite{Bhattacharjee:2022ave} 
\begin{align}
    \kappa_n := \partial_{\theta}^n G(\theta,t)|_{\theta = 0}\,. \label{kcum}
\end{align}
This implies that the Krylov complexity and the Krylov variance are simply the first and the second cumulants of the distribution, i.e., $\kappa_1 \equiv K(t)$ and $\kappa_2 \equiv \Delta K(t)^2$. The higher cumulants encode additional information about the distribution (e.g., regarding its skewness, kurtosis, etc.) and are computed straightforwardly, given knowledge of the cumulant-generating function. In a similar spirit, the generalized notion of Krylov complexity has also been introduced \cite{Fan_generalized_Kcomp}.

\begin{figure*}[t]
{\includegraphics[width=0.32\textwidth]{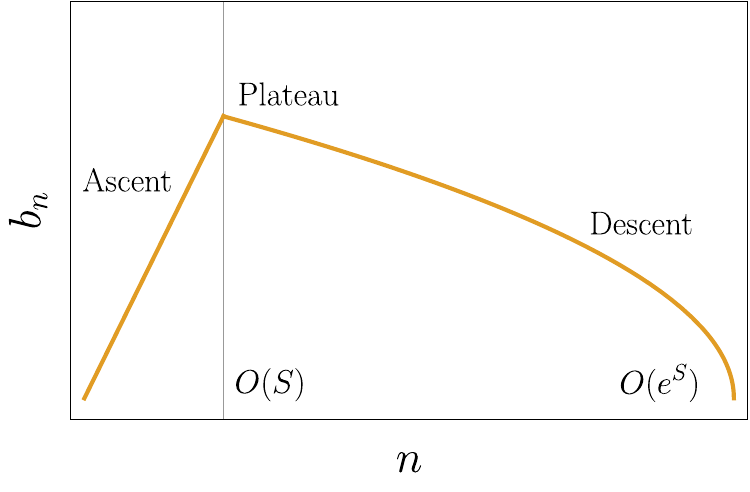}}
\hfil
{\includegraphics[width=0.32\textwidth]{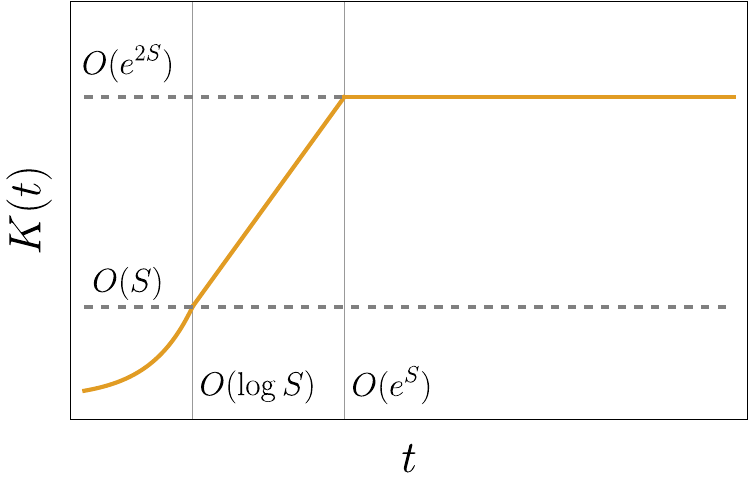}}
\hfil
{\includegraphics[width=0.32\textwidth]{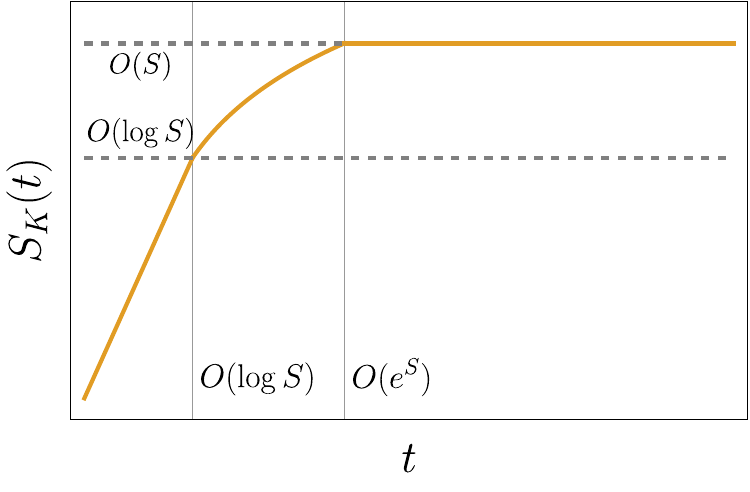}}
\caption{Schematic diagrams of the generic behavior of Lanczos coefficients (left), the associated Krylov complexity (middle), and Krylov entropy (right). The region of Lanczos ascent, Lanczos plateau, Lanczos descent, and the relevant timescales for Krylov complexity and Krylov entropy are shown. Here, $S \sim O(\log N)$ is the thermodynamic entropy of the system, directly proportional to the degrees of freedom $N$. The plots are schematic and do not have a proper numerical scale. For example, the entropy plots are zoomed in, hence the linear region seems longer compared to the Krylov complexity. Also, the difference between the plateau and the descent regime of the Lanczos coefficients is often subtle and may not be separated clearly. The generic structure of these figures is adapted from \cite{barbon2019, Kar2022, Bhattacharjee:2022vlt, Rabinovici:2022beu, Rabinovici:2020ryf}.} \label{fig:genricbnkt}
\end{figure*}

\subsection{Krylov entropy}

The probabilistic interpretation of the Krylov wavefunction allows us to define various quantum information-theoretic quantities \cite{barbon2019, caputa2021, Patramanis:2021lkx}. This includes Shannon entropy, negativity \cite{Horodecki09}, and the capacity of entanglement \cite{Boer19, Nakaguchi2016, Nandy:2021hmk}. We  focus on \emph{Krylov entropy}, which is the Shannon entropy of the probability distribution $P(n,t)$, 
\begin{align}
    S_K (t) = -\sum_{n=0}^{D_K-1}  \abs{\varphi_{n} (t)}^2 \log{\abs{\varphi_{n} (t)}^2}\,.
\end{align} 
The authors of \cite{barbon2019} investigated the behavior of Krylov complexity in various quantum systems, analyzing time scales much longer than the scrambling period. Their study provided evidence that Krylov complexity for operators conforming to the Eigenstate Thermalization Hypothesis \cite{DAlessio16} exhibits a characteristic pattern, growing exponentially during the scrambling period and transitioning to linear growth between the scrambling and the Heisenberg time, at which it saturates. %for exponentially long periods. 
The Krylov entropy shows logarithmic growth in the post-scrambling period. It was shown that the Krylov complexity and the Krylov entropy fulfill the logarithmic relation \cite{Fan:2022xaa}
\begin{align}
    S_K(t) \sim \log K(t)\,, ~~~~~ t > t_{*}\,, \label{sck}
\end{align}
at late times. This late time corresponds to the timescale beyond the scrambling time $t_{*} \sim O(\log N)$ for systems with $N$ degrees of freedom. The relation \eqref{sck} is reminiscent of the equation for Boltzmann's entropy and hints at the appearance of an irreversible thermodynamic-like behavior in the post-scrambling regime.  However, at early times, Eq. \eqref{sck} does not hold, as $S_K(t)$ and $K(t)$ exhibit a product-logarithmic relation instead \cite{barbon2019, Fan:2022xaa}.
%for more details.

The generic behavior of the full profile of Lanczos growth in relation to the Krylov complexity and Krylov entropy can be divided into three distinct regimes. The initial growth of Lanczos coefficients, known as \emph{Lanczos ascent} \cite{barbon2019, Kar2022} persists until $n \sim O(S)$, where $S$ is the thermodynamic entropy of the system. Correspondingly, the Krylov complexity exhibits exponential growth till the \emph{scrambling time} $t_{*} \sim O(\log S)$ with a value $K \sim O(S)$. Beyond the post-scrambling regime, the Lanczos coefficients plateau for $n > O(S)$ is associated with the linear growth of Krylov complexity till the \emph{Heisenberg time} $t_H \sim O(e^S)$, reaching a value at $K \sim O(e^{2S})$. The regime is known as the \emph{Lanczos plateau} \cite{barbon2019, Kar2022}. In this regime, the Krylov wavefunction is uniformly distributed over the Krylov space \cite{Rabinovici:2020ryf}
\begin{align}
    |\varphi(t > t_H)|^2 \sim \frac{1}{D_K}\,. \label{wdec}
\end{align}
In the plateau, the Krylov complexity and Krylov entropy read \cite{Rabinovici:2020ryf}
\begin{align}
    &K(t > t_H) \sim \frac{1}{D_K} \frac{D_K (D_K -1)}{2} \sim \frac{D_K}{2}\,,  \label{wdec2}\\
    &S_K (t > t_H) \sim -D_K \frac{1}{D_K} \log(1/D_K) \sim \log (D_K)\,.
\end{align}
In chaotic systems, the Krylov dimension saturates the bound \eqref{Kbound} and hence, $D_K \sim d^2 \sim e^{2S}$, where $S$ is the entropy of the system and $d \sim e^S$. Thus, the logarithmic relation between them is evident according to \eqref{sck}. After the saturation regime, the $b_n$ decreases at exponentially large $n \sim O(e^S)$, where the Krylov complexity stays at the plateau value $K \sim O(e^{2S})$. This regime is known as \emph{Lanczos descent} \cite{barbon2019, Kar2022}. A schematic behavior of Lanczos growth, Krylov complexity, and Krylov entropy are shown in Fig.\,\ref{fig:genricbnkt}.

\section{Lanczos algorithm: Monic version and orthogonal polynomials}\label{secLAMonic}

%\begin{widetext}
%\begin{center}

\begin{table*}[t]
\resizebox{\columnwidth}{!}{
    \begin{tabular}{ | l | l | l | p{5cm} |}
    \hline
    ~ & ~~~~~Lanczos algorithm (orthonormal version)  & ~~~~~~~~~Lanczos algorithm (monic version) \\ \hline
    Initialization & (0a). $| \mathcal O  _{0}) = |\mathcal O)$ and $| \mathcal O _{-1}) = 0$, ~ $b_0 = 0$. & (0a) $|\mathsf{O}_0) = |\mathcal{O})$ and $|\mathsf{O}_{-1}) = 0$, ~ $\Delta_0 = 0$ \\ \hline
    $n = 1$ & \makecell{(1a). Compute $a_0 = ( \mathcal O _0|\mathcal{L}| \mathcal O _0)$,\\ (1b). $|\mathcal{A}_1) = \mathcal{L}| \mathcal O _0) - a_0 | \mathcal O _0)$\,,\\  (1c). Compute $b_1 = \|\mathcal{A}_1\|$, \\(1d). If $b_1 = 0$, stop, else $| \mathcal O _1) = b_1^{-1} |\mathcal{A}_1)$} &  \makecell{(1a). Compute $a_0 = (\mathsf{O}_0|\mathcal{L}|\mathsf{O}_0)$, \\ (1b). $|\mathsf{O}_1)=\mathcal{L} |\mathsf{O}_0) - a_0 |\mathsf{O}_0)$. \\ (1c). Compute $\Delta_1 = \|\mathsf{O}_1\|^2$.\\ (1d). If $\Delta_1 =0$, stop.} \\ \hline
    $n > 1$ & \makecell{(2a). Compute $a_{n-1} = ( \mathcal O _{n-1}|\mathcal{L}| \mathcal O _{n-1})$, \\ (2b). $|\mathcal{A}_n) = \mathcal{L}| \mathcal O _{n-1}) -a_{n-1} | \mathcal O _{n-1}) - b_{n-1} | \mathcal O _{n-2})$.\\ (2c). Compute $b_n = \| \mathcal{A}_n \|.$\\(2d). If $b_n=0$, stop, else $| \mathcal O _n) = b_n^{-1} |\mathcal{A}_n)$.}   & \makecell{(2a). Compute $a_{n-1} = \frac{(\mathsf{O}_{n-1}|\mathcal{L}|\mathsf{O}_{n-1})}{\|\mathsf{O}_{n-1}\|^2}$,\\
    (2b). $|\mathsf{O}_n) = \mathcal{L}|\mathsf{O}_{n-1}) -a_{n-1} |\mathsf{O}_{n-1}) - \Delta_{n-1} |\mathsf{O}_{n-2})$.\\ (2c). Compute $\Delta_n = \frac{\|\mathsf{O}_n\|^2}{\|\mathsf{O}_{n-1}\|^2}$.\\(2d). If $\Delta_n =0$, stop, else return to (2b).\\
    }  \\
    \hline
    \end{tabular}}
    \caption{Relations between the orthonormal and monic versions of the Lanczos algorithm. The table is adapted from \cite{viswanath1994recursion, Muck:2022xfc} with few changes of notations.}
\label{TableMonic}
%\end{center}
%\end{widetext}
\end{table*}
For completeness, we discuss an alternate version of the Lanczos algorithm, known as the monic version \cite{viswanath1994recursion, Muck:2022xfc}. This version is equivalent to the orthonormal version discussed in the previous section. Given an operator $\mathcal{O}$,  normalized as $\|\mathcal{O}\| = 1$, the algorithm goes as follows:
\begin{enumerate}
    \item $|\mathsf{O}_0) = |\mathcal O)$.
    \item $|\mathsf{O}_1)=\mathcal{L} |\mathsf{O}_0)$. Compute $\Delta_1 = (\mathsf{O}_1|\mathsf{O}_1) = \|\mathsf{O}_1\|^2$. If $\Delta_1 =0$, stop. Otherwise, proceed to step 3.
    \item For $n>1$: $|\mathsf{O}_n) = \mathcal{L}|\mathsf{O}_{n-1}) - \Delta_{n-1} |\mathsf{O}_{n-2})$. Compute $\Delta_n = \frac{(\mathsf{O}_n|\mathsf{O}_n)}{(\mathsf{O}_{n-1}|\mathsf{O}_{n-1})} = \frac{\|\mathsf{O}_n\|^2}{\|\mathsf{O}_{n-1}\|^2}$. If $\Delta_n=0$, stop. Otherwise, repeat the procedure.
\end{enumerate}
The crucial difference of the monic version is that the operators are not normalized at each step. Hence, the operators $\mathsf{O}_n$ (to be distinguished from $\mathcal{O}_n$) are orthogonal but not orthonormal to each other. Note that $\mathcal{A}_n$ in the orthonormal version plays the same role as $\mathsf{O}_n$ in the monic version, while the coefficients in these two versions are related by \cite{viswanath1994recursion}
\begin{align}
    \Delta_n = b_n^2\,.
\end{align}
Table \ref{TableMonic} shows the comparison between the two methods. This table is an extended version of the table provided in \cite{Muck:2022xfc}. Here, an additional coefficient $a_n$ is introduced so that the initial operator $\mathcal{O}$ need not be Hermitian. For the Hermitian initial operator, $a_n$ coefficients identically vanish.
For $n \geq 0$, the action of the Liouvillian translates to 
\begin{align}
       \mathcal{L} | \mathcal O _n) &= a_n | \mathcal O _n) + b_{n+1} | \mathcal O _{n+1}) + b_n | \mathcal O _{n-1})\,, \label{lortho}\\
       \mathcal{L} |\mathsf{O}_n) &= a_n |\mathsf{O}_n) + |\mathsf{O}_{n+1}) + b_n^2 |\mathsf{O}_{n-1})\,, \label{lmonic}
\end{align} 
which imply
\begin{align}
 &a_n = ( \mathcal O _n|\mathcal{L}| \mathcal O _n)\,,~~
    b_{n} = ( \mathcal O _{n-1}|\mathcal{L}| \mathcal O _n)\,, \\
&a_n = \frac{(\mathsf{O}_n|\mathcal{L}|\mathsf{O}_n)}{\|\mathsf{O}_n\|^2}\,,~~
    b_{n} = \frac{\|\mathsf{O}_n\|}{\|\mathsf{O}_{n-1}\|}\,,
\end{align}
in the orthonormal and monic versions, respectively.  This leads to exactly the same values for the Lanczos coefficients in the two versions. For example, in the orthonormal version, for any generic operator $ \mc O _n$, one finds
\begin{align*}
    a_n = ( \mathcal O _n|\mathcal{L}|  \mathcal O _n) \propto \mathrm{Tr}( \mathcal O _n^{\dagger}[H,  \mathcal O _n]) = \mathrm{Tr}([ \mathcal O _n,   \mathcal O _n^{\dagger}] H)\,.
\end{align*}
The Hermitian conjugates of the Krylov basis operators obey, $\mathcal{O} _n^{\dagger} = (-1)^n   \mathcal{O} _n$, i.e., the even Krylov basis elements are Hermitian and the odd are anti-Hermitian. Thus $[ \mathcal O _n,   \mathcal{O}_n^{\dagger}] = (-1)^n [ \mathcal{O}_n,   \mathcal{O}_n] =0$, implying that the $a_n$ coefficients vanish. This property also holds for the monic version, and in general, when the Krylov basis elements $\mc O_n$ are \textit{normal} matrices, i.e., they commute with their Hermitian conjugate.

Both versions fall under a general class of orthogonal polynomials \cite{viswanath1994recursion, Muck:2022xfc, Muck:2024fpb, Sasaki:2024puk}. The recursion method is equivalent to the repeated application of the  Liouvillian to the initial operator. In particular, the $n$-th Krylov basis (either in the orthonormal or monic version) can be represented by
\begin{align}
    |\mathscr{O}_n) = \mathscr{P}_n (\mathcal{L}) | \mathcal O )\,,
\end{align}
where $ \mathcal O $ is the initial operator and $\mathscr{P}_n (\mathcal{L})$ denotes any polynomial of the Liouvillian $\mathcal{L}$ of index $n$. Both the orthonormal and monic Krylov basis are compactly denoted by $|\mathscr{O}_n) \in \{ | \mathcal O _n), |\mathsf{O}_n)\}$, as well as in the polynomial $\mathscr{P}_n (\mathcal{L}) \in \{\mathcal{P}_n (\mathcal{L}), \mathsf{P}_n (\mathcal{L})\}$. It is instructive to note that the leading  coefficients of $\mathcal{P}_n(\mathcal{L})$ is a function of the Lanczos coefficients, while the leading coefficient of $\mathsf{P}_n(\mathcal{L})$ is unity, 
\begin{align}
\mathcal{P}_n(\mathcal{L}) &= \Big(\prod_{i= 1, \cdots, n} \frac{1}{b_i}\Big) \,\mathcal{L}^n + \cdots \,, \\
  \mathsf{P}_n(\mathcal{L}) &= \mathcal{L}^n + \cdots\,,
\end{align}
where the dots denote the lower order monomials in $\mathcal{L}$. This follows from the Lanczos algorithm's recursion relation, which gives the name monic polynomial \cite{chiharabook} and makes the Lanczos algorithm a monic version. Furthermore, in terms of these polynomials, Eqs. \eqref{lortho}-\eqref{lmonic} can be rewritten  as
\begin{align}
       \mathcal{L} \mathcal{P}_n (\mathcal{L}) &= a_n \mathcal{P}_n (\mathcal{L}) + b_{n+1} \mathcal{P}_{n+1} (\mathcal{L}) + b_n \mathcal{P}_{n-1} (\mathcal{L})\,, \label{lortho1}\\
       \mathcal{L} \mathsf{P}_n (\mathcal{L}) &= a_n \mathsf{P}_n (\mathcal{L}) + \mathsf{P}_{n+1} (\mathcal{L}) + b_n^2 \mathsf{P}_{n-1} (\mathcal{L})\,. \label{lmonic1}
\end{align}
This is a three-term recurrence relation satisfied by the orthonormal and the monic polynomial, respectively. They are orthogonal with respect to some measure $\mu(\mathcal{L})$ such that
\begin{align}
    \int d\mu(\mathcal{L}) \, \mathscr{P}_m (\mathcal{L}) \mathscr{P}_n (\mathcal{L}) \propto \delta_{m,n}\,,
\end{align}
where the proportionality constant depends on the chosen polynomial. This is a consequence of Favard's theorem \cite{chiharabook}; see \cite{Muck:2022xfc} for more details.

While both versions of the Lanczos algorithm are equivalent, we preferentially follow the orthonormal version to make use of the associated orthonormal Krylov basis set. However, the monic version is useful in connection with the Toda chain method \cite{dymarsky2020a},  discussed in Sec.\,\ref{todadiscussion}.

\section{The universal operator growth hypothesis}\label{secOGH}

\subsection{Statement of the hypothesis}

The universal operator growth hypothesis identifies different classes of physical systems according to the specific laws governing the growth of the Lanczos coefficients, which in turn dictate the dynamics of operators and complexity measures in Krylov space \cite{parker2019}. It builds on the accumulated numerical evidence regarding the growth of Lanczos coefficients in many-body systems \cite{viswanath1994recursion}
and the relation between chaos and spectral properties of the observables \cite{elsayed2014signatures}.
Consider a many-body system in the thermodynamic limit in $d$ dimensions, governed by a time-independent Hamiltonian $H$ that is non-integrable or chaotic.  The hypothesis regards an initial local operator $\mathcal{O}_0$ that does not commute with any conserved quantity of the system. According to it, the Lanczos coefficients, which capture the dynamics of operator evolution and information spreading within the system, should exhibit maximal growth in the asymptotic limit. This maximal growth is linear for a generic chaotic system in $d$ dimensions, with an additional logarithmic correction in one-dimensional systems. The hypothesis asserts that for the generic $H$ and $O_0$ the asymptotic behavior of $b_n$ is   \cite{parker2019}
\begin{align}
b_n =
  \begin{cases}
    A \frac{n}{\log n} + o(n/\log n) \,,      & ~~d = 1\,,\\
    \alpha n + \gamma + o(1) \,,   & ~~d > 1\,,  \label{uogh}
  \end{cases}
\end{align}
with constants $A$, $\alpha$ and $\gamma$. 
The constants $A$ and $\alpha$ dictate the slope of the growth, while $\gamma$ accounts for any linear shift that is irrelevant in the asymptotic limit of $n$. The values of $A$ and $\alpha$ are not arbitrary; they possess energy dimensions and are constrained by the bandwidth for local Hamiltonians, see \cite{parker2019}. They are specific to the Hamiltonian and the local initial operator. The growth coefficients change for different initial operators, but the asymptotic linear growth should remain unaltered. This is a generic conclusion of the Lanczos algorithm, as the initial vector has lesser effects on the eigenvalues of the tridiagonal matrix. Note that \eqref{uogh} assumes that $b_n$ varies smoothly,  disregarding even and odd parity effects, so that the index $n$ can be treated as a continuous variable.

The hypothesis concerns thermodynamic systems in the asymptotic limit. In finite-dimensional systems, however, Lanczos coefficients eventually terminate at the end of the Krylov space and may display saturation due to the finite dimensionality of the Hilbert space \cite{Rabinovici:2020ryf}. Hence, one must judiciously select an appropriate growth regime before finite-size effects become significant. Notably, linear growth occurs in dimensions greater than one, with a logarithmic correction in one dimension. Distinguishing between linear growth and its logarithmic correction is a formidable numerical task with conventional algorithms \cite{parker2019, Heveling:2022hth, JDNohoperatorising}, a feat only recently achieved with specialized Monte Carlo methods \cite{Caostochastic}. Curiously, the Lanczos coefficients have been found to show faster than linear growth in deep Hilbert space \cite{CaodeepH}, where the operator growth hypothesis does not hold.  Efforts have been made to extend the recursion method in two dimensions \cite{parker2019, PhysRevB.109.L140301}. Therefore, we adopt linear growth as the hallmark of the universal operator growth hypothesis across all dimensions, including all-to-all systems like the SYK model.

\subsection{Does linear growth of Lanczos coefficients always imply chaos?}\label{sec:linearGrowth_chaos?}

\begin{figure}
\centering
\includegraphics[width=0.94\textwidth]{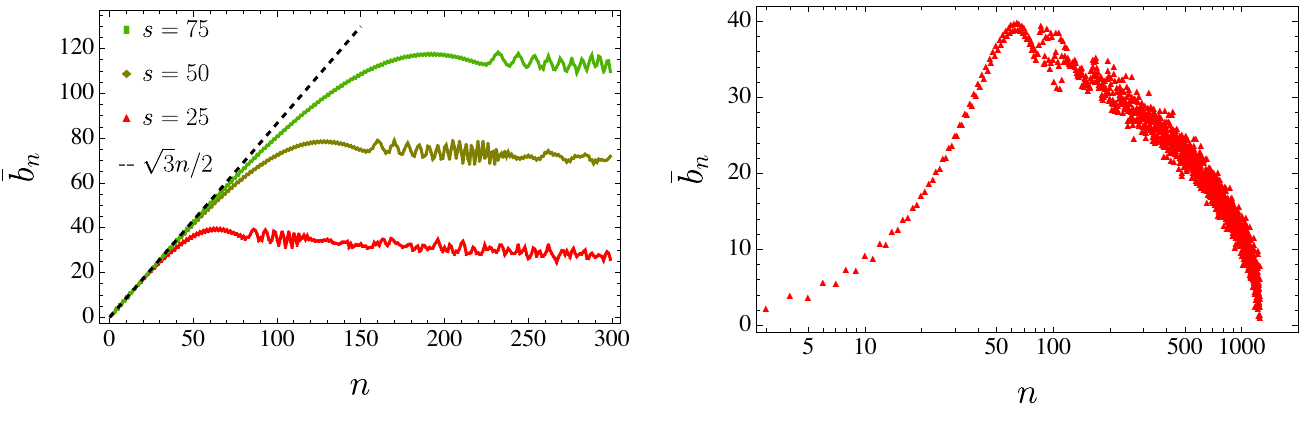}
    \caption{(Left) Behavior of the Lanczos coefficients $\overline{b}_n = b_n s$ in the LMG model for the initial operator $\hat{z}$ for spins $s = 25$, $50$, and $75$ respectively. The black dashed line indicates the linear growth with coefficient $\alpha \simeq \sqrt{3}/2$. (Right) The entire Lanczos spectrum for $s = 25$. The Krylov dimension $D_K \sim 1250$ is much smaller than the maximum dimension $d^2 - d + 1 = 2551$, where the Hilbert space dimension is $d = 2s+1 = 51$. Adapted from \cite{Bhattacharjee:2022vlt}.}
    \label{fig:saddle}
\end{figure}

The hypothesis also has a bearing on integrable systems \cite{Rabinovici:2021qqt, Rabinovici:2022beu}. In them, the Lanczos coefficients  \emph{typically} exhibit sublinear growth and a power-law growth of Krylov complexity
\begin{align}
    b_n \sim \alpha\, n^{\delta}, ~~\Leftrightarrow~~ K(t) \sim (\alpha t)^{\frac{1}{1-\delta}}\,, ~~~ 0< \delta < 1\,. \label{intbe}
\end{align}
By contrast, non-interacting systems like free-fermionic models exhibit a bounded sequence, $b_n \sim O(1)$ \cite{parker2019}, by extending $\delta = 0$ in \eqref{intbe}. In such cases, an operator does not grow
%; the one-body initial operator remains one-body, 
and the operator size remains constant. Table\,\ref{TableTwo} shows the generic behavior of Lanczos coefficients, Krylov complexity, and Krylov entropy for a variety of systems, including two unknown growth models yet to be found in a physical system. Nonetheless, the hypothesis \eqref{uogh} primarily addresses non-integrable systems, sidestepping the nuances of integrable system behaviors. Some special integrable systems may display 
%such nuanced 
singular growth patterns. This includes integrable systems with saddle-dominated scrambling \cite{Bhattacharjee:2022vlt} or systems showing many-body localization \cite{Trigueros:2021rwj}. We consider an example of the former case, the Lipkin-Meshkov-Glick (LMG) model \cite{LIPKIN1965188, MESHKOV1965199, GLICK1965211}. The Hamiltonian in this model is constructed by the following SU(2) operators $\{\hat{x}, \hat{y}, \hat{z}\} = \{\hat{s}_x/s, \hat{s}_y/s, \hat{s}_z/s\}$,  obeying $[\hat{x}, \hat{y}] = i \hbar_{\mathrm{eff}} \hat{z}$, with other cyclic commutators. The Hamiltonian is given by \cite{Xu:2019lhc, Bhattacharjee:2022vlt}
\begin{align} \label{H_LMG}
    H = \hat{x} + 2 \hat{z}^2\,.
\end{align}
Here, $\hat{s}_{i} = \hat{\sigma}_i/2$, with $i = x,y,z$ are the SU(2) spin operators of spin $s$, and $\hat{\sigma_i}$ being the Pauli matrices. The effective Planck constant depends on the spin as $\hbar_{\mathrm{eff}} = 1/s$. The dimension of the Hilbert space is $d = 2s+1$. In other words, the large spin limit $s \rightarrow \infty$ effectively implies the classical limit $\hbar_{\mathrm{eff}} \rightarrow 0$. The LMG model is classically integrable, and the naive expectation is that its integrability is preserved as it transitions to the quantum regime through semiclassical approximation. Upon rescaling the Hamiltonian $\overline{H} = H/\hbar_{\mathrm{eff}}$, the Lanczos 
coefficients acquire a spin factor \cite{Xu:2019lhc, Bhattacharjee:2022vlt}
\begin{align}
    \overline{b}_n = b_n/\hbar_{\mathrm{eff}} = b_n s\,,
\end{align}
which directly comes from the Lanczos algorithm. The growth of $\overline{b}_n$ is shown in Fig.\,\ref{fig:saddle} (left) for the initial operator $\hat{z}$ and the spin values $s = 25$, $50$, and $75$. It exhibits linear growth $\overline{b}_n \sim \alpha n$, even though the LMG model is classically integrable. The black dashed line shows the growth coefficient computed as $\alpha \simeq \sqrt{3}/2$. The entire Lanczos spectrum is shown in Fig.\,\ref{fig:saddle} (right) for $s=25$. The Krylov dimension shows the integrable behavior and is much lower than the saturation bound given in Eq. \eqref{Kbound}. Nevertheless, the linear growth of the Lanczos coefficients and the associated exponential growth of Krylov complexity are unexpected for an integrable model. This peculiar behavior arises due to the presence of an unstable saddle point $(x,y,z) = (1,0,0)$ in its classical phase space, a phenomenon referred to as \emph{saddle-dominated scrambling} \cite{Bhattacharjee:2022vlt}. Indeed, the OTOC of this system also shows exponential growth with the Lyapunov exponent $\lambda_{\rm OTOC}=\sqrt{3}$ given by the underlying saddle point \cite{Xu:2019lhc}, see Fig.\,\ref{fig:OTOC_LMG}. A modified version of such exponent was also proposed \cite{TruninLyap1, TruninLyap2}. It is also interesting to note that the infinite-temperature Lyapunov exponent obeys the bound $\lambda \leq 2 \alpha$ \cite{parker2019}, see Secs.\,\ref{sykexample} and \ref{sec:SYK_finiteT}, and using the computed $\alpha$ for LMG, we find a saturation of the bound $\lambda_{\rm OTOC} = 2 \alpha$. This appears to be an apparent violation of the universal operator growth hypothesis. However, such a special case of saddle-dominated scrambling does not invalidate the universal operator growth hypothesis \eqref{uogh}, which asserts that {\it generic} chaotic systems should demonstrate maximal growth of Lanczos coefficients. The latter is thus a \emph{necessary}, yet  \emph{not sufficient}, condition for chaos, only if the choice of the Hamiltonian and the initial operator is sufficiently generic. Aligning with the previous study with OTOC \cite{PhysRevE.101.010202, Xu:2019lhc}, Eq. \eqref{uogh} suggests that the linear growth of Lanczos coefficients \emph{at best} provides necessary and sufficient conditions for scrambling, but not for chaos \cite{Bhattacharjee:2022vlt}. In fact, the \emph{necessary} conditions for scrambling can also be proved rigorously using operator entanglement \cite{dowling_scrambling_2023}. The saddle-dominated scrambling also occurs in classically chaotic systems, an example being the Feingold-Peres model \cite{FeingoldPeres1, FeingoldPeres2, Xu:2019lhc}. Being classically chaotic, in these models, the Lanczos coefficients exhibit linear growth as expected \cite{parker2019, Bhattacharjee:2022vlt}.

%\begin{figure}[t]
%\hspace*{-0.37 cm}
%{\includegraphics[height=3.2cm,width=0.51\linewidth]{bnsaddle.pdf}}
%\hfill
%{\includegraphics[height=3.2cm,width=0.51\linewidth]{saddle25.pdf}}
%\caption{(Left) Behavior of the Lanczos coefficients $\overline{b}_n = b_n s$ in the LMG model for the initial operator $\hat{z}$ for spins $s = 25$, $50$, and $75$ respectively. The black dashed line indicates the linear growth with coefficient $\alpha \simeq \sqrt{3}/2$. (Right) The entire Lanczos spectrum for $s = 25$. The Krylov dimension $D_K \sim 1250$ is much smaller than the maximum dimension $d^2 - d + 1 = 2551$, where the Hilbert space dimension is $d = 2s+1 = 51$. Adapted from \cite{Bhattacharjee:2022vlt}.} \label{fig:saddle}
%\end{figure}

%\begin{widetext}
%\begin{center}
\begin{table*}[t]
    \begin{tabular}{ | l | l | l | l| l| l| p{3.5cm} |}
    \hline
    Model & Chaotic  & $1d$ Chaotic & Integrable & Bounded & Unknown & Unknown \\ \hline
    $b_n$ & $\alpha n$ & $\alpha n/\log n$ & $\alpha n^{\delta}$ & $b$ & $\alpha \ln n$ & $\alpha n^{\delta} (\log n)^{\pm}$ \\ 
    $K(t)$ & $e^{2 \alpha t}$ &  $e^{\sqrt{4 \alpha t}}$ & $(\alpha t)^{\frac{1}{1-\delta}}$ & $ b t$ & $ 2 \alpha t \log(2 \alpha t)$ & $(2 \alpha t)^{\frac{1}{1-\delta}} \log^{\pm \frac{1}{1-\delta}}(2 \alpha t) $\\ 
    $S_K(t)$ & $2 \alpha t$  & $\sqrt{4 \alpha t}$ & $\log (\alpha t)$ & $\log (2 bt)$ & $\log(2 \alpha t)$ & $\log(2 \alpha t)$\\
    \hline
    \end{tabular}
        \caption{Predictions of the operator growth hypothesis, regarding the growth of the Lanczos coefficients and the associated dynamics of the Krylov complexity and Krylov entropy, for different classes of physical systems. The table is taken from \cite{Fan:2022xaa}.}
\label{TableTwo}
\end{table*}    
%\end{center}
%\end{widetext}

Finally, let us briefly mention some consequences of the hypothesis in quantum field theory (QFT) and conformal field theory (CFT).  The hypothesis \eqref{uogh}, initially posited for discrete quantum many-body systems \cite{parker2019}, encounters significant challenges when extended to continuous models, e.g., in  QFT or CFT. 
In this case, constructing a Krylov basis by means of the Lanczos algorithm would require introducing a finite temperature; see Section \ref{secKtemp}.
One can circumvent this step and determine Lanczos coefficients directly from the two-point autocorrelation function (see Secs.\,\ref{sec:moment1} and \ref{QFT}) has been applied \cite{Dymarsky:2021bjq}. In this framework, Lanczos coefficients demonstrate linear growth in free theories, barring the introduction of a UV cutoff \cite{Avdoshkin:2022xuw, Camargo:2022rnt}. Intriguingly, the presence of an IR scale can lead to splitting of the Lanczos coefficients into even and odd sequences $b_n \in \{b_{\mathrm{even}}, b_{\mathrm{odd}}\}$, challenging the smoothness assumption presupposed in \eqref{uogh}. Hence, the results in continuum field theories go beyond the hypothesis \eqref{uogh} and might seek an extension of the same in such theories.

% Relation to autocorrelation function and moments
With the above caveats, the operator growth hypothesis has consolidated the Krylov complexity as a diagnostic tool for scrambling and quantum chaos. In this context, the Krylov complexity is under exhaustive investigation and has been analyzed in random matrix theory \cite{Rabinovici:2022beu, Tang:2023ocr}, elementary models in quantum mechanics \cite{Hashimoto23, Camargo:2023eev, KrylovCalabiYau, Guo:2022hui}, regular graphs \cite{Xia:2024xfd}, Bethe lattice \cite{Avdoshkin:2019trj, ZotosKrylovBethe}, quantum reservoirs \cite{Domingo:2023kjr}, toy models of gauge theories dual to AdS black holes \cite{magan2020, Jian:2020qpp, Mohan2023, Kar2022, iizuka2023,iizuka2023b}, inflationary cosmology \cite{Li2024inflationKcomp, Li:2024iji}, and quantum metrology \cite{Chu:2024cah}. Further, it has been emphasized that its effectiveness in probing chaos strongly depends on the chosen initial operator \cite{Espanyol23}.

In addition, 
%it is worth mentioning that 
several works have promoted its use with a complementary scope. 
For instance, it has been shown to act as an order parameter in certain settings in dynamical phase transitions \cite{bento2023krylov}. Likewise, its behavior reflects the confinement/deconfinement
phase transitions at large $N$ \cite{anegawa2024krylov}. The Krylov complexity can also be used to probe transitions occurring as a function of the duration of the Trotter time-step in Floquet circuits associated with the Trotter decomposition of unitary dynamics \cite{suchsland2023krylov}. Further, the Krylov complexity has been used to characterize the charging power of quantum batteries utilizing SYK models formulated on a graph \cite{KimRosa22}.

\subsection{Computation of Lanczos coefficients} \label{Lanc_analytic}

This section presents two analytic methods to compute the Lanczos coefficients. The first method is known as the ``moment method'' and relies on computing the moments of the spectral function, which relates to the Taylor series expansion of the autocorrelation function. A special case of the method was presented in \cite{parker2019} while the more general method was developed in \cite{Bhattacharjee:2022lzy, Bhattacharjee:2023uwx}. Both rely on the generic recursion method \cite{viswanath1994recursion}. The second method is based on the Toda chain flow in Krylov space and is often known as the ``Toda chain technique'' \cite{dymarsky2020a, Dymarsky:2021bjq}. Both methods are equivalent and provide the same Lanczos coefficients, which can be further matched with the results of numerical algorithms (see Sec.\,\ref{kcomponum}). 

\subsubsection{Lanczos coefficients via the moment method: Pole structure of autocorrelation function} \label{sec:moment1}

The primary motivation is based on calculating the moments of the spectral function $\Phi (\omega)$, defined below. Given the spectral function $\Phi (\omega)$, the moments are given by
\begin{align}
    m_{n} = \frac{1}{2\pi} \int_{-\infty}^{\infty} d \omega\, \omega^{n} \,\Phi (\omega)\,, ~~~ n = 0, 1, 2, \cdots\,. \label{mom1}
\end{align}
This computes a set of numbers $\{m_n\}$. They can be separated into even $m_{2n}$ and odd $m_{2n+1}$ moments (for $n = 0, 1, 2, \dots$). Conventionally,  $\Phi (\omega)$ is normalized such that $m_0 = 1$. For some special cases, odd moments vanish, although we keep considering both odd and even moments for the general discussion.  Further, the spectral function is the Fourier transform of the autocorrelation function $\mathcal{C}(t)$, given by
\begin{align}
    \Phi (\omega) = \int_{-\infty}^{\infty} d t \,e^{- i \omega t} \, \mathcal{C}(t)\,, ~~~~~~~
    \mathcal{C}(t) = \frac{1}{2\pi} \int_{-\infty}^{\infty} d \omega \,e^{ i \omega t} \,\Phi (\omega) \, \,. \label{autocomega}
\end{align}
Hence, $\mathcal{C}(t)$ is the inverse Fourier transform of $\Phi (\omega)$. Taking the $n$-th derivative of $\mathcal{C}(t)$ as well as the  limit $t \rightarrow 0$, and using \eqref{mom1}, we obtain \cite{parker2019}
\begin{align}
    m_{n} = \frac{1}{i^{n}} \lim_{t \rightarrow 0} \frac{d^n \mathcal{C} (t) }{d t^n} \,. \label{mom4}
\end{align}
The first moment $m_0 = \mathcal{C}(0)$ is the autocorrelation function evaluated at the initial time. Conveniently, the autocorrelation function is normalized to unity at $t = 0$, so by definition, $m_0 = \mathcal{C}(0) = 1$. This is consistent with the normalization of $\Phi(\omega)$. Here, the autocorrelation function is assumed to have no pole in the real axis. However, it might have poles in the imaginary axis, including complex infinity. An example is $\mathcal{C}_1 (t) = \mathrm{sech}(\alpha t)$ or $\mathcal{C}_2 (t) = \exp(-\alpha t^2)$. On the other hand, for functions that have a pole in the real axis, e.g., $\mathcal{C} (t) = \mathrm{sec}(\alpha t)$, one needs to rotate $t \rightarrow it$ to transform it into an imaginary axis and then apply \eqref{mom4}. Equation \eqref{mom4} suggests an expansion of the autocorrelation function in a Taylor series form \cite{Bhattacharjee:2022lzy}
\begin{align}
    \mathcal{C}(t)\, \stackrel{t>0} = \, \sum_{n=0}^{\infty} m_n \,\frac{(it)^n}{n!}\,, \label{autodef}
\end{align}
with the expansion coefficients being the moments \eqref{mom1}. The mutual conversion between the moments and the autocorrelation function (and spectral function $\Phi (\omega)$ from Eq. \eqref{mom1}) is subtle and often regarded in the realm of the Hamburger moment problem \cite{parkerthesis, Bhattacharjee:2022ave, chiharabook}.

\begin{figure}[t]
   \centering
\includegraphics[width=0.45\textwidth]{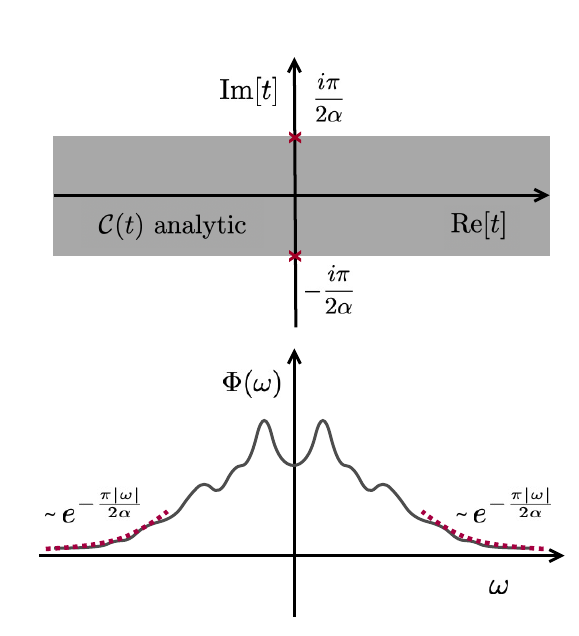}
\caption{Diagrammatic illustration of the analytical structure of the autocorrelation function alongside the spectral function. (above) The autocorrelation function is analytic within a gray-shaded area. The closest poles lie at $t_{\pm} = \pm i \pi/2 \alpha$ in the imaginary axis, corresponding to the linear growth $b_n = \alpha n$.   (below) This is equivalent to the exponential tail of the spectral function. Adapted from \cite{parker2019}.} \label{fig:autofig}
\end{figure}

Given the moments, the Lanczos coefficients can be obtained by the following recursive algorithm
~\cite{viswanath1994recursion, Bhattacharjee:2022lzy}:
\begin{align}
\mathsf{M}_{k}^{(0)} &= (-1)^k m_k \,,\, \mathsf{L}_{k}^{(0)} = (-1)^{k+1} m_{k+1}\,, \nonumber \\ 
\mathsf{M}_{k}^{(n)}  &= \mathsf{L}_{k}^{(n-1)} - \mathsf{L}_{n-1}^{(n-1)} \frac{\mathsf{M}_{k}^{(n-1)}}{\mathsf{M}_{n-1}^{(n-1)}}\,, \nonumber\\
\mathsf{L}_{k}^{(n)}  &= \frac{\mathsf{M}_{k+1}^{(n)}}{\mathsf{M}_{n}^{(n)}} - \frac{\mathsf{M}_{k}^{(n-1)}}{\mathsf{M}_{n-1}^{(n-1)}}\,,\,~ k \ge n \,,
\nonumber \\
b_n &= \sqrt{\mathsf{M}_{n}^{(n)}}\,, ~~~ a_n = - \mathsf{L}_n^{(n)} \,. \label{mombn}
\end{align}
Given the Lanczos coefficients, the moments can also be evaluated. The problem boils down to a fully combinatorial problem of evaluating {\it Dyck paths} \cite{parker2019}, also see below. Figure \ref{fig:diagmoment} shows a diagrammatic picture for evaluating such moments. The first four moments read \cite{viswanath1994recursion}:
\begin{align}
        m_0 &= 1\,, \nonumber \\
    m_1 &= a_0\,,  \nonumber \\
    m_2 &= a_0^2 + b_1^2\,, \\
    m_3 &= a_0^3 +2 a_0 b_1^2 + a_1 b_1^2\,, \nonumber \\
    m_4 &= b_1^2 (a_0^2 + a_1^2 + a_0 a_1 + b_1^2 +b_2^2) + a_0 (a_0^3 +2 a_0 b_1^2 + a_1 b_1^2)  \,. \nonumber
\end{align}
%and so on. 

In general, this diagrammatic approach can be used to evaluate any general matrix element of powers of the Liouvillian $(\mc O_j|\mc L^n |\mc O_k)$, starting and finishing at any general position. The result is given by the sum over all the possible paths that connect sites $k$ and $j$ in $n$ steps. These paths are known as \emph{Motzkin paths} \cite{Stanley_Fomin_1999, Motzkinpaths, KatsuraMotzkin}. The weight of each path is given by the product of Lanczos coefficients $a_n, \; b_n$ associated with the path. At each application of the Liouvillian, the paths can move at most one site; this means that all matrix elements in which $j>k+n$ or $j<k-n$ identically vanish $(\mc O_j|\mc L^n|\mc O_k) = 0$. The first non-zero matrix element reads $(\mc O_{k+n}|\mc L^n|\mc O_k)= b_k b_{k+1} \dots b_{k+n}$, which corresponds to the direct path connecting sites $k$ and $k+n$.

If the autocorrelation function is an even function of $t$, only the even moments survive and the coefficients $a_n$ vanish, i.e., $m_0 = 1, m_2 = b_1^2, m_4 = b_1^4 +b_1^2 b_2^2, \cdots$. The scaling of such coefficients was shown to indicate some universal feature of the autocorrelation function at early times \cite{Zhangauto2024}. This is generically true for unitary evolution with a Hermitian operator $\mc O^\dagger = \mc O$, which reduces the above recursion to a simpler form \cite{parker2019}. The corresponding Dyck paths are the Motzkin paths without the side movement. Hence, Dyck paths only allow to move up and down, not going below the level of the initial point \cite{Motzkinpaths}. At the end of this section, we provide a path integral method to evaluate such Dyck paths in the asymptotic limit of large $n$. In generic non-unitary evolution, or when the initial operator is non-Hermitian $\mc O^\dagger \neq \mc O$, however, both even and odd moments exist, giving rise to two sets of Lanczos coefficients $\{a_n\}$ and $\{b_n\}$. They can be complex numbers in general. For example, the previously introduced autocorrelation functions $\mathcal{C}_1 (t) = \mathrm{sech}(\alpha t)$ and $\mathcal{C}_2 (t) = \exp(-\alpha^2 t^2/2)$ provide $b_n^{(1)} = \alpha n$ and $b_n^{(2)} =  \alpha \sqrt{n}$,  respectively. A natural question is whether they come from any orthonormalization algorithm like the Lanczos algorithm or its variants. We will see in later sections that they do not necessarily do so. In fact, the direct generalization to Arnoldi iteration \cite{Arnoldipaper} (see Sec.\,\ref{Arnoldisection}) only gives these coefficients \emph{approximately} \cite{Bhattacharjee:2022lzy}. However, they are the \emph{exact} output of a bi-orthonormalization procedure, obtained by a different generalization of the Lanczos algorithm known as the bi-Lanczos algorithm (see  Sec.\,\ref{bidisc}), and expressed in the form of the tridiagonal Lindbladian matrix \eqref{ld} \cite{Bhattacharjee:2023uwx}.

\begin{figure}[t]
   \centering
\includegraphics[width=0.55\textwidth]{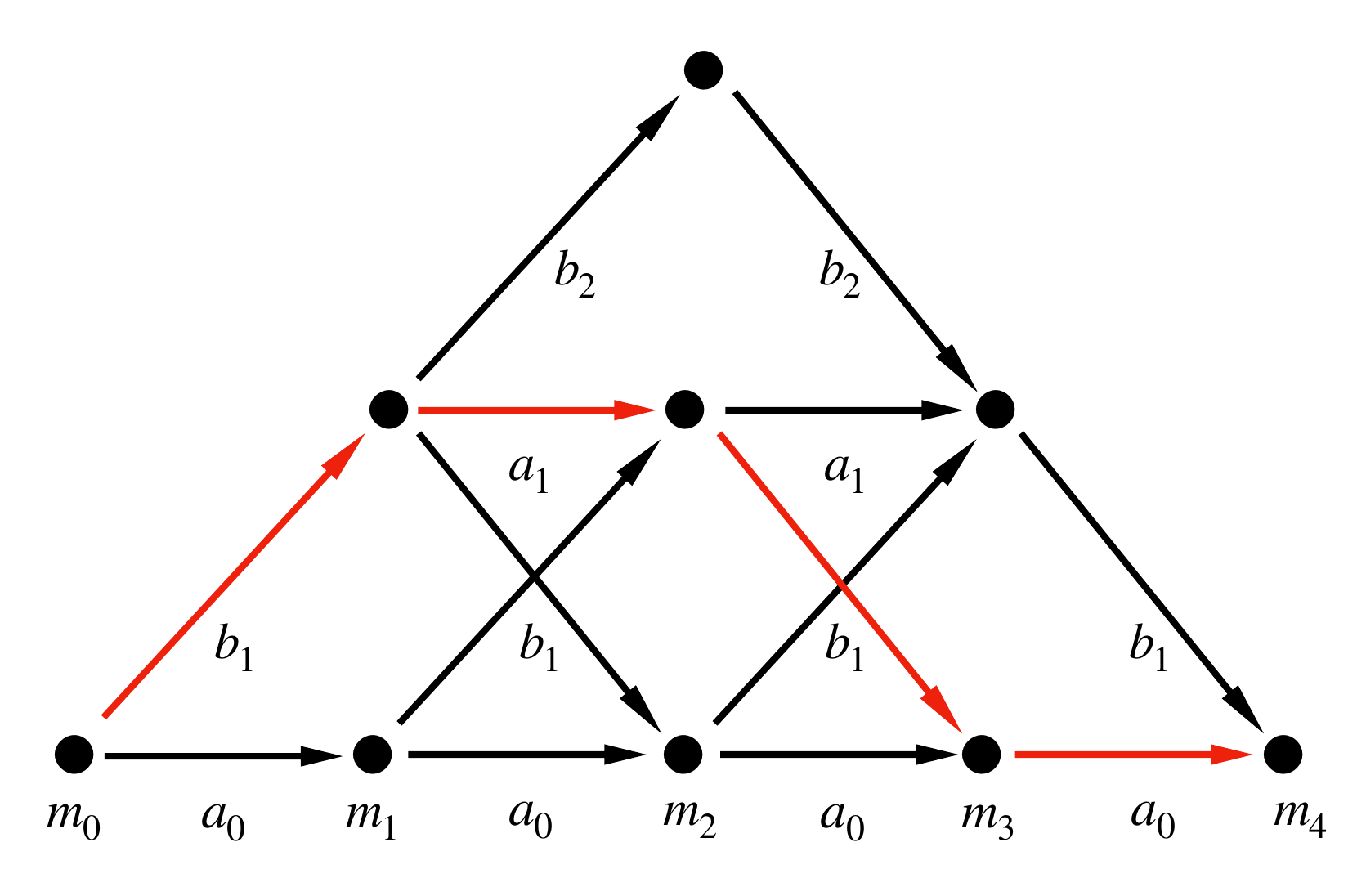}
\caption{Pictorial illustration of the evaluation of moments from the Lanczos coefficients. To evaluate any $m_n$, they are represented as dots starting from $m_0$ and building the triangle. Each horizontal row represents fixed $a_n$, which the $b_n$ forwards one step as one climbs up to one step top. Then, discrete paths are to be tracked (by multiplying each step's contribution) and summed up that leads from $m_0$ from $m_n$. 
 For example, this picture evaluates the moment $m_4$. We start from $m_0$, track all discrete paths that lead to $m_4$, and sum up all the paths. Here, the red arrow shows one such path, which evaluates to $b_1 a_1 b_1 a_0 = a_0 a_1 b_1^2$. For unitary evolution, all $a_n$ vanish, thus we are only restricted to moving upwards and downwards (but not sideways). This significantly restricts the number of terms in the moments. The figure is adapted from the spread complexity picture in \cite{balasubramanian2022} with the necessary modifications in the Lanczos coefficients.} \label{fig:diagmoment}
\end{figure}

The linear growth is special,
in all known physical systems it is
related to the singularity of the autocorrelation function $\mathcal{C}$ in the complex $t$ plane, and thereby, the decay of the spectral function at high frequencies \cite{parker2019}. Yet, mathematically, linear growth of $b_n$ may not require $\mathcal{C}(t)$ to be singular, see an example in Eq. \eqref{mockauto} below. However, the opposite statement is mathematically rigorous \cite{lubinsky1987survey}: the singularity of $\mathcal{C}(t)$ would necessarily require unbounded linear growth of $b_n$, as will be evident from the Dyck path formalism discussed below. 

In most cases, when $b_n$ grows linearly, the distance to the closest pole of $\mathcal{C}(t)$ along the imaginary axis from the origin determines the asymptotics of the Lanczos coefficients.  For example, the autocorrelation $\mathcal{C}_1 (t) = \mathrm{sech}(\alpha t)$ has poles in the imaginary axis at $t_{\pm} = \pm i \pi/(2 \alpha)$; see Fig.\,\ref{fig:autofig}. This determines $b_n^{(1)} = \alpha n$, thereby the exponential growth of Krylov complexity $K(t) \sim e^{2 \alpha t}$. Alternately, the linear growth of the Lanczos coefficients results in the exponential decay of the spectral function. For the above case, we find $\Phi_1 (\omega) = \frac{\pi}{\alpha} \text{sech}\left(\frac{\pi  \omega }{2 \alpha }\right) \sim e^{-\frac{\pi  |\omega| }{2 \alpha }}$. The exponential decay is the slowest possible decay, which in turn provides the fastest growth (linear) of the Lanczos coefficients (see Fig.\,\ref{fig:autofig}). In other words, the higher moments control the asymptotic growth and the high-frequency regime, usually associated with late-time physics \cite{parker2019}. The sublinear growth of the Lanczos coefficients is usually associated with a faster decay. For example, $b_n^{(2)} \sim  \alpha \sqrt{n}$ implies $\Phi_2 (\omega) \sim e^{-\omega^2/(2 \alpha^2)}$. However, the fastest decay in 1D local systems is $\Phi(\omega) \sim |\omega|^{-|\omega|} \equiv e^{-|\omega| \log |\omega|}$ \cite{parker2019,Avdoshkin:2019trj}, which imposes a logarithmic correction on the maximal possible growth of Lanczos coefficients $b_n \sim n/\log n$. %This logarithmic correction cannot be explained from the pole structure since the autocorrelation function is \emph{entire} or \emph{holomorphic} in the whole complex plane for any local systems in $1d$ \cite{parker2019, Araki1969, Abanin2015}.
% In 1D local systems, the autocorrelation function $C(t)$ is an \textit{entire} function, i.e.,  \textit{holomorphic} in the whole complex plane \cite{Araki1969, Abanin2015}. This restricts the growth of the Lanczos coefficients to be sublinear, bounded by the logarithmic correction \cite{parker2019,Avdoshkin:2019trj}.

As we saw above, evaluating Lanczos coefficients starting from the moments, or the other way around, is a complicated task. It is nevertheless important, especially in the light of the operator growth hypothesis, to relate the asymptotic behavior of $b_n$ and $\mu_n$. This can be accomplished using the approach of the Dyck path formalism, as we now explain. For simplicity we assume that only the even moments are present; including the odd moments is a straightforward exercise.
The tridiagonal form of the Liouvillian directly yields the moments \cite{parker2019}
\begin{align}
    m_{2n} = (\mathcal{O}|\mathcal{L}^{2n}_H|\mathcal{O})\,, \label{evenmom}
\end{align}
where the suffix $H$ in $\mathcal{L}_H$ indicates the case of a Hermitian Liouvillian, corresponding to $a_n = 0$. The moments are given by the sum over weighted \emph{Dyck paths} \cite{parker2019, Avdoshkin:2019trj}
\begin{align}
    m_{2n} = \sum_{h_0 \cdots h_{2n}} \prod_{i = 1}^{2n} b_{(h_i + h_{i-1})/2}\,, \label{eqm}
\end{align}
where the set $\{h_1, \cdots, h_{2n}\}$ denotes the \emph{height} of the Dyck path of length $2n$, with $h_i \geq 1/2$ and $h_0 = h_{2n} = 1/2$, i.e., the height ultimately reaches to the same height of the initial starting point at the end of the path. We wish to evaluate the sum in the asymptotic limit of $n$. To this end, two approaches have been outlined in \cite{parker2019, Cao:2020zls}. However, we take an alternative approach that evaluates the number of weighted Dyck paths using a saddle point approximation. This amounts to express \eqref{eqm} as a path integral over a smooth function $f(\mathsf{t})$ with  $\mathsf{\mathsf{t}} \in [0,1]$, such that $h_i = \frac{1}{2} + 2n f(i/(2n))$. This puts the boundary condition on $f(\mathsf{t})$ such that $f(0) = f(1) = 0$. In other words, the derivative of $f(\mathsf{t})$ denotes the slope of moving \emph{up} and \emph{down} of a \emph{microscopic} Dyck path at the index $i = 2 n \mathsf{t}$, such that $f'(\mathsf{t}) = +1$ and $f'(\mathsf{t}) = -1$ for \emph{up} and \emph{down} jumps respectively. Further, if the probability associated with such \emph{up} and \emph{down} jumps are $p$ and $(1-p)$, we can write $2p(\mathsf{t}) - 1 = f'(\mathsf{t})$. Since this is considered an asymptotic limit, we further assume $b_n$ is a smooth function of $n$, i.e., $b_n \equiv b(n)$. The asymptotic form of the moment is given by considering the total number of weighted Dyck paths \cite{Avdoshkin:2019trj}
\begin{align}
    m_{2n} \sim \int D f(\mathsf{t}) \,e^{\mathsf{S}(n)} \,,~~~~~ 
    \mathsf{S}(n) = 2n \int_{0}^{1} d\mathsf{t} \,[\mathsf{H}(p(\mathsf{t})) + \mathrm{log}(b(2 n f(\mathsf{t}))) ]\,, \label{msadd}
\end{align}
where $\mathsf{t} \in [0,1]$ is just a parameter with no relation with real time $t$. Here, $p(\mathsf{t}) = (1 + f'(\mathsf{t})/2)$ with the prime indicating the derivative with respect to $\mathsf{t}$, and $\mathsf{H}(x) = -x \log x - (1-x) \log(1-x)$ is the \emph{microscopic} Shannon entropy of the associated variable $x$. Therefore $\mathsf{S}(n)$ evaluates the total contribution of the \emph{microscopic} entropy weighted by the Lanczos coefficients. Alternatively, it can be considered as the saddle point action that gives the total entropy, i.e., the number of total weighted Dyck paths for the moments \eqref{msadd}.

The idea of the approach is to evaluate the path integral \eqref{msadd} using saddle point approximation, by finding a classical trajectory that maximizes the effective action $\mathsf{S}[f(\mathsf{t})]$. We skip technical details and only give the results. For the generic growth of Lanczos coefficients $b(n) = \alpha n^{\delta}$ with $\delta=1$, we find
\begin{align}
f(\mathsf{t}) =   \begin{cases}
    \sin (\pi \mathsf{t})/\pi      & ~~\delta = 1\,,\\
    \mathsf{t}(1-\mathsf{t})    & ~~\delta = 1/2\,, \\
    0 &~~\delta = 0\,. \label{ftleading}
  \end{cases}
\end{align} 
The saddle point action readily gives the  moments \cite{Avdoshkin:2019trj}
\begin{align}
m_{2n} \sim
  \begin{cases}
    \left(\frac{4 n \alpha}{\pi e}\right)^{2n}      & ~~\delta = 1\,,\\
    \big(\frac{2 n \alpha^2}{e}\big)^{n}    & ~~\delta = 1/2\,, \\
    4^n &~~ \delta = 0\,.\label{m2nleading}
  \end{cases}
\end{align}
In the case ($\delta = 0$), the Dyck paths are not weighted \cite{parker2019, Avdoshkin:2019trj}, which gives rise to the asymptotic behavior of the Catalan number, $\mathrm{Cat}(n) \sim 4^n$ \cite{BERNHART199973}. An extension of  this result, establishing the relation between constant $\gamma$ in \eqref{uogh} and the order of the singularity on the imaginary axis of $\mathcal{C}(t)$ 
is discussed in \cite{Dymarsky:2021bjq}.
The extension to the case in which $b_n$ split into two approximately continuous branches is developed in \cite{Avdoshkin:2022xuw}.

The case of ``chaotic'' behavior in one spatial dimension is a bit subtle. Since the Lanczos coefficients show a logarithmic correction $b_n \sim \alpha\, n/\log n$, the function $f(\mathsf{t})$ acquires a subleading term which is logarithmic \cite{Avdoshkin:2019trj}
\begin{align}
    f(\mathsf{t}) =  \frac{\sin (\pi \mathsf{t})}{\pi} + O\left(\frac{1}{\mathrm{log}(2n)} \right)\,, \label{ft1d}
\end{align}
where the first term equals that in \eqref{ftleading} for the linear growth. The corresponding correction term $\Delta \mathsf{S}$ to the action $\mathsf{S}$ is given by \cite{Avdoshkin:2019trj}
\begin{align}
    \Delta \mathsf{S} = 2n \int_{0}^{1} d\mathsf{t} \left(\delta -1\right) \mathrm{log}(f(\mathsf{t}))\,,
\end{align}
where $f(\mathsf{t})$ is given by \eqref{ft1d}. The saddle point solution is readily evaluated, and the moments are given by \cite{Avdoshkin:2019trj}
\begin{align}
    m_{2n} \sim \left(\frac{4 n \alpha}{\pi e \, \mathrm{log}(2n)}\right)^{2n} (2\pi)^{2n/\mathrm{log}(2n)}\,.
\end{align}
Hence, for the maximal growth of the Lanczos coefficients, the following statements are equivalent (we take systems with $a_n = 0$)
    \begin{align}
     &\Phi(\omega) \sim e^{-\frac{\pi  |\omega| }{2 \alpha }}\,~~~\Leftrightarrow ~~~b_n \sim \alpha n  ~~~~~~\Leftrightarrow ~~~m_{2n} \sim \left(\frac{4 n \alpha}{\pi e}\right)^{2n} ~~~~\Leftrightarrow ~~~ K(t) \sim e^{2 \alpha t} ~~~~~~~~~~~ (d > 1)\,,\\
     & \Phi(\omega) \sim |\omega|^{-|\omega|}~~~\Leftrightarrow ~~~b_n \sim \frac{\alpha n}{\log n}  ~~~\Leftrightarrow ~~~m_{2n} \sim \left(\frac{\alpha n}{\log n}\right)^{2n}~~~\Leftrightarrow ~~~ K(t) \sim e^{\sqrt{4 \alpha t}} ~~~~~~~ (d = 1)\,.
\end{align}
We reiterate that these relations hold in an asymptotic sense.

\begin{figure}[t]
\centering
\includegraphics[width=0.43\textwidth]{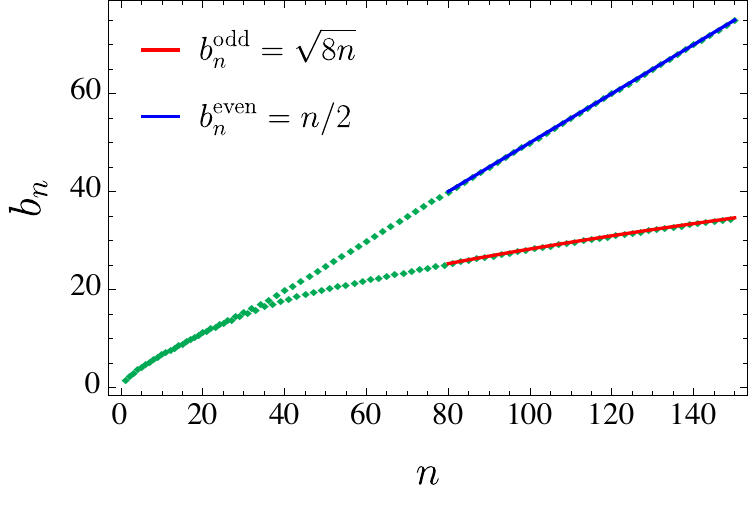}
\caption{Growth of $b_n$ for the mock autocorrelation function \eqref{mockauto}. The Lanczos coefficients split into two branches $b_n \in \{b_{\mathrm{even}}, b_{\mathrm{odd}}\}$ after certain $n_{*} \approx 30$. The odd and even coefficients are fitted by the respective smooth functions, according to the individual growth \eqref{mockbn}. The figure is adapted from \cite{Avdoshkin:2019trj} with associated fitting.} \label{fig:mockbnplot}
\end{figure}

This discussion assumes a smooth behavior of the Lanczos coefficients. By this, we mean $b_n \sim \alpha n$ with $n$ varies continuously, although it is a discrete index. However, the smoothness of the moments $m_n$ does not guarantee the smoothness of the Lanczos coefficients. A particular example is given by the following mock auto-correlation function of the form \cite{Avdoshkin:2019trj} 
\begin{align}
    \mathcal{C}(t) = \frac{1}{2} \big[e^{ (e^{i t} -1)}+ e^{(e^{-i t} -1)}\big]\,, \label{mockauto}
\end{align}
such that $\mathcal{C}(0) = 1$. The moments can be straightforwardly computed using Eq. \eqref{mom4}. Since the above autocorrelation is even $ \mathcal{C}(-t) = \mathcal{C}(t)$, all odd moments vanish and even moments are given by $m_{2n} = B_{2n}$, where $B_{2n}$ are even Bell numbers \cite{A000110}. It is easy to see that the moments are smooth functions of $n$. Applying \eqref{mombn}, we find $a_n = 0$, while $b_n$ coefficients split into two branches $b_n \in \{b_{\mathrm{even}}, b_{\mathrm{odd}}\}$, after certain $n_{*} \approx 30$, see Fig.\,\ref{fig:mockbnplot}. While the overall growth of $b_n$ is not smooth, the odd and even $b_n$ separately show smooth behavior in the asymptotic sense \cite{Avdoshkin:2019trj}
\begin{align}
    b_n^{\mathrm{odd}} \sim \sqrt{n}\,, ~~~~~~~~ b_n^{\mathrm{even}} \sim n\,. \label{mockbn}
\end{align}
Hence, the odd coefficients are sublinear, while the even coefficients are linear. Moreover, the autocorrelation function is periodic in $t$ in the real axis. In other words, the autocorrelation function does not decay to zero. This property is presumably responsible for such oscillation-the even and odd splitting can be mathematically expressed as \cite{PhysRevB.102.195419, Bhattacharjee:2022vlt}
\begin{align}
    b(n) = f(n) + (-1)^n g(n)\,,
\end{align}
with slowly varying function $f(n)$ and $g(n)$ such that $g(n)$ decays with $n$, and $f(n) \gg g(n)$ in the asymptotic limit of $n$. This gives two distinct branches, oscillating between $f(n) \pm g(n)$. If $f(n)$ is assumed to be linear, i.e., $f(n) \sim \alpha n$ for some $\alpha >0$, then an asymptotic analysis suggests that $g(n)$ has the form \cite{Bhattacharjee:2022vlt}
\begin{align}
    g(n) \sim (\log n)^{-\mathsf{a}}\,, ~~~ \mathsf{a} \geq 0\,,
\end{align}
on top of the linear growth. This gives rise to an autocorrelation function of the form \cite{Bhattacharjee:2022vlt}
\begin{align}
    \mathcal{C}(t) \sim t^{-\mathsf{a}}\,, ~~~ \mathsf{a} \geq 0\,.
\end{align}
Thus, a power law decay of the autocorrelation results from the logarithmic decay of $g(n)$ on top of the linear growth of $f(n)$. However, constant $\mathsf{a}=0$ implies $g(n) \sim \mathrm{const}$, which results in autocorrelation that does not decay over time and can be periodic. Such non-decaying behavior of autocorrelation usually gives rise to the unusual splitting of the Lanczos coefficients \cite{Avdoshkin:2019trj, PhysRevB.102.195419, Bhattacharjee:2022vlt}. An alternate point of view from the continuity of the spectral function is also discussed \cite{viswanath1994recursion, Camargo:2022rnt}.

Similar odd and even separation (with different growth) was observed in the next-to-leading order expansion in the large $q$ SYK model \cite{Bhattacharjee:2022ave}, and around the saddle point solution of the Lipkin-Meshkov-Glick (LMG) model \cite{Bhattacharjee:2022vlt}, as well as in systems mimicking the inhomogeneous  Su-Schrieffer-Heeger (SSH) model \cite{PhysRevLett.124.206803, PhysRevB.102.195419}. More specifically, such cases are prevalent in quantum field theory, especially when an explicit IR cutoff is present (see Sec.\,\ref{QFT}). In such a case, however, the pole structure of the autocorrelation function and the asymptotic decay of the spectral function cannot be used to deduce the generic growth of the Lanczos coefficients \cite{Dymarsky:2021bjq, Avdoshkin:2022xuw}. The full understanding of such splitting is still under investigation.

We can express the recursive relation \eqref{mombn} in a suggestive way, using a continued fraction \cite{parker2019, viswanath1994recursion,Bhattacharjee:2022lzy}
\begin{align}
  G(z) = \sum_{n=0}^{\infty} \frac{m_n}{z^{n+1}}  = \dfrac{1}{z - a_0 - \dfrac{b_1^2}{z- a_1 - \dfrac{b_2^2}{z - a_2 - \dots }}} \,, \label{confrac}
\end{align}
where $G(z)$ is the Green's function, which is related to the autocorrelation function $\mathcal{C} (t)$ by a Laplace-like transform
\begin{align}
    G(z) =  i \int_{0}^{\infty} d t \, e^{-i z t} \,\mathcal{C} (t)\,. %= \llangle p_1|(z-\mathcal{L}_o^{\dagger})^{-1}|p_1\rrangle\,.
\end{align} 
This can be verified using \eqref{autodef}. For a Hermitian Liouvilian $\mathcal{L}_H$, we have an orthonormal basis and the Green's function can be written as $G_H(z) =  (\mathcal{O}|(z-\mathcal{L}_H)^{-1}|\mathcal{O})$ with the even moments given by \eqref{evenmom}, which is connected to the inverse scattering operator \cite{parker2019}. Furthermore, $G(z)$ is associated with paths starting in the first site and returning to it after propagating over the chain. Applying this idea recursively allows for an intuitive understanding of the continuous fraction expansion \eqref{confrac}. The continued fraction representation of Green's function is closely related to the generating function of the \emph{Motzkin polynomials} \cite{Motzkinpaths}.

\subsubsection{An example: large $q$ SYK model} \label{sykexample}

In principle, the above formalism works for any system. For illustration, we choose the paradigmatic Sachdev-Ye-Kitaev (SYK) model as our system, as done in \cite{parker2019, Bhattacharjee:2022lzy, Bhattacharjee:2023uwx}. A similar construction has also been considered in the double-scaled SYK model \cite{Bhattacharjee:2022ave, Rabinovici2023DSSYK, Aguilar-Gutierrez:2024nau}. The motivation for choosing the SYK is twofold - it is numerically amenable \cite{garcia-garcia_SYK_2017, Kobrin_many_body_SYK} and analytically tractable \cite{Maldacena:2016hyu} for the desired computation. Further, the model is also important from the holographic considerations due to its similarity with Jackiw-Teitelboim (JT) gravity in the low-energy limit. We consider the $q$-body SYK model with $N$ Majorana fermions, given by the Hamiltonian 
\cite{PhysRevLett.70.3339, Kittu}
\begin{align}
    H = i^{q/2} \sum_{1 \leq i_1 < \cdots < i_q \leq N} J_{i_1 \cdots i_q} \psi_{i_1} \cdots \psi_{i_q}\,, \label{sykh}
\end{align}
where the fermionic operators $\psi_i$ obey the  Clifford algebra $\{\psi_a, \psi_b\} = \delta_{ab}$, and $J_{i_1 \cdots i_q}$ are random couplings drawn from a Gaussian ensemble with zero mean and variance given by
\begin{align}
    \langle{J_{i_1 \cdots i_q}\rangle} = 0\,, ~~ \langle{J^2_{i_1 \cdots i_q}\rangle} = \frac{(q-1)! J^2}{N^{q-1}} = 2^{q-1} \frac{(q-1)! \mathcal{J}^2}{q N^{q-1}}\,, \label{sykparamters}
\end{align}
where $\mathcal{J}^2 = 2^{1-q} q J^2$ is the convenient energy scale in the large $q$ limit. The limit $N\to\infty$ is already implied and facilitates analytical tractability, especially computing correlation functions in the $1/q$ expansion. However, for numerical purposes, we choose finite $N$ and thus focus on finite $q$ results. See \cite{SachdevSYKreview, Trunin_2021, Rosenhaus_2019} for a review of the SYK model and its connection to holography.

Traditionally, the growth of operator size in this model has been studied using the melon diagrams technique in Pauli spin-basis \cite{Roberts:2018mnp} and epidemic models \cite{Qi:2018bje}. Let us study the growth of a single Majorana operator, say $\psi_1 := \sqrt{2} \psi_1$, and the Krylov complexity, evolved by the Hamiltonian \eqref{sykh}. The two-point autocorrelation function
$\mathcal{C} (t) = (\psi_1(t) |\psi_1(0))_{\beta}$ at finite temperature $1/\beta$ is known
\cite{Maldacena:2016hyu}. However, in this section, we only focus on the infinite-temperature case. The autocorrelation function can be expanded as \cite{Maldacena:2016hyu} 
\begin{align}
    \mathcal{C} (t) &= 1 + \frac{1}{q}\, g(t)  +  O(1/q^2)\,. \label{autoq}
\end{align}
Here, we have considered the leading order term in $O(1/q)$ expansion. The subleading order has also been considered \cite{Tarnopolsky:2018env, Bhattacharjee:2022ave}, but we ignore that for our discussion.
The function $g(t)$ satisfies the Liouville differential equation \cite{parker2019}
\begin{align}
    \partial_{t}^2 g(t) &=  -2 \mathcal{J}^2 e^{g(t)}\,. \label{ode1}
\end{align}
With the boundary conditions $g(0) = 0$ and $g'(0) = 0$, we obtain the solution $g(t) = 2 \ln (\sech (\mathcal{J} t ))$. Hence, the auto-correlation function in the leading order is given by
\begin{align}
    \mathcal{C} (t) &= 1 + \frac{2}{q}\, \ln (\sech (\mathcal{J} t ))  + + O(1/q^2)\,. \label{autoqsyk1}
\end{align}
This form of autocorrelation directly allows us to compute the moments using \eqref{mom4}. Since the autocorrelation is an even function in time, the odd moments vanish $m_{2n+1} = 0$, while the even moments are given by 
\begin{align}
    m_{2n} = \frac{1}{q}  \mathcal{J}^{2n} T_{n-1} + O(1/q^2)\,, ~~~~~~ n \geq 1\,, \label{momsyk}
\end{align}
expressed in terms of the tangent numbers $\{T_{n-1}\}_{n=1}^{\infty} = \{1, 2, 16, 272, 7936, \cdots \}$ \cite{oeis}. Applying the recursive algorithm \eqref{mombn}, we obtain the Lanczos coefficients \cite{parker2019}
\begin{align}
b_n =
  \begin{cases}
    \mathcal{J}\sqrt{2/q} + O(1/q) \,      & ~~n = 1\,,\\
    \mathcal{J}\sqrt{n(n-1)} + O(1/q) \,   & ~~n > 1\,,  \label{bnleading}
  \end{cases}
\end{align}
implying an asymptotic growth $b_n \sim \alpha n$, with $\alpha = \mathcal{J}$. The growth is set by the energy scale of the problem. Plugging \eqref{bnleading} into the differential equation \eqref{discschr}, we obtain the Krylov basis wavefunctions as \cite{parker2019}
\begin{align}
\varphi_n(t) =
  \begin{cases}
    1 + (2/q) \ln (\sech (\mathcal{J} t)) + O(1/q^2) \,      & ~~n = 0\,,\\
    \sqrt{\frac{2}{nq}}\, \tanh^n (\mathcal{J} t) + O(1/q^2) \,   & ~~n \geq 1\,. \label{diss}
  \end{cases}
\end{align}

\begin{figure*}[t]
\centering
{\includegraphics[width=0.32\textwidth]{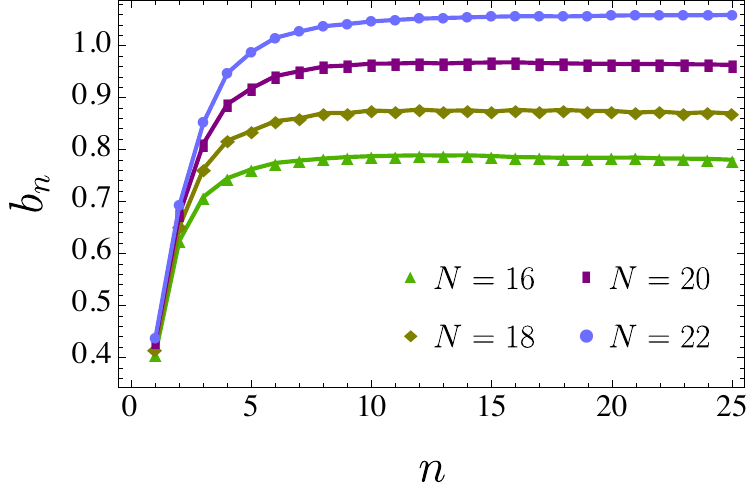}}
\hfil
{\includegraphics[width=0.32\textwidth]{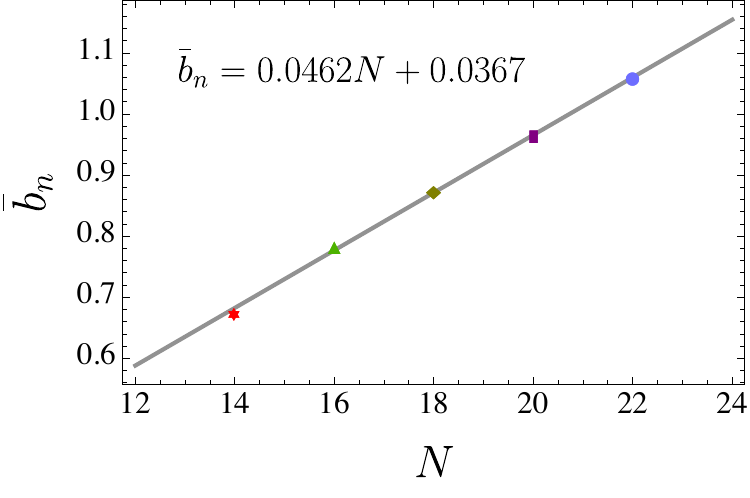}}
\hfil
{\includegraphics[width=0.32\textwidth]{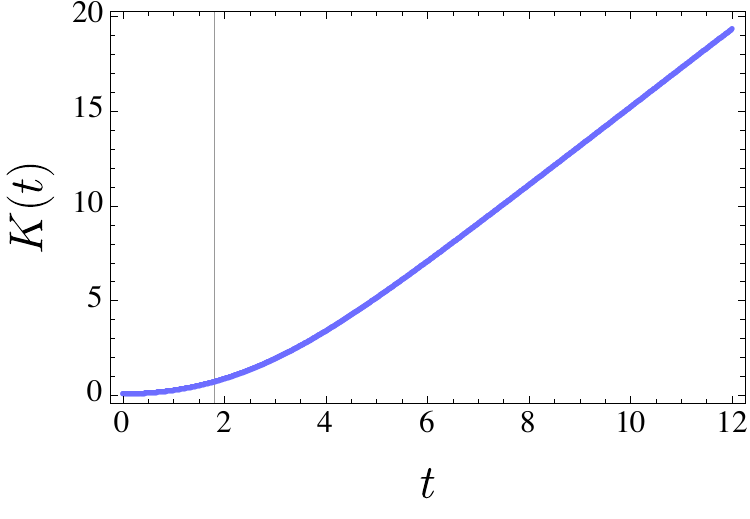}}
\caption{(Left) The growth of Lanczos coefficients of SYK$_4$ model at infinite temperature for different system sizes (number of fermions) $N$. The coefficients exhibit linear growth until they saturate to a system size-dependent value. (Middle) The finite size scaling of the saturation value $\overline{b}_n$. For the average, we take coefficients from $n=15$ to $n=25$. The fitted function is $\overline{b}_n = 0.0465 N + 0.0337$ up to four significant digits, shown by the gray line. The ensemble average of Hamiltonians taken are $100\, (N=14)$, $100\, (N=16)$, $50\, (N=18)$, $20\, (N=20)$ and $10 \,(N=22)$ respectively. In the left figure, $N=14$ data is not shown, while it is evaluated in the middle figure. (Right) The exponential growth of Krylov complexity is followed by a linear growth in the SYK$_4$ model at infinite temperature. Our system size is $N =22$, and we have taken $10$ ensemble averages of the Hamiltonian. The gray solid line indicates the scrambling time $t_{*} \propto \log N/q$, i.e.,  the boundary between the exponential and the linear growth regime. Due to the small system size ($N = 22$), the exponential regime is transient. We set $\mathcal{J} = 1/\sqrt{2}$ in all computations. Adapted from \cite{Jian:2020qpp} with different variance. To match with the results of \cite{Jian:2020qpp}, the coefficients will be multiplied by a factor of $\sqrt{2}$.} \label{fig:SYKfiniteT1}
\end{figure*}

It can be easily checked that the total probability is unity $\sum_{n=0}^{\infty}|\varphi_n(t)|^2 = 1$ up to the $O(1/q)$, as expected. Further, we compute the Krylov complexity as \cite{parker2019}
\begin{align}
    K (t) = \frac{2}{q} \sinh^2 (\mathcal{J} t) + O(1/q^2)\,,\label{cc} 
\end{align}
which grows exponentially with Krylov growth coefficient $\lambda_K = 2 \alpha = 2 \mathcal{J}$. It is also straightforward to compute the Krylov variance \eqref{kvar0}, which is given by
\begin{align}
    \Delta K(t)^2 = \frac{1}{2q} \sinh^2(2 \mathcal{J} t) + O(1/q^2)\,. \label{sykv0}
\end{align}
This grows exponentially with coefficient $4 \mathcal{J}$.

The infinite-temperature Lyapunov exponent obtained from the OTOC is also available \cite{Roberts:2018mnp}. In particular, it is upper bounded \cite{parker2019}
\begin{align}
    \lambda_{\mathrm{OTOC}} \leq 2 \alpha \,. \label{kbound1}
\end{align}
The bound is tight, and no tighter bound can be obtained. It is saturated only in the large $q$ limit and remains valid near saturation for finite $q$. However, this inequality is proved only in the infinite-temperature limit for the large $q$ limit. In the finite temperature case, the inequality is only a conjecture. In particular, it was speculated that the bound can only be saturated for all-to-all systems. A recent study also formulates the OTOC on the Krylov basis in the SYK model \cite{Chen:2024imd}. Further, the bound \eqref{kbound1} is shown to be valid for classical systems and becomes tight for classically chaotic systems, including systems that exhibit saddle-dominated scrambling \cite{parker2019, Bhattacharjee:2022vlt}.

In contrast to our analytic computation, we can also numerically determine the Lanczos coefficients from the Lanczos algorithm. In principle, we can choose any fermionic operator which is \emph{local} in the sense of full Hamiltonian. This includes single or two-body fermionic operators, for example. However, for our numerical analysis, we take $\mathcal{O} = \sqrt{2} \psi_1$ as the normalized initial operator, for which we have analytically computed the autocorrelation function in \eqref{autoq}.  Figure \ref{fig:SYKfiniteT1} shows the behavior of the Lanczos coefficients $b_n$ for $q =4$ with different system sizes (number of fermions) $N$. Since the system is closed, all $a_n$ vanish. The behavior of $b_n$ is typical for chaotic systems, it grows linearly followed by saturation at $n \gtrsim N/q$ to a system-size dependent value. At $n \sim e^{N}$, the Lanczos coefficients begin to decrease. One needs to exhaust the full Krylov space to see such a full regime of the \emph{Lanczos descent} of $b_n$. See \cite{Rabinovici:2020ryf} for the full profile of the Lanczos coefficients in SYK. However, in our case, increasing $N$ increases the saturation linearly, which is shown in the right figure in Fig.\,\ref{fig:SYKfiniteT1}. In the thermodynamic limit, the saturation is pushed to infinity, making it irrelevant physically. Hence, in the thermodynamic limit, only the slope of $b_n$ is important, not its saturation value. The slope is linear and given by \eqref{bnleading}.  However, our numerical model consists of $q=4$ body interaction with system sizes up to $N = 22$. Hence, both $q$ and $N$ are not large enough to compare our numerical result with the analytical large $q$ (and large $N$) result.

Finally, from the numerical results of the Lanczos coefficients, we directly compute the Krylov complexity by solving the Eq. \eqref{discschr}. An easy way to achieve this will be discussed in \eqref{dsevec}. Figure \ref{fig:SYKfiniteT1} shows the behavior of the Krylov complexity. Here, only transient exponential and linear growth are observed. The crossover happens at the scrambling time $t_{*} \propto \log N/q$, corresponding to the saturation of $b_n$. At exponentially large times, the Krylov complexity saturates, which is missing in our analysis. This is because we have not taken the full profile of Lanczos coefficients; the large $n$ Lanczos coefficients are responsible for the late-time value of the Krylov complexity. See \cite{Rabinovici:2020ryf} for the full profile of Krylov complexity in the SYK model.

\subsection{Revisiting Lanczos coefficients: a Toda chain method} \label{todadiscussion}

In this section, we derive an alternate method to compute Lanczos coefficients. This can be obtained from the monic version of the Lanczos algorithm \cite{dymarsky2020a}, with Euclidean time $\tau = it$. The method goes as follows: given the autocorrelation $\mathcal{C}(\tau)$, we construct the $(n+1)$-dimensional square Hankel matrix $\mathscr{M}$ such that
\begin{align}
    \mathscr{M}_{jk}^{(n)} (\tau) = \mathcal{C}^{(j+k)} (\tau)\,,~~~~~ j, k = 0, 1, \cdots, n\,,
\end{align}
where the $(j,k)$-th element is denoted by the $(j+k)$-th derivative of the autocorrelation function, i.e., $\mathcal{C}^{(j+k)}(\tau) = d^{j+k} \mathcal{C}(\tau)/d \tau^{j+k}$. In other words, the Hankel matrix is formed by the $n$-th moments. Then the determinant of this matrix is given by the Toda function
\begin{align}
    \uptau_n (\tau) = \det \mathscr{M}^{(n)} (\tau)\,,
\end{align}
where $\uptau_0 (\tau) = \mathcal{C}(\tau)$ is the autocorrelation function. The Toda function satisfies the \emph{Hirota's bilinear form} \cite{Hirota_2004}, 
\begin{align}
      \uptau_n \Ddot{\uptau}_n - \dot{\uptau}_n^2= \uptau_{n+1} \uptau_{n-1}\,,~~~~~~ \uptau_{-1} := 1\,, \label{todaeq1}
\end{align}
where the dot denotes the derivative with respect to $\tau$ and we have suppressed the $\tau$ dependence in $\uptau_n \equiv \uptau_n (\tau)$. This equation is equivalent to the Toda chain equation $\Ddot{\mathsf{q}}_n = e^{\mathsf{q}_{n+1} - \mathsf{q_n}} - e^{\mathsf{q}_n - \mathsf{q}_{n-1}}$ for $ n= 0, 1, \cdots$ \cite{toda1, todabook}, with $\mathsf{q}_{-1} = -\infty$, which is obtained upon the substitution with Toda variables $\mathsf{q}_n = \log(\uptau_n/\uptau_{n-1})$ with $\uptau_0 (\tau) = e^{\mathsf{q}_0 (\tau)}$. Given the Toda function $\uptau_n (\tau)$, the Lanczos coefficients are obtained as\footnote{Our notation uses a shift in $n \rightarrow n+1$ in the expression of $b_n^2(\tau)$ in \eqref{bndef} compared to \cite{dymarsky2020a}. This is consistent with the Toda equation and produces the exact result obtained from the moment method.} \cite{dymarsky2020a}
\begin{align}
    a_n (\tau)  = \frac{d}{d \tau} \log \left(\frac{\uptau_n (\tau)}{\uptau_{n-1} (\tau)}\right)\,, ~~~~~
    b_{n+1}^2 (\tau) = \frac{\uptau_{n+1} (\tau)\,\uptau_{n-1} (\tau)}{\uptau_n^2 (\tau)}\,, \,~~~~ n \geq 0\,,  \label{bndef}
\end{align}
which depend on the parameter $\tau$. The parameter-dependent Lanczos coefficients are known as \emph{Flaschka variables}. They are related to the Toda variables as $ a_n (\tau)  = \dot{\mathsf{q}}_n (\tau)$ and $b_{n+1} (\tau) = e^{\frac{1}{2}(\mathsf{q}_{n+1} (\tau) - \mathsf{q}_n(\tau))}$ for $n= 0, 1, \cdots$. In other words, the Toda equation can alternatively written in the following Flaschka form \cite{Flaschka}
provided the $a_n(\tau)$ and $b_n(\tau)$ satisfy 
\begin{align}
        \dot{a}_n (\tau)  = b_{n+1}^2  (\tau) - b_{n}^2 (\tau)\,,~~~~
    \dot{b}_{n+1} (\tau) = \frac{1}{2} b_{n+1} (\tau)\big(a_{n+1} \big(\tau) - a_n (\tau)\big)\,,~~ n \geq 0\,, \label{f1}
\end{align}
with $b_0 = 0$. Now we choose an appropriate cutoff $\tau = \tau_0$, similar to the moment method. The actual Lanczos coefficients are then given by
\begin{align}  
    a_n = a_n (\tau)\big|_{\tau_0} \,, ~~~~~~ b_{n+1} = b_{n+1} (\tau)\big|_{\tau_0} \,,~~~~~~~~ n\geq 0\,.
\end{align}
For most cases, we consider $\tau_0 = 0$, but for certain systems (e.g., in CFT), we may need to consider $\tau_0 \neq 0$. This yields another powerful method to obtain the Lanczos coefficients from the autocorrelation function.

There is a very simple \emph{Ansatz}: 
\begin{align}
b_n^2(\tau)=b(\tau)^2 p(n)\,,    \label{TodaAnsatz}
\end{align}
which satisfies \eqref{f1} provided $p(n)$ is a polynomial of degree no higher than two. The corresponding solutions were explicitly found in \cite{dymarsky2020a}. In the next section, we will encounter exactly the same sequences of $b_n$ again when discussing the algebraic approach.

Both the ``moment method'' and ``Toda chain method'' are useful in many situations, especially when we know the analytic form of the autocorrelation function. A particular example is the SYK model \cite{parker2019} (see Sec.\,\ref{sykmomentmethod}) and its large $q$ expansions \cite{Bhattacharjee:2022ave}. 
The moment method has been successfully applied to compute the Lanczos coefficients, which have been numerically verified.
Another specific interesting case is that of quantum field theories (QFT) and their conformal limit, described by conformal field theories (CFT) \cite{Dymarsky:2021bjq}. Because of the infinite degrees of freedom, we cannot generate an orthonormal (or bi-orthonormal) basis. However, many theories allow us to compute the autocorrelation function exactly, mainly due to the conformal symmetry. In such cases, the Lanczos coefficients can be exactly computed using either the moment method or the Toda chain method. See Sec.\,\ref{QFT} for details. 

\section{Coherent states, complexity algebra, and dispersion bound in Krylov space}\label{secKdyn}

\subsection{Coherent states}
\label{sec:cohStates}
From the vast array of complex quantum systems, we narrow down our study to those that are symmetrical. We focus on systems where the Liouvillian operator belongs to the Lie algebra of a specific symmetry group. For these systems, the representation of the Liouvillian in the Krylov basis decomposes into two components according to \cite{caputa2021}
\begin{align}
    \mathcal{L} = \alpha (\mathcal{L}_{+} + \mathcal{L}_{-})\,, \label{lgroup1}
\end{align}
where $\mathcal{L}_{+}$ and $\mathcal{L}_{-}$ represent the \emph{raising} and \emph{lowering} parts of the Liouvillian, akin to ladder operators. While their representations vary across different symmetry groups, their operational essence mirrors that of the creation and annihilation operators, which we will examine shortly. The coefficient $\alpha$ is not constrained by symmetry considerations and must be calibrated based on the system's specifics and according to the chosen norm. Considering the scenario where $a_n = 0$, the action of $\mathcal{L}_{\pm}$ on the Krylov basis is given by \eqref{lortho}:
\begin{align}
    \alpha \mathcal{L}_{+} |\mathcal{O}_n) = b_{n+1} |\mathcal{O}_{n+1})\,, ~~~~~~~\alpha \mathcal{L}_{-} |\mathcal{O}_n) = b_{n} |\mathcal{O}_{n-1})\,.
\end{align}
Systems with non-vanishing $a_n$ can be %straightforwardly generalized 
also considered \cite{balasubramanian2022}. Notably, the temporal evolution dictated by the Liouvillian \eqref{lgroup1} can be equivalently described through the evolution of coherent states, which are constructed via the exponential action of the ladder operators \cite{perelomov_generalized_2012, caputa2021}, 
\begin{align}
    D(\xi) := e^{\xi \mathcal{L}_{+} - \overline{\xi} \mathcal{L}_{-}}\,, \label{disp}
\end{align}
where $\xi$ denotes a complex number, with $\overline{\xi}$ being its complex conjugate. The operator $D(\xi)$, known as the \emph{displacement operator}, corresponds to the Lie group generated by the ladder operators $\mathcal{L}_{\pm}$. Similar displacement operators frequently arise when studying coherent states in quantum optics. A prototypical coherent state emerges from the application of $D(\xi)$ to a reference state $\ket{\psi}$ \cite{perelomov_generalized_2012}.
%\begin{align}
%    \ket{z,\xi} = D(\xi) \ket{\xi} \,, \label{coh1}
%\end{align}
We next explore how a judicious selection of $\xi$ and $\ket{\psi}$ facilitates the study of operator growth in Krylov space governed by the Liouvillian \eqref{lgroup1} with specific symmetry groups. In particular, we present results for the three closed algebras: $\mathrm{SL}(2, \mathbb{R})$, Heisenberg-Weyl, and $\mathrm{SU}(2)$, which describes different physical systems including the SYK model and spin chains \cite{caputa2021}.

%the identification of the resultant coherent state \eqref{coh1} as the time-evolved state 
%governed by the Liouvillian \eqref{lgroup1}, in alignment with the associated symmetry group.

\subsubsection{$\mathrm{SL}(2,\mathbb{R})$ algebra} \label{subsubsec:SL2R}

Our first example is the algebra of SL$(2,\mathbb{R})$, which is locally isomorphic to SU$(1,1)$ \cite{perelomov_generalized_2012}. The generators of this group are the set $\{L_0, L_{\pm 1}\}$, where they satisfy the following Lie algebra \cite{perelomov_generalized_2012}
\begin{align}
    [L_0, L_{\pm 1}] = \mp L_{\pm 1}\,, ~~~~~ [L_{1},L_{-1}] = 2 L_0\,.
\end{align}
The above generators construct the complete representation of the Lie algebra \cite{perelomov_generalized_2012, caputa2021}
\begin{align}
\begin{split}
      L_{0} \ket{h, n} &= (h+n) \ket{h, n}\,,\\
    L_{-1} \ket{h, n} &= \sqrt{(n+1)(2h+n)} \ket{h, n+1}\,, \\
     L_{1} \ket{h, n} &= \sqrt{n(2h+n-1)} \ket{h, n-1}\,. \label{lopp}
\end{split}
\end{align}
Here, $h$ and $n \geq 0$ are two indices for the corresponding states $\ket{h, n}$, usually known as the conformal weight and the excitation index, respectively. The states obey the condition $\langle h,m|h,n \rangle = \delta_{m,n}$, ensuring orthonormality. Furthermore, we can introduce the Casimir operator $\mathsf{C}_2 = L_0^2 - \frac{1}{2}(L_{-1} L_1 + L_1 L_{-1})$ of the algebra, which acts invariantly on the state $\ket{h,n}$  \cite{perelomov_generalized_2012}
\begin{align}
    \mathsf{C}_2 \ket{h, n} = h(h-1)\ket{h, n}\,. \label{cas}
\end{align}
Since the Casimir operator commutes with $L_0$, the states $\ket{h, n}$ are the simultaneous eigenstates of $\mathsf{C}_2$ and $L_0$. This fact is evident from \eqref{lopp} and \eqref{cas}.

The states $\ket{h, n}$ are obtained by the repeated action ($n$-fold) of the generator $L_{-1}$ on the highest-weight state $\ket{h}$, expressed as \cite{perelomov_generalized_2012, caputa2021}
\begin{align}
    \ket{h, n} = \sqrt{\frac{\Gamma (2h)}{n! \, \Gamma(2h+n)}}  L_{-1}^n \ket{h} \,.
\end{align}
Interestingly, the  generalized coherent states of the SL$(2, \mathbb R)$ algebra $\ket{z,h}$ are obtained by the action of the displacement operator $D(\xi)$ \cite{perelomov_generalized_2012, caputa2021}:
\begin{align}
    \ket{z,h} = D(\xi) \ket{h}\,, ~~~~ D(\xi) = e^{\xi L_{-1} - \overline{\xi} L_1}\,.
\end{align}
The parametrization between $z$ and $\xi$ is given by $z = \frac{\xi}{|\xi|} \tanh(|\xi|)$, where $|\xi|^2 = \xi \overline{\xi}$. Thus, a generic coherent state aligned with the SL$(2,\mathbb{R})$ symmetry can be written explicitly in the eigenbasis $\{\ket{h, n}\}$ as \cite{perelomov_generalized_2012, caputa2021, Patramanis:2021lkx}
\begin{align}
    \ket{z,h} = (1-|z|^2)^{h} \sum_{n=0}^{\infty} z^n \sqrt{\frac{(2h)_n}{n!}} \ket{h, n}\,.
\end{align}
Here, $(a)_n = \Gamma(a + n)/\Gamma(a)$ is the Pochhammer symbol. With the above formalism in hand, we define the Liouvillian as a combination of $L_{\pm 1}$ according to \cite{caputa2021}
\begin{align}
    \mathcal{L} = \alpha (L_{1} + L_{-1})\,, \label{lsp}
\end{align}
where $\alpha$ is a non-zero real coefficient. Although we do not impose any constraint on it, we take it independent of both $h$ and $n$. In Eq. \eqref{lsp}, we can also include $L_0$ and an identity operator. However, for the discussion, we stick to the simplest choice in \eqref{lsp}. To appreciate the generic structure of the operator evolution   \cite{caputa2021}
\begin{align}
    |\mathcal{O}(t)) \equiv e^{i \mathcal{L} t}  |\mathcal{O}) \equiv e^{i  \alpha (L_{1} + L_{-1}) t} |\mathcal{O}) \,,
\end{align} 
note that the time-evolution operator generated by the Liouvillian \eqref{lsp} is nothing but the displacement operator of the algebra with $\xi = i \alpha t$. Once we identify the initial operator as the highest-weight state of the representation, we can immediately identify the time-evolved state and the Krylov basis
\cite{caputa2021, Patramanis:2021lkx},
\begin{align}
    |\mathcal{O}(t)) = \ket{z,h}\,, ~~~|\mathcal{O}) = \ket{h}\,, ~~~ |\mathcal{O}_n) = \ket{h,n}\,, 
\end{align}
with the corresponding Lanczos coefficients. In particular, in this case, it is easy to see that they are given by the eigenvalues of the SL$(2,\mathbb{R})$ generators \cite{caputa2021}
\begin{align}
    b_n = \alpha \sqrt{n(n-1 + 2 h)} \,. \label{bns0}
\end{align}
Thus, the corresponding operator evolution is governed by the SL$(2,\mathbb{R})$ symmetry. 
This completely furnishes the structure of the operator evolution corresponding to the underlying symmetry algebra.

We note that the behavior \eqref{bns0} is one of the examples of the exactly solvable Toda equations \eqref{f1} with the factorizable $b_n^2(\tau)$ \eqref{TodaAnsatz}. The time dependence, which we discuss below, can be extended to Toda formalism using Wick rotation $\tau=i t$.

\begin{figure}[t]
\centering
%\subfigure[]
{\includegraphics[width=0.43\textwidth]{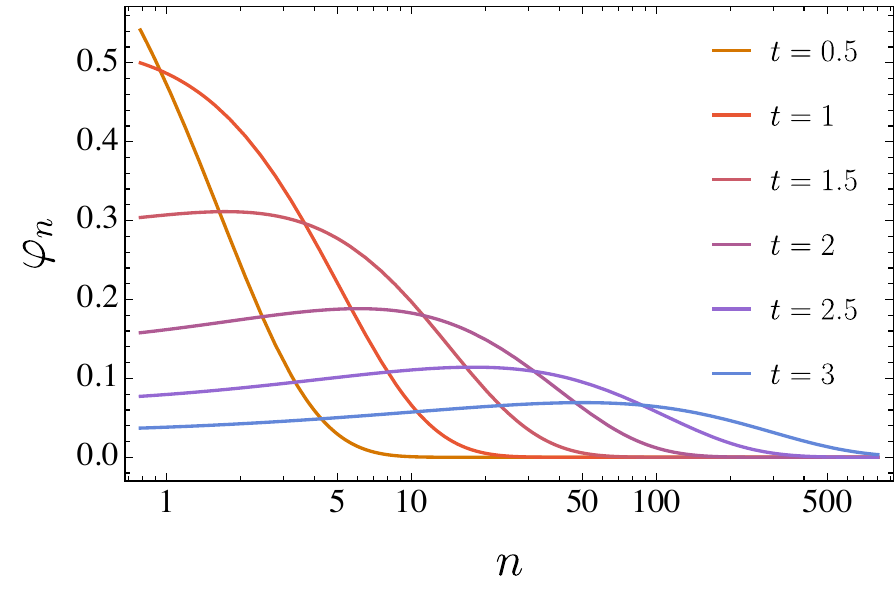}}
\hfil
%\subfigure[]
{\includegraphics[width=0.43\textwidth]{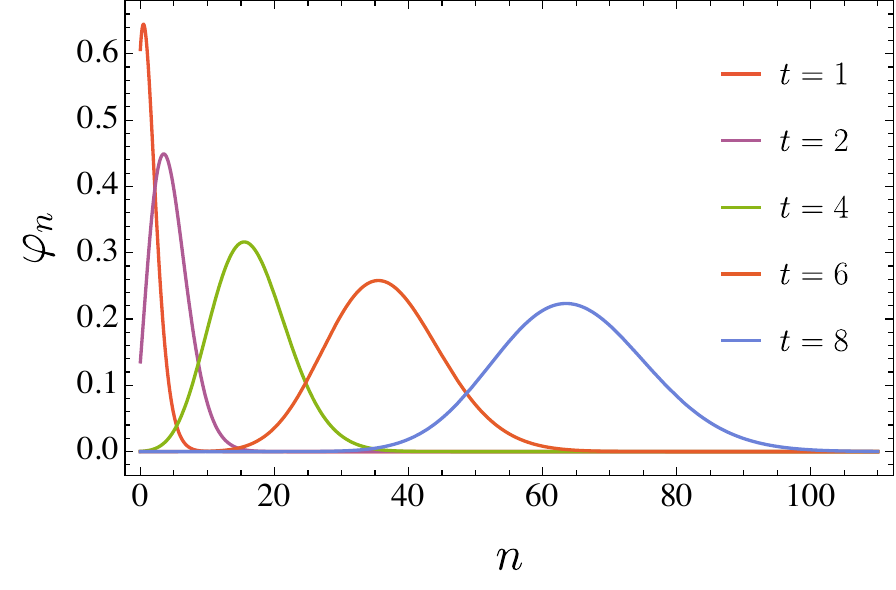}}
\caption{(Left) Snapshots (each curve shows a fixed time) of the Krylov basis wavefunction \eqref{alvphi}, with $\alpha = 1$ and $\eta = 1.5$. While $n$ is a discrete index, we treat it as a continuum value for visual display. At late times, the large $n$ wavefunctions become more dominant. The tail decays exponentially (not shown in the above figure) according to \eqref{del1} for large $n$, with the exponentially increasing delocalization length. (Right) Snapshots of Krylov basis wavefunctions \eqref{su2w} for the HW algebra ($\alpha = 1$). Adapted from \cite{caputa2021}.} \label{fig:wavesnap}
\end{figure}

So far, the discussion is completely general and restrained by the SL$(2,\mathbb{R})$ symmetry only. However, to make contact with our previous discussion, we identify $\xi = i \alpha t$. This readily gives $z = i \tanh \alpha t$ and the time-evolved operator is given by the coherent state \cite{caputa2021, Patramanis:2021lkx}
\begin{align}
    |\mathcal{O}(t)) &= \ket{z,h}|_{z= i \tanh \alpha t,\, h= \eta/2} = \sum_{n=0}^{\infty} \sqrt{\frac{(\eta)_n}{n!}} \sech^{\eta}(\alpha t)\tanh^n(\alpha t) |\mathcal{O}_n)\,, \label{pa1}
\end{align}
where $\eta$ is related to the parameters of the operator growth hypothesis \eqref{uogh} by $\eta = 2 \gamma/\alpha + 1$ \cite{parker2019}. Hence, the Krylov basis function is easily obtained by the expansion coefficients of the above time-evolved operator, associated with the Lanczos coefficients. They read \cite{parker2019}
\begin{align}
    b_n &= \alpha \sqrt{n(n-1 + \eta)}\,, \label{bns}\\
    \varphi_n (t) &=  \sqrt{\frac{(\eta)_n}{n!}} \sech^{\eta}(\alpha t)\tanh^n(\alpha t)\,, \label{alvphi}
\end{align}
with the same choice of parameters $z$ and $h$, as in \eqref{pa1}. One can check that \eqref{alvphi} satisfies the recurrence relation \eqref{discschr} with the corresponding Lanczos coefficients \eqref{bns}. Each $\varphi_n(t)$ shows an exponentially decreasing behavior $\sim e^{-\eta \alpha t}$ at late times, independent of $n$. For a finite time, Fig.\,\ref{fig:wavesnap} (left) shows snapshots of the wavefunction at each time. As time progresses, the contribution of the higher $\varphi_n$ becomes more dominant. To find the asymptotic tail of such behavior, consider the asymptotic limit of \eqref{alvphi}. We find \cite{parker2019}
\begin{align}
     \varphi_n (t)  \simeq  n^{\frac{\eta -1}{2}} \tanh^n (\alpha t) \sim e^{-n/\xi(t)} \,  n^{\frac{\eta -1}{2}}\,, \label{del1}
\end{align}
where $\xi(t)^{-1} \sim 2e^{-2 \alpha t}$. This expression follows from the asymptotic expansion $\Gamma(n+\eta)/\Gamma(n) \sim n^{\eta}$ for $n \rightarrow \infty$, ignoring any $n$-independent terms, which are irrelevant in the asymptotic limit. The delocalization length $\xi(t)$ grows exponentially, i.e., the operator delocalizes \cite{Avdoshkin:2019trj}. Later, we will see a dissipative version \cite{Bhattacharjee:2022lzy} of this delocalization in Eq. \eqref{del2}.

The probability is conserved $\sum_{n=0}^{\infty} |\varphi_n (t)|^2 = 1$ and the Krylov complexity reads \cite{parker2019}
\begin{align}
    K(t) = \sum_{n=0}^{\infty} n |\varphi_n(t)|^2 = \eta \sinh^2 (\alpha t) \sim \eta e^{2 \alpha t}\,,
\end{align}
which is exponential with an arbitrary prefactor $\eta$ which \emph{does not} scale with $n$. Let us consider two special values of $\eta$. For the simplest case, with $\eta = 1$, 
the Lanczos coefficients become linear, and the Krylov basis functions are given by \cite{parker2019, barbon2019}
\begin{align}
    b_n = \alpha n \,, ~~~~~~~~~\varphi_n (t) = \sech(\alpha t)\tanh^n(\alpha t)\,. \label{bb}
\end{align}
The autocorrelation function is $\varphi_0(t) = \sech(\alpha t) \equiv \mathcal{C}_1(t)$, whose property, pole structure, and corresponding spectral function is already discussed in detail, see Sec. \ref{sec:moment1}. The Krylov complexity grows exponentially $K(t) = \sinh^2 (\alpha t) \sim e^{2 \alpha t}$.

The second choice concerns the SYK model. The above result correctly reproduces the analytic expressions in this model. Identifying $\alpha = \mathcal{J}$, $\eta = 2/q$ in \eqref{bns}, and performing the $1/q$ expansion readily gives
\begin{align*}
     b_n = \alpha \sqrt{n(n-1 + 2/q)} = \mathcal{J} \sqrt{n(n-1)} + O(1/q),
\end{align*}
which is exactly \eqref{bnleading} for $n>1$. Both the Krylov wavefunction and Krylov complexity also reduce to \eqref{diss} and \eqref{cc} upon the same identification. This is intimately tied with the underlying SL$(2,\mathbb{R})$ symmetry of the Liouvillian generator, which we have exploited in the corresponding melon diagrams in \eqref{eq:Lpm}. Later, we see this generic structure of algebra is also retained when we include an addition coefficient $a_n$ for open systems (see Eq. \eqref{eq:anbnexact}).

\subsubsection{Heisenberg-Weyl (HW) algebra}\label{subsubsec:HW}

The Heisenberg-Weyl (HW) algebra is generated by the four generators $\{a, a^{\dagger}, \id, \hat{n}\}$. Here $\id$ is the identity operator, and $\hat{n} = a^{\dagger} a$ is the number operator. They satisfy the following algebra
\begin{align}
    [a,a^{\dagger}] = \id\,,~~~~~ [n,a] = -a \,, ~~~~~ [n,a^{\dagger}] = a^{\dagger}\,. \label{hwalgebra}
\end{align}
The Hilbert space is infinite-dimensional and spanned by the number basis set $ \ket{n} = \frac{1}{\sqrt{n!}} (a^{\dagger})^n \ket{0}$,
where $\ket{0}$ is the lowest-weight state in the representation corresponding to $n=0$, i.e., $\langle 0|\hat{n}|0\rangle = 0$. In other words, it is annihilated by the operator $a$, i.e., $a \ket{0} = 0$. The basis states are orthonormal $\langle m|n \rangle = \delta_{m,n}$. The creation ($a^{\dagger}$) and annihilation ($a$) operators act on the number state according to
\begin{align}
    a^{\dagger} \ket{n} = \sqrt{n+1} \ket{n+1} \,, ~~~~~ a \ket{n} = \sqrt{n} \ket{n-1}\,.
\end{align}
They are also known as the raising and the lowering operator, respectively, as obvious from their action. The generic coherent states is given by \cite{perelomov_generalized_2012, caputa2021}:
\begin{align}
    \ket{z} = D(z) \ket{0}\,, ~~~~ D(z) = e^{z a^{\dagger} - \overline{z} a}\,, \label{cohhw}
\end{align}
with $z = e^{i \phi}$ being a complex number. The operator $D(z)$ is the standard displacement operator. Using the algebra \eqref{hwalgebra}, we obtain the generic form of the coherent state
\begin{align}
    \ket{z} = e^{-|z|^2/2} \sum_{n=0}^{\infty} \frac{z^n}{\sqrt{n!}} \ket{n}\,.
\end{align}
To find the operator evolution, we write the Liouvillian in terms of the creation and annihilation operator \cite{caputa2021}
\begin{align}
    \mathcal{L} = \alpha (a^{\dagger} + a)\,.~~ 
\end{align}
Following the identification $z = i \alpha t$, the Krylov basis and the time-evolved operator can readily be obtained as a coherent state \cite{caputa2021}
\begin{align}
    |\mathcal{O}(t)) = \ket{z = i \alpha t} = e^{i  \alpha (a^{\dagger} + a)t} \ket{0}\,,~~~ |\mathcal{O}_n) = \ket{n}\,.
\end{align}
Similarly, we identify the Lanczos coefficients and the Krylov basis wavefunctions \cite{caputa2021}
\begin{align}
    b_n = \alpha \sqrt{n}\,, ~~~~
    \varphi_n (t) =  e^{-\alpha^2 t^2/2} \,\frac{(\alpha t)^n}{\sqrt{n!}}\,, \label{bnwavehwalgebra}
\end{align}
such that the wavefunction is appropriately normalized, i.e., $\sum_{n=0}^{\infty}|\varphi_n(t)|^2 = 1$. The snapshots of the wavefunctions are shown in Fig.\,\ref{fig:wavesnap} (right). The Krylov complexity is given by \cite{caputa2021}
\begin{align}
    K(t) = \sum_{n=0}^{\infty} n |\varphi_n (t)|^2 = \alpha^2 t^2 \,,
\end{align}
which grows quadratically over time. This algebra is thus an example of a system where the Lanczos coefficients grow sub-linearly, and therefore the Krylov complexity grows sub-exponentially, in particular quadratically. Various generalizations and extensions of the above algebra are possible, like the $q$-deformed version. An interesting extension was done for the Schr\"odinger group in \cite{Patramanis:2023cwz}. 

\subsubsection{SU(2) algebra} \label{subsubsec:SU2}

The SU(2) algebra is defined by the set of three generators $\{J_i\}_{i=1}^3$, obeying the following commutation rule 
$[J_i, J_j] = i \epsilon_{ijk} J_{k}$ with $i,j,k = 1,2,3$, where $\epsilon_{ijk}$ is the Levi-Civita symbol. Introducing the raising and lowering operators $J_{\pm} = J_{1} \pm i J_{2}$, and relabeling $J_3 \equiv J_0$, the SU(2) algebra is written as
\begin{align}
    [J_0, J_{\pm}] = \pm J_{\pm}\,, ~~~~~ [J_{+}, J_{-}] = 2 J_0\,. \label{su2a}
\end{align}
The Casimir operator of this algebra is $J^2 := J_1^2 + J_2^2 + J_3^2 = J_{+}J_{-} + J_0^2$. Since $[J^2,J_0] = 0$, they have a common eigenbasis
\begin{align}
    J^2 \ket{j, n} = j(j+1) \ket{j, n}\,,~~~~
    J_0 \ket{j, n} = n \ket{j, n}\,.
\end{align}
The states are orthonormal $\langle j, n| j', n' \rangle = \delta_{j j'} \delta_{n n'}$. The quantum numbers $j = 0, 1/2, 1, \cdots$ and $n$ with $-j \leq n \leq j$ are spin quantum numbers. For convenience, we shift $n \rightarrow -j+n$, such that $0 \leq n \leq 2j$. The lowest-weight state is $\ket{j,-j}$ which is annihilated by $J_{-}$, i.e, $J_{-}\ket{j,-j} = 0$. A similar action on the highest-weight state is $J_{+}\ket{j,j} = 0$. However, repeated action of $J_{+}$ on the lowest-weight state $\ket{j,-j}$ builds the corresponding orthonormal basis states
\cite{perelomov_generalized_2012}
\begin{align}
    \ket{j, -j+n} = \sqrt{\frac{\Gamma (2j-n+1)}{n! \,\Gamma(2j+1)}} \, J_{+}^n \ket{j,-j} \,, \label{js1}
\end{align}
which we denote with another index $n$. Alternatively, we could reach the same state from the highest-weight state $\ket{j, j}$ and the repeated action of $J_{-}$. The action of the generators $\{J_{\pm}, J_0\}$ on this state \eqref{js1} is the following \cite{perelomov_generalized_2012, caputa2021}
\begin{align}
\begin{split}
      J_{0} \ket{j, -j+n} &= (-j+n) \ket{j, -j+n}\,,\\
    J_{+} \ket{j, -j+n} &= \sqrt{(n+1)(2j-n)} \ket{j, -j+n+1}\,, \\
     J_{-} \ket{j, -j+n} &= \sqrt{n(2j-n+1)} \ket{j,-j + n-1}\,. \label{jopp}
\end{split}
\end{align}
Similar to the $\mathrm{SL}(2,\mathbb{R})$ and SU$(2)$ cases, the coherent states $\ket{z,j}$ are obtained by the action of the displacement operator $D(\xi)$ \cite{perelomov_generalized_2012, caputa2021}:
\begin{align}
    \ket{z,j} = D(\xi) \ket{j,-j}\,, ~~~~ D(\xi) = e^{\xi J_{+} - \overline{\xi} J_{-}}\,, \label{jcoherent}
\end{align}
where $z = \tan(\theta/2) e^{i \phi}$. As a result, the SU$(2)$ coherent state is written in the eigenbasis as \cite{perelomov_generalized_2012, caputa2021, Patramanis:2021lkx}
\begin{align}
    \ket{z,j} = (1+|z|^2)^{-j} \,\sum_{n=0}^{\infty} z^n \sqrt{\frac{\Gamma(2j+1)}{n! \,\Gamma(2j-n+1)}} \ket{j, -j + n}\,. \label{jcoh}
\end{align}
This completes the general discussion of the familiar SU(2) algebra. To connect with our discussion, we split the Liouvillian according to \cite{caputa2021}
\begin{align}
    \mathcal{L} = \alpha (J_{+} + J_{-})\,. \label{jsp}
\end{align}
The operator evolution is given by \cite{caputa2021}
\begin{align}
    |\mathcal{O}(t)) \equiv e^{i \mathcal{L} t}  |\mathcal{O}) \equiv e^{i  \alpha (J_{+} + J_{-}) t} |\mathcal{O}) \,.
\end{align}
With the initial operator $|\mathcal{O}) = \ket{j,-j}$, given that the effective evolution is described by the displacement operator \eqref{jcoherent}, the time-evolved operator is given by the coherent state \cite{caputa2021}, i.e.,
\begin{align}
        |\mathcal{O}(t)) := \ket{z = i \tan(\alpha t),j}\,,~~~~~~~~
    |\mathcal{O}_n) := \ket{j,-j+n}\,, ~~ n = 0,\cdots, 2j\,.
\end{align}

\begin{figure}[t]
\centering
%\subfigure[]
{\includegraphics[width=0.43\linewidth]{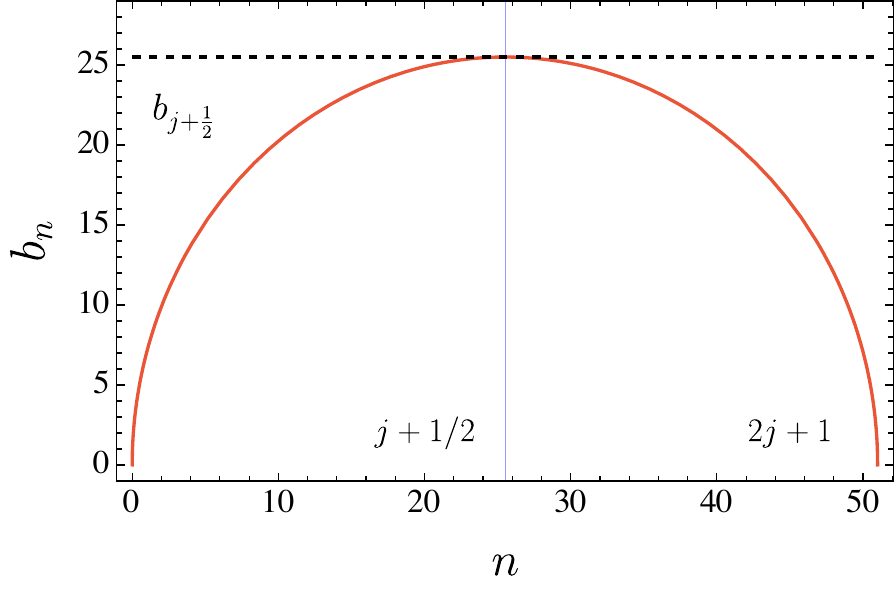}}
\hfil
%\subfigure[]
{\includegraphics[width=0.43\linewidth]{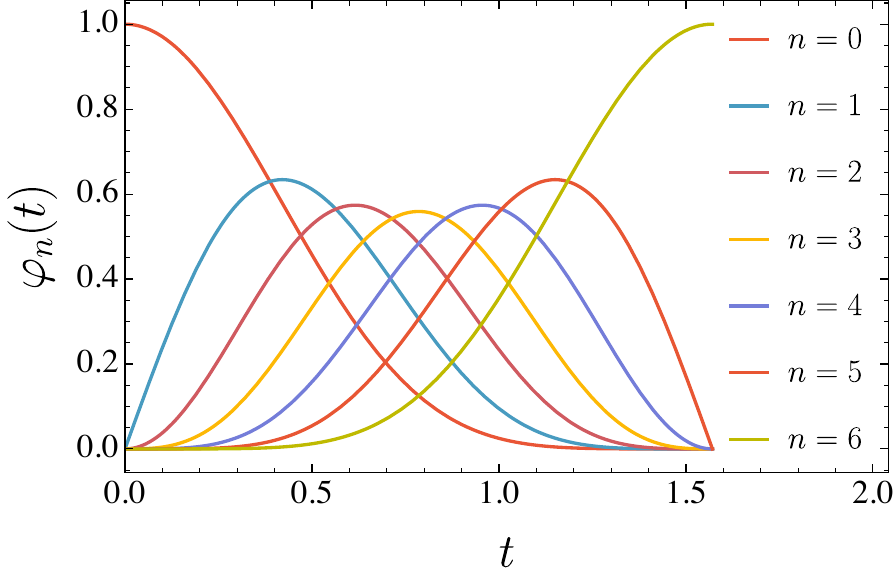}}
\caption{(left) Variation of Lanczos coefficients for SU(2) algebra for $j=25$. The structure is generic for any spin $j$. The coefficient reaches at maximum for $n = j+1/2$ (shown by opaque and vertical blue line) with $b_{j+1/2}$ (shown by dashed black line), while it vanishes at the end of the Krylov space $b_{2j+1} = 0$. A comparison between the $\mathrm{SL}(2,\mathbb{R})$ and HW algebra is shown in Fig.\,\ref{simplicityplot}(b). (right) Evolution of the Krylov basis wavefunctions \eqref{su2w} for SU(2) algebra for $j=3$ for range $t \in [0,\pi/2]$ (we choose $\alpha = 1$). The dimension of the Hilbert space is $2j +1= 7$. Adapted from \cite{caputa2021}.} \label{fig:wavesu2plot}
\end{figure}

In other words, we have identified the Krylov basis states with the orthonormal spin states. Hence the Lanczos coefficients and the Krylov basis wavefunctions are readily identified \cite{caputa2021}
\begin{align}
    b_n &= \alpha \sqrt{n(2j - n +1)}\,, \label{bnsj}\\
    \varphi_n (t) &= \sqrt{\frac{\Gamma(2j+1)}{n! \,\Gamma(2j-n+1)}} \sec^{-2j}(\alpha t)\tan^{n}(\alpha t)\,. \label{su2w}
\end{align}
The Krylov dimension $D_K$ is identified when $b_{D_K} = 0$, which implies $D_K = 2j+1$, the same as the Hilbert space dimension. Due to the finite structure of the Hilbert space, $b_n$ furnishes the symmetry
\begin{align}
    b_1 = b_{2j} = \alpha \sqrt{2j}\,, ~~~~ b_{\mathrm{max}} = \alpha \left(j+\frac{1}{2}\right)\,.
\end{align}
The variation of the Lanczos coefficients is shown in Fig.\,\ref{fig:wavesu2plot} (left). The coefficients peak at $n = j+1/2$ with $b_{j+1/2} = b_{\mathrm{max}}$ and vanish at the end of the Krylov space.
The corresponding evolution of a set of wavefunctions is shown in Fig.\,\ref{fig:wavesu2plot} (right). The symmetry structure of $b_n$ reflects the symmetric profile of the wavefunctions. They are normalized, i.e., $\sum_{n=0}^{2j} |\varphi_n (t)|^2 = 1$.  Hence, the Krylov complexity is given by \cite{caputa2021}
\begin{align}
    K(t) = \sum_{n=0}^{2j} n |\varphi_n (t)|^2 = 2j \sin^2(\alpha t) \,.
\end{align}
Since $\alpha\in \mathbb R_+$, the complexity is periodic. The average complexity is $\overline{K} =  j$, which is directly proportional to the spin. This fact has been exploited in the computation of the Krylov state (spread) complexity (see Sec.\,\ref{secStates}) for spin $j=N/2$ representation of SU(2) algebra for the paramagnetic Hamiltonian, where $N$ represents the number of lattice sites \cite{Bhattacharjee:2022qjw}.

A particular extension of the above algebra is known as the $\mathtt{q}$-deformed SU(2) algebra \cite{Chaichian1996} (denoted as SU$_{\mathtt{q}}$(2)), which was first studied in Ref.~\cite{Bhattacharjee:2022qjw} in the context of quantum many-body scars. This amounts to define $[x]_{\mathtt{q}}$, the $\mathtt{q}$-deformed number of $x$, such that \cite{Chaichian1996}
\begin{align}
    [x]_{\mathtt{q}} = \frac{\mathtt{q}^x - \mathtt{q}^{-x}}{\mathtt{q} - \mathtt{q}^{-1}}\,,~~~~ \lim_{\mathtt{q}\rightarrow 1} [x]_{\mathtt{q}} = x\,.
\end{align}
The SU(2) algebra \eqref{su2a} is modified according to 
\begin{align}
    [J_0, J_{\pm}] = \pm J_{\pm}\,, ~~~~~ [J_{+}, J_{-}] = [2 J_0]_{\mathtt{q}}\,.
\end{align}
where $[2 J_0]_{\mathtt{q}}$ is the $\mathtt{q}$-deformed version of the operator $2 J_0$, with the associated Lanczos coefficients \cite{Bhattacharjee:2022qjw}
\begin{align}
    b_n^{(\mathtt{q})} = \alpha \sqrt{[n]_{\mathtt{q}} [2j - n +1]_{\mathtt{q}}}\,,~~~~ \lim_{\mathtt{q}\rightarrow 1} b_n^{(\mathtt{q})} = b_n\,,
\end{align}
where $b_n$ is the SU(2) Lanczos coefficients \eqref{bnsj}. For a more detailed application, see \cite{Bhattacharjee:2022qjw}.

\subsection{Complexity algebra: the simplicity hypothesis} 

The particular tridiagonal structure of the Liouvillian \eqref{triLiouvillian} and the Krylov operator \eqref{kop} in the Krylov basis enables us to define a particular notion of algebra in Krylov space. To see this, we define the \textit{anti-Liouvillian} \cite{caputa2021,Hornedal2022}
\begin{align}
\mathcal{M} := 
	\begin{pmatrix}
0 & - b_1 &  &  &  \\
b_1 & 0 & -b_2 &   &  \\
 & b_2 & 0   & & \\
 &  & &  \ddots  \\
 &  &  &  & 0 &  -b_{D_K-1}   \\
 &  &  &  &  b_{D_K-1}  & 0  \\
	\end{pmatrix}\,, 
\end{align}
with the property $\mathcal{M}^{\dagger} = \mathcal{M}^{\intercal} = -\mathcal{M}$, and a real vector $|\varphi(t) ) := \big( \varphi_0 (t), \varphi_1 (t), \varphi_2 (t), \dots , \varphi_{D_K-1} (t) \big)^{\intercal}$, where $\intercal$ denotes the transpose. Note that, analogously to the Liouvillian in \eqref{lgroup1}, the anti-Liouvillian may be decomposed into two components simply as $\mathcal M \propto \mathcal L_+ - \mathcal L_-$. In the Krylov basis, the anti-Liouvillian is expressed as 
\begin{align}
    \mathcal{M} = \sum_{n=0}^{D_K-1} b_{n+1} \Big[ |\mathcal{O}_{n+1})(\mathcal{O}_{n}| - |\mathcal{O}_{n})(\mathcal{O}_{n+1}| \Big]\,,
\end{align}
The normalization \eqref{dsenorm} corresponds to the unit norm $(\varphi(t)| \varphi(t) ) = 1$. This allows one to write Eq. \eqref{discschr} in the form of a linear dynamical system of the form   
\begin{align}
\partial_t |\varphi(t) ) = \mathcal{M} |\varphi(t))\,, \label{dsevec}
\end{align}
with initial condition in the Krylov basis $|\varphi(0) ) = \big(1,  0, 0,  \dots,  0 \big)^\intercal$. This is simply the (imaginary-time) Schr\"odinger equation of $|{\varphi}(0))$ with an effective Hamiltonian $\mathcal{M}$.
% This comment is wrong: \footnote{Note that the Heisenberg evolution reads $\partial_t | \mathcal{O} (t) ) = i \mathcal{L} | \mathcal{O} (t) )$. Hence, Eq.  \eqref{dsevec} is equivalent to the Heisenberg evolution following the identification $\mathcal{M} \equiv i \mathcal{L}$ and $| \vec{\varphi} (t)\rangle \equiv | \mathcal{O} (t) ) $.}
In particular, the definition of the vector $|{K}(t) ) := \sqrt{\mathcal{K}} |\varphi(t)) = \big( \varphi_0 (t), \sqrt{1}\varphi_1 (t), \sqrt{2}\varphi_2 (t),  \dots , \sqrt{D_K-1}\varphi_{D_K-1} (t) \big)^\intercal$, allows to directly solve \eqref{dsevec} with the appropriate initial condition. The Krylov complexity is thus equivalent to the norm $ K(t) = ({K}(t)|{K}(t))$.
This form is particularly suitable for the numerical evaluation of Krylov complexity from the numerical form of the Lanczos coefficients. See \cite{Bhattacharya:2023zqt} for the detailed numerical implementation.

The Liouvillian, the anti-Liouvillian, and the Krylov complexity operator always obey the commutation relations \cite{caputa2021}
\begin{equation}
	[\mathcal{K}, \mathcal{M}] = \mathcal{L}\,, ~~~~~ [\mathcal{K}, \mathcal{L}] = \mathcal{M} . \label{comKLM}
\end{equation}
However, the commutator between the Liouvillian and the anti-Liouvillian 
$[\mathcal{L}, \mathcal{M}]$ is not universal. Yet, it is diagonal in the Krylov basis \cite{caputa2021}
\begin{align}
	[\mathcal{L}, \mathcal{M}] =  2 \sum_{n=0}^{D_K-1} \left( b_{n+1}^2 - b_n^2 \right) |\mathcal{O}_n)(\mathcal{O}_n| \,, \label{comLM}
\end{align}
with the diagonal coefficients given by the difference between the squared Lanczos coefficients. Since $\mathcal{K}$ is diagonal in this basis, the commutator $[\mathcal{K},[\mathcal{L}, \mathcal{M}]] = 0$ always holds. This also follows from the Jacobi identity
\begin{align}
    [\mathcal{K} , [ \mathcal{L} , \mathcal{M} ]] + [\mathcal{L},[ \mathcal{M}, \mathcal{K}]] + [ \mathcal{M},[ \mathcal{K}, \mathcal{L}]] = 0\,, \label{jacobiKLM}
\end{align}
using \eqref{comKLM}. In principle, any polynomial of $\mathcal{K}$ will satisfy $[\mathcal{K},[\mathcal{L}, \mathcal{M}]] = 0$. However, a particular choice \cite{caputa2021}
\begin{align}
    [\mathcal{L}, \mathcal{M}] = \upalpha \mathcal{K} + \upgamma \id \,, ~~~~~ \upalpha , \upgamma \in \mathbb{R}  \label{clm}
\end{align}
is often referred to as the \textit{simplicity hypothesis} \cite{caputa2021}. Here,  the identity matrix is denoted by $\mathbb{I}$. The commutator $[\mathcal{L}, \mathcal{M}]$ is directly proportional to the Krylov complexity operator $\mathcal K$, which is otherwise non-trivial. This amounts to the redefinition $\tilde{\mathcal{K}} = \upalpha \mathcal{K} + \upgamma \mathbb{I}$, such that the modified operator $\tilde{\mathcal{K}}$ closes the complexity algebra \cite{Hornedal2022}, i.e.,
\begin{equation}
	[\tilde{\mathcal{K}}, \mathcal{M}] = \upalpha \mathcal{L}, ~~~~~~ [\tilde{\mathcal{K}}, \mathcal{L}] = \upalpha \mathcal{M}\,, ~~~~~~ [\mathcal{L}, \mathcal{M}] = \tilde{\mathcal{K}} . \label{comKLMnew}
\end{equation}
It is straightforward to see that Eq. \eqref{comLM} satisfies \eqref{clm}, provided the Lanczos coefficients take the following form \cite{caputa2021, Hornedal2022}
\begin{align}
     b_n = \sqrt{\frac{1}{4} \mathfrak{\upalpha} n(n-1) + \frac{1}{2} \upgamma n}\,, \label{bkl}
\end{align}
with $n \geq 0$. In other words, the simplicity hypothesis restricts the growth of Lanczos coefficients, with linear growth being the maximal. This particular form completely specifies the behavior of the Krylov complexity for the associated algebra. To understand this, consider the first-time derivative of Krylov complexity
\begin{align} \label{Kderiv}
    \partial_t K(t) &= \partial_t ({\varphi} (t) |  \mathcal{K} | {\varphi} (t)) \rangle = ({\varphi} (t) |  [\mathcal{K}, \mathcal{M}] | {\varphi} (t) )\,,
\end{align}
where to deduce the second equality, we used \eqref{dsevec} with $\mathcal{M}^{\dagger} = \mathcal{M}^{\intercal} = -\mathcal{M}$. Each additional time derivative of the Krylov complexity brings a commutator with $\mathcal{M}$. The $\ell$-th time derivative gives
\begin{align}
 \partial_t^\ell K(t) &= ({\varphi} (t) | \big[\dots \big[[ \mathcal{K}, \mathcal{M} \underbrace{ ], \mathcal{M} \big],  \dots , \mathcal{M} \big]}_{\ell-\text{times}} | {\varphi} (t) )\,, \label{krycomjdot}
\end{align}
the nested commutator being evaluated in the state $|{\varphi}(t))$. The behavior of these nested commutators provides the differential equation for its time evolution and thus fully characterizes the dynamics of Krylov complexity. For example, the second derivative reads \cite{Hornedal2022, Erdmenger2023}
\begin{align}
    \partial_t^2 K(t) = ({\varphi} (t) |  [\mathcal{L}, \mathcal{M}] | {\varphi} (t) )\,,
\end{align}
where we have used $[\mathcal{K}, \mathcal{M}] = \mathcal{L}$. Provided that the simplicity algebra is fulfilled \cite{Hornedal2022}
\begin{align}
    \partial_t^2 K(t) = \upalpha K(t) + \upgamma \,. \label{keq}
\end{align}
This equation has also been termed as the \emph{Ehrenfest theorem} for the Krylov complexity \cite{Erdmenger2023}. Three special cases can be distinguished:

{\it Case 1: Linear growth \& $\mathrm{SL}(2,\mathbb{R})$ algebra.} This particular case corresponds to $\upalpha = 4 \alpha^2$ and $\upgamma = \upalpha/2 = 2 \alpha^2$. Then, $b_n$ grows linearly in the asymptotic limit of $n$, as $b_n = \alpha n$. 
%for $n \geq 0$.
The complexity algebra is $[\mathcal{L}, \mathcal{M}] = 2 \alpha^2 (2 \mathcal{K} + \mathbb{I})$. Since $\upalpha, \upgamma > 0$, the solution of the Eq. \eqref{keq}, with the initial condition $K(0) = 0$ and $K(-t) = K(t)$, is given by
\begin{align}
     K(t) = \frac{2\upgamma}{\upalpha} \sinh^2(\sqrt{\upalpha}\, t/2) = \sinh^2(\sqrt{\upgamma}\, t/\sqrt{2}) \sim e^{2 \alpha t}\,,
\end{align}
in the asymptotic limit. However, $\upgamma = \upalpha/2$ is not necessarily required - arbitrary independent $\upgamma$ and $\upalpha$ also give the linear growth in the asymptotic limit. Such growth of Lanczos coefficients gives rise to the exponential growth of the Krylov complexity. This growth corresponds to $\mathrm{SL}(2,\mathbb{R})$ algebra \cite{caputa2021, Hornedal2022}; see Sec.\,\ref{subsubsec:SL2R}.

%%%
\begin{figure}
\centering
\includegraphics[scale=0.65]{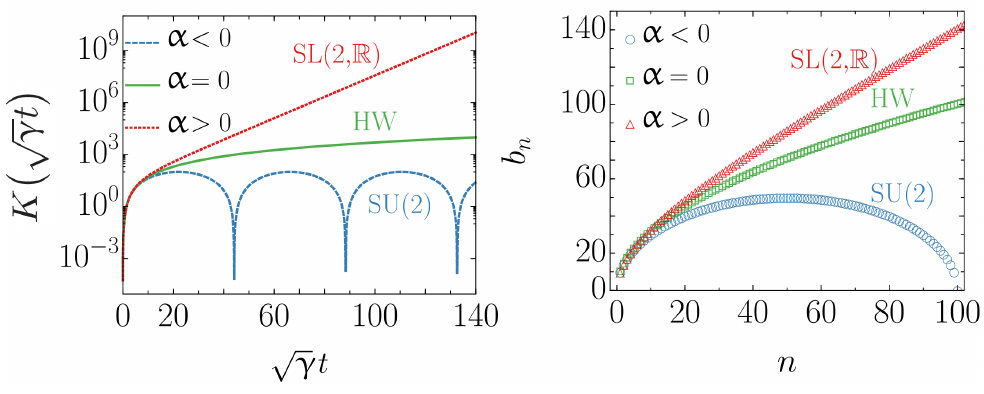}
\caption{Examples of the three different cases of the simplicity hypothesis. (Left) Time evolution of Krylov complexity $K\big(\sqrt{\upgamma}t\big)$, is plotted against the rescaled time $\sqrt{\upgamma} t$. For the $\mathrm{SL}(2,\mathbb{R})$ case we choose $\upalpha = 4$ and $\upgamma = 202$. (Right) The behavior of the Lanczos coefficients. The particular case of SU(2) is shown in detail in Fig.\,\ref{fig:wavesu2plot}.  Adapted from \cite{Hornedal2022}.}
\label{simplicityplot}
\end{figure} 
%%%

{\it Case 2: Sublinear growth \&  HW algebra.} This particular case corresponds to $\upalpha = 0$ and $\upgamma = 2 \alpha^2$, when the simplicity algebra closes with $[\mathcal{L}, \mathcal{M}] = 2 \alpha^2 \mathbb{I}$. The growth of the Lanczos coefficients $b_n$ is sublinear in the asymptotic limit of $n$, $b_n = \alpha \sqrt{n}$ for $n \geq 0$. The solution of \eqref{keq} is
\begin{align}
    K(t) = \frac{1}{2} \upgamma t^2 = \alpha^2 t^2\,,
\end{align}
i.e., the Krylov complexity grows quadratically in time. The growth corresponds to the Heisenberg-Weyl (HW) algebra; see Sec.\,\ref{subsubsec:HW} \cite{caputa2021, Hornedal2022}.

{\it Case 3: Finite dimensions \& $\mathrm{SU(2)}$ algebra.}  For finite dimensions, the Lanczos sequence must terminate. Hence, the Lanczos coefficient must vanish at the Krylov dimension $D_K > 1$, i.e., $b_{D_K} = 0$. This implies from \eqref{bkl} that
$\upalpha = -2 \upgamma/(D_K-1)$. The solution of Krylov complexity \eqref{keq} becomes \cite{Hornedal2022}
\begin{align}
    K(t)  = (D_K-1) \sin^2 \omega t\,,
\end{align}
where $\omega = \left(\frac{\upgamma }{2 (D_K-1)}\right)^{\frac{1}{2}}$, with the corresponding Lanczos coefficients   
$b_n = \omega \sqrt{n(D_K-n)}$. The Krylov complexity is thus periodic in time and associated with the SU(2) algebra, see Sec.\,\ref{subsubsec:SU2} \cite{caputa2021, Hornedal2022}.

Figure \ref{simplicityplot} presents a representative example of the Krylov complexity and the Lanczos coefficients for each of the three aforementioned cases. The classification of the complexity algebras is highly useful since, as shown in the previous section, it allows us to solve the time evolution of operators analytically in terms of coherent states. More general cases of complexity algebra have also been proposed recently \cite{Patramanis:2023cwz, Haque:2022ncl}.

%%%
\begin{figure}[t]
\centering
\includegraphics[scale=0.55]{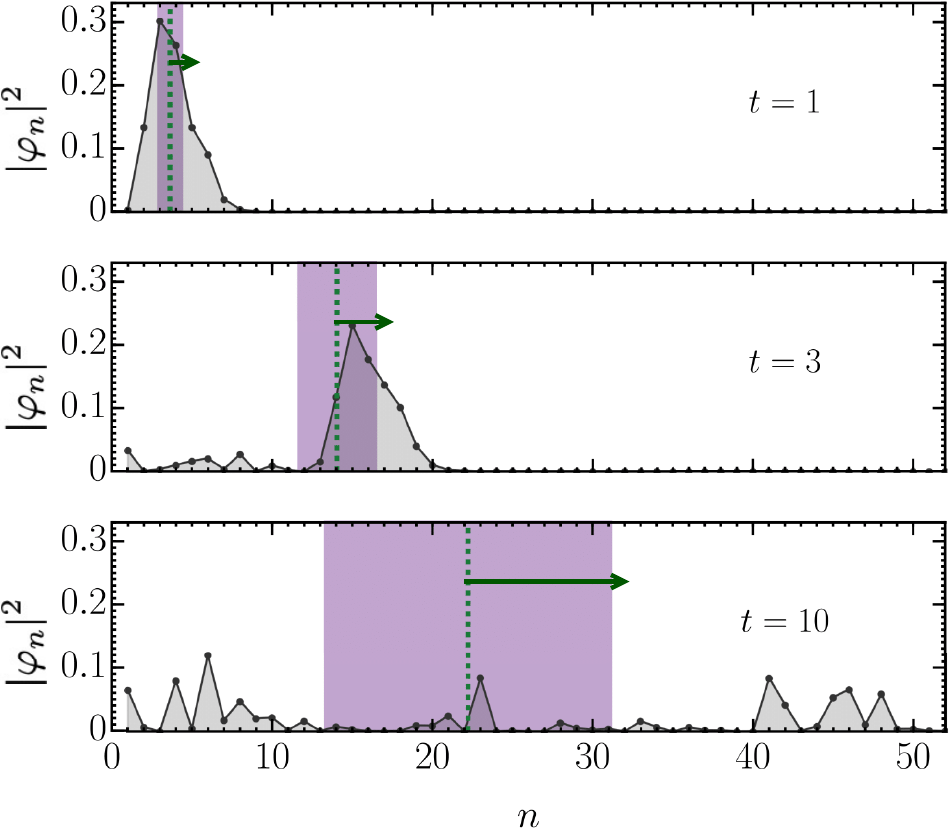}
\caption{Distribution of the operator amplitudes on the Krylov chain for a single random matrix Hamiltonian, sampled by the Gaussian orthogonal ensemble, and an initial operator equal to the constant matrix. The Krylov dimension is $D_K=91$. The green line shows the average position on the Lanczos chain, i.e., the Krylov complexity $K(t)$.  The width of the purple area corresponds to the dispersion of the complexity operator $\Delta K (t)$. According to the dispersion bound \eqref{dispbound}, the larger the width of the purple area, the faster the Krylov complexity can change.}
\label{lanczosfailplot}
\end{figure} 
%%%

\subsection{Dispersion bound and quantum speed limits to the complexity growth rate}

The authors of \cite{Hornedal2022} introduced a universal bound to the growth of Krylov complexity through a Robertson uncertainty relation involving the Krylov complexity operator and the Liouvillian as the generator of time evolution.
Specifically, within the Krylov space, which constitutes an inner product space, when $\mathcal{L}$ and $\mathcal{K}$ are self-adjoint superoperators, the following uncertainty relation holds: $\Delta K\Delta \mathcal{L} \geq \frac{1}{2}\left|\left\langle[\mathcal{K},\mathcal{L}]\right\rangle\right|$. Here, $\Delta K^2 = \left\langle\mathcal{K}^2\right\rangle-\left\langle\mathcal{K}\right\rangle^2$ denotes the Krylov variance \eqref{kvar0}, or equivalently the squared dispersion, relative to some operator $|\mathcal{O}(t))$. The dispersion of the Liouvillian reduces to $\Delta \mathcal{L} = b_1$. From Eq. \eqref{Kderiv}, one obtains the \textit{dispersion bound} \cite{Hornedal2022}
%%%
\begin{equation}
\label{dispbound}
    \abs{\partial_t K(t)} \leq 2b_1 \Delta K\,.
\end{equation}
%%%%
The form of this bound is reminiscent of the celebrated Mandelstam-Tamm time-energy uncertainty relation for the minimum time required for the mean value of an observable  %evolving in the Heisenberg picture, 
to vary by an amount comparable to its variance \cite{Mandelstam45,Busch2008,Hornedal2022}.  However, it is formulated in Krylov space and is derived for the Krylov complexity superoperator, with the first Lanczos coefficient providing an upper bound to the speed of evolution in Krylov space.

Figure \ref{lanczosfailplot} illustrates the dispersion bound for a simple numerical example of a random Liouvillian and random initial operator. The dispersion bound \eqref{dispbound} is saturated when the following three equivalent conditions are satisfied \cite{Hornedal2022}:
\textit{(i)} the superoperators $\tilde{\mathcal{K}}$, $\mathcal{L}$, and $\mathcal{M}$ close the algebra \eqref{comKLMnew}, 
\textit{(ii)} the Lanczos coefficients are given by \eqref{bkl}, 
\textit{(iii)} the Krylov complexity satisfies the differential equation \eqref{keq}.

We note that in isolated systems, every operator evolves under the Heisenberg equation, and the superoperators that act on them obey the generalized uncertainty principle. Therefore, the dispersion bound applies to every superoperator, not only the Krylov complexity operator. Recently, it has been used to bound the rate of change of the Krylov entropy \cite{Fan2022}.

As Eq.  \eqref{bkl} can be satisfied without the linear growth of Lanczos coefficients, the saturation of the dispersion bound is not tied with chaos in general. However, the linear growth of Lanczos coefficients is a \emph{sufficient} but \emph{not necessary} condition for the saturation. The saturation is intimately tied with the closure of the algebra, which provides \emph{necessary} and \emph{sufficient} conditions for the saturation of the dispersion bound \cite{Hornedal2022}.

Without assuming the complexity algebra,  a short-time asymptotic analysis of the Krylov complexity yields \cite{Carabba2024}
\begin{align*}
K(t) = b_1^2 t^2 + \frac{1}{6}b_1^2 (b_2^2-2b_1^2) \,t^4 + \frac{1}{180}b_1^2(8 b_1^4 - 7 b_2^4  + b_1^2 b_2^2 + 3 b_2^2 b_3^2) \,t^6 +\mathcal{O}(t^8) \,.
\end{align*}
From it, it can be shown that the generic system saturates the dispersion bound at short times of $\mathcal{O}(t^4)$ and generally deviates from it when contributions of $\mathcal{O}(t^6)$ become relevant. This occurs at the characteristic time \cite{Carabba2024}
\begin{eqnarray}
 \tau_d=\left|\frac{20(b_2^2-2b_1^2)}{8b_1^4-7b_2^4+b_1^2b_2^2+3b_2^2b_3^2}\right|^{\frac{1}{2}},
\end{eqnarray}
which corrects the estimate in \cite{Hornedal2022}.

An RMT Hamiltonian provides an interesting example of a typical system \cite{Hornedal2022, Tang:2023ocr, Loc:2024oen}. Eigenvalue repulsion is a key property of quantum chaos, and yet, the RMT Hamiltonians neither saturate the dispersion bound nor lead to an exponential growth of the Krylov complexity \cite{Hornedal2022}. 
%, with a non-necessarily local operator
This further suggests that the exponential growth of the Krylov complexity is a signature of scrambling rather than of quantum chaos. However, the choice of an initial local operator for testing the operator growth hypothesis is hard to realize within the standard ensembles in RMT. As an alternative, one may build in the required structure of many-body composite systems by considering Hamiltonians and operators with a tensor product structure \cite{CaoXu2022}, or described by random banded matrices \cite{Casati1990,Fyodorov1991,Fyodorov1993,Shepelyansky1994,Fyodorov1995,Borgonovi2016}.

Beyond the dispersion bound and its generalizations, the ultimate limits to the complexity growth rate can also be analyzed from a complementary point of view in the framework of quantum information geometry \cite{Amari2000,Bengtsson2006}. Bounds known as quantum speed limits identify the minimum time in which a process can unfold. They provide a refinement of the conventional time-energy uncertainty relations \cite{Busch2008} by introducing a distance in state space and an upper bound to the speed of evolution. Their use ranges from foundations of physics to quantum technologies, including quantum metrology and quantum computation.  As such, their study has motivated a large body of literature \cite{Deffner2017,Gong2022}. Recently, quantum speed limits have been generalized to characterize operator flows \cite{Carabba2022,Hornedal2023}, by introducing a distance in operator space, and identifying the corresponding maximum operator flow rate; see \cite{hamazaki2023limits,Hamazaki24} for other extensions. In particular, it has been shown that the saturation of the Krylov complexity growth rate is equivalent to the saturation of the geometric operator quantum speed limit \cite{Hornedal2023}.  
Further developments involve the generalization of the Krylov complexity to open quantum systems, discussed in Sec.\,\ref{secOpen}.

\section{Operator size concentration}\label{secOpSize}

So far, our primary focus has been on the Lanczos coefficients, a key output of the Lanczos algorithm. However, the Krylov basis elements $|\mathcal{O}_n)$ have not been as extensively examined. Remarkably, the linear growth of the Lanczos coefficients, a characteristic feature in all-to-all systems such as the SYK model, as illustrated in Eq. \eqref{bnleading}, is a broader attribute of the Krylov basis itself. This trait, referred to as ``operator size concentration'', was first identified in \cite{Bhattacharjee:2022lzy}. It has since proven to be of significant utility in both closed and open quantum systems, as detailed in \cite{Bhattacharya:2023zqt, Bhattacharjee:2023uwx}.  Notably, the growth of Lanczos coefficients in the SYK model is a consequence of the operator size concentration.

In this section, we provide a combinatorial derivation of the linear growth of the Lanczos coefficients in the large $q$ SYK model, Eq. \eqref{bnleading}. To this end, we make use of the diagrammatic approach of ``open'' melon diagrams \cite{Bhattacharjee:2022lzy}, which is a generalized version of the melon diagrams introduced in \cite{Roberts:2018mnp} and related the ``size'' of the operator. The essence is to describe the operators in a graph, where the Heisenberg time evolution is depicted quantum evolution of a particle moving on the graph. The approach is a consequence of a very special property of the Krylov basis, operator size concentration: the $n$-th Krylov basis element $\mathcal{O}_n$ is formed by a linear combination of the Majorana strings of the same size \cite{Bhattacharjee:2022lzy}. Mathematically, we write
\begin{align}
      \mathcal{O}_n = \sum_{i_1< \dots < i_s} c_{i_1, \dots, i_s} \psi_{i_1} \dots \psi_{i_{s}} + O(1/q) \,, \label{eq:OSC}
\end{align}
where $s = n(q-2) + 1$. Correspondingly, $n$ is the step size such that at the first step, $n = 1$, ensuing $s = q-1$. In other words, the $n$-th Krylov basis is concentrated on the same size of Majorana strings. The integer $n$ is the index of the Krylov basis dictating the number of nested commutators, often referred to as the generation index \cite{Roberts:2018mnp}. We outline the proof provided in \cite{Bhattacharjee:2022lzy} and consider the large $q$ SYK model, setting $\mathcal{J} = 1/\sqrt{2}$ for convenience. 

We initiate with a normalized operator $ \mathcal{O}_0 \equiv \mathcal{O} = \sqrt{2} \psi_1$ of size one, denoted by the following line \cite{Roberts:2018mnp}
\begin{align}
     \mathcal{O} = \sqrt{2}\psi_1 =\includegraphics[scale=.15,valign=c]{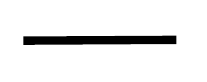}\,.
\end{align}
Next, we split the closed system Liouvillian into the following two parts \cite{caputa2021, Bhattacharjee:2022lzy}
\begin{align}
    \mathcal{L}_H = \mathcal{L}_+ + \mathcal{L}_-\,, \label{eq:Lpm}
\end{align}
where $\mathcal{L}_{+}$ is the increasing part of $\mathcal{L}_H$, and it increases the size of the operator. The decreasing operator $\mathcal{L}_{-}$ has the reverse effect. Given a size one operator, the operation of $\mathcal{L}_{+}$ can be diagrammatically written as
\begin{align} 
\mathcal{L}_+ \psi_1  \propto \includegraphics[scale=.5,valign=c]{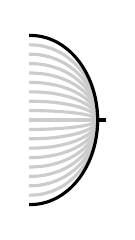} = \includegraphics[scale=.5,valign=c]{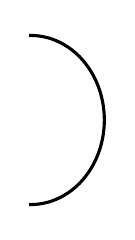} \,.
\end{align} 
This is the output at the first step. Here, the first diagram consists of $q$ lines and denotes a $(q-1)$ body operator formed by the action of $\mathcal{L}_{+}$. In the large $q$ limit, we neglect the intermediate gray lines and compactly denote it as a single arc (known as melon), represented by the second diagram on the right side. Further actions of  $\mathcal{L}_{+}$ leads to the following diagrams \cite{Bhattacharjee:2022lzy}
\begin{align*}
   &  \mathcal{L}_+^2 \psi_1 \propto \includegraphics[scale=.5,valign=c]{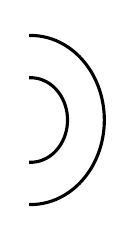}  \,, ~~~~
      \mathcal{L}_+^3 \psi_1 =  c_3\includegraphics[scale=.5,valign=c]{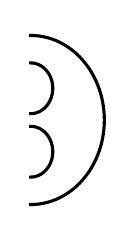} + c_4 \includegraphics[scale=.5,valign=c]{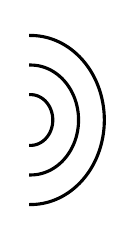}  \,, \\
    & \mathcal{L}_+^4 \psi_1 =  c_5\includegraphics[scale=.5,valign=c]{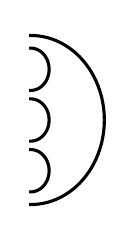} + c_6 \includegraphics[scale=.5,valign=c]{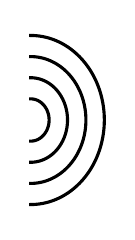} + c_7 
\includegraphics[scale=.5,valign=c]{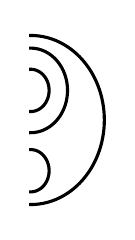} + c_8
\includegraphics[scale=.5,valign=c]{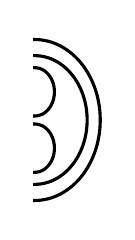} ,
\end{align*}
and the process goes continuously. Every action of  $\mathcal{L}_+$ creates a ``child'' arc of size $(q-1)$ within its ``parent'' arc. Hence, the full action $ \mathcal{L}_{+}^n \psi_1$ can be presented as unmarked $n$ arcs (vertices). These diagrams are not disorder-averaged and consist of the leading order diagrams. However, the observables can be constructed after closing the melon diagrams and taking the disorder-averaging. This, however, neglects any subleading contributions that have negligible effect in any disorder-averaged observables \cite{Bhattacharjee:2022lzy}. 

Next, we focus on the precise evaluation of the prefactors. As an example, we consider the number of ways the diagram with prefactor $c_7$ can be rearranged
\begin{align}
        c_7 
 \includegraphics[scale=.5,valign=c]{sykq-op7.pdf} =  \includegraphics[scale=.5,valign=c]{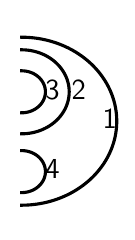} +
\includegraphics[scale=.5,valign=c]{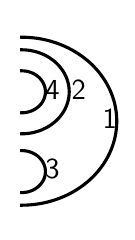} +
\includegraphics[scale=.5,valign=c]{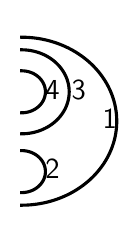} \,. \label{marked}
\end{align}
These are the only possible ways to construct these diagrams since a child diagram can only appear after its parent diagram. These diagrams in \eqref{marked} are known as \emph{unmarked ordered} diagrams. More precisely,  $\mathcal{L}_+^n \psi_1$ enumerates the number of possible ways a diagram with arc $n$ can be constructed with respective amplitude (multiplicity), i.e.,  
\begin{equation}
    \mathcal{L}_+^n \psi_1 = \sum \left[ \text{ordered diagrams of $n$ arcs} \right] \,.  \label{eq:Lnismarkedtreee}
\end{equation}
Similarly, the action of $\mathcal{L}_{-}$ removes an arc from the parent, known as a childless arc \cite{Bhattacharjee:2022lzy}. As an example, consider the following diagram:
\begin{align}
    \mathcal{L}_- \mathcal{L}_+^{3+1} \psi_1 = \,\mathcal{L}_-  \includegraphics[scale=.5,valign=c]{sykq-op7m2.pdf} + \dots  
    = \includegraphics[scale=.5,valign=c]{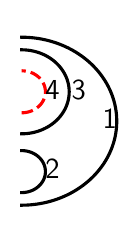} + \includegraphics[scale=.5,valign=c]{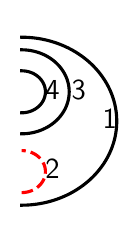} + \dots \,,
\end{align}
The diagram on the top is \emph{unmarked ordered} while the two diagrams on the bottom are called \emph{marked ordered} diagrams. Thus, the removal is \emph{marked} and is denoted by the dashed red line.  Specifically, they are \emph{ordered} diagrams with one \emph{marked} child. For example, consider the second diagram in the above example. The removal of the marked child produces the following \emph{unmarked ordered} diagram \begin{align}
\includegraphics[scale=.5,valign=c]{sykq-op7m2c1.pdf} ~~\mapsto~~ \includegraphics[scale=.5,valign=c]{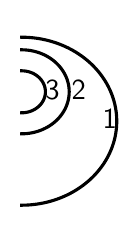}  \,. \label{marked2}
\end{align}
Since an \emph{unmarked ordered} diagram can be constructed in several ways from \emph{marked ordered} diagrams, this removal is not unique. In other words, the map is many-to-one: removal of $\sum n = n(n+1)/2$ \emph{marked ordered} diagrams with $(n+1)$ arcs gives rise to a single \emph{unmarked ordered} diagram with $n$ arcs. However, given the \emph{datum} of the parent and the childless arc $(p,c)$, the construction of the \emph{unmarked ordered} diagram is unique. For example, the left diagram of \eqref{marked2} has $(p,c) = (1,2)$. Another example is the following:
\begin{align}
    \left[ \includegraphics[scale=.5,valign=c]{sykq-op7m3.pdf}, p=2, c=3 \right]  \mapsto   \includegraphics[scale=.5,valign=c]{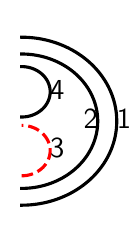} \,. 
\end{align}
Of course, it is easy to see that the left-hand side diagram is obtained after removing the red arc from the right-hand side diagram. Thus, we propose the following statement: For any $n \ge 1$, the action of the Liouvillian gives \cite{Bhattacharjee:2022lzy}
\begin{align}
        \mathcal{L}_- \mathcal{L}_+^{n+1} \psi_1 = \frac12 n (n + 1) \mathcal{L}_+^{n} \psi_1 \,. \label{eq:central}
\end{align}
While the left-hand side represents the sum of \emph{ordered} diagrams with one \emph{marked} child, the right-hand side represents $n(n+1)/2$ \emph{unmarked ordered} diagrams with $n$ arcs. Hence, the removal map is $n(n+1)/2$-to-one.

The identity \eqref{eq:central} is central to our discussion. Following the action of the Liouvillian $\mathcal{L}_H$, this directly leads to the following. For $n \geq 2$
\begin{align}\label{eq:LhLp}
    \mathcal{L}_H \mathcal{L}_+^n \psi_1 = 
   (\mathcal{L}_+ + \mathcal{L}_-) \mathcal{L}_+^n  \psi_1 =  \mathcal{L}_+^{n+1} \psi_1 + \frac12 n (n-1) \mathcal{L}_+^{n-1} \psi_1 \,,
\end{align}
where the second term in the second line uses \eqref{eq:central}. However, the specific terms for $n = 0, 1$ have to be evaluated independently, and are given by
\begin{align}
       \mathcal{L}_H \psi_1 = \mathcal{L}_+ \psi_1 \,,\,~~ \mathcal{L}_H \mathcal{L}_+ \psi_1  = \mathcal{L}_+^{2} \psi_1 + \frac1q  \psi_1\,.
\end{align}
See \cite{Bhattacharjee:2022lzy, Bhattacharjee:2023uwx} for the explicit evaluation in these cases. In other words, the consequence of \eqref{eq:central} provides the Krylov basis \cite{Bhattacharjee:2022lzy}
\begin{align}
         \mathcal O _n \propto \mathcal{L}_+^n \psi_1 \,, \label{eq:OnisLplus}
\end{align}
i.e., the Krylov basis formed by the Liouvillian $\mathcal{L}_H$ is effectively similar to the action of $\mathcal{L}_{+}$. The multiplicity factors are simply obtained by \eqref{eq:LhLp}, which are the Lanczos coefficients $b_ 1 = \sqrt{1/q}$ and $b_n = \sqrt{n(n-1)/2}$, and exactly match those in \eqref{bnleading} (we set $\mathcal{J} = 1/\sqrt{2}$) to the leading order in $q$. Equation ~\eqref{eq:OnisLplus} is equivalent to the statement of the operator size concentration \eqref{eq:OSC}. Since the Krylov basis \eqref{eq:OnisLplus} produces the operator size $s = n(q-2) + 1$, Eq. \eqref{eq:OSC} immediately follows. For further details, see \cite{Bhattacharjee:2022lzy, Bhattacharjee:2023uwx}.

As highlighted in Ref.\,\cite{Roberts:2018mnp}, the radial direction in the operator graph can be interpreted similarly to the radial direction in the bulk gravitational theory. This analogy not only provides an intuitive understanding of the index $n$, at least for all-to-all systems but also suggests a strong resemblance of a particle propagating deeper into the graph with a particle falling into the bulk (black hole). Although this has not been rigorously explored, it may offer an intuitive picture of the holographic nature of operator growth and Krylov complexity and its relation to black holes.

\section{Krylov complexity at finite temperature}\label{secKtemp}

In the preceding discussions, our attention was centered on the infinite temperature inner product. Nonetheless, many relevant studies of thermalization, quantum field theory, and black hole physics in the AdS/CFT correspondence incorporate finite temperatures. For example, a black hole is known to be a maximal scrambler, satisfying  Maldacena-Shenker-Stanford (MSS) bound on chaos \cite{Maldacena2016}. Hence, it is imperative to include finite temperatures in our analysis. We shall commence the discussion by delineating the inner product at finite temperatures, followed by the corresponding Lanczos algorithm. Our exploration will reveal the changes to the Krylov complexity bound induced by finite temperatures and ascertain whether these modifications can enhance the universal MSS bound in a stricter sense. As a concrete example, we consider the
%will particularly focus on the 
% Sachdev-Ye-Kitaev (SYK) 
SYK model through both numerical and analytical methods. %at finite temperatures. 

\subsection{Finite temperature inner product and Lanczos algorithm} \label{sec:finiteT}
Incorporating the thermal density matrix $\rho_{\beta} = e^{-\beta H}/Z_{\beta}$ at inverse temperature $\beta = 1/T$, we define the inner product as \cite{viswanath1994recursion, parker2019}
\begin{equation}
    (A|B)_\beta^g := \frac{1}{Z_\beta}\int_0^\beta d \lambda \,g(\lambda)\, \mathrm{Tr}( e^{- (\beta - \lambda) H}A^\dagger e^{-\lambda H}B)\, , \label{ipT}
\end{equation}
for two operators $A$ and $B$, where $g(\lambda)$ is an even function on the interval $[0, \beta]$, and $Z_\beta = \mathrm{Tr}(e^{-\beta H})$ is the thermal partition function corresponding to the Hamiltonian $H$. Note that the inner product is defined through an integral over a continuous parameter $\lambda \in [0, \beta]$. In particular, the function $g(\lambda)$ must satisfy the following conditions \cite{viswanath1994recursion}
\begin{align}
    g(\lambda)\geq 0\,,~~~~ g(\beta - \lambda) = g(\lambda)\,,~~~ \frac{1}{\beta}\int_0^\beta d \lambda \, g(\lambda) = 1\,. \label{cong}
\end{align}
%Operators are typically considered within the subspace where they have zero thermal average $\mathrm{Tr}(\rho_{\beta} A)=0$. Consequently, it is unnecessary to subtract the disconnected term \cite{parker2019}. \AD{[I don't think the condition $\mathrm{Tr}(\rho_{\beta} A)=0$ is important; everything works fine even if it is not satisfied.]}
The chosen inner product \eqref{ipT} dictates the autocorrelation function, which is expressed as:
\begin{equation}
    \mathcal{C}^g_\beta(t) = (\mathcal{O}|\mathcal{O}(t))^g_\beta = \int_0^\beta  d\lambda \, g(\lambda) \, \mathrm{Tr}(\rho_\beta\, \mathcal{O}^\dagger \mathcal{O}(t+i\lambda))\,.  \label{autoT}
\end{equation}
Given the inner product \eqref{ipT}, the Lanczos algorithm is applied to construct the Krylov basis. The process is outlined as follows \cite{viswanath1994recursion, parker2019}:
\begin{align}
    |\mathcal{O}_{-1})_{\beta}^g &:= 0\,,~~~ b^{(g)}_{0,T} :=0\,, ~~~|\mathcal{O}_{0})_{\beta}^g := |\mathcal{O})\,, \\
    |\mathcal{A}_n)_{\beta}^g &= \mathcal{L} |\mathcal{O}_{n-1})^g_\beta - b^{(g)}_{n-1, T} |\mathcal{O}_{n-2})^g_\beta\,, \\
    b_{n, T}^{(g)} &= \sqrt{(\mathcal{A}_n|\mathcal{A}_n)^g_\beta}\,, \\
    |\mathcal{O}_n)^g_\beta &= (b^{(g)}_{n,T})^{-1}|\mathcal{A}_n)_{\beta}^g\,.
\end{align}
It is important to note that the only modification to the Lanczos algorithm is the inner product; the definition of the Krylov subspace remains unchanged as the span of %$\{ |\mathcal{O})^g_\beta, \mathcal{L} |\mathcal{O})^g_\beta, \dots, \mathcal{L}^n|\mathcal{O})^g_\beta\}$
$\{ |\mathcal{O}), \mathcal{L} |\mathcal{O}), \dots, \mathcal{L}^n|\mathcal{O})\}$. The concept of orthogonality is simply redefined to align with the temperature of the system. Consequently, the Lanczos coefficients and the Krylov basis acquire a temperature dependence. 
%in their respective expressions.

The Krylov construction comes with freedom in the possible choices of $g(\lambda)$, constrained only by the conditions in Eq. (\ref{cong}). 
%Admittedly, the ambiguities of the definition \eqref{ipT} is not an escaped notice-apparently any function $g(\lambda)$ satisfying \eqref{cong} is a valid function. 
The Lanczos coefficients may differ depending on the function $g(\lambda)$. 
However, the two most common choices of $g(\lambda)$ are associated with the ``standard'' and ``Wightman'' inner products, denoted by the superscripts ``S'' and ``W''. For the standard inner product, $g(\lambda) = \frac{1}{2}(\delta(\lambda)+ \delta(\lambda-\beta))$ \cite{viswanath1994recursion, parker2019}
\begin{equation}
    (A|B)^{(S)}_\beta := \frac{1}{2 Z_{\beta}} \mathrm{Tr}(e^{-\beta H}A^\dagger B + A^\dagger e^{-\beta H}B)\,, \label{standard}
\end{equation}
which gives the standard thermal correlation function. In quantum field theory, the Wightman inner product is often preferred and corresponds to $g(\lambda) = \delta(\lambda - \beta/2)$ \cite{viswanath1994recursion, parker2019}
\begin{equation}
    (A|B)^{(\mathrm{W})}_\beta := \frac{1}{Z_{\beta}}\mathrm{Tr}(e^{-\beta H/2}A^\dagger e^{-\beta H/2} B)\,. \label{wight}
\end{equation}
The relation between Lanczos coefficients, defined with the help of \eqref{standard} and \eqref{wight}, is given by the Toda equations \eqref{f1} discussed in subsection \ref{todadiscussion} above. Note that another set of inner products can be defined by choosing $g(\lambda) = \delta(\lambda)$, in which case the diagonal coefficients $a_n$ of the Liouvillian may be non-vanishing for Hermitian initial operators; see \cite{Kundu:2023hbk} for such example in $2d$ CFTs.

 In the high-temperature limit, both the standard \eqref{standard} and the Wightman \eqref{wight} inner product reduces to the infinite-temperature inner product \eqref{IP}, which is uniquely defined. We focus on the Wightman inner product, which we denote by a superscript $(\mathrm{W})$. Note that the regularized finite temperature OTOC introduced in Sec.\,\ref{sec:ChaosOTOCs} can be naturally expressed in terms of the Wightman inner product as $\mr{OTOC}_\beta(t) = \left([W(t), V(0)]|[W(t), V(0)]\right)^\mr{(W)}_\beta$.

All definitions associated with the Lanczos algorithm and the autocorrelation function generalize naturally to finite temperatures. Consequently, the universal operator growth hypothesis at finite temperature suggests that for chaotic systems in $d>1$, the Lanczos coefficients asymptotically exhibit linear growth \cite{parker2019},
\begin{align}
    b^{(\mathrm{W})}_{n,T} = \alpha^{(\mathrm{W})}_T n + \gamma + o(1)\,, \label{uoghfiniteT}
\end{align}
where $\alpha^{(W)}_T$ is specific to the Wightman inner product, $\gamma$ is a $n$ independent constant. It is important to note that for inner products different from \eqref{wight}, the hypothesis has to be generalized to accommodate necessarily non-zero Lanczos coefficients $a_n$'s.
%this hypothesis \eqref{uoghfiniteT} may not hold under other inner product definitions, such as the standard inner product \eqref{standard} mentioned earlier.

When the temperature is small, such that $J/T\gg 1$, where $J$ is a characteristic local coupling of a lattice system, the thermal correlation length is much larger than the lattice spacing, and the system can be regarded as approximately continuous. In this limit, Lanczos coefficients will exhibit linear growth with the slope $\pi, T$, as in the continuous field theory,
\begin{align}
    b^{(\mathrm{W})}_{n,T} \approx  \pi\, T\, n + O(1)\,\quad n\gtrsim 1, \label{uoghsmallT}
\end{align}
which will persist until $n\lesssim J/T$ \cite{Avdoshkin:2022xuw}.

\begin{figure}[t]
\centering
%\subfigure[]
{\includegraphics[width=0.45\linewidth]{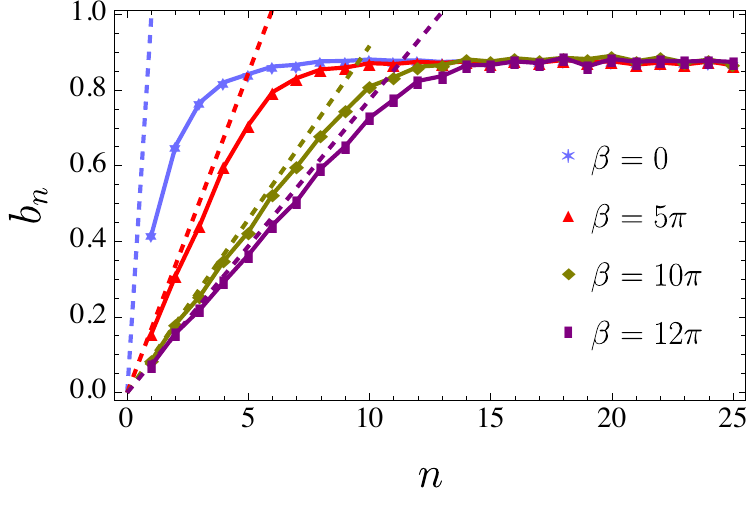}}
\hfil
%\subfigure[]
{\includegraphics[width=0.47\linewidth]{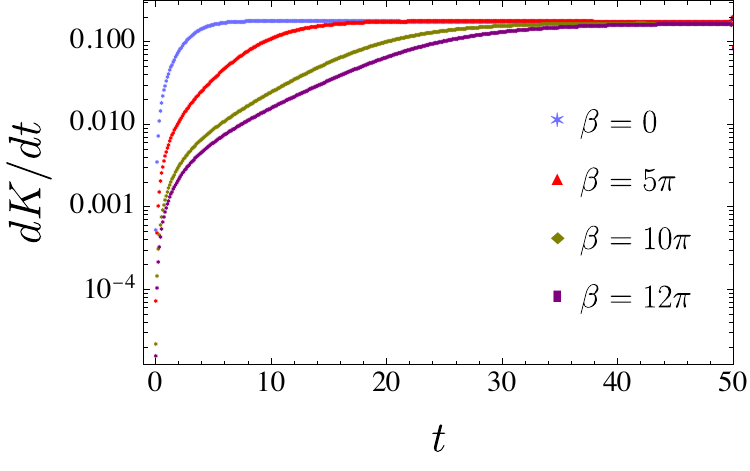}}
\caption{The growth of (left) the Lanczos coefficients (right) the rate of Krylov complexity in SYK$_4$ model at different inverse temperatures $(\beta = 1/T)$. The dashed lines in the left figure represent respective growth rates in Eq. \eqref{alphafiniteT}. The parameters are $N =18$ and $\mathcal{J} = 1/\sqrt{2}$ with $10$ Hamiltonian realizations. The figures are adapted from \cite{Jian:2020qpp} with different system parameters.} \label{fig:LanczosfiniteTSYK}
\end{figure}

\subsection{SYK at finite temperature} \label{sec:SYK_finiteT}

In Sec.\,\ref{sykexample}, we computed the Lanczos coefficients and associated Krylov cumulants at infinite temperature where the norm used was unique since it does not differentiate between the Wightman and the standard norms. Now, we turn our attention to the finite temperature scenario. For this purpose, we adopt the Wightman inner product to compute the Lanczos coefficients. The temperature is typically parameterized as follows \cite{Maldacena:2016hyu},
\begin{align}
    \frac{T}{\mathcal{J}} = \frac{\cos(\frac{\pi v}{2})}{\pi v}\,, \label{temppa}
\end{align}
 where $v \in (0,1)$ is parameter and $\mathcal{J}$ is the coupling constant defined in \eqref{sykparamters}. The high and low-temperature limits correspond to
\begin{align}
    T \rightarrow \infty ~ \Leftrightarrow ~ v \rightarrow 0\,,~~~~~~ T \rightarrow 0 ~ \Leftrightarrow ~ v \rightarrow 1\,,
\end{align}
respectively. The autocorrelation function under the Wightman inner product is expressed as \cite{parker2019, Jian:2020qpp}
\begin{align}
    \mathcal{C}^{(\mathrm{W})}(t) = 1 + \frac{2}{q} \log \big(\sech (\pi v T t)\big) + O(1/q^2)\,,
\end{align}
where the superscript $``\mathrm{W}$'' stands for ``Wightman''. As in the infinite temperature case,  the moment method can be used to determine the Lanczos coefficients, which are given by \cite{parker2019, Jian:2020qpp}
\begin{align}
b^{(\mathrm{W})}_{n,T} =
  \begin{cases}
    \pi v T \sqrt{2/q} + O(1/q) \,      & ~~n = 1\,,\\
    \pi v T  \sqrt{n(n-1)} + O(1/q) \,   & ~~n > 1\,.  \label{bnleadingfiniteT}
  \end{cases}
\end{align}
These coefficients acquire a temperature dependence, which we encapsulate by introducing the parameter \cite{parker2019}
\begin{align}
    \alpha^{(\mathrm{W})}_T = \pi v T\,, \label{finitealpha}
\end{align}
which dictates the growth rate of the Lanczos coefficients \eqref{bnleadingfiniteT}. Thus, one obtains the linear growth \eqref{uoghfiniteT} in the asymptotic limit of $n$. In terms of $\alpha^{(\mathrm{W})}_T$, the parametrization of temperature Eq. \eqref{temppa} translates to \cite{Jian:2020qpp}
\begin{align}
    \alpha^{(\mathrm{W})}_T = \mathcal{J} \cos\bigg(\frac{\alpha^{(\mathrm{W})}_T \beta}{2}\bigg) \rightarrow \begin{cases}
    \mathcal{J} \,      & ~~ \beta \mathcal{J} \ll 1\,,\\
    \pi/\beta   & ~~\beta \mathcal{J} \gg 1\,,  \label{alphafiniteT}
  \end{cases}
\end{align}
where $\beta = 1/T$ is the inverse temperature. In the standard inner product, however, the temperature dependence on 
$\alpha^{(\mathrm{W})}_T$ is different and shows the opposite behavior of \eqref{alphafiniteT} \cite{parker2019}. The wavefunctions acquire a similar form of \eqref{diss}, also with a temperature dependence. The growth of Krylov complexity is given by \cite{parker2019, Jian:2020qpp}
\begin{align}
    K(t) = \frac{2}{q} \sinh^2 (\alpha^{(\mathrm{W})}_T t) + O(1/q^2)\,.
\end{align}
Higher moments of the Krylov operator can also be straightforwardly computed \cite{Jian:2020qpp, Bhattacharjee:2022ave}. The comparison with the finite-temperature Lyapunov exponent reveals that
\begin{align}
    \lambda_{\mathrm{OTOC}} (T) = 2 \pi  v T  = 2 \alpha^{(\mathrm{W})}_T \rightarrow \begin{cases}
    2\mathcal{J} \,      & ~~ \beta \mathcal{J} \ll 1\,,\\
    2\pi/\beta   & ~~\beta \mathcal{J} \gg 1\,.
      \end{cases} \label{otocfin}
\end{align}
This reduces to the growth \eqref{cc} at the infinite-temperature limit and satisfies the Maldacena-Shenker-Stanford bound \cite{Maldacena2016} at low temperature. Moreover, \eqref{otocfin} results in the saturation of the Krylov bound at all temperatures, including the infinite-temperature bound \eqref{kbound1}.

\begin{figure}[t]
\centering
%\subfigure[]
{\includegraphics[width=0.45\textwidth]{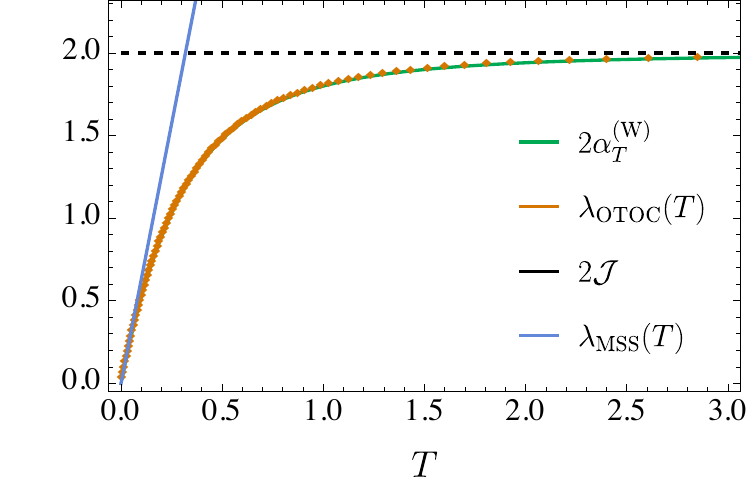}}\label{fig:mss}
\hfil
%\subfigure[]
{\includegraphics[width=0.44\textwidth]{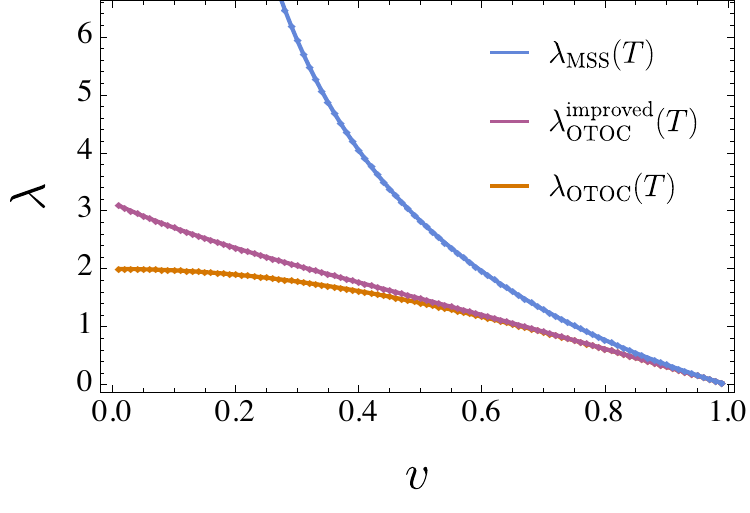}}
\caption{(Left) Illustration of the bound \eqref{kbound1}, and its finite temperature version \eqref{boundT} in the large $q$ limit. In both cases, the bound saturates, while the MSS bound is saturated at a low-temperature limit. Adapted from \cite{parker2019}. (Right) Illustration of the improved OTOC bound $\lambda_{\mathrm{OTOC}}^{\mathrm{improved}} = \frac{2\pi T}{1+4\beta^{*} T}$. We take $2\beta^{*} = 1$. The figure is adapted from \cite{Avdoshkin:2019trj}.} \label{fig:mssbound}
\end{figure}

%\begin{figure}[t]
%   \centering
%\includegraphics[width=0.385\textwidth]{mss.pdf}
%\caption{Illustration of the bound \eqref{kbound1}, and its finite temperature version \eqref{boundT} in the large $q$ limit. In both cases, the bound saturates, while the MSS bound is saturated at a low-temperature limit. Adapted from \cite{parker2019}.} \label{fig:mss}
%\end{figure}

In addition to the above analytic results in the large $q$ and large $N$ limit, Fig.\,\ref{fig:LanczosfiniteTSYK} illustrates the numerical computation of Lanczos coefficients and the growth rate of Krylov complexity at varying temperatures within the SYK$_4$ model. %for a system size of $N=18$ and $10$ Hamiltonian realizations. 
The Lanczos coefficients exhibit linear growth before reaching a plateau at a value determined by the system size $N$, which remains constant across different temperatures. In contrast, the growth rate, as defined in Eq. \eqref{alphafiniteT}, is temperature-dependent, decreasing as the temperature lowers. The dashed lines in Fig.\,\ref{fig:LanczosfiniteTSYK} (left) represent the growth rate, i.e., the solution of Eq. \eqref{alphafiniteT}, constrained by $\pi/\beta \leq \alpha \leq \mathcal{J}$. This temperature-dependent behavior also influences the early-time growth rate of Krylov complexity, causing it to decelerate. Nonetheless, its exponential growth to the linear growth regime is still prominent. The eventual saturation of the Lanczos coefficients is mirrored in the late-time behavior of the Krylov complexity growth rate, $dK(t)/dt$, which converges to a temperature-independent constant, as depicted in Fig.\,\ref{fig:LanczosfiniteTSYK}.  Similar results for $T \Bar{T}$-deformed SYK model have also been reported \cite{HeSYK2022}. The appropriate timescale of this growth rate \cite{Jian:2020qpp} shares similar timescales in holographic complexity \cite{Stanford2014}.

The numerical computation of Lanczos coefficients, as depicted in Fig.\,\ref{fig:LanczosfiniteTSYK}, does not reveal the saturation point of Krylov complexity due to the finite set of coefficients evaluated. A comprehensive exploration of the entire Krylov space is necessary to observe such saturation. Reference \cite{Rabinovici:2020ryf}  provides an illustrative example of a full Krylov space analysis.

\subsection{Krylov exponent at finite temperature} \label{secmssbound}

In Ref.~\cite{parker2019}, the finite temperature Krylov exponent $\lambda_K(T) = 2\alpha^{(\mathrm{W})}_T$ was conjectured to be the upper bound of the Lyapunov exponent, capturing the growth of the out-of-time-ordered correlator (OTOC):
\begin{align}   \lambda_{\mathrm{OTOC}}(T) \leq 2\alpha^{(\mathrm{W})}_T \leq 2\pi T\,. \label{boundT}
\end{align}
This bound on $\lambda_{\mathrm{OTOC}}$ can be tighter than the universal Maldacena-Shenker-Stanford (MSS) bound  \eqref{mssbound} as we will see below.

The large $q$ SYK model saturates the left bound in \eqref{boundT} at all temperatures, we discuss explicit calculation in the later section. 
The right bound is tight only at low temperatures; see Fig.\,\ref{fig:mss}. 
At high temperatures, the bound simply reduces to \eqref{kbound1}.

In fact, the right inequality in \eqref{boundT} can be improved. Ref.~\cite{Avdoshkin:2019trj} put forward a tighter inequality,
\begin{align}
\lambda_{\mathrm{OTOC}}(T) \leq \lambda_K (T)=2\alpha^{(\mathrm{W})}_T \leq \frac{2\pi T}{1+4\beta^{*} T}\,, \label{boundT1}
\end{align}
where $\beta^{*}$ bounds the location of the finite temperature autocorrelation function \eqref{autoT} in an infinite strip, see \cite{Avdoshkin:2019trj, GuKitaevZhang2022} for a proof.  In the continuum field theory limit  $(\beta^{*})^{-1} \sim O(\Lambda)$, where $\Lambda \rightarrow \infty$ is the UV-cutoff. Hence, Eq.~\eqref{boundT1} reduces to \eqref{boundT}. However, $\beta^{*}$ retains a significant, non-trivial value for discrete lattice non-integrable systems. The refined bound \eqref{boundT1} reverts to the MSS bound \eqref{boundT} at low temperatures and remains applicable across the entire temperature spectrum, including at infinite temperatures; see Fig.\,\ref{fig:mssbound}. Consequently, this improved bound offers a more comprehensive constraint than its predecessor, applicable under a broader range of conditions.

It is also important to mention that $\lambda_K(T)$ is not always equal to $2\alpha^{(\mathrm{W})}_T$. This leads to a generalization of \eqref{boundT}, proposed in \cite{Avdoshkin:2022xuw},
\begin{align}
    \lambda_{\mathrm{OTOC}}(T) \leq \lambda_K(T) \leq 2\pi T\,. \label{boundTextended}
\end{align}
There are several non-trivial cases exemplifying \eqref{boundTextended}, including the large $q$ SYK model discussed above, as well as different models of quantum field theory \cite{Avdoshkin:2022xuw,Camargo:2022rnt}. Free massive field theory in 4D exhibits no exponential growth of OTOC, $\lambda_{\mathrm{OTOC}}=0$, and less than maximal growth of Krylov complexity $0<\lambda_K < 2\pi T$. A free massless field theory placed on a sphere exhibits no exponential growth of Krylov, rendering $\lambda_{\mathrm{OTOC}}=\lambda_K=0$.
There are also holographic examples \cite{Dymarsky:2021bjq,Avdoshkin:2022xuw} when $\lambda_{\mathrm{OTOC}}=\lambda_K$  either both vanish or are both equal to $2\pi T$, depending on whether $T$ is above the point of Hawking-Page transition. These examples provide arguments to support  \eqref{boundTextended} and suggest that both inequalities are non-trivial.

%\begin{figure}[t]
%   \centering
%\includegraphics[width=0.385\textwidth]{MSSbound.pdf}
%\caption{Illustration of the improved OTOC bound $\lambda_{\mathrm{OTOC}}^{\mathrm{improved}} = \frac{2\pi T}{1+4\beta^{*} T}$. We take $2\beta^{*} = 1$. The figure is adapted from \cite{Avdoshkin:2019trj}.} \label{fig:mssbound}
%\end{figure}

%%%%%%%%%%%%%%%%   KRYLOV SPACE OF PURE STATES  %%%%%%%%%%%%%%%%%%%%
\section{Krylov Space of Pure States} \label{secStates}

\subsection{Krylov space and spread complexity}

Let us consider a Hermitian Hamiltonian $ H$  and the corresponding Schr\"odinger equation, $i  \partial_t \ket{\Psi(t)} =  H \ket{\Psi(t)}$ (setting $\hbar=1$), governing the evolution of a pure initial quantum state $\ket{\Psi_0} \equiv \ket{\Psi(0)}$ in a $d-$dimensional Hilbert space $\mathscr{H}$. The time evolution admits the expansion
%%%
\begin{equation}
    \ket{\Psi(t)} = e^{-it  H} \ket{\Psi_0} = \sum_{n=0}^\infty \frac{(-i t)^n}{ n!} H^n \ket{\Psi_0}\,,
\end{equation}
%%%
and is thus contained in the Krylov space spanned by the powers of the Hamiltonian acting on the initial state $\mathrm{span}\{\ket{\Psi_0},  H \ket{\Psi_0},  H^2 \ket{\Psi_0}, \dots \}$. One can construct an orthonormal basis for a Krylov space of pure states in the same way as for operators, using the Hamiltonian as the generator of time evolution instead of the Liouvillian (an alternative basis construction uses the time-evolved states \cite{Cindrak:2024exj}). The Gram–Schmidt procedure \cite{gram1883,schmidt1908} applied to the set of vectors $\{H^n \ket{\Psi_0}\}_{n=0}^{\infty}$ yields an orthonormal basis set $\{|K_n\rangle\}_{n=0}^{D_K-1}$. Here, $D_K$ is the corresponding dimension of the Krylov space, whose maximum value is set by the dimension of the Hilbert space itself
\begin{align}
    D_K \leq d\,.
\end{align}
The basis elements $|K_n\rangle$ are known as the Krylov basis for the corresponding Hamiltonian $H$ with the initial state $\ket{\Psi(0)}$. They  are orthonormal
%\begin{align}
$\langle K_m |K_n\rangle = \delta_{mn},$
%\end{align}
and the first element is the initial state $|K_0\rangle = | \Psi(0)\rangle$.
One can find the Krylov basis in this space $\{\ket{K_n}\}$ using the Lanczos algorithm \cite{liesenbook,viswanath1994recursion}. Setting $\ket{K_{-1}} = \mathsf{a}_{-1} = \mathsf{b}_0 = 0$, one performs the following steps for $n \geq 0$:
\begin{enumerate}
    \item  Compute the diagonal coefficient $\mathsf{a}_n = \bra{K_n} H \ket{K_n}$, and
    $\ket{A_{n+1}} = H \ket{K_n} -\mathsf{a}_n \ket{K_n} - \mathsf{b}_{n} \ket{K_{n-1}}$.
    \item Compute $\mathsf{b}_{n+1} =\sqrt{\braket{A_{n+1}}}$.
    If $\mathsf{b}_{n+1}=0$, stop the algorithm. Otherwise define $\ket{K_{n+1}} = \mathsf{b}_{n+1}^{-1} \ket{A_{n+1}}$, and repeat the procedure 1.
\end{enumerate}
Along with the Krylov basis set, this generates two sets of Lanczos coefficients $\{\mathsf{a}_n, \mathsf{b}_n\}$, to be distinguished from $\{a_n, b_n\}$ in the operator picture.
%\footnote{The Lanczos coefficients in the state complexity picture are denoted by $\{\mathsf{a}_n, \mathsf{b}_n\}$ to distinguish from $\{a_n, b_n\}$ in the operator picture.} 
%We distinguish them in Greek letters instead of Roman letters used for the operator basis. 
In particular, the algorithm yields the following recurrence relation
\begin{equation}
\label{lanczosH}
     H \ket{K_n} = \mathsf{b}_{n} \ket{K_{n -1}} + \mathsf{a}_n \ket{K_n} + \mathsf{b}_{n + 1}\ket{K_{n +1}} .
\end{equation}

Thus, the Hamiltonian in the Krylov basis takes the following tridiagonal form
\begin{align}
H = 
	\begin{pmatrix}
\mathsf{a}_0 & \mathsf{b}_1 &  &  &  \\
\mathsf{b}_1 & \mathsf{a}_1 & \mathsf{b}_2 &   &  \\
 & \mathsf{b}_2 & \mathsf{a}_2 & & \\
 &  & &  \ddots  \\
 &  &  & &  &  \mathsf{b}_{D_K-1}   \\
 &  &  &  &  \mathsf{b}_{D_K-1}  & \mathsf{a}_{D_K-1}  
	\end{pmatrix}\,. \label{triHamiltonian}
	\end{align}
Unlike the operator case, where the Liouvillian is tridiagonal with zero diagonal elements (for a Hermitian Hamiltonian) in the Krylov basis, the Hamiltonian's diagonal elements $\mathsf{a}_n$ are in general finite.
The use of Householder reflections \cite{householder1958} to tridiagonalize the Hamiltonian and obtain Lanczos coefficients is a technique often employed in numerical linear algebra, making it more amenable to analysis and numerical methods. 
Nevertheless, to employ Householder reflections one has to start from a basis in which the Hamiltonian is not already tridiagonal. This method is often pursued under the name \emph{Hessenberg decomposition} \cite{Hessenberg}. The procedure casts a matrix in an upper-triangular form (known as upper-Hessenberg form), and for the special case of Hermitian matrix, it reduces to complete tridiagonal form. The method is faster than the Lanczos algorithm and can be implemented using the command $\mathtt{HessenbergDecomposition}[m]$ in $\emph{Mathematica}$ where $m$ is the matrix under consideration \cite{MathematicaHessenberg}. While in the Lanczos algorithm, we choose the initial state at our disposal, the Hessenberg decomposition picks up a special initial state $(1, 0, 0, \cdots )^{\intercal}$. Thus, the Lanczos coefficients computed by these two methods are different. Hence, the Lanczos algorithm is preferred for any physical Hamiltonian since one chooses the initial state according to the given problem (e.g., as in a quantum quench protocol). On the other hand, in random matrix theory (RMT), choosing any initial state is sufficient to capture the properties in RMT. Hence, to study the statistics of the RMT \cite{balasubramanian2022}, Hessenberg decomposition is usually employed.

%%%

The Krylov basis provides a framework within the Hilbert space for the evolution of the wavefunction, which can be expressed as
\begin{align}
\label{krylovstateH}
    \ket{\Psi(t)}= \sum_{n = 0}^{D_K-1} \uppsi_n(t) \ket{K_n},
\end{align}
where $\uppsi_n (t)$ represents the wavefunction amplitudes within the Krylov chain, and $D_K$ denotes the Krylov dimension. It is important to note that $D_K$ may be finite, even in an infinite-dimensional Hilbert space. 

The Schr\"odinger equation, in conjunction with the recursion relation \eqref{lanczosH}, dictates that these amplitudes follow the equation
\begin{align}
\label{amplitudesH}
    i  \partial_t \uppsi_n (t) = \mathsf{b}_{n} \uppsi_{n-1}(t) + \mathsf{a}_{n}\uppsi_n(t) + \mathsf{b}_{n +1} \uppsi_{n +1}(t) \,.
\end{align}
Here the imaginary unit $i$ is an integral part of the equation, indicating that the amplitudes $\uppsi(t)$ belong to the complex plane $\mathbb C$. Furthermore, we define the Krylov operator as
\begin{align}
\label{krycompH}
    \mathcal{K}_S = \sum_{n=0}^{D_K-1} n \ketbra{K_n}\,.
\end{align}
The expectation value of this operator, with respect to the state \eqref{krycompH}, yields
\begin{align}
    K_S(t) &= \bra{\Psi(t)} \mathcal{K}_S \ket{\Psi(t)}  = \sum_{n=0}^{D_K-1} n \abs{\uppsi_n (t)}^2\,,
\end{align}
reflecting the mean position on the Krylov chain. This definition is often referred to as the \emph{Krylov state complexity} or \emph{spread complexity} 
%(or spread complexity for short) 
and has found a variety of applications such as probing quantum scars in PXP model \cite{Bhattacharjee:2022qjw, Nandy:2023brt}, topological states in quantum matter \cite{CaputaLiu22, Caputa:2022yju} in the Su-Schrieffer-Heeger (SSH) model \cite{SSH79}, quench protocols \cite{Pal:2023yik, Gautam:2023bcm, Gautam:2023pny}, random matrix theory \cite{Bhattacharyya:2023grv, Balasubramanian:2022dnj, Balasubramanian:2023kwd}, PT-symmetric quantum mechanics \cite{Beetar:2023mfn}, 
localization and thermalization phenomena \cite{Alaoui:2023cdo, Alishahiha:2024rwm, Menzler:2024atb, Cohen:2024ngg}, evolution of modular Hamiltonian and modular chaos \cite{CaputamodularH24},
LMG model \cite{Afrasiar:2022efk, bento2023krylov, huh2023, Zhou:2024rtg}, quantum measurements \cite{Gill:2023umm, Bhattacharya:2023yec},  open quantum systems \cite{Bhattacharya:2023zqt, Carolan:2024wov}, high-energy quantum chromodynamics \cite{Caputa:2024xkp}, and the characterization of networks for quantum walks \cite{jeevanesan2023krylov}.

The spread complexity offers a measure that captures the dynamical spreading of states through the Hilbert space. The significance of the Krylov basis in this context arises from its ability to capture the spread of this state effectively. While it is true that any basis could theoretically be employed for this purpose, the Krylov basis is special. Consider a basis $\mathcal{B} := \{\ket{B_n}: n= 0,1, \cdots\}$, alongside a cost functional defined as \cite{balasubramanian2022}
\begin{align}
    C_\mathcal{B}(t) = \sum_n \mathsf{c}_n \abs{\braket{\Psi(t)}{B_n}}^2 \,,
\end{align}
where the coefficients $\mathsf c_n$ are both positive and monotonically increasing. Given the completeness of the basis $\mathcal{B}$ and the unitarity constraint, it follows that $\sum _n |\mathsf{c}_n|^2 = 1$. By minimizing this cost functional across all possible bases $\mathcal{B}$, and specifically choosing $\mathsf{c}_n = n$, we arrive at the minimum value representing the spread complexity \cite{balasubramanian2022}
\begin{align}
    K_S(t) := \min_\mathcal{B} C_\mathcal{B}(t)\,.
\end{align}
This minimization process is a \emph{functional minimization}, which identifies the Krylov basis as the optimal basis for a finite duration of time. In scenarios of discrete-time evolution, commonly analyzed in unitary circuits \cite{Nahum:2016muy, PhysRevX.8.041019, PhysRevX.9.031009}, the Krylov basis consistently minimizes the cost functional at all times. In conclusion, the Krylov basis provides a natural and computationally efficient basis where the spreading of the initial wave function is minimal. Ref.~\cite{balasubramanian2022} provides a detailed proof of the above statement.

\subsection{Survival amplitudes and thermofield double state}
The spread of the wavefunction, akin to operator complexity, is encapsulated by the function\footnote{Our notation is aligned with the definition of the autocorrelation function  \eqref{defauto}, and differs from \cite{balasubramanian2022} with a complex conjugate. Hence, according to \cite{balasubramanian2022}, the survival amplitude is $S(t)^{*}$.}
\begin{align}
    S(t) := \langle \Psi(0)|\Psi(t) \rangle = \uppsi_0 (t)\,,
\end{align}
which is the overlap between the initial state and its temporal evolution. This quantity is known as the \emph{survival amplitude} of the initial state  $\ket{\Psi(0)}$, and plays a crucial role in quantum dynamics, e.g., in the context of quantum speed limits \cite{Deffner2017}, Loschmidt echoes \cite{Gorin06}, and quantum decay \cite{Fonda1978}. Given the amplitude $S(t)$, the moments are computed as 
\begin{align}
    \mu_k := \lim_{t \rightarrow 0} \,\frac{d^n S(t)}{d t^n} \,. \label{statemom}
\end{align}
Note the absence of factor $i^n$ compared to \eqref{mom4}, a fact which can be traced to the hopping equation \eqref{amplitudesH} for the state. The corresponding Lanczos coefficients can be computed either through the recursive algorithm \eqref{mombn} or using the pictorial diagram in Fig.\,\ref{fig:diagmoment} with an additional factor of $i$ in both $\mathsf{a}_n$ and $\mathsf{b}_n$) \cite{balasubramanian2022}.

For the specific case we are considering, we choose our initial state as the thermofield double (TFD) state \cite{Maldacena:2001kr, Nishiokarmp}
\begin{align}
    \ket{\Psi(0)} := \ket{{\rm TFD}(\beta)} = \sum_{n} \frac{ e^{- \beta E_n/2}}{\sqrt{Z(\beta)}} \ket{n}_1 \otimes\ket{n}_2\,, \label{tfd}
\end{align}
where $\beta$ is the inverse temperature $\beta = 1/T$, where $Z(\beta) = \mathrm{Tr}(e^{-\beta H}) = \sum_n e^{-\beta E_n}$ is the partition function. Here, ``1'' and ``2'' indicate the first and the second copies of the system, which is formed by doubling the Hilbert space. $E_n$ and $\ket{n}_{1,2}$ are the eigenvalues and eigenstate of the Hamiltonian $H_{1,2}$ under consideration, obeying $H_{1,2} \ket{n}_{1,2} = E_n \ket{n}_{1,2}$. Tracing out one system produces a thermal mixed state, which is indistinguishable from the pure TFD state in the double Hilbert space. The TFD state can be understood as the purification of the thermal Gibbs state \cite{Nishiokarmp}. Note that the state \eqref{tfd} is written in a doubled Hilbert space with two identical copies of the same state. Such state in conformal field theory (CFT) is thus considered as holographically dual to the two-sided eternal black holes in anti-de-Sitter (AdS) space, with two boundaries of the asymptotic spacetime dual to the two copies of CFT, denoted as ``1'' and ``2'' respectively \cite{Maldacena:2001kr}. In this context, the preparation of TFD state \cite{Cottrell2019}, its entanglement and complexity structure \cite{chapmantfd, YangTFD}, and applications to wormhole geometry \cite{Maldacenawormhole, Gao2017} and quantum teleportation protocols in quantum circuits \cite{QGlab1, QGlab2, Schusterteleportation} constitute the foundation of the slogan ``entanglement builds spacetime'' \cite{VanRaamsdonk2010, Swingle2012}.

\subsection{Spread complexity in RMT}
Initializing with an arbitrary state, random matrices can be tridiagonalized using the Lanczos algorithm. For an ensemble of random matrices, there will be an ensemble of Lanczos coefficients. If the ensemble is Gaussian, then the tridiagonal representation is known analytically. Even beyond Gaussianity, the statistics of the Lanczos coefficients can be found numerically.

The density of states (DOS) is a critical aspect of RMT, describing the distribution of eigenvalues. For an $N \times N$ random matrix, an approximate relation between the density of states $\rho (E)$ and the \emph{mean} or \emph{average} Lanczos coefficients $\mathsf{a}(x) \equiv \mathsf{a}_{xN}$ and $\mathsf{b}(x) \equiv \mathsf{b}_{xN}$, with $x = n/N$ is in the large $N$ limit can be found. The relation reads
\begin{align}
    \rho (E) \approx \int_{0}^1 dx\, \frac{\Theta(4 \mathsf{b}(x)^2 - (E - \mathsf{a}(x))^2)}{\pi \sqrt{4 \mathsf{b}(x)^2 - (E - \mathsf{a}(x))^2}}\,, \label{den0}
\end{align}
where $\Theta(z)$ is the Heaviside theta function taking values $\Theta(z) = 1$ for $z \geq 0$ and $\Theta(z) = 0$ for $z < 0$. This is an integral equation involving the energy on either side. The density of states $\rho(E)$ has compact support over an interval $[-E_{\mathrm{min}}, E_{\mathrm{max}}]$ in the large $N$ limit. For non-compact cases, an explicit cut-off can be put at the tail of the DOS. Thus, the above integral equation can be solved using the bi-section method \cite{Balasubramanian:2022dnj}. Eq.\,\eqref{den0} has deep historical roots, originating from the enumeration of graphs on Riemann surfaces \cite{BESSIS1980109} and extending to the recursion relations in classical orthogonal polynomials \cite{Marino:2004eq}. Recently, it has gained renewed interest through the study of quantum chaos in Krylov space \cite{Balasubramanian:2022dnj}.

\begin{figure}[t]
\centering
\includegraphics[width=0.48\textwidth]{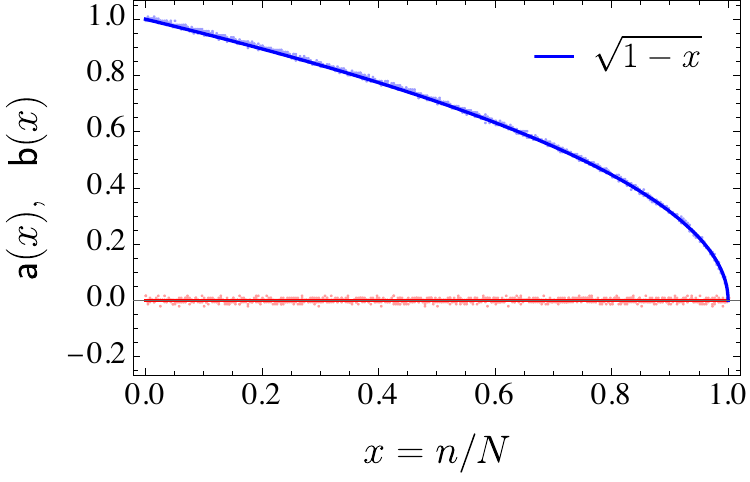}
\caption{The Lanczos spectrum for GUE. We choose $N = 1024$ (total of 20 realizations). The coefficients $\mathsf{a}(x)$ fluctuate around $\mathsf{a}(x)=0$ (shown by the red line), while the $\mathsf{b}(x)$ gradually decreases and terminates at $n = N$ (shown by the blue line). They match with the analytic result \eqref{RMTa}. The figure is adapted from \cite{Balasubramanian:2022dnj} with different system parameters. The initial state is chosen as $(1, 0, \cdots, 0)^{\intercal}$.} \label{fig:RMTstate}
\end{figure}

As an example, consider the GUE with the potential $V(H) = H^2$, given by \eqref{RMTdiffpot}. In this case, the density of states follows Wigner's semi-circle law
\begin{align}
    \rho(E) = \frac{1}{2\pi}\sqrt{4-E^2}\,.
\end{align}
The integral equation is solved exactly with the \emph{mean} Lanczos coefficients given by \cite{Balasubramanian:2022dnj}
\begin{align}
    \mathsf{a}(x) = 0\,,~~~~~ \mathsf{b}(x) = \sqrt{1-x}\,. \label{RMTa}
\end{align}
Figure \ref{fig:RMTstate} shows the numerical results for Lanczos coefficients with GUE random matrices of size $N =1024$ (20 realizations), which match the analytic expressions exactly. Here we take the variance $\sigma^2 = 1/N$, i.e., the real diagonal elements (since the matrix is Hermitian, the diagonal elements are real) have the zero mean, and the variance $1/N$ while the off-diagonal elements are complex numbers with both the real and the imaginary parts have individual variances $1/(2N)$. This choice of variance ensures that the Lanczos coefficients remain independent of the system size $N$, which otherwise scales as $\mathsf{b}_n \sim \sqrt{N}$ given that $n \sim o(1)$ if the variance is taken as unity. For the numerical implementation, we used the $\mathtt{HessenbergDecomposition}$ command in $\emph{Mathematica}$, which offers computational advantages over the traditional Lanczos algorithm, making it particularly well-suited for handling large matrices and ensembles. However, unlike the Lanczos algorithm, the Hessenberg decomposition chooses a fixed initial state $(1, 0, \cdots, 0)^{\intercal}$. In addition, in the Hessenberg decomposition, the Lanczos coefficients can be negative. As discussed in  \cite{Balasubramanian:2022dnj}, its source is due to the phase factors from the initial state and can be avoided by taking the modulus of the coefficients. A similar approach can be applied to the Liouvillian \cite{Loc:2024oen}. The non-Gaussian potentials with $V(H)$ being a polynomial of $H$ are discussed in \cite{Balasubramanian:2022dnj}.

A particular point regarding Eqn.\,\eqref{den0} is worth mentioning. The left-hand side (LHS) of \eqref{den0} is independent of the choice of the initial state, while the Lanczos coefficients are state-dependent. Another way to interpret this is that different states can be related by a unitary transformation that changes the Lanczos coefficients but not the eigenvalues, thus not affecting the DOS, i.e., the LHS of \eqref{den0}. Therefore, one might question the validity of \eqref{den0}. The answer is as follows: For a fixed Hamiltonian, different initial states produce different Lanczos coefficients. Although these coefficients can be arbitrary, they are distributed (for different initial states) around a \emph{mean} value. The Lanczos coefficients on the right-hand side (RHS) of \eqref{den0} represent this mean value, not individual realizations. Hence, \eqref{den0} implicitly accounts for an average over the initial states for any \emph{fixed} Hamiltonian.

For random matrices, each realization corresponds to a different Hamiltonian. Thus, a random matrix with a fixed initial state is analogous to a fixed Hamiltonian with different initial states, as previously described. A random matrix with TFD states introduces an additional random state for each realization, thereby justifying \eqref{den0} automatically. Consequently, the bulk distribution of the \emph{mean} Lanczos coefficients does not depend on the choice of a large class of initial states; it only depends on the DOS. For different universality classes with the same DOS, the mean Lanczos coefficients are the same, given by \eqref{RMTa}. This can be explicitly verified for the tenfold Altland-Zirnbauer classes \cite{AZsymmtery}, including the three Dyson classes: GOE, GUE, and GSE matrices.

%%%%%%%%%%%%%%%%%%%%%%%%%%%%%%%%%%%%%%%%%%%%%%%%%%%%%%%
\subsection{Spread complexity in the thermofield double state} \label{SCandSFF}

\begin{figure} 
\centering
\includegraphics[width=0.77\linewidth]{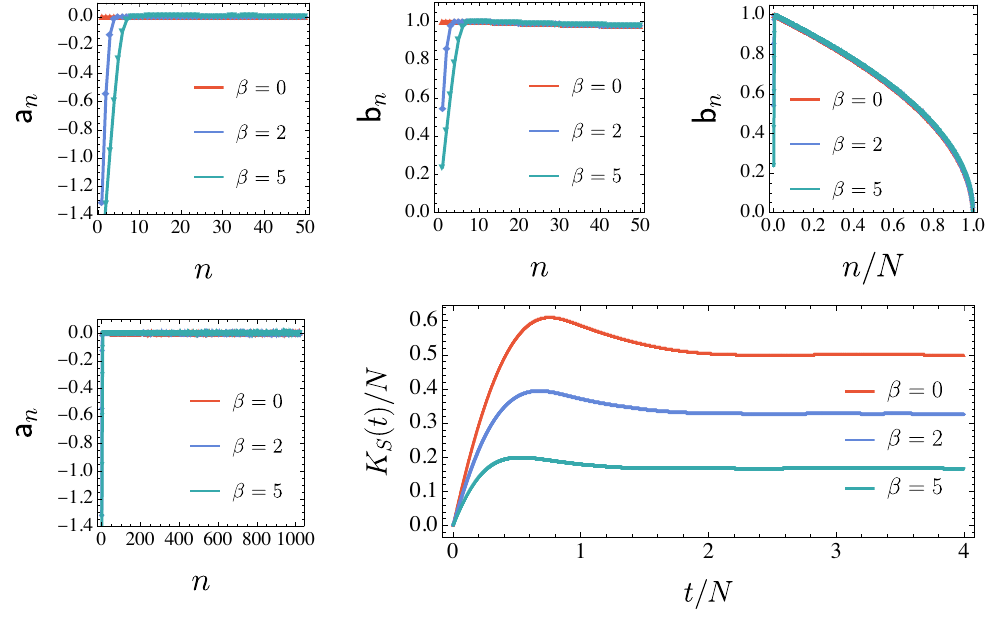}
\caption{Mean Lanczos coefficients and spread complexity for a single Hamiltonian sampled from  $\mathrm{GUE}$, with size $N=1024$ and variance $\sigma^2=1/N$. Diagonal Lanczos coefficients (left, two rows) and off-diagonal (top, center, and right) are shown for different temperatures in the TFD state. (Lower right) Spread complexity averaged over $100$ matrices drawn from the same ensemble. The initial state is the coherent Gibbs state (one copy of the TFD state) at inverse temperature $\beta=0$ (red), $\beta = 2$ (blue) and $\beta = 5$ (green). The saturation value at infinite temperature matches with \eqref{infsat}. Details are explained in the text. The figures are adapted from \cite{balasubramanian2022} with different system parameters.}
\label{GOERMTfig}
\end{figure}

The TFD state \eqref{tfd} is time-invariant under the time evolution of $H_{\mathrm{tot}} = H_1 - H_2$, which corresponds to the boost symmetry in the bulk spacetime \cite{Maldacena:2001kr}. However, it evolves with a Hamiltonian acting on one side. The time-evolved state, under such condition, is given by
\begin{align}
    \ket{\Psi(t)} = e^{-i H t} \ket{{\rm TFD}(\beta)} = \ket{{\rm TFD}(\beta + i t)}\,.
\end{align}
Here the time evolution is generated by the Hamiltonian of one of the copies, and compactly denoted as $H \equiv H \otimes \id$. Alternatively, the time evolution can also be generated with $\bar{H}_{\mathrm{tot}} = (H \otimes \id + \id \otimes H)/2$, which has the same effect as the single-side Hamiltonian. As a result, the time evolution shifts the inverse temperature as $\beta \rightarrow \beta + it$. The survival amplitude is calculated as \cite{del_campo_scrambling_2017}
\begin{align}
    S(t) =  \langle {\rm TFD}(\beta)| e^{-i H t} |{\rm TFD}(\beta) \rangle =  \langle {\rm TFD}(\beta)| {\rm TFD}(\beta + i t)  \rangle = \frac{Z (\beta + it)}{Z(\beta)}\,,
\end{align}
which is the ratio of the analytical continuation of the partition function $Z(\beta + i t)$ and the standard partition function $Z(\beta)$.
Therefore,  the survival probability in the TFD state equals the SFF in Eq. (\ref{SFFdef}) \cite{del_campo_scrambling_2017,Xu2021}.

In the TFD state, the moments \eqref{statemom} of the survival amplitude are conveniently expressed as \cite{balasubramanian2022}
\begin{align}
    \mu_n = \frac{1}{Z(\beta)} \mathrm{Tr}(e^{-\beta H}   (i H)^n )\,. \label{statemom2}
\end{align}
An important difference between $\mu_n$ and the moments $m_n$ computed in Sec.\,\ref{sec:moment1} is that here the moments are given by the Hamiltonian moments, while in the operator complexity picture, the moments $m_n$ are the Liouvillian moments \eqref{evenmom}. The Lanczos coefficients can be computed straightforwardly from \eqref{statemom2}. 

Figure \ref{GOERMTfig} shows the behavior of mean $\mathsf{a}_n$, $\mathsf{b}_n$ in the TFD state when the Hamiltonian is sampled from the Gaussian unitary ensemble (GUE) with dimension $N = 1024$ (averaged over $100$ instances) at different temperatures. Here,  the Hessenberg decomposition is used to compute the Lanczos coefficients rather than  \eqref{statemom2}. Depending on the temperature,  $\mathsf{a}_n$ increases with a different slope and saturates at $\mathsf{a}_n \approx 0$. The saturation occurs for $n \ll N$, which is much lower than the dimension of the matrix \cite{balasubramanian2022}. On the other hand, the coefficients $\mathsf{b}_n$ show a similar trend as $\mathsf{a}_n$ for $n \ll N$, but terminate at $n =N$ due to the finite dimensions of the matrices. The slope of the growth increases with decreasing temperature, a trend also observed for the Krylov complexity in the operator picture in the SYK model. See Fig.\,\ref{fig:LanczosfiniteTSYK} for a comparison.

\begin{figure}[t]
\centering
\includegraphics[width=0.9\textwidth]{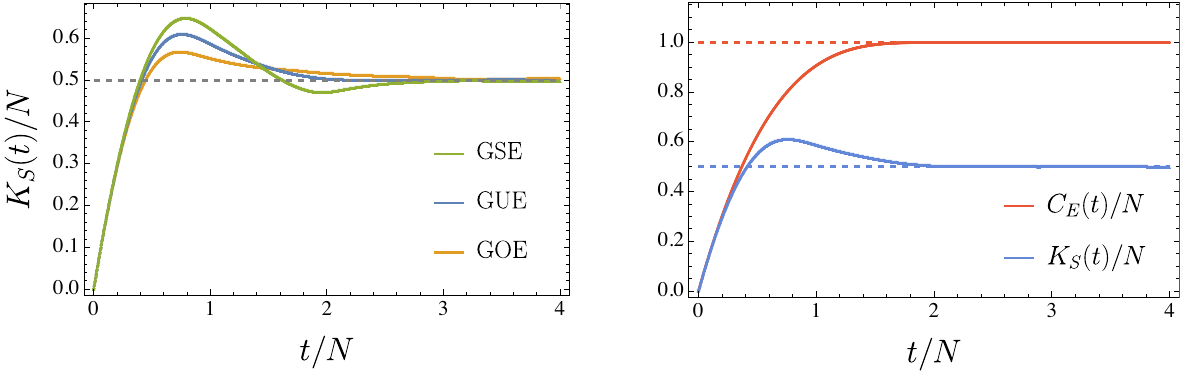}
\caption{Left: Krylov (spread) complexity for GOE (red), GUE (blue), and GSE (green) matrices respectively. Notice the peak height is maximum for GSE, followed by GUE and GOE. Right: Comparison between spectral complexity (red) and the spread complexity (blue) at infinite temperature in the GUE. The spectral complexity is given the analytic expression \eqref{specCan}, while the spread complexity is computed in the maximally entangled state (i.e., the TFD state at infinite temperature). The red and blue dashed lines are the spectral and spread complexity saturation values given by $(N-1)/(2N)$ (gray dashed line for the left figure) and $(N-1)/N$, respectively. For all the plots, the system size is $N = 512$ with an average taken on $100$ samples. The variance of GOE, GUE, and GSE matrices are taken as $\sigma^2 = 2/N, 1/N$, and $1/(2N)$ respectively, with a non-degenerate block chosen for the GSE case. The right figure is adapted from \cite{Erdmenger2023} with different system parameters.} \label{fig:specvsspread}
\end{figure}

The corresponding behavior of the Krylov complexity is also shown (bottom right). A key finding is the identification of four distinct regimes in the time-evolution of the complexity measure for TFD states in chaotic systems: an initial linear increase, a peak, a decrease, and finally a plateau \cite{balasubramanian2022, Erdmenger2023}. In the infinite temperature limit, the saturation value is given by \cite{Erdmenger2023}
\begin{align}
   \bar{K}_S := K_S (t \rightarrow \infty) = \frac{N - 1}{2}\,, \label{infsat}
\end{align}
while the ratio $\bar{K}_S/N \rightarrow 1/2$ in the large $N$ limit. The wavefunction becomes completely delocalized, resulting in $|\uppsi_n (t \rightarrow \infty)|^2 = 1/N$, akin to the behavior observed in the operator case; see \eqref{wdec} and \eqref{wdec2} for instance. For a fixed dimension $N$, the peak value and the saturation plateau decrease with temperature, as evident from Fig.\,\ref{GOERMTfig}. This ramp-peak-slope-plateau behavior of the Krylov state (spread) complexity is reminiscent of the slope-dip-ramp-plateau structure observed in the SFF, as introduced in Sec.\,\ref{sec:ChaosSFF}, indicating a deep connection between spectral properties and quantum state complexity. The peak value of the complexity indicates the strength of the eigenvalue correlations. It is highest for GSE, followed by GUE and GOE matrices; see Fig\,\ref{fig:specvsspread} (left). A noticeable decrease in complexity is observed for GSE, which is not present in GOE or GUE. This behavior is likely similar to the sharp peak observed in the SFF for GSE, whereas GOE and GUE exhibit a smooth corner and a kink, respectively \cite{Guhr1998, Cotler2017, Liu2018}. It is important to note that neither the linear growth nor the saturation value of the spread complexity indicates chaos. However, for uncorrelated energy levels, the peak in the complexity disappears \cite{balasubramanian2022}, which is a potential indicator of chaos. 
A more exhaustive discussion between the SFF and the Krylov complexity can be found in \cite{Erdmenger2023, Tan:2024kqb}.

\subsection{A digression: Spectral Complexity}

In systems characterized by a large-dimensional Hilbert space, the initial increase in spread complexity is analogous to the behavior of \emph{spectral complexity} \cite{Iliesiu2022}. The latter is defined as follows
\cite{Iliesiu2022, Erdmenger2023}
\begin{align}
    C_E(t) := \frac{1}{Z(2\beta) N} \sum_{E_i \neq E_j} \left(\frac{\sin (t (E_i - E_j)/2)}{(E_i - E_j)/2}\right)^2 e^{-\beta (E_i + E_j)}\,, \label{specC}
\end{align}
where the temperature is denoted by $T = 1/\beta$. This measure is conjectured to be holographically dual to the length of the Einstein-Rosen bridge, exhibiting an initial quadratic growth, transitioning into a linear regime, and ultimately reaching saturation. The late-time average value of spectral complexity is expressed as \cite{Camargo:2023eev}
\begin{align}
    \bar{C}_{E} =  \frac{2}{Z(2\beta) N} \sum_{E_i \neq E_j} \frac{e^{-\beta (E_i + E_j)}}{(E_i - E_j)^2}\,, \label{satce}
\end{align}
which, at infinite temperature, only depends on the eigenvalue separation. Notably, the spectral complexity relies entirely on the eigenspectrum of the boundary Hamiltonian, allowing for its computation across any Hamiltonian system. This has been explicitly demonstrated for random matrices and the SYK model \cite{Iliesiu2022}. Curiously, the spectral complexity is related to the SFF by \cite{Iliesiu2022, Erdmenger2023}   
\begin{align}
    \frac{d^2}{dt^2}  C_{E}(t) = \frac{2}{N} \frac{Z(\beta)^2}{Z(2 \beta)} \,\mathrm{SFF}(t) - \frac{2}{N}\,,
\end{align}
where the SFF is defined in \eqref{sffdef} with the Boltzmann factor serving as the eigenvalue filter. Although the time scales governing the SFF do not straightforwardly translate into the time scales in $C_E$, the late time saturation value of $C_E$ \eqref{satce} is proposed to be an indicator of chaos. Chaotic systems reach a lower saturation value than integrable systems, which is attributed to the eigenvalue repulsion present in chaotic systems.  An explicit example of such a case is shown for quantum billiards \cite{Camargo:2023eev}. This feature usually differs from the late-time behavior of the spread complexity \eqref{infsat}, whose saturation does not generically reflect the chaotic properties of the system \cite{Erdmenger2023}.

We consider the GUE at infinite temperature to understand the spectral and spread complexity in more detail. The analytic form of the spectral complexity can be derived from the expression of the SFF. It is given by \cite{Erdmenger2023}
\begin{align}
    C_E (t) &= \, _1F_2\left(-\frac{1}{2};1,2;-4 t^2\right)-1+N -\frac{16 t}{3 \pi } \nonumber \\
    &+ \Re\bigg[\frac{t}{6 \pi } \left(\frac{t^2}{N^2}+26\right) \sqrt{1-\frac{t^2}{4 N^2}} -\frac{2 N}{\pi }  \left(\frac{t^2}{N^2}+1\right) \sin ^{-1}\left(\sqrt{1-\frac{t^2}{4 N^2}}\right) \bigg]\,. \label{specCan}
\end{align}
Figure\,\,\ref{fig:specvsspread} (right) compares the analytic spectral complexity \eqref{specCan} and the numerically-evaluated spread complexity for a maximally entangled state in the GUE ensemble. While the initial growth is the same for both complexities, the late-time saturation differs. The spectral complexity reaches a saturation point that is twice as high as that of the spread complexity without any indication of a peak. One obtains three distinctive regimes of spectral complexity \cite{Erdmenger2023} 
\begin{align}
C_E(t) \rightarrow
  \begin{cases}
    t^2 \,      & ~~t \ll 1\,,\\
    \frac{16 t}{3 \pi} \,   & ~~1 \ll t \ll N\,,  \\ 
     N-1 \,      & ~~N \ll t\,.\label{CElimit}
  \end{cases}
\end{align}
These regimes represent the early quadratic growth, followed by linear growth and saturation. For analysis at finite temperatures and an extensive discussion on GOE and GSE, readers are referred to  \cite{Erdmenger2023}.

%%%%%%%%%%%%%%%%   KRYLOV SPACE OF DENSITY OPERATORS  %%%%%%%%%%%%%%%%%%%%
\section{Krylov Space of Density Operators} \label{secDensity}

The Krylov complexity was originally studied for observables evolving in Heisenberg picture \cite{parker2019}, as discussed in Sec.\,\ref{secObservable}. It was soon after extended to the case of pure quantum states evolving in Schr\"odinger picture \cite{balasubramanian2022}, reviewed in Sec.\,\ref{secStates}. However, the most general quantum state need not be pure and can be represented by a classical statistical mixture of pure states, which is modeled by a \textit{density matrix}. 
General mixed states 
%are not just a theoretical concept but instead 
occur naturally in the description of open quantum systems, especially after decoherence acts on pure states, making them effectively classical \cite{zurek_decoherence_2003}. The Krylov complexity for density matrices has been studied recently \cite{Alishahihakcomp1, caputa2024krylov}. Here, we present an alternative formulation of the Krylov space for density matrices, focusing on the constraints on the evolution in the Krylov chain imposed by the defining properties of density matrices, which we recall now.

A generic quantum state $\rho = \sum_j p_j \ket{\psi_j}\bra{\psi_j}$ is an operator that: \textit{(i)} has unit trace $\Tr(\rho) = 1$
%, this property generalizes the normalization condition of pure states $\braket{\Psi}{\Psi}=1$, 
%\textit{(ii)} is Hermitian 
and \textit{(ii)} is positive semidefinite $\rho\geq 0$, which implies hermiticity $\rho^\dagger = \rho$. These properties imply that \textit{(i)} $\sum_j p_j = 1$, \textit{(ii)} $ p_j\geq 0, \; p_j \in \mathbb R$ so that the eigenvalues of the density matrix can be associated with a probability distribution. 

The unitary evolution of a general quantum state $ \rho(t)$ evolving under the Hamiltonian $ H$ is described by the Liouville-von Neumann equation
\begin{equation}
    \partial_t \rho(t) = -i [ H,  \rho(t)] = - i \mathcal L \rho(t)\,,
\end{equation}
with the \textit{Liouvillian superoperator} $\mathcal L \bullet = [ H, \bullet]$. The solution of this equation gives the evolution of the density matrix as
\begin{equation}\label{evol_rhot}
    \rho(t) = e^{-i H t} \rho(0) e^{i H t} = e^{-i\mathcal L t}[\rho(0)] = \sum_{n=0}^\infty \frac{(-i t)^n}{n!} \mathcal L^n[\rho(0)]\,. 
\end{equation}
Using vectorization, the Liouville-von Neumann equation can be expressed as a linear differential equation as
\begin{equation} 
\label{vonneumann}
	\partial_t | \rho(t) ) = - i \mathcal{L} |\rho(t) )\,,
\end{equation}
in terms of the vectorized Liouvillian  $\mathcal{L} =  H \otimes \id  - \id \otimes H^\intercal$ and the vectorized density matrix $|\rho) = \frac{1}{\sqrt{d}}{\rm vec} (\rho) = \frac{1}{\sqrt{d}}\sum_{m,n} \rho_{mn}\ket{n}\otimes \ket{m}^*$. Expressing the vectorized Liouvillian in the Hamiltonian eigenbasis $H =  \sum_{n=1}^{d-1} E_{n} \ketbra{n}{n}$ gives
$\mathcal{L} = \sum_{n,m=1}^{d} \omega_{nm} |\omega_{nm}) (\omega_{nm}| ,$
where $\omega_{nm} \equiv E_n - E_m$ are all the energy differences, which are the eigenvalues of the Liouvillian,  and $ |\omega_{nm}) = \ket{n} \otimes \ket{m}^* $ are their associated eigenvectors. 

From an initial density operator $|\rho_0)$, in analogy to the formalism developed for operators in Sec.\,\ref{secObservable}, we can define the \textit{Krylov space} generated by the repeated application of the Liouvillian as $\mathrm{span} \{ |\rho_0), \mathcal{L} |\rho_0), \mathcal{L}^2 |\rho_0), \dots\}$. The Krylov space spans the subspace of the total Hilbert space in which the evolution of the initial state $|\rho_0)$ occurs, which is most clearly seen in \eqref{evol_rhot}. This process yields a linearly independent set of a certain dimension $D_K$, the \textit{Krylov Dimension}. %Generally, this set is not orthogonal. 

% Lanczos algorithm
The set obtained by repeated application of the Liouvillian is not orthogonal 
and we can construct the orthogonal Krylov basis $\{ |\rho_0), |\rho_1), \dots  |\rho_{D_K-1}) \}$by a procedure closely resembling that in the case of operators. To proceed, we choose the \textit{Hilbert-Schmidt} inner product, introduced previously. Let us detail the Lanczos algorithm we follow
\begin{enumerate}
    \item Define the starting density operator to be $|\rho_0)$.
    \item Compute $|\mathcal A_1) = \mathcal L|\rho_0)$. If $(\mathcal A_1|\mathcal A_1)\neq 0$ define $b_1 = \sqrt{(\mathcal A_1|\mathcal A_1)}$ and $|\rho_1) = |\mc A_1)/b_1$.
    \item For $n>1$ compute $|\mc A_n) = \mc L|\rho_{n-1})-b_{n-1}|\rho_{n-2})$. If $(\mathcal A_n|\mathcal A_n)\neq 0$ define $b_n = \sqrt{(\mathcal A_n|\mathcal A_n)}$ and $|\rho_n) = |\mc A_n)/b_n$.
\end{enumerate}
The Liouvillian has the following recurrence relation in the Krylov basis
\begin{equation}
\label{lanczos}
   \mathcal{L} |\rho_{n-1}) =	b_n|\rho_n) + b_{n-1}|\rho_{n-2})\,.
\end{equation}
The Lanczos algorithm introduced here has a key difference with respect to the formalism in \cite{caputa2024krylov}. In particular, it does not normalize the initial density matrix by $\sqrt{(\rho_0|\rho_0)}= (\Tr(\rho_0^2)/d)^{1/2}= \sqrt{P(0)/d}$. In doing so, the algorithm keeps the first element of the Krylov chain a physical density matrix with unit trace. Note that if the initial density matrix was normalized, the trace of an initially mixed state $|\rho_0)$ becomes $\Tr(\rho_0) = d \; (\id|\rho_0) = 1/ \sqrt{P(0)}$, which is different from unity if the state is mixed $P(0)<1$.\footnote{The Lanczos coefficients obtained by the two methods (here denoting $\tilde b_n$ as the definition in \cite{caputa2024krylov}) are not the same but are easily related through the purity by $\tilde b_n = b_n/\sqrt{P(0)}$. The Krylov basis elements $|\rho_n)$ are the same for $n\geq 1$, since $|\tilde{\mc A}_n)=|A_n)/\sqrt{P(0)}$ which cancels out with the factor in $\tilde b_n$, and only differ in $n=0$ by the factor $\sqrt{P(0)}$.} With this definition, the Krylov basis is orthogonal but not necessarily normalized for the first element $(\rho_0|\rho_0)=P(0)/d$%, which is not unity for a mixed state
, although for $n, m \geq 1$, the standard orthonormality condition $(\rho_n|\rho_m) = \delta_{nm}$ holds.

\begin{figure} 
\centering
\includegraphics[width=0.47\linewidth]{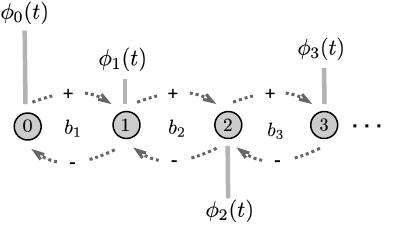}
\caption{Illustration of the Krylov chain for density matrices. The illustration displays the different Krylov basis elements, $|\rho_n)$ with their associated amplitudes $\phi_n(t)$ characterizing the probability of the state to be in each site. The hopping rates in the chain are given by the Lanczos coefficients $b_n$. For the evolution of the amplitudes, these coefficients carry a positive sign in hoppings to the right and a negative sign for hoppings to the left, as can be seen in the anti-Liouvillian \eqref{eq:antiLiouvillian}.  }
\label{fig:KrylovChainDM}
\end{figure}

% The Krylov basis
In the Krylov basis, the Liouvillian is tridiagonal, with all diagonal elements being zero,
\begin{equation}
	\mathcal{L} = \sum_{n=1}^{D_K-1} b_n \big( |\rho_{n})(\rho_{n-1}| + |\rho_{n-1})(\rho_{n}| \big)\,.
\end{equation}
The density matrix on this basis is
\begin{equation} \label{rho_krylov}
	|\rho(t) ) = \sum_{n=0}^{D_K-1} (-i)^n \phi_n(t)|\rho_n)\,,
\end{equation}
where the \textit{density matrix amplitudes} $\phi_n(t)$ are given by	$ \phi_n(t) = i^n (\rho_n|\rho(t)) $. Note that the density matrix amplitudes contain information equivalent to the full density matrix, conditioned to a particular initial state. It is thus possible to recast the Liouville-von Neumann equation \eqref{vonneumann} into a dynamical equation for the amplitudes
\begin{equation}
\label{diffeq}
	\partial_t \phi_n(t) = b_n \phi_{n-1} (t) - b_{n+1}\phi_{n+1}(t)\,.
\end{equation} 
This difference equation, sometimes called a \textit{discrete Schr\"odinger equation} \cite{parker2019},  may be written in the compact form
\begin{equation}
	\partial_t | \phi (t) ) = \mathcal{M} | \phi (t) )\,,
\end{equation}
where it is convenient to introduce the vector of density matrix amplitudes $|\phi (t) ) = \sum_{n=0}^{D_K-1} \phi_n(t)|\rho_n)$, and the \textit{anti-Liouvillian} superoperator
\begin{equation}
	\mathcal{M} = \sum_{n=1}^{D_K-1} b_n \big( |\rho_{n})(\rho_{n-1}| - |\rho_{n-1})(\rho_{n}| \big)\,. \label{eq:antiLiouvillian}
\end{equation}
The density matrix amplitudes are illustrated in Fig.\,\ref{fig:KrylovChainDM} along with the roles of the Lanczos coefficients as hopping amplitudes and the corresponding signs entering the anti-Liouvillian, positive for right jumps and negative for left jumps. This allows us to think of the anti-Liouvillian $\mc M$ as a current operator on the Krylov chain, carrying information on the sign of the current.

\subsection{Properties of the elements of the density matrix Krylov chain}

% Trace of Krylov elements

From Eq. (\ref{lanczos}), an interesting expression for the trace of the Krylov elements, $\{ \rho_n\}_{n=0}^{D_K-1}$, can be found. Noting that $\mathrm{Tr}(\mathcal{L}  \rho_n)=\mathrm{Tr}([H, \rho_n])=0$ for all $n$, we find that
\begin{eqnarray}
    \mathrm{Tr}( \rho_n)= -\frac{b_{n-1}}{b_n} \,\mathrm{Tr}( \rho_{n-2}) \,.
\end{eqnarray}
Hence, if $\mathrm{Tr}(\rho_0)=1$ as appropriate for a density matrix,  even and odd Krylov elements differ by their trace
\begin{subequations}
\label{eq:tr_rhon}
\begin{eqnarray}
    \mathrm{Tr}( \rho_{2n}) &=& (-1)^n \frac{b_{2n-1}}{b_{2n}} \frac{b_{2n-3}}{b_{2n-2}} \cdots \frac{b_1}{b_2}\,,\\
    \mathrm{Tr}( \rho_{2n+1}) &=& 0 \,. 
\end{eqnarray}
\end{subequations}
This condition means that the Krylov elements for density matrices, $|\rho_n),\; n\geq 1$ are never density matrices on their own (since they do not satisfy $\mathrm{Tr}( \rho_n)=1)$. These traces can also be understood as the inner product with the identity $\mathrm{Tr}(\rho)= d \; (\id|\rho)$. Therefore, Eqs.  \eqref{eq:tr_rhon} specify the components of the identity operator in the Krylov basis, which reads
\begin{equation}
    |\id) = \frac{1}{\sqrt{d}}\sum_{n=1}^{D_K-1} (-1)^n \prod_{j=1}^{n} \frac{b_{2j-1}}{b_{2j}} |\rho_{2n})\,. 
    \label{Id}
\end{equation}
The identity operator is proportional to the \textit{maximally mixed state} $ \rho_\mathrm{MM} =  \id /d$. Note that this state has support over all the even Krylov basis elements. Interestingly, 
this expression seems to be closely related to the null state of the Liouvillian \cite{Rabinovici:2022beu}, and its norm is related to the area under the autocorrelation function ${\cal C}(t)$ \cite{Bartsch:2023kjz,wang2023diffusion}.
The relation \eqref{Id}  suggests that the Lanczos algorithm, at least in the absence of degeneracies, collapses the zero eigenvalue subspace of dimension $N$ to the maximally mixed state. 

%Hermiticity of the Krylov basis elements

The commutator of two Hermitian operators $ A^\dagger =  A, \;  B^\dagger =  B$ is anti-Hermitian, i.e., $[ A,  B] =  C$ where $ C^\dagger = -  C$. Therefore, when building the Krylov space by repeated application of the Liouvillian, the even powers will be Hermitian $(\mathcal L^{2n} \rho_0)^\dagger=\mathcal L^{2n} \rho_0$  and the odd powers will be anti-Hermitian $(\mathcal L^{2n+1} \rho_0)^\dagger=-\mathcal L^{2n+1} \rho_0$. Due to the Lanczos algorithm building the Krylov basis by a real linear combination of only even or only odd elements, the resulting Krylov basis is composed of Hermitian operators for the even basis elements and anti-Hermitian operators for the odd ones.

\subsection{Constraints on the evolution}

%Trace Preservation

Any physical density matrix $ \rho$ must be unit trace to represent a properly normalized quantum state, i.e., $\mathrm{Tr}( \rho(t))=1$. This condition can be expressed for the vectorized density matrix as
\begin{equation}
    \mathrm{Tr}( \rho(t))= d (\id|\rho(t))=1\,,
\end{equation}
where $|\id)= \frac{1}{\sqrt{d}}\sum_{m=1}^d \ket{m}\otimes \ket{m}^*$ is the vectorized identity matrix.
Any physical quantum dynamics, mapping physical quantum states to physical quantum states, must preserve the trace of the density matrix $\mathrm{Tr}( \rho(t)) = d  (\id|\rho(t))= 1$ at all times.  Substituting \eqref{eq:tr_rhon} yields the constraint for the even amplitudes
\begin{equation}
    \mathrm{Tr}( \rho(t)) = \sum_{n=0}^{D_K/2} \phi_{2n}(t) \frac{b_{2n-1}}{b_{2n}}\frac{b_{2n-3}}{b_{2n-2}}\cdots \frac{b_1}{b_2}=1\,.
\end{equation}
This constraint involves only even density matrix amplitudes $\phi_{2n}(t)$, and ratios of odd and even Lanczos coefficients, which can show drastically different scalings as discussed in Sec.\,\ref{sec:moment1}. It thus shows that the dynamics of the density matrix amplitudes will differ in even and odd sites, providing further insight into the dynamics of the amplitudes in the Krylov chain.

%Purity conservation

Furthermore, since the dynamics is unitary, the purity of the initial density matrix will be preserved $P(t) = \mathrm{Tr}( \rho^2(t))= d (\rho(t)|\rho(t))=\mathrm{Tr}( \rho^2_0) = d (\rho_0|\rho_0)=P (0)$, which is unity if the initial state is pure $ \rho_0 = \ket{\psi_0}\bra{\psi_0}$. When written in terms of the amplitudes, the purity preservation condition implies
\begin{equation}
    \sum_{n=0}^{D_K-1} \phi_n^2(t)=P (0)\,.
\end{equation}

The density matrix of a physical state needs to be Hermitian and positive semi-definite, i.e., $ \rho^\dagger =  \rho$ and $ \rho \geq 0$. The former condition can be easily applied to the density matrix written in the Krylov basis \eqref{rho_krylov}. It is easy to see that all the elements of the sum are Hermitian.
%, the even ones are Hermitian directly, and the odd ones are multiplied by the imaginary unit, which makes them Hermitian. 
This implies that for the full density matrix to be Hermitian, the density matrix amplitudes must be real $\phi_n^*(t) = \phi_n(t)$. In addition, positive semi-definiteness implies that the eigenvalues of the density matrix must be positive semidefinite since $ \rho$ is Hermitian. Determining the constraints imposed by positive semi-definiteness on the density matrix amplitudes constitutes an interesting open problem.

%The probability amplitude for propagation between any two sites $n, \; m$ of the Krylov chain in time $t$ is given by 
%\begin{equation}
%    \phi_{nm}(t) = (\rho_n|e^{-i \mathcal{L} t}|\rho_m).
%\end{equation}
%For simplicity we denote $\phi_{0n}(t) \equiv \phi_{n}(t)$.

%\subsection{Chapman-Kolmogorov equation for the density operator amplitudes}

%The $0$-th operator amplitude, or return probability to the 0-th site, reads
%\begin{equation}
%    \phi_{00}(t) = (\rho_0|\rho(t)) = (\rho_0|e^{-i \mathcal{L} t}|\rho_0),
%\end{equation}
%the time evolution may be split in two, from $0$ to $t_1$ and from $t_1$ to $t$. Introducing the identity $\id = \sum_{k=0}^{D_K-1} |\rho_k)(\rho_k|$ we find
%\begin{align} \notag
%    \phi_{0,0}(t) &= \sum_{k=0}^{D_K-1} (\rho_0|e^{-i \mathcal{L} (t-t_1)}|\rho_k)(\rho_k|e^{-i \mathcal{L} t_1}|\rho_0)\\ 
%     &= \sum_{k=0}^{D_K-1} \phi_{0,k}(t-t_1) \phi_{k,0}(t_1).
%\end{align}

%The most general form of the Chapman-Kolmogorov equation can be found splitting the propagation from $0$ to $t$ in $J$ intermediate step, i.e.,$(0, t_1),  (t_1, t_2), \dots, (t_J, t)$. Introducing the resolution of the identity in terms of the Krylov basis $\id = \sum_{k=0}^{D_K-1} |\rho_k)(\rho_k|$ $J$ times yields
%\begin{align} 
%    \phi_{mn}(t) = \sum_{k_1, \dots, k_J=0}^{D_K-1}&\phi_{m k_{J}}(t - t_{J})\\ \notag  \times &\prod_{j=2}^{J} \phi_{k_j k_{j-1}}(t_j - t_{j-1})\phi_{k_1 n}(t_1).
%\end{align}

%%%%%%%%%%%%%%%%   OPEN QUANTUM SYSTEMS  %%%%%%%%%%%%%%%%%%%%
\section{Open Quantum Systems} \label{secOpen}

\subsection{An introduction to the Lindblad master equation}

The dynamics in Krylov space discussed so far have only focused on closed quantum systems. Any realistic treatment of a quantum system needs to include the effects of decoherence and noise caused by the surrounding environment, it is thus of key importance to extend the Krylov formalism to open quantum dynamics. The theory of Open Quantum Systems (OQS) offers a powerful description of dissipative quantum dynamics, here we review some of the key results, in particular the Lindblad master equation, for the extension of the Krylov formalism to the open case. For a more thorough study of OQS, we refer the reader to \cite{breuer_theory_2002, rivas_open_2012}. 

The description of OQS considers a bipartite Hilbert space composed of two key constituents: the system $S$, with Hilbert space  $\mathscr H_S$ and Hamiltonian $H_S$, which includes the relevant degrees of freedom, and the environment $E$, with Hilbert space $\mathscr H_E$ and Hamiltonian $H_E$, which models the effect of the surroundings. The full system is thus composed of System and Environment  $\mathscr H = \mathscr H_S \otimes \mathscr H_E $, and importantly, is a closed system, with Hamiltonian 
\begin{align}
 H_{S+E} = H_S \otimes \id_E + \id_S \otimes H_E + H_{\mathrm{int}}\,,
\end{align}
where $H_\mr{int}$ describes the interaction between system and environment.  The full system plus environment thus evolves unitarily. The description based on the full $S+E$ Hamiltonian is too complicated since in relevant situations the environment is composed of extremely many, or even infinite, degrees of freedom. Therefore, one of the key tools in the theory of OQS is master equations that describe the dissipative evolution of the system alone. There exists a plethora of master equations, each one valid and useful under particular conditions. Here we are interested in the most general Markovian dissipative evolution, which is generated by the Lindblad equation, also known as the Gorini-Kossakowski-Sudarshan-Lindblad equation \cite{lindblad_generators_1976, gorini_completely_1976}. 
The original derivation of this equation starts from postulating the most general evolution of the system that sends physical states to physical states. Consider that the evolution is generated by a dynamical map $\rho(t)= \mc E_t(\rho_0)$, then the most general Completely-Positive and Trace-Preserving (CPTP) evolution admits the Kraus decomposition \cite{Kraus83,Wilde13,Holevo19} $\rho(t) = \mc E_t(\rho_0)= \sum_k E_k \rho_0 E_k^{\dagger}$ where $E_k$ are the Kraus operators subject to the normalization condition $\sum_k E_k^\dagger E_k = \id$. A map is \textit{positive} iff it sends positive operators to positive operators $\rho\geq 0 \rightarrow \mc E(\rho) \geq 0$. A map is \textit{completely positive} iff the map $ \mc E \otimes \id_n$ is positive for all $n$, where $\id_n$ is the $n$-dimensional identity map \cite{CHOI1975285, lindblad_generators_1976}. Physically this requirement may be understood as the map extended to act on the system and any general ancilla being positive. The condition of complete positivity guarantees that the states generated by the evolution are always Hermitian and positive semidefinite $ \rho_t^\dagger = \rho_t, \;\rho_t \geq 0, \;  \forall \; t$, and trace preservation implies that the density matrix remains normalized $\Tr(\rho_t) = 1, \;  \forall \; t$. Thus the dynamics generated by a CPTP map sends physical quantum states to physical quantum states. If the dynamical map obeys the semigroup property $\mc E_{t_1}\mc E_{t_2} = \mc E_{t_1 + t_2}$, the dynamics described by the system is Markovian and its generator admits the Lindblad form
\begin{align}
	\dot{\rho}(t) = - i [H, \rho(t)] +  \sum_k \upmu_k \left[L_k \rho(t) L_k^{\dagger} - \frac{1}{2} \{L_k^{\dagger} L_k, \rho(t)\}\right] =- i \mc L_o \rho_t = - i \mc L_H \rho_t - i \mc L_D \rho_t\,,\label{st4}
\end{align}
where $\upmu_k \geq 0$ are the dissipation rates, $L_k$ are the \textit{jump operators} describing the dissipative evolution and $H$ is the generator of the unitary part of the evolution and in general is not equal to the system Hamiltonian $H_S$. $\mc L_o $ represents the \textit{Lindbladian superoperator} where $\mc L_H, \; \mc L_D$ characterize, respectively, the unitary and dissipative part of the dynamics.  If the rates are negative $\upmu_k < 0$, the evolution is, in general, non-Markovian \cite{Rivas_2014_NonMarkov, breuer_2016_NonMarkov, Li_2021_NonMarkov}. The dynamical map can be written in terms of the Lindbladian superoperator as $\mc E_t = e^{-i \mc L_o t}$, note that the inverse of this map $\mc E_t^{-1}$ is not a CPTP map unless $\mc L_o$ describes unitary evolution \cite{rivas_open_2012}.

\begin{figure}[t]
\centering
\includegraphics[width=0.48\textwidth]{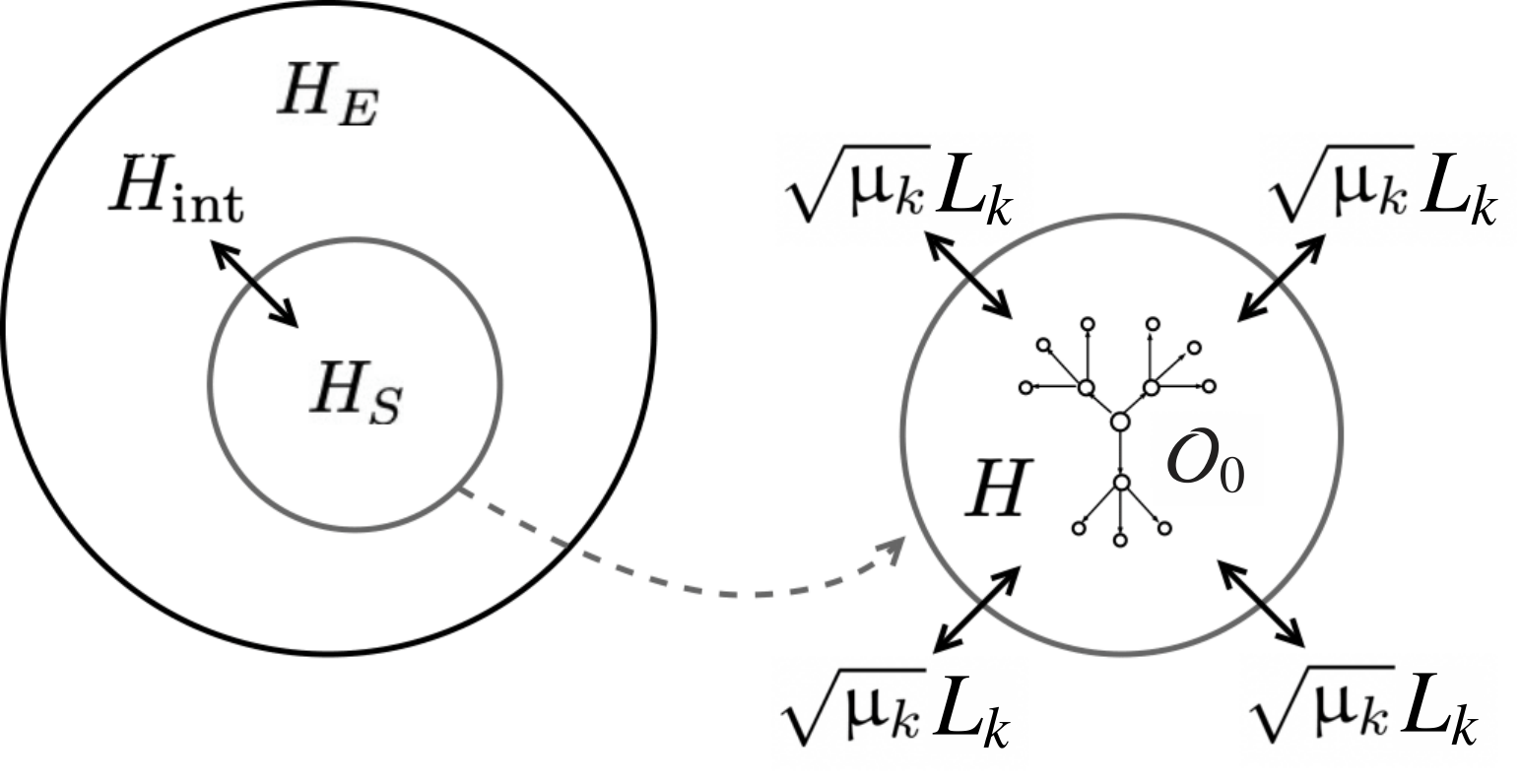}
\caption{The above schematic diagram illustrates the open system dynamics, where the system interacts with the environment (left) and evolves on its own (right). The jump operators $\{L_k\}$ with their associated dissipation rates $\{\upmu_k\}$ characterize the system-bath interaction. An operator $|\mathcal{O}_0) \equiv |\mathcal{O})$ grows due to the system's action and loses information to the environment.} \label{fig:opendiagram}
\end{figure}

The Lindblad equation can also be derived from a ``microscopic'' point of view \cite{breuer_theory_2002, rivas_open_2012}, in particular, one starts from the Liouville-von Neumann equation for the full system and environment and traces over the degrees of freedom of the environment, for this several conditions and approximations need to be imposed. Firstly, a Lindblad equation can usually be derived only in the weak system-bath coupling, or in the singular coupling limits. Secondly, the \textit{Born}, \textit{Markov}, and \textit{rotating-wave} approximations need to be imposed. These require: the system and environment state to be in a product form $\rho_S(t) \otimes \rho_E$ at all times, the characteristic time scale of the system $\tau_S$ to be much larger than the characteristic time-scale of the environment $\tau_E$ and, given a spectral decomposition of the jump operators $L(\omega)$, the terms involving different frequencies $\omega \neq \omega'$ to be negligible, respectively. The microscopic derivation provides the specific relation between the full Hamiltonian $H_{S+E}$ and the jump operators $\{L_k\}$, dissipation rates $\{\upmu_k\}$ and Hamiltonian $H$ appearing in the master equation \eqref{st4}. Figure \ref{fig:opendiagram} illustrates a schematic diagram of the connection.
%where $\upmu_k \geq 0$ are dissipation rates. From now on, we omit the subscript $S$ for the system and use $\rho$ and $H$ to denote $\rho_S$ and $H_S$ unless otherwise stated. Note that in Eq. (\ref{st4}), $H$ includes any Lamb shift term arising from the system-environment interaction. If the rates are negative $\upmu_k < 0$, the evolution is, in general, non-Markovian; see \cite{Rivas_2014_NonMarkov, breuer_2016_NonMarkov, Li_2021_NonMarkov} for a review on the topic. The Lindblad jump operators $L_k$'s are constructed from the system operators and encode the information of interaction  $H_{\mathrm{int}}$ assuming the coupling between the system and environment is weak. Figure \ref{fig:opendiagram} illustrates a schematic diagram of the open system dynamics. If the interaction is strong or the environment has a specific structure, then this approximation breaks down. This is the simplest approximation, and we can obtain more information if we know the exact form of the environment. The first term in Eq. \eqref{st4} generates unitary evolution, while the second term accounts for dissipation. As a result, the system evolution is no longer unitary in the presence of the environment.
The Lindblad equation \eqref{st4} can be interpreted from a quantum measurement point of view \cite{ZhouMIPT, cresser_measurement_2006, Wiseman_Milburn_2009}. For illustration, let us consider a closed system in a state $\rho(t)$ at time $t$. Assume the system evolves unitarily for a time interval $\delta t$ and then undergoes a quantum measurement. The measurement is characterized by a probability $P(s)$ at time $t = s$ and a set of measurement operators $\{M_k\}$. The density matrix of the system at time $t + \delta t$ can be expressed as
\begin{align}
    \rho(t+\delta t) = \rho(t) + \dot{\rho} (t) \delta t + O(\delta t^2)
    =\rho(t) - i [H, \rho (t)] \delta t + O(\delta t^2)\,,
\end{align}
where we have used the unitary evolution equation $\dot{\rho} (t) = - i [H, \rho (t)]$ in the second line. The measurement at this state changes the state to (we use a subscript $M$ to indicate that the system is measured)
\begin{align}
     \rho^M(t+\delta t) = [1 - P(t+\delta t)] \rho(t+\delta t) + P(t+\delta t) \sum_k M_k \,\rho(t+\delta t)  M_k^{\dagger}\,, \label{ms}
\end{align}
where $P(t+\delta t)$ is the probability of the measurement at time $t + \delta t$. The first term on the RHS of \eqref{ms} represents the probability that the system stays in the same state $\rho(t+\delta t)$ after the measurement (i.e., the measurement does not affect the system). The second term represents the effect of the measurement with the Kraus operators $M_k$ satisfying $\sum_k M_k^{\dagger} M_k = \id$. Expanding the probability as $P(t+\delta t) = P(t) + \eta (t) \delta t + O(\delta t^2)$, where $\eta (t) = \delta P(t)/\delta t$ is the measurement rate, we can rewrite \eqref{ms} as
\begin{align}
    \dot{\rho}^M (t) = - i [H, \rho(t)] + \eta (t) \sum_k \left[M_k \rho(t) M_k^{\dagger} - \frac{1}{2} \{M_k^{\dagger} M_k, \rho(t)\}\right]\,. \label{ms2}
\end{align}
This equation has the same form as a Lindblad equation for open quantum system \eqref{st4} if we identify the jump operators with the measurement operators $L_k \equiv M_k$ and with all dissipation rates equal and equal to the measurement rate $\eta(t) = \upmu \geq 0$. This is no coincidence, as the open system dynamics can be interpreted as that of a system continuously monitored by the environment \cite{Wiseman_Milburn_2009, carmichael2007statistical, Jacobs06, Jacobs_2014}. The measurement is a non-unitary process that disrupts the unitary dynamics of the system. The higher the measurement rate, the more the system deviates from the unitary evolution. Therefore, the stronger dissipation drives the system away from its unitary evolution.

The evolution of the density matrices \eqref{st4} is given in the Schr\"odinger picture. An analogous equation can be derived for operators in the Heisenberg picture. Recall that, in Schr\"odinger picture, the states (density matrices) evolve, and the operators stay constant while in the Heisenberg picture, the operators evolve while the states are fixed. The expectation value of an operator is the same in both pictures and thus,
$\mathrm{Tr}(\rho (t)\,   \mathcal{O}) = \mathrm{Tr} (\rho(0) \,  \mathcal{O} (t))\,, %\label{sh}
$
where $\mathcal{O} \equiv \mathcal{O}(t=0)$ is a normalized operator. Note that the LHS is in the Schr\"odinger picture while the RHS is in the Heisenberg picture. To derive the operator evolution, we differentiate both sides  with respect to time and get
$\mathrm{Tr}(\dot{\rho} (t) \,  \mathcal{O}) = \mathrm{Tr} (\rho(0) \, \dot{\mathcal{O}}(t))\,. %\label{sh1}
$
Using \eqref{st4} %and rearranging the terms (using the cyclic property of trace), 
we can write the expression in the Schr\"odinger picture as
\begin{align*}
    \mathrm{Tr}(\dot{\rho} (t)\,  \mathcal{O}) = \mathrm{Tr} \bigg[ \Big(-i [H, \rho(t)] +  \sum_k \upmu_k \left[ L_k  \rho(t) L_k^{\dagger} - \frac{1}{2} \{L_k^{\dagger} L_k,  \rho(t)\}\right] \Big)\mathcal{O}\bigg]\,.
\end{align*}
This expression can be recast into the Heisenberg picture by using the cyclic property of the trace
\begin{align*}
    \mathrm{Tr}(\rho(0)\,  \dot{\mathcal{O}}(t)) = \mathrm{Tr} \bigg[ \rho(0) \Big(i [H,  \mathcal{O} (t)] +  \sum_k \upmu_k \left[ L_k^{\dagger}  \mathcal{O} (t) L_k - \frac{1}{2} \{L_k^{\dagger} L_k, \mathcal{O} (t)\}\right] \Big)\bigg]\,.
\end{align*}
Therefore, the \textit{adjoint master equation}, which characterizes the dissipative evolution of operators in the Heisenberg picture, can be written as 
\begin{align}
	\dot{\mathcal{O}}(t) &=  i [H,  \mathcal{O} (t)] +  \sum_k \upmu_k \left[\pm L_k^{\dagger}  \mathcal{O} (t) L_k - \frac{1}{2} \{L_k^{\dagger} L_k,  \mathcal{O} (t)\}\right]\,, \notag \\  &= i \mc L_o^\dagger \mc O(t)  = i \mc L^\dagger_H \mc O(t) + i  \mc L ^\dagger_D \mc O(t)\,, \label{ot0}
\end{align}
where $\mathcal{L}^{\dagger}_o$ is the \textit{adjoint} Lindbladian superoperator, with the corresponding unitary $\mc L_H^\dagger$ and dissipative $\mc L_D^\dagger$ parts, and the sign $\pm$ accounts for the fermionic operators as well. The minus sign is used when the jump operators and initial operator $\mathcal{O}$ are both fermionic, i.e., they have odd parity \cite{Liu:2022god, Clark_exact_2010}. The evolution of any operator $\mc O(t)$ can formally be written as
\begin{align}
 \mathcal{O}(t) = e^{i \mathcal{L}_o^{\dagger}t} \, \mathcal{O}\,.\label{oeq}
\end{align}
% given by
%\begin{align}
	%\mathcal{L}^{\dagger}_o  \mathcal{O} (t) &= [H,  \mathcal{O}(t)] \nonumber \\
% &-i  \sum_k \upmu_k \left[\pm L_k^{\dagger}  \mathcal{O}(t) L_k - \frac{1}{2} \{L_k^{\dagger} L_k,  \mathcal{O} (t)\}\right]\,. \label{ot1}
%\end{align}
%The equation \eqref{oeq} satisfies the semigroup property, namely $\Lambda_{t_2} \Lambda_{t_1} = \Lambda_{t_2 + t_1}$, where the map $\Lambda_t = e^{i \mathcal{L}_o^{\dagger}t}$ is the semigroup generator. The inverse of this map exists but is not a CPTP map unless $\Lambda_t$ is unitary \cite{rivas_open_2012}. The map is also completely positive and trace-preserving (CPTP) \cite{breuer_theory_2002, breuer_2016_NonMarkov}. These properties ensure that the Lindblad generator has the form \eqref{ot1}.

The unitary and dissipative contributions to the adjoint Lindbladian can be written explicitly as \cite{Bhattacharya:2022gbz, Bhattacharjee:2022lzy}
\begin{align}
	 \mathcal{L}_H^{\dagger}  \mathcal{O}  = [H,  \mathcal{O} ] \,,~~~~~~
	  \mathcal{L}_D^{\dagger}  \mathcal{O}  =  -i \sum_k \upmu_k \left[\pm L_k^{\dagger}  \mathcal{O}  L_k- \frac{1}{2}\{L_k^{\dagger} L_k,  \mathcal{O} \} \right]\,. \label{negpos}
\end{align}
We also focus on the infinite-temperature Gibbs state $\rho_{\infty} =\id/\tr \id$, which is always a steady state of \eqref{ot0} since $\mc L_o \id = 0$, note that the identity $\id$ is not a fermionic operator and thus the correct sign in the adjoint Lindbladian is $+$. This property, preservation of the identity,  is called \textit{unitality} and provides the analog of trace preservation in the Heisenberg picture. It allows for a simplification of the problem by endowing a specific and unique inner product $\left( A | B \right):= \mathrm{Tr}(\rho_{\infty} A^\dagger B)$, matching that in \eqref{IP}.

For the Krylov construction, resorting to vectorization is convenient; recall Sec.\,\ref{SecVec}.  Making use of it, the normalization of the quantum state reads $\mathrm{Tr} \rho = (\mathrm{vec} \, \id)^{\dagger} \,\mathrm{vec} \, \rho =1$.
%
%An obvious application of this identity is $\mathrm{tr} \rho = (\mathrm{vec} \, \id)^{\dagger} \,\mathrm{vec} \, \rho =1$.
%
To express the Lindbladian superoperator in vectorized form,  making use of the identity (\ref{EqAOB}) 
in \eqref{ot0} yields
\begin{align}
    \mathcal{L}_o^{\dagger} \equiv ( H \otimes  \id-  \id \otimes H^{\intercal} ) - i \sum_k \Big[\pm  L_{k}^{\dagger} \otimes L_k^{\intercal} - \frac{1}{2} \left(L_{k}^{\dagger} L_k  \otimes \id + \id \otimes L_k^{\intercal} L_k^{*} \right) \Big]\,, \label{lindsup}
\end{align}
where for convenience the jump operators have been rescaled as $\sqrt{\upmu_k}L_k \rightarrow L_k$. The advantage of vectorization is that it transforms the superoperator Lindbladian of dimension $d$ (i.e., a map between matrices of dimension $d \times d$) into an operator of dimension $d^2$ (i.e., $d^2 \times d^2$ matrix) that acts on vectors ($\mathrm{vec} \,  \mc O $) of length $d^2$.  This is essential to compute the spectrum of the Lindbladian.

\section{Krylov complexity in open systems: Different approaches} \label{kcomponum}

%We are interested in how information propagates in a generic system. So far, we have seen that inserting an operator as input information and observing its time evolution gives us some insight into the information flow. In particular, the operator ``grows'' over time, and we can measure this growth using the Krylov complexity. However, this growth is affected if the system interacts with the environment. For a chaotic system, the information spreads exponentially fast until the scrambling time $t_s \sim \log N$, where $N$ is the degrees of freedom of the system \cite{Sekino:2008he, Lashkari:2011yi}. Dissipation inherently changes the dynamics by introducing additional timescales, which depend on the balance between the system's inherent scrambling and the dissipation rate. Therefore, we ask the precise question: how does an operator grow in a system that is coupled to a dissipative environment? Can we find some universal quantities that describe this growth? What are the appropriate timescales in such a situation? We will show in later sections that several answers are positive, at least for a large class of chaotic systems, such as the SYK model.

\subsection{Numerical approaches}

Several numerical methods extend the Krylov construction to open quantum systems. We present them in the following sequence. An analytic approach has been presented in Sec.\,\ref{Lanc_analytic}.

\subsubsection{Arnoldi iteration} \label{Arnoldisection}

The first study of Krylov construction in open systems was initiated in Ref.~\cite{Bhattacharya:2022gbz}, where a generalization of the Lanczos algorithm was proposed. The algorithm is known as Arnoldi iteration \cite{Arnoldipaper} where an orthonormal basis set $\{ \mathcal{V}_0, \dots \mathcal{V}_n, \dots \}$ is constructed using the full open-system Lindbladian:
\begin{align}
    \mathrm{span}(\mathcal{V}_0,\dots, \mathcal{V}_n) = \mathrm{span}(  \mathcal{O}, \mathcal{L}_o^{\dagger}   \mathcal{O}, \dots, (\mathcal{L}_o^{\dagger})^n   \mathcal{O} ) \,. 
\end{align}
The algorithm proceeds as follows. By initializing with a normalized vector $\mathcal{V}_0 \propto  \mathcal{O}$, an iterative construction yields
\begin{align}
    |\mathcal{U}_k )= \mathcal{L}_o^{\dagger} \,|\mathcal{V}_{k-1})\,.
\end{align}
$k=1, 2, \ldots$. Then, for $j=0$ to $n-1$, the algorithm works as follows \cite{Bhattacharya:2022gbz, Bhattacharjee:2022lzy}:
\begin{align}
    &1. ~h_{j,k-1} = ( \mathcal{V}_j|\mathcal{U}_k )\,.~~~~~~~~~~~~~~~~~~~~~~~~ ~~~~~~~~~~~~ ~~~~~~~~~~~~~~~~~~~~~~~~~   \nonumber \\
    &2. ~|\tilde{\mathcal{U}}_{k} ) =|\mathcal{U}_{k} )-\sum_{j=0}^{k-1} h_{j,k-1} |\mathcal{V}_{j} )\,. \nonumber \\
    &3. ~ h_{k,k-1} =  \sqrt{({\tilde{\mathcal{U}}_k|\tilde{\mathcal{U}}_k})}\,.
\end{align}
If $h_{k,k-1}=0$, stop; otherwise, define $\mathcal{V}_k$ as
\begin{align}
   | \mathcal{V}_k )= \frac{|\tilde{\mathcal{U}}_k )}{h_{k,k-1}}\,.
\end{align}
If the operators are vectorized, the appropriate inner product \eqref{IP} should be used. The Arnoldi iteration transforms the Lindbladian into an upper Hessenberg form (see also \cite{Minganti2022arnoldilindbladtime}) in the Arnoldi basis (or Krylov basis, keeping in mind that the basis is generated by the full Lindbladian $\mathcal{L}_o^{\dagger}$) \cite{Bhattacharya:2022gbz, Bhattacharjee:2022lzy},
\begin{align}\label{Arnoldimatrix}
    \mathcal{L}_o^{\dagger} \equiv \begin{pmatrix}
h_{0,0} & h_{0,1} & h_{0,2} & \cdots & \cdots & h_{0,n}\\
h_{1,0} & h_{1,1} & h_{1,2} & \cdots & \cdots & h_{1,n}\\
0 & h_{2,1} & h_{2,2} & h_{2,3} & \cdots & \cdots\\
\cdots & 0 & h_{3,2} &\cdots & \cdots & \cdots\\
0 & \cdots & 0 &\cdots &\cdots & h_{n-1,n}\\
0 & 0 & \cdots &0 & h_{n,n-1} & h_{n,n}\\
\end{pmatrix}\,, 
\end{align}
with the Arnoldi coefficients $h_{m,n}= (\mathcal{V}_m|\mathcal{L}_o^{\dagger}|\mathcal{V}_n)$. Without dissipation, $\mathcal{L}_o^{\dagger}$ reduces to the Hermitian counterpart $\mathcal{L}_H$, and the Arnoldi iteration reduces to the usual Lanczos algorithm. The matrix becomes tridiagonal with the non-zero primary off-diagonal elements given by the Lanczos coefficients $b_n$.

The construction is closely related to the Hessenberg decomposition, where any matrix $\mathcal{A}$ can be written as $\mathcal{U} \mathcal{A}_{\mathrm{Hess}} \,\mathcal{U}^{\dagger}$, where $\mathcal{U}$ is a unitary matrix and $\mathcal{A}_{\mathrm{Hess}}$ is of Hessenberg form, with all elements below the first subdiagonal vanishing. The determinant, trace, and eigenvalues of $\mathcal{A}$ are the same as those of $\mathcal{A}_{\mathrm{Hess}}$. Finding the Hessenberg form of any matrix is useful because it makes the matrix sparser. However, the Hessenberg decomposition is not unique and does not ensure that the diagonal and subdiagonal elements are positive. The Arnoldi iteration, on the other hand, finds a Hessenberg matrix that has positive diagonal and subdiagonal elements. Therefore, the Arnoldi iteration gives a very specific Hessenberg decomposition among the many possible ones.

\subsubsection{Closed Krylov basis}

In \cite{Liu:2022god}, the second method was proposed, where the Krylov basis is generated by the \textit{closed-system Liouvillian} $\mathcal{L}_H^{\dagger}$, instead of the full Liouvillian $\mathcal{L}_o^{\dagger} \equiv \mathcal{L}_H^{\dagger} + \mathcal{L}_D^{\dagger}$. This leads to a Krylov basis that confines the operator dynamics to the subspace:
\begin{align}
\mathrm{span}(\mathcal{O}_0,\dots,   \mathcal{O}_n) = \mathrm{span}(\mathcal{O}, \mathcal{L}_H^{\dagger}  \mathcal{O}, \dots, (\mathcal{L}_H^{\dagger})^n  \mathcal{O}) \,.
\end{align}
Alternatively, this converts the operator dynamics into a non-Hermitian tight-binding model of particles hopping between sites \cite{Liu:2022god}
\begin{align}
\partial_t \varphi_{n}(t) = b_{n} \varphi_{n-1} (t)  - b_{n+1} \varphi_{n+1} (t) +  i \sum_m a_{m, n} \varphi_{m} (t) \,, \label{linH}
\end{align}
where $ n \geq 1$, and $a_{m,n}$ are additional coefficients. These coefficients resemble Arnoldi coefficients, and the diagonal ones $a_n \equiv a_{n,n}$ are dominant. Ref.~\cite{Liu:2022god} provides a method to calculate them, but the reason for the dominance of the diagonal coefficients by the dissipative part of the Lindbladian is unclear. Ref.~\cite{Liu:2022god} also conducted numerical simulations in an one-dimensional interacting spinless fermionic model and the finite-fermion Sachdev-Ye-Kitaev (SYK) model. The diagonal elements are consistent with $h_{n,n}$ from the Arnoldi iteration, increasing linearly before a finite saturation. However, the behavior has no analytical support, although the general growth is observed to be in agreement with \cite{Bhattacharya:2022gbz}.

\subsubsection{Bi-Lanczos algorithm} \label{bidisc}

The third and final approach is a particularly convenient one. It creates a bi-orthonormal basis set instead of an orthonormal one, starting from initial vectors $|p_0\rrangle$ and $|q_0\rrangle$, evolved by the adjoint Lindbladian $\mathcal{L}_o^{\dagger}$ and the Lindbladian $\mathcal{L}_o$ respectively. Thus, it yields two separate bases \cite{Bhattacharya:2023zqt, Bhattacharjee:2023uwx}
\begin{align} \mathrm{Kry}^j(\mathcal{L}_{o}^\dagger,\,|p_0\rrangle)&= \{|p_0\rrangle, \mathcal{L}_{o}^\dagger \, |p_0\rrangle, (\mathcal{L}_{o}^\dagger)^2 \,|p_0\rrangle, \ldots\}\,,   \\
	\mathrm{Kry}^j(\mathcal{L}_{o},\,|q_0\rrangle)&= \{|q_0\rrangle, \mathcal{L}_{o} \, |q_0\rrangle, \mathcal{L}_{o}^2 \, |q_0 \rrangle, \ldots \}\,,
\end{align}
and imposes the bi-orthonormality condition 
\begin{equation}\label{biortho}
	\llangle q_m|p_n \rrangle =\delta_{m,n}\,.
\end{equation}
The ``double braces'' notation indicates the bi-Lanczos vectors \cite{Bhattacharya:2023zqt}, derived by the vectorization principle satisfying the inner product \eqref{IP}.  Such a bi-orthonormality condition is typically encountered in the context of non-Hermitian Hamiltonians. In such scenarios, the eigenvectors corresponding to the non-Hermitian Hamiltonian do not exhibit orthogonality with respect to one another \cite{Brody_2014}. Unlike the Arnoldi iteration, the bi-orthonormality condition is imposed - the vector spaces are no longer individually orthonormal. This renders the Lindbladian in a tridiagonal form \cite{Bhattacharya:2023zqt, Bhattacharjee:2023uwx}
\begin{align}
	\mathcal{L}_{o}^{\dagger} \equiv \begin{pmatrix} a_{0}&b_{1}&&&&0\\c_{1}&  a_{1}& b_{2}&&&\\&c_{2}&\ddots&\ddots &&\\&&\ddots &a_{m-1} &b_{m}&\\&&&c_{m}&\ddots&
		\ddots\\0&&&&\ddots&\ddots\\\end{pmatrix}\,,
\end{align}
in contrast with the upper Hessenberg form obtained from the Arnoldi iteration. A similarity transformation also implies that $d_n := \sqrt{b_n c_n}$ can be regarded as generalized Lanczos coefficients for open systems \cite{Bhattacharya:2023zqt, NSSrivatsa:2023qlh}. This is seemingly equivalent to different versions of the bi-Lanczos algorithm, which provides a non-unique basis \cite{gruning, biLcode, sogabe2023krylov}. These three sets of coefficients $\{a_j\}, \{b_j\}$ and $\{c_j\}$ are recursively related by the following 
two sets of three-term recurrence relations \cite{gruning, Bhattacharya:2023zqt, Bhattacharjee:2023uwx}
\begin{align}
	\mathcal{L}_{o}^\dagger |p_{j}\rrangle &= b_{j} |p_{j-1}\rrangle + a_j |p_j \rrangle  + c_{j+1} |p_{j+1}\rrangle\,, \label{bilanczosbasic} \\
	\mathcal{L}_{o} |q_{j}\rrangle &= c^*_{j} |q_{j-1}\rrangle + a^*_j |q_j\rrangle + b^*_{j+1} |q_{j+1}\rrangle\,, \label{bilanczosbasic2}
\end{align}
where $*$ denotes complex conjugation. The bi-Lanczos algorithm produces both these coefficients and the two sets of bi-orthogonal vectors ${|p_j\rrangle}$ and ${|q_j\rrangle}$, which we describe below \cite{Bhattacharya:2023zqt, Bhattacharjee:2023uwx, biLcode}:

\begin{enumerate}
	\item \textbf{Initialization.}
	
	Let $|p_{-1}\rrangle = |q_{-1}\rrangle = 0$ and $a_{-1} = b_0 = c_0 = 0$. Also, let  $|p_0\rrangle = |q_0\rrangle \equiv |\mathcal{O})$, where $\mathcal{O}$ is the initial normalized operator.
	\item \textbf{Lindbladian action and bi-Lanczos coefficients.}
	
	For $j = 0, 1, \ldots$,  perform the following iterations:
	\begin{enumerate}
		\item Compute:  $|r_j\rrangle =\mathcal{L}_{o}^\dagger  |p_j\rrangle$, and $|s_j\rrangle =\mathcal{L}_{o} |q_j\rrangle$.
		
		\item Redefine the vectors:
		
		$|r_j\rrangle := |r_j\rrangle - b_{j} |p_{j-1}\rrangle$, and $|s_j\rrangle := |s_j\rrangle - c_{j}^{*} |q_{j-1}\rrangle$.
		
		\item  Evaluate the inner product: $a_j = \llangle q_j|r_j \rrangle$.
		
		\item Again, redefine the vectors:
		
		$|r_j\rrangle := |r_j\rrangle - a_j |p_{j}\rrangle$, and $|s_j\rrangle := |s_j\rrangle - a_j^{*} |q_{j}\rrangle$.
		
		\item Evaluate the inner product: $\omega_{j}=\llangle r_{j}|s_{j} \rrangle$.
		
		\item Evaluate the norm:  $c_{j+1}= \sqrt{|\omega_{j}|}$, and $b_{j+1} = \omega^*_{j}/c_{j+1}$.
		
		\item If $b_{j+1}\neq0$, then define the vectors:
		\begin{align*}
			|p_{j+1}\rrangle = \frac{|r_{j}\rrangle}{c_{j+1}}\,,~~ \mathrm{and} ~~ |q_{j+1}\rrangle =\frac{|s_{j}\rrangle}{b^*_{j+1}}\,.
		\end{align*}
		
		\item If required, perform the full orthogonalization (FO) procedure.
	\end{enumerate}
	\item Stop, if $b_k=0$ for some $k$.
\end{enumerate}

To summarize, the bi-Lanczos algorithm differs from the Arnoldi iteration in that the Krylov spaces are bi-orthonormal to each other, not orthonormal. This leads to a tridiagonal Lindbladian, unlike the Arnoldi iteration. However, both methods are equivalent in capturing the non-Hermiticity of the Lindbladian
and reduces to the usual Lanczos algorithm when dissipation is absent, and the Lindbladian becomes Hermitian \cite{parker2019}.

\begin{figure}[t]
   \centering
\includegraphics[width=0.65\textwidth]{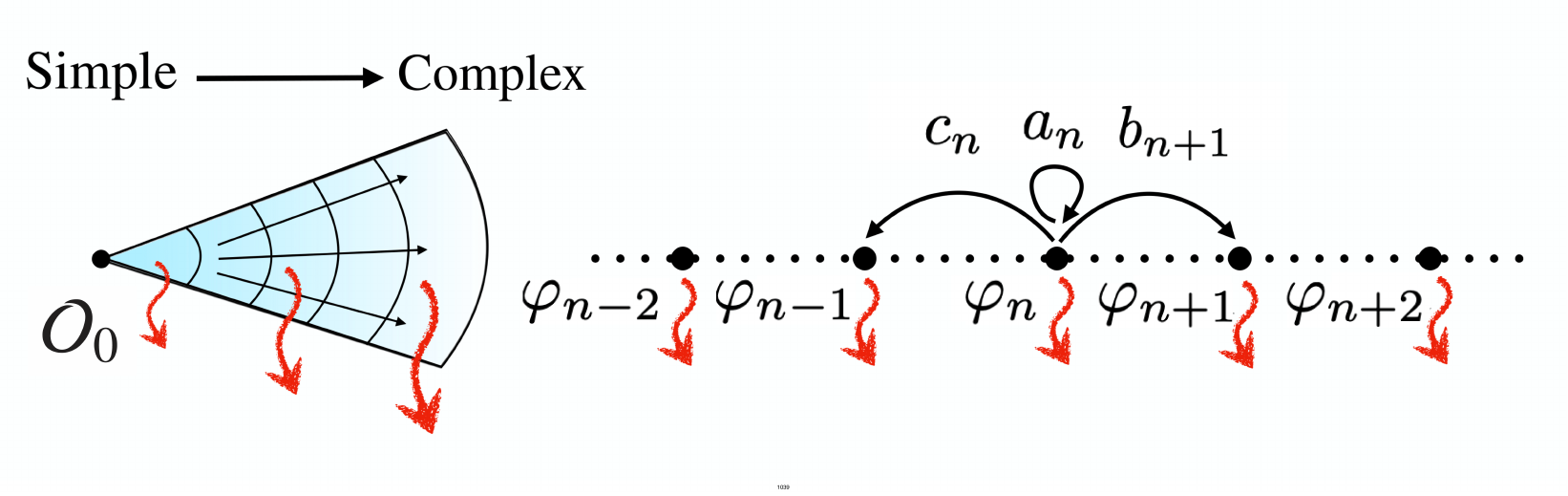}
\caption{A schematic diagram of the operator growth of the initial operator $\mathcal{O}_0$ in dissipative systems. The dissipative model is mapped to a non-Hermitian Krylov chain. The model parallels a particle-hopping problem from $n$-th side to the adjacent $(n+1)$-th and $(n-1)$-th sites with hopping amplitudes $b_{n+1}$ and $c_n = b_n$ respectively. Further, $a_{n}$ denotes the amplitude of staying at site $n$. The dissipation acts stronger as the operator grows, marked by the red arrows. The figure is taken from \cite{Bhattacharjee:2023uwx}.} \label{fig:nonHchain}
\end{figure}

\subsection{Structure of the Lindbladian}

Equipped with the generic sets of bi-Lanczos coefficients, we discuss the generic structure of the Lindbladian. The elements of the Lindbladian can be completely general complex numbers. However, the bi-Lanczos algorithm and the generic properties of Lindbladian (i.e., the eigenvalues of $ i \mathcal{L}_{o}^{\dagger}$ can either be non-positive real elements or complex conjugate in pairs \cite{Lucas2023PRX, albert_symmetries_2014}, with non-positive real parts) force the Lindbladian to take the following form \cite{Bhattacharya:2023zqt, Bhattacharjee:2023uwx}
\begin{align}
	\mathcal{L}_{o}^{\dagger} \equiv \begin{pmatrix} i|a_{0}|&b_{1}&&&&0\\b_{1}& i |a_{1}| & b_{2}&&&\\&b_{2}&\ddots&\ddots &&\\&&\ddots &i |a_{m-1}| &b_{m}&\\&&&b_{m}&\ddots&		\ddots\\0&&&&\ddots&\ddots\\\end{pmatrix}\,, \label{ld}
\end{align}
where $b_n = c_n \in \mathbb{R}_{+}$. In a more general setting, the off-diagonal coefficients can differ by a phase factor \cite{Bhattacharya:2023zqt}. The Lindbladian is written in the bi-orthonormal basis of the form $\llangle q_i | \mathcal{L}_{o}^{\dagger} | p_j \rrangle$. These off-diagonal coefficients are the same as the Lanczos coefficients for a closed system. The diagonal coefficients of \eqref{ld} break the Hermiticity of the Lindbladian $\mathcal{L}_{o}^{\dagger} \neq (\mathcal{L}_{o}^{\dagger})^{\dagger}$ (it is neither Hermitian nor anti-Hermitian), which is otherwise true in the absence of $a_n$. Further,
the imaginary part of any eigenvalue $\lambda_{\mathcal{L}}$ of the (adjoint) Lindbladian $\mathcal{L}_{o}^\dagger$ satisfies \cite{Bhattacharya:2023zqt}
\begin{align}
    \underset{n}{\mathrm{min}} \,\mathrm{Im}(a_n) \,\leq \,\mathrm{Im}(\lambda_{\mathcal{L}})\, \leq\, \underset{n}{\mathrm{max}} \,\mathrm{Im}(a_n)\,, \label{theorem}
\end{align}
where the equality trivially holds for a closed system. 

We briefly explain why the elements have this specific form \cite{Bhattacharya:2022gbz}. Consider the eigenvalue of the matrix $i \mathcal{L}_{o}^{\dagger}$, where $\mathcal{L}_{o}^{\dagger}$ is given by \eqref{ld}, with real $b_n$. The eigenvalue equation is
\begin{align}
       Q_n = \mathrm{det} \left(i \mathcal{L}_{o}^{\dagger} - \lambda \id\right) = 0\,, \label{co1}
\end{align}
where $Q_n$ is known as a specific version of the continuant of dimension $n$ \cite{muir2003treatise}, which is the determinant of a tridiagonal matrix of dimension $n$. The $n$-dimensional continuant can be obtained by its lower dimensional continuant from the following form \cite{Bhattacharya:2022gbz, muir2003treatise}
\begin{align}
    Q_n=(-|a_{n-1}|-\lambda)\, Q_{n-1}+b_{n-1}^2 Q_{n-2}\,,
\end{align}
with an initial condition $Q_0=1$ and $Q_1=-|a_0|-\lambda$. Setting $Q_n = 0$ from \eqref{co1}, we get a polynomial of $\lambda$ as
\begin{align*}
    \lambda^n + f_1 (\{|a_n|, b_n\}) \lambda^{n-1}  + \cdots +  f_{n} (\{|a_n|, b_n\}) = 0\,,
\end{align*}
where $f_k(\{|a_n|, b_n\})$ is a set of real functions in $\{|a_n|, b_n\}$. The complex conjugate roots theorem says for a polynomial equation with all real coefficients, if $x+iy$ is a root, then $x-iy$ is also a root. This includes real eigenvalues ($y = 0$), so the theorem means that eigenvalues $\lambda$ are
either real or come in complex conjugate pairs. This generically holds for any physical Lindbladian. Any real part in $a_n$ or any complex part in $b_n$ will violate this theorem. In other words, if the Lindbladian takes the form of \eqref{ld}, then $b_n$ has to be real, and $a_n = i |a_n|$ has to be purely imaginary. This provides a generic argument for the real and imaginary nature of the bi-Lanczos coefficients.

A time-evolved operator (evolved by $\mathcal{L}_o^{\dagger}$) can be written in the bi-Lanczos basis as
\begin{align}
    |\mathcal{O}(t)) = \sum_n i^n \varphi_{n} (t) \,|p_n\rrangle\,,
\end{align}
with $\varphi_n (t)$ denoting the Krylov basis wavefunctions. The Heisenberg equation of motion for the operator $d |\mathcal{O}(t))/d t = i \mathcal{L}_o^{\dagger} |\mathcal{O}(t))$ translates to \cite{Bhattacharya:2023zqt, Bhattacharjee:2022ave}
\begin{align}
    \partial_t \varphi_{n} (t) = b_{n} \varphi_{n-1} (t) + i a_n \varphi_{n} (t) - b_{n+1} \varphi_{n+1} (t)\,, \label{nonHtight}
\end{align}
for $n \geq 1$, with the boundary condition $\varphi_{-1}(t) =0$ and $\varphi_{n}(0) = \delta_{n,0}$. Here, $\varphi_0(t)$ is the standard autocorrelation function, defined as $\varphi_0(t) \equiv \mathcal{C}(\{\mu\}, t) = \frac{1}{2^N} \mathrm{Tr}(\mathcal{O}(t) \mathcal{O})$ with the time-evolved operator \eqref{oeq} for a system consisting of $N$ two level systems, in a similar way to \eqref{defauto}. Let us collectively denote $\{\mu\}$ as the set of the dissipative parameters. The similarity with Eq. \eqref{linH} is apparent, especially Eq. \eqref{nonHtight} is exactly equal to Eq. \eqref{linH} if only the diagonal terms are present in Eq. \eqref{linH}. In other words, the bi-Lanczos algorithm transforms the evolution dynamics to a non-Hermitian tight-binding model given by Eq. \eqref{nonHtight}; see Fig.\,\ref{fig:nonHchain}. Where now the dissipation is purely local on-site $n$ and not approximately local as in \eqref{linH}.
Starting from a particular site $n$, a particle hops to the $(n-1)$-th and $(n+1)$-th site with hopping rates given by $b_{n}$ and $b_{n+1}$, respectively. Moreover, the particle has the probability to stay at site $n$ with amplitude $a_n$. As the coefficients $a_n$ are purely imaginary, the tight-binding equation \eqref{nonHtight}  reduces to \cite{Bhattacharya:2023zqt, Bhattacharjee:2022ave}
\begin{align}
    \partial_t \varphi_{n}(t) = b_{n} \varphi_{n-1}(t) -  |a_n| \varphi_{n} (t) - b_{n+1} \varphi_{n+1} (t)\,,
\end{align}
with $n \geq 1$. Since the evolution is non-unitary and the basis $|p_n\rrangle$ is not orthonormal, the total probability $\sum_n  |\varphi_n (t) |^2 < 1$ is not conserved. However, we can define a modified probability $\tilde{\varphi}_n (t) = \varphi_n (t)/ \sqrt{\mathcal{Z}(t)}$, where $\mathcal{Z}(t) =  \sum_n  |\varphi_n (t) |^2$ is the total probability, acting as a normalization constant. This modified probability is conserved by definition ($\sum_n  |\tilde{\varphi}_n (t) |^2 = 1$) and the Krylov complexity in such setting equals  the average position of the particle in the non-Hermitian Krylov chain, i.e.,
\begin{align}
    K (t) =   \sum_n n |\tilde{\varphi}_n (t)|^2 = \frac{1}{\mathcal{Z}(t)} \sum_n n |\varphi_n (t)|^2 \,. \label{kcompopen}
\end{align}
Similarly, higher orders of Krylov cumulants can be defined, for example, the normalized variance~\cite{Bhattacharjee:2022ave, Bhattacharjee:2022lzy}, which is the second cumulant. We define it with respect to the normalized wavefunction
\begin{align}
    \Delta K (t)^2 :=
    \sum_n n^2 |\tilde{\varphi}_n (t)|^2 - \Big(\sum_n n |\tilde{\varphi}_n (t)|\Big)^2 
    =
    \frac{1}{\mathcal{Z}(t)} \sum_n |\varphi_n (t)|^2 (n - K(t))^2 \,. \label{kvar}
\end{align}
For any generic system, we are mostly interested in the first two cumulants - Krylov complexity and the Krylov variance. 

%In the next section, we focus on a particular class of models, the SYK model and its dissipative variants.

\subsection{Constraints imposed by Trace Preservation}

As discussed in Sec.\,\ref{secOpen}, the Lindbladian is the generator of a CPTP dynamical semigroup. This structure constrains the evolution so that the dynamics remain physical. The way that these constraints manifest in the Krylov representation of the Lindbladian poses a formidable question. Here we detail how Trace-Preservation poses constraints for the dynamics in the Krylov space. 

Trace preservation of Lindbladians imply that $\partial_t \Tr(\rho) = \Tr(\dot{\rho}) = \Tr(\mc L_o(\rho)) = 0$. In the Heisenberg picture it manifests as \emph{unitality}, i.e. $\Tr(\id \mc L_o(\rho)) = \Tr(\mc L_o^\dagger( \id)\rho)=0 \Leftrightarrow \mc L_o^\dagger (\id) = 0$. This condition can be written in the bi-orthonormal basis as $\mc L_o^\dagger | \id \rrangle = 0$, where the adjoint Lindbladian is now in the tridiagonal form \eqref{ld}. Using the bi-orthonormal resolution of the identity superoperator $\mc I = \sum_j |p_j \rrangle \llangle q_j| $ the expression of the vectorized identity matrix can be expressed in the bi-orthonormal basis as 
\begin{equation}
    |\id \rrangle = \frac{1}{d}\sum_j \Tr(q_j) |p_n \rrangle\,, 
\end{equation}
therefore the coefficients of the identity matrix in the bi-orthonormal basis are simply the traces of the left Krylov basis $\{|q_j \rrangle\}$.  Leveraging the recurrence relation \eqref{bilanczosbasic2} and the trace preservation of the Lindbladian $\Tr(\mc L_o (q_j))=0$ we find the recurrence relation for the traces 
\begin{equation}
    c_j^* \Tr(q_{j-1}) + a_j^* \Tr(q_j) + b_{j+1}^* \Tr(q_{j+1}) = 0.
\end{equation}
From the recurrence relation and the fact that $|q_0\rrangle = |\mc O)$ the traces of the first elements follow as
\begin{subequations}
\begin{align}
    \Tr(q_0) &= \Tr(\mc O), \\
    \Tr(q_1) &= - \frac{a_0^*}{b_1^*}\Tr(\mc O),\\
    \Tr(q_2) &= \left(\frac{a_1^* a_0^*}{b_2^* b_1^*} - \frac{c_1^*}{b_2^*}\right)\Tr(\mc O), \\
    \Tr(q_3) &= \left( \frac{c_2^* a_0^*}{b_3^* b_1^*} + \frac{a_2^* c_1^*}{b_2^* b_3^*} - \frac{a_2^* a_1^* a_0^*}{b_3^* b_2^* b_1^*} \right) \Tr(\mc O).
\end{align}
\end{subequations}
These expressions are very cumbersome. However, if a traceless operator $\Tr(\mc O)=0$ is chosen as a starting operator for the bi-Lanczos algorithm, the identity is the zero vector and the dynamics is \emph{unital}, and thus trace-preserving. In the following section, a traceless operator is always chosen. The restrictions imposed by complete positivity on the structure of the Lanczos coefficients and the associated dynamics in the bi-orthonormal Krylov basis remain an interesting open problem.

\section{Examples of open quantum dynamics in Krylov space} \label{Secopex}

\subsection{Dissipative SYK Model}

For illustration, we choose the SYK model \eqref{sykh} and its dissipative variants \cite{Kulkarni:2021gtt, sa_lindbladian_2022, KawabataDQPTSYK, Kawabata2023}. Such systems exhibit a profound parallel with non-Hermitian physics, wherein the SYK model is generalized to a non-Hermitian version \cite{DarioPTSYK, DarioSYK2, Garcia-Garcia:2023yet, Garcia-Garcia:2023uwh, GarciaGarcianonHSYK2}. A potential gravity dual has also been discussed \cite{GarciaKeldysh1}. The Arnoldi iteration and bi-Lanczos construction in spin chains in their respective integrable and chaotic limits \cite{AkemannProsen19} were also studied in detail \cite{Bhattacharya:2022gbz, Bhattacharya:2023zqt}. We consider the following two classes of dissipators.

\emph{Class 1: Linear dissipator.} When each fermion dissipates at an equal rate, the evolution is characterized by a linear Lindblad operator of the form~\cite{Kulkarni:2021gtt}:
\begin{align}
    L_{i} = \sqrt{\lambda} \,\psi_{i}\,, ~~~~~~ i = 1, 2, \cdots, N\,, \label{jumpop}
\end{align}
where $\lambda \geq 0$ is the dissipation strength associated with the interaction between the system and the environment. This model with the Hamiltonian \eqref{sykh} is analytically solvable in the large $q$ limit.\\

\emph{Class 2: Non-linear dissipator.} The 
most generic non-linear dissipators involve $p$-body Lindblad operators with a structure similar to the SYK Hamiltonian that can be written in the form~\cite{Kawabata2023}
\begin{align}
    L_a = \sum_{1 \leq i_1 < \cdots < i_p \leq N} V_{i_1 i_2 \ldots i_p}^a \,\psi_{i_1} \psi_{i_2} \cdots \psi_{i_p}\,,  \label{jpp0}
\end{align}
with $a =1, 2, \cdots, M$ and the random interaction $V_{i_1 i_2 \ldots i_p}$ satisfies the following distribution:
\begin{align}
    \langle V_{i_1 i_2 \ldots i_p}^a \rangle = 0\,, ~~ \langle |V_{i_1 i_2 \ldots i_p}^a|^2 \rangle = \frac{p!}{N^p}  V^2\,,~~ \forall i_1, \cdots, i_p, a\,, \label{jp}
\end{align}
with $V \geq 0$. The random average has to be taken for the ensemble of the interaction strength. It reduces to the linear dissipator without the random average and for $p =1$. For numerical purposes, we specifically consider the $p=2$ case. Let $M$ denote the number of jump operators in \eqref{jpp0}.  We take a special double-scaling limit $N, M \rightarrow \infty$ with $R = M/N$ being held constant for analytical purposes.

Due to the special structure of the large $q$ SYK model, the operator size concentration \eqref{eq:OSC} plays a key role. It is interesting to see that the strings of Majorana fermions act as an eigenstate of the dissipative part of the Lindbladian \eqref{negpos}, i.e.,
\begin{align}
  \mathcal{L}_D^{\dagger} (\psi_{i_1} \cdots  \psi_{i_s}) =
  \begin{cases}
    i \lambda s \,(\psi_{i_1} \cdots \psi_{i_s})       & ~~p = 1\,,\\
    i RV^2 \frac{ps}{2^{p-1}} (\psi_{i_1} \cdots \psi_{i_s}) \,   & ~~p > 1\,,  \label{dissact}
  \end{cases}
\end{align}
where the $p > 1$ limit is strictly valid in the large $q$ and large $N$ limit. Hence, the rate of annihilation by $\mathcal{L}_D^{\dagger}$ is proportional to the size $s$ of the operator, defined in Sec.\,\ref{secOpSize}, and the rate of dissipation $\propto \lambda$ or $\propto RV^2$ for single and $p>1$-body dissipator respectively. This distinctive characteristic, attributed to the operator size concentration, is a unique feature of the large $q$ SYK model.  It will be manifest in the diagonal Lanczos coefficients within the Lindbladian matrix, which we will explore in the forthcoming analysis.

\subsection{Analytical approach: moment method} \label{sykmomentmethod}

Section \ref{sykexample} outlined the use of the moment method in the SYK model.
Next, we present its generalization to the dissipative SYK. The autocorrelation function is given by \cite{Kulkarni:2021gtt}
\begin{align}
  &  \mathcal{C}(\tilde{\lambda},t) =  1 + \frac{1}{q} g(\tilde{\lambda},t) + O(1/q^2) \,, \label{eq:Ct} \\
  & g(\tilde{\lambda},t) =  \log \left[\frac{\alpha^2}{\mathcal{J}^2\cosh^2(\alpha t + \aleph)}\right] \,,\, ~~t > 0 \,, \label{ect1}
\end{align}
where $\alpha$ and $\aleph$ read
\begin{align}
     \alpha = \sqrt{(\tilde{\lambda}/ 2)^2 +  \mathcal{J}^2}\,, ~~~~ \aleph = \arcsinh (\tilde{\lambda}/ (2 \mathcal{J}))\,. 
\label{pardef}
\end{align}
Here, $\tilde{\lambda} = \lambda q$ is the dissipative parameter in the large $q$ limit. With no dissipation, $\lambda = 0$, the autocorrelation function reduces to \eqref{autoqsyk1}. The function $g(\tilde{\lambda},t)$ satisfies the following Liouville equation \cite{Kulkarni:2021gtt}
\begin{align}
    \partial_t^2 g(\tilde{\lambda},t)  = - 2 \mathcal{J}^2 e^{g(\tilde{\lambda},t) }\,,
\end{align}
with the boundary condition \cite{Kulkarni:2021gtt}
\begin{align}
    g(\tilde{\lambda},0) = 0\,,~~ ~~ g'(\tilde{\lambda},0)  = - \tilde{\lambda}\,.
\end{align}
They include and generalize the known result \cite{parker2019} with zero dissipation.
However, the autocorrelation function \eqref{eq:Ct} is not an even function as in \cite{parker2019}, and hence both even and odd moments exist. Specifically, one finds  in a $1/q$ expansion the moments
\begin{align}
    m_n = \frac{2}{q} \tilde{m}_n + O(1/q^2) \,,\,~~ n \ge 1\,,
\end{align}
where $\tilde{m}_n$ is a polynomial of $w := i \tilde\lambda$. For example, the leading moments are given by  \cite{Bhattacharjee:2022lzy}
\begin{align}
      \tilde m_1 &= w/2\,, \nonumber \\
      \tilde m_2 &= 1\,, \nonumber \\
      \tilde m_3 &= w \,, \nonumber \\
      \tilde m_4 &= w^2 + 2\,, \nonumber  \\
      \tilde m_5 &= w^3 + 8 w \,, \nonumber \\
      \tilde m_6 & = w^4 + 22 w^2 + 16 \,, \nonumber \\
       \tilde m_7 & = w^5 + 52 w^2 + 136 \,, \nonumber \\
        \tilde m_{8} & = w^6+114 w^4+720 w^2+272 \,.
\end{align}
For $n>1$, the moments are associated with the triangle ``$T(n,k)$'', and generated according to the following sequence \cite{A101280}
\begin{align*}
    T(n,k) = (k+1)T(n-1,k) + (2n-4k)T(n-1,k-1)\,,
\end{align*}
with $\lfloor \frac{n-1}{2} \rfloor \geq k \geq 0$ for each $n \geq 1$.  Alternatively, the moment $\tilde{m}_n$ is given by the number of Motzkin paths of length $n$ where $k$ of them are upsteps, 
\begin{align}
    \tilde{m}_n = \sum _{k=0}^{\lfloor \frac{n}{2}-1\rfloor} T(n-1,k)\, w^{n -2 k-2}\,,
\end{align}
for $n \geq 2$. Further, they can be written in terms of the continued fraction of the form \eqref{confrac} \cite{Bhattacharjee:2022lzy}. Applying the recursive algorithm \eqref{mombn}, the moments provide us two sets of Lanczos coefficients \cite{Bhattacharjee:2022lzy}
\begin{align}
&a_n = i \tilde{\lambda} n + O(1/q) \,,\,~~ \tilde{\lambda} := \lambda q \,,\, \label{eq:an-main} \\ &b_n =  \begin{cases} \mathcal{J}\sqrt{2/q}  \,      & \,n = 1\,,\\
    \mathcal{J}\sqrt{n(n-1)} + O(1/q) \,   & \,n > 1\,. \label{eq:bn-main}
  \end{cases} 
 \end{align}
Note that the coefficients $b_n$ are exactly equal to their closed-system counterparts and do not depend on the dissipation. In contrast, the coefficients $a_n$ are purely imaginary and linearly depend on the dissipation. Further, both coefficients grow linearly in $n$. We will come back to this point in detail.

%%%%%%%%%%%%%%%%%%% FIGURE  %%%%%%%%%%%%%%%%%%%%%%%%%%

\begin{figure}[t]
\centering
%\subfigure[]
{\includegraphics[width=0.48\textwidth]{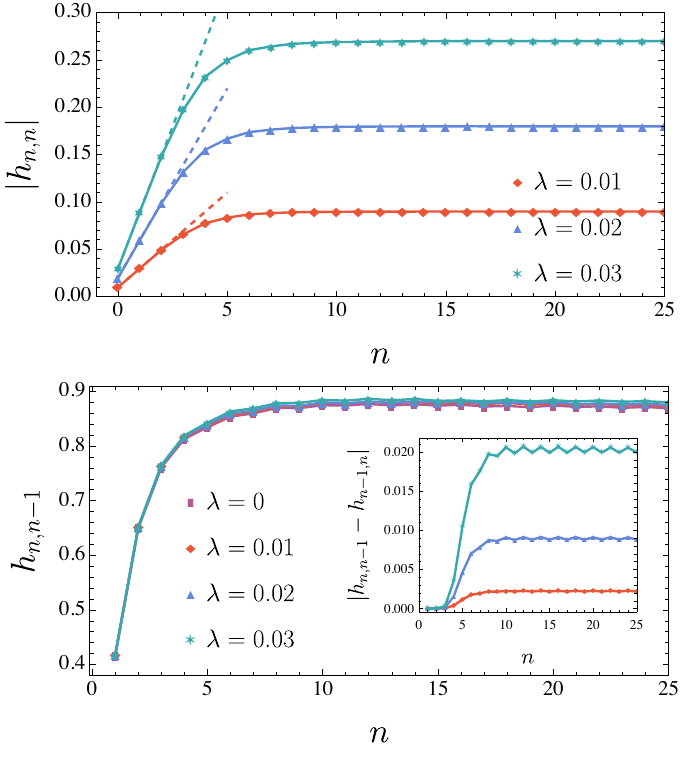}}\label{fig:Arnoldi}
\hfil
%\subfigure[]
{\includegraphics[width=0.48\textwidth]{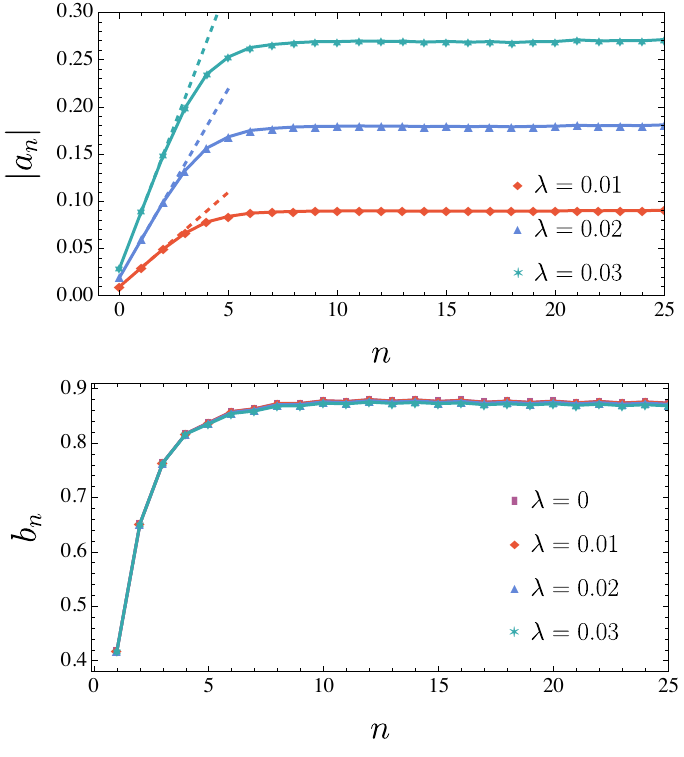}}
\caption{(Left, Top) The behavior of the diagonal coefficients $|h_{n,n}|$ from the Arnoldi iteration. The dashed line is the linear fit given by \eqref{asb}. (Left, Bottom) the lower primary off-diagonal coefficients $h_{n,n-1}$. The inset shows the difference between the magnitude of the lower and the upper primary off-diagonal elements. The total number of fermions is $N = 18$ (100 Hamiltonian realizations), the linear dissipator \eqref{jumpop}, and the initial operator $\mathcal{O} = \sqrt{2} \psi_1$. The figure is adapted from \cite{Bhattacharjee:2022lzy} with the change of parameters. (Right, Top) The behavior of the (top) diagonal coefficients $|a_{n}|$ for different dissipative strengths in the SYK$_4$ model with $N = 18$ fermions (100 Hamiltonian realizations), using the bi-Lanczos algorithm. The dashed line is the linear fit given by \eqref{asb2}. (Right, Top) The behavior of the primary off-diagonal coefficients $b_{n}$ for the same. We choose the linear dissipators given by \eqref{jumpop}.  We choose the initial operator as $\mathcal{O} = \sqrt{2} \psi_1$. The figure is adapted from \cite{Bhattacharjee:2023uwx} with the change of parameters.} \label{fig:biLanc}
\end{figure}

%\begin{figure}[t]
%   \centering
%\includegraphics[width=0.46\textwidth]{ArnoldiSYKN18.pdf}
%\caption{(Top) The behavior of the diagonal coefficients $|h_{n,n}|$ from the Arnoldi iteration. The dashed line is the linear fit given by \eqref{asb}. (Bottom) the lower primary off-diagonal coefficients $h_{n,n-1}$. The inset shows the difference between the magnitude of the lower and the upper primary off-diagonal elements. The total number of fermions is $N = 18$ (100 Hamiltonian realizations), the linear dissipator \eqref{jumpop}, and the initial operator $\mathcal{O} = \sqrt{2} \psi_1$. The figure is adapted from \cite{Bhattacharjee:2022lzy} with the change of parameters.} \label{fig:Arnoldi}
%\end{figure}

%%%%%%%%%%%%%%%%%%% FIGURE  %%%%%%%%%%%%%%%%%%%%%%%%%%

\subsubsection{Arnoldi iteration}

To appreciate the analytical findings in the previous section, let us implement the Arnoldi iteration in the dissipative SYK model with the Hamiltonian \eqref{sykh} and the linear dissipator \eqref{jumpop}. We vectorize the Lindbladian according to \eqref{lindsup} and choose the vectorized initial operator $\mathcal{O} = \sqrt{2} \psi_1$. For its numerical study, we choose $q=4$ and $N=18$ fermions. Figure \ref{fig:biLanc} (left) shows the behavior for the diagonal and primary off-diagonal elements of the Lindbladian in Arnoldi (Krylov) basis. The (imaginary) diagonal elements $|h_{n,n}|$  depend on the dissipation strength and grow linearly before saturation at $n \lesssim N/q$, which is dependent on the system size. The linear fit gives 
\begin{align}
    |h_{n,n}| = \lambda (2n + 1 )=  2 n \lambda + O(1)\,, \label{asb}
\end{align}
which is shown by the dashed line in Fig.\,\ref{fig:biLanc} (left, top). By contrast, the primary off-diagonal elements $h_{n,n-1}$ and $h_{n-1,n}$ are \emph{almost} independent of the dissipation and closely overlap with the closed system counterparts, see  Fig.\,\ref{fig:biLanc} (left, bottom). However,  they are not equal, i.e., $h_{n-1,n} \neq h_{n,n-1}$, primarily due to the presence of other off-diagonal elements $h_{m,n}$. Their small relative differences are shown in the inset of Fig.\,\ref{fig:biLanc} (left, bottom). Although the other off-diagonal elements $h_{m,n}$ are much smaller in magnitude compared to the diagonal and the primary off-diagonal elements, which dominate the Lindbladian matrix \eqref{Arnoldimatrix}, their presence makes it difficult to compute the Krylov complexity in general.

%%%%%%%%%%%%%%%%%%% FIGURE  %%%%%%%%%%%%%%%%%%%%%%%%%%

\subsection{Numerical approaches}

\subsubsection{Bi-Lanczos algorithm}

In this section, we apply the bi-Lanczos algorithm for the  Hamiltonian \eqref{sykh}, with the linear dissipator \eqref{jumpop}. We keep all the parameters the same as in the Arnoldi iteration. The bi-Lanczos algorithm generates two sets of coefficients, which are shown in Fig.\,\ref{fig:biLanc} (right). The diagonal coefficients increase linearly and are proportional to the dissipative parameter. The linear fit gives
\begin{align}
    |a_n| = \lambda \,(2n + 1) = 2 n \lambda + O(1)\,. \label{asb2}
\end{align}
This property is similar to the diagonal Arnoldi coefficients $|h_{n,n}|$, barring any $O(1)$ numbers, if any, which are insignificant in the asymptotic limit of $n$. All the upper and lower off-diagonal elements are the same and equivalent to the closed system counterparts; see Fig.\,\ref{fig:biLanc} (right, bottom). This contrasts with the Arnoldi iteration, where upper and lower off-diagonal elements differ. The upshot is that the Lindbladian in the bi-Lanczos basis is expressed in the purely tridiagonal form \eqref{ld} as discussed in Sec.\,\ref{bidisc}.

%\begin{figure}[t]
%   \centering
%\includegraphics[width=0.46\textwidth]{BiLanczosSYKN18.pdf}
%\caption{The behavior of the (top) diagonal coefficients $|a_{n}|$ and (bottom) the primary off-diagonal coefficients $b_{n}$ for different dissipative strength in the SYK$_4$ model with $N = 18$ fermions (100 Hamiltonian realizations), using the bi-Lanczos algorithm. We choose the linear dissipators given by \eqref{jumpop}. The dashed line in (top) is the linear fit given by \eqref{asb2}. We choose the initial operator as $\mathcal{O} = \sqrt{2} \psi_1$. The figure is adapted from \cite{Bhattacharjee:2023uwx} with the change of parameters.} \label{fig:biLanc}
%\end{figure}

Finally, we comment on the numerical stability of both approaches. Although Arnoldi iteration appears to be more stable than the bi-Lanczos algorithm, both show numerical instability in Lanczos coefficients in small system sizes (i.e., small $N$). Ideally, $N \geq 18$ shows reasonably stable Lanczos coefficients. Further, since the coupling of the SYK model is chosen randomly, ideally, a random average (multiple realizations) is required. We performed 100 Hamiltonian realizations in both the Arnoldi iteration and the bi-Lanczos method. Comparing these two methods, we find perfect agreement in the slope and the saturation. Multiple realizations are considerably significant for smaller system sizes due to the numerical instability of the Lanczos coefficients. Partial re-orthogonalization methods \cite{VANDERVEEN1995605} can also be employed.

\subsection{Krylov complexity in chaotic open quantum systems} \label{marchao}

Motivated by the extensive analytical and numerical studies in the SYK model, we propose that both sets of Lanczos coefficients show asymptotic linear growth
~\cite{Bhattacharjee:2022lzy}
\begin{align}
    a_n \sim i \chi \mu n\,, ~~~~ b_n = c_n \sim \alpha n\,. \label{asymab}
\end{align}
Here,  $\mu$ denotes the generic dissipative parameter, with $\mu \propto \lambda$ for the linear dissipator (class 1) and $\mu \propto R V^2$ for the generic $p$-body dissipator (class 2). The proportionality directly follows from \eqref{dissact}, as a consequence of the operator size concentration in the large $q$ SYK model. The parameter $\chi$ is independent of $n$ and the dissipative parameter. This provides a more generic operator growth hypothesis, at least for chaotic systems, which includes \cite{parker2019} as a specific case for the unitary systems.

\subsubsection{Continuum limit: large $n$ result}

To understand the behavior of the Krylov complexity, we first take a heuristic approach. We take the continuum limit by elevating the index $n$ to a parameter and denoting $\varphi_n (t) \equiv \varphi(n,t)$ as an Ansatz. The equation \eqref{nonHtight} with \eqref{asymab} can be written as \cite{Bhattacharjee:2022lzy}
\begin{align}
    \partial_t \varphi(n,t) + n \left(\chi \mu \varphi(n,t) + 2 \alpha \partial_n \varphi(n,t)\right) = 0\,,
\end{align}
where we have elevated $\varphi_{n-1}(t) = \varphi (n-1,t)$ and  $\varphi_{n+1}(t) = \varphi (n+1,t)$, and further used $\varphi (n+1,t) -  \varphi (n-1,t) = \varphi (n+1,t) - \varphi (n,t) + \varphi (n,t) -  \varphi (n-1,t) = 2 \partial_n \varphi(n,t)$. We further assume $b_{n+1} = b_n \equiv b(n)$, which is true in the asymptotic limit.\footnote{Otherwise, we have an extra $\partial_n b(n)$ term which can be added to $a(n)$ since both terms are proportional to $\varphi(n,t)$. For the linear growth $b(n) \propto n$, this extra term will add a constant to $a(n)$, which can be ignored in the asymptotic limit.} We look for a stationary solution (where $\partial_t \varphi = 0$) at $t \rightarrow \infty$, which is given by
\begin{align}
    \varphi_{*} (n, t \rightarrow \infty) \propto e^{-n/\xi}\,, ~~~~ \xi := \frac{2 \alpha}{\chi \mu}\,. \label{an2}
\end{align}
Using the above wavefunction, the late-time (stationary) Krylov complexity (after normalization using \eqref{kcompopen}) saturates to
\begin{align}
    K(t) = \frac{\xi}{2} + O(1/\xi) = \frac{\alpha}{\chi \mu} + O(\mu)\,, ~~~ t \rightarrow \infty\,,
\end{align}
which is valid in the weak dissipation regime $\xi \gg 1$ and inversely proportional to the dissipation strength. However, the early-time growth is exponential $K(t) \sim e^{2 \alpha t}$, with the time-scale for saturation $t_{*}$ estimated as 
\begin{align}
    e^{2 \alpha t_{*}} = \frac{\alpha}{\chi \mu}\,~~~ \Rightarrow ~~~ t_{*} = \frac{1}{2 \alpha} \ln \left(\frac{\alpha}{\chi \mu}\right)\,.
\end{align}
%We refer to it as a dissipative time scale. In other words, 
This dissipative time scale varies logarithmically with the inverse of the dissipation strength.

\subsubsection{Exact results}

%We try to see how it matches with the exact calculation of Krylov complexity. 

Equipped with the Lanczos coefficients \eqref{asymab}, the non-Hermitian tight-binding model \eqref{nonHtight} is solved to obtain the basis wavefunctions. To do this, we assume a specific form of the coefficients \cite{Bhattacharjee:2022lzy, Bhattacharjee:2023uwx}
\begin{equation}\label{eq:anbnexact}
    b_n^2 = \gamma^2(1-u^2) n (n-1+\eta) \,, ~~~~a_n = i u \gamma (2n + \eta) \,,
\end{equation}
where $u \in (0,1)$ and $\eta \sim O(1)$ number. Equation \eqref{asymab} is recovered as a particular choice
\begin{align}
    \alpha^2 = \gamma^2(1-u^2) \,,\,~~~~ \chi \mu = 2 \gamma u \,,  \label{ag}
\end{align}
in the asymptotic limit of $n$. Using \eqref{eq:anbnexact}, the non-Hermitian tight-binding model \eqref{nonHtight} can be exactly solved, and the solution is given by
\cite{Bhattacharjee:2022lzy}
\begin{align}
    \varphi_n(t) =   \frac{ \sech(\gamma t)^\eta }{(1 + u \tanh(\gamma t))^\eta}  
     (1 - u^2)^{\frac{n}2}  \sqrt{\frac{(\eta)_n}{n!}}    \left( \frac{\tanh (\gamma t)}{1 +u \tanh(\gamma t)} \right)^n \,. \label{kwave}
\end{align}
Although the appearance of $u$ in both $b_n$ and $a_n$ is more subtle, curiously, the form of the wavefunction is controlled by the underlying $\mathrm{SL}(2,\mathbb{R}$) symmetry \cite{caputa2021, balasubramanian2022}. It can be easily checked that the probability is not conserved in general
\begin{align}
   \mathcal{Z}(t) = \sum_n |\varphi_n (t)|^2 = \left(u (u \cosh (2 \gamma t)+\sinh (2 \gamma t))-u^2+1\right)^{-\eta } \,.
\end{align}
It is interesting to perform an early and late-time asymptotic analysis of the wavefunction \eqref{kwave}. 
%We first consider at late times $t \rightarrow \infty$, and later at finite time $\gamma t \gg 1$. 
At late times $t \rightarrow \infty$, 
%\eqref{kwave} becomes
\begin{align}
    \varphi_n(t \rightarrow \infty) \simeq \left(\frac{\sqrt{1-u^2}}{1+u}\right)^n n^{\frac{\eta-1}{2}}\,, 
\end{align}
where we have only kept terms involving terms with $n$ and neglected any other terms, such as those involving $\eta$. This is justified since we focus on the asymptotic limit of $n$. We also used the asymptotic expansion $\Gamma(n+\eta)/\Gamma(n) \sim n^{\eta}$ for $n \rightarrow \infty$. Since $\log((1+u)/\sqrt{1-u^2}) = u + O(u^3)$, we can readily see \cite{Bhattacharjee:2022lzy}
\begin{align}
    \varphi_n(t \rightarrow \infty) \sim e^{-n/\xi(u)}\, n^{\frac{\eta-1}{2}}\,, ~~ \xi(u)^{-1} = u + O(u^2)\,.  \label{as1}
\end{align} 
This correctly reproduces the stationary state ansatz solution in \eqref{an2}, with $\eta = 1$ and $\xi(u) = 1/u \propto \mu/\gamma$. 

The finite time limit with  $\gamma t \gg 1$ is more involved. One can repeat the analysis and find \cite{Bhattacharjee:2022lzy}
\begin{align}
\begin{split}
     &\varphi_n(t) \sim e^{-n/\xi(u,t)} \,n^{\frac{\eta-1}{2}}\,, \\
     &\xi(u,t)^{-1} = u + 2 e^{-2 \gamma t} + O(e^{-4 \gamma t}, e^{-2 \gamma t} u, u^2)\,, \label{del2}
\end{split}
\end{align}
which agrees with the previous result \eqref{as1} as well as the zero dissipation result in \eqref{del1}. The delocalization length $\xi(u,t)$ captures both the spreading of the operator and the dissipation. Equating its first and second terms provides the saturation timescale $t_{d} \sim \log(1/u)$, which grows logarithmically.

\begin{figure}[t]
\centering
%\subfigure[]
{\includegraphics[height=0.3\textwidth,width=0.44\textwidth]{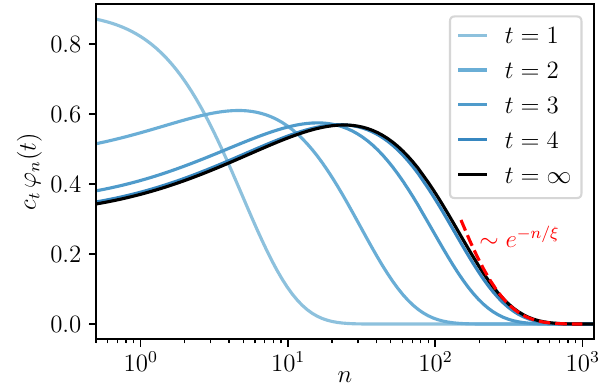}}\label{fig:wavefunc}
\hfil
%\subfigure[]
{\includegraphics[height=0.305\textwidth,width=0.44\textwidth]{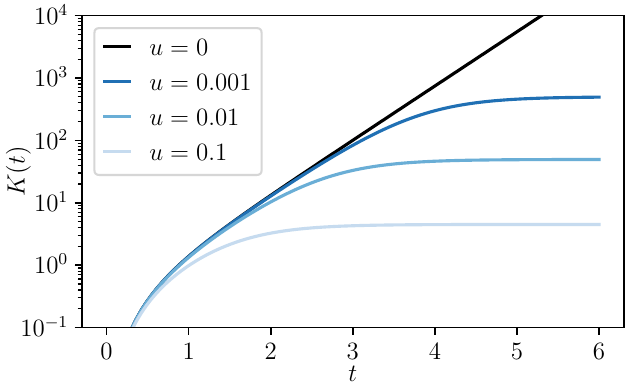}}
\caption{(Left) Snapshots of the Krylov wavefunctions \eqref{kwave}, obtained from the exact solution of the Lanczos coefficients \eqref{eq:anbnexact}, with rescaling of the wavefunctions by a $t$-dependent constant $c_t$. We choose $\eta = 1.5$ and $u = 0.01$. Although the wavefunctions are defined for integer values of $n$ only, they are interpolated for non-integer values of $n$. At late times, $t \rightarrow \infty$, the wavefunction profile reached a stationary exponential tail $\propto e^{-n/\xi(u)}$, where $\xi(u) = 1/u \propto \mu/\gamma$. This is indicated by the red dashed curve. The figure is taken from \cite{Bhattacharjee:2022lzy}. (Right) The behavior of Krylov complexity \eqref{kcplot} for different dissipation strengths. The black dashed line indicates the behavior of $\sim e^{2t}$ (for $u=0$) of a closed system. We choose $\eta = 1.5$. The figure is taken from \cite{Bhattacharjee:2022lzy}.} \label{fig:kcomp}
\end{figure}

The exact wavefunction \eqref{kwave} allows us to compute the Krylov complexity \eqref{kcompopen} exactly. It is given by
\cite{Bhattacharjee:2022lzy, Bhattacharjee:2023uwx}
\begin{align}
    K(t) = \frac{\eta  \left(1-u^2\right) \tanh ^2(\gamma t)}{1+2 u \tanh (\gamma t)-\left(1-2 u^2\right) \tanh ^2(\gamma t)}\,. \label{kcplot}
\end{align}
and shown in Fig.\,\ref{fig:kcomp} for different dissipation strengths. In the weak coupling limit, it reduces to
\begin{align}
    K(t) = \eta \left[ \sinh^2(\gamma t)-2 u  \sinh^3(\gamma t) \cosh (\gamma t) + O(u^2)\right] \,. \label{kweak}
\end{align}
Without dissipation (i.e., $\mu = 0~\mathrm{or}~u = 0$), when $\gamma = \alpha$, the exponential growth $K(t) \sim \eta e^{2 \alpha t}$ is recovered. The higher order terms in $u$ in \eqref{kweak} are responsible for weak dissipation, and the approximation becomes invalid when the first and second terms are comparable. In other words, it gives a time-scale $t_d$ when the dissipative regime begins. This happens at \cite{Bhattacharjee:2022lzy}
\begin{align}
    t_d = \frac{1}{2 \gamma} \sinh^{-1}(1/u) \sim \frac{1}{2 \alpha} \log\left(\frac{4 \alpha}{\chi \mu}\right) + O(\mu)\,,
\end{align}
where in the second inequality, we have used $\sinh^{-1}(1/u) = \log(2/u) + O(u)$ and kept terms up to $O(u)$. This allows us to set $\gamma = \alpha$ and $2/u = 4 \alpha/(\chi \mu)$ at $O(u)$ using \eqref{ag}. Thus, we recover the dissipative scale $t_d = t_{*}$ from the general argument and also from the generic delocalization length $\xi(u,t)$.

%\begin{figure}[t]
%\includegraphics[width=0.44\textwidth]{wavefunc.pdf}
%\caption{Snapshots of the Krylov wavefunctions \eqref{kwave}, obtained from the exact solution of the Lanczos coefficients \eqref{eq:anbnexact}, with rescaling of the wavefunctions by a $t$-dependent constant $c_t$. We choose $\eta = 1.5$ and $u = 0.01$. Although the wavefunctions are defined for integer values of $n$ only, they are interpolated for non-integer values of $n$. At late times, $t \rightarrow \infty$, the wavefunction profile reached a stationary exponential tail $\propto e^{-n/\xi(u)}$, where $\xi(u) = 1/u \propto \mu/\gamma$. This is indicated by the red dashed curve. The figure is taken from \cite{Bhattacharjee:2022lzy}.} \label{fig:wavefunc}
%\end{figure}

On the other hand, the late-time value for the Krylov complexity at fixed dissipation $u > 0$ is given by \cite{Bhattacharjee:2022lzy}
\begin{align}
    K(t\to\infty) = \frac{\eta}{2 u} - \frac{\eta}{2}  \,,
\end{align}
which is a constant independent of the initial growth parameter $\gamma$ (or $\alpha$) but only depends on the dissipation $u$.

Hence, the generic arguments are consistent with the calculation. We also obtain two quantities, namely the dissipative timescale $t_d$ and the saturation value of the Krylov complexity $K_{\mathrm{sat}}$, which show universal aspects of this behavior \cite{Bhattacharjee:2022lzy, Bhattacharjee:2023uwx}
\begin{align}
t_{d} \sim \frac{1}{\gamma} \log(1/u)\,,~~~~~~ K_{\mathrm{sat}} \sim 1/u\,. \label{kfinal}
\end{align}
The dissipative timescale resembles the logarithmic timescale of scrambling, implying that the dissipative strength acts as an effective degree of freedom. The saturation value is independent of the system size and thus generically holds in the thermodynamic limit. The saturation plateau appears to be generic in other notions of operator growth, namely operator size and OTOC \cite{schuster_operator_2022, Bhattacharjee:2022lzy, Liu:2024stj}. We propose these quantities are robust for any generic all-to-all quantum chaotic systems.

It is interesting to compute the normalized variance \eqref{kvar}, which is given by \cite{Bhattacharjee:2022lzy, Bhattacharjee:2023uwx}
\begin{align}
   \Delta K (t)^2 = \,\frac{\eta\left(1-u^2\right) \tanh^2(\gamma t) (u \tanh (\gamma t)+1)^2}{\left(1+ 2 u \tanh (\gamma t)- \left(1-2 u^2\right) \tanh^2(\gamma t)\right)^2} \,.
\end{align}
While it behaves as $(\eta/4) \sinh^2(2 \gamma t) \sim \eta e^{4\gamma t}$ in the growth regime, it saturates at $\eta / (4u^2)$ at late times. In either case, we find
\begin{align}
    \Delta K (t) \sim K(t)\,, \label{kck}
\end{align}
i.e., the standard deviation of the Krylov complexity is comparable to its average, indicating a broad distribution. This is due to the ``operator size concentration'',  discussed earlier. Incidentally, this property was also found to be true in all-to-all random unitary circuits \cite{schuster_operator_2022}. It is further interesting to note the identity
\begin{align}
    \partial_t \log \mathcal{Z}  (t) = - 2 u \gamma (2 K(t) + \eta)\,,
\end{align}
resembling the equality of the Loschmidt fidelity and operator size \cite{schuster_operator_2022}. For a closed system, $\mathcal{Z}(t) = 1$, and $u =0$, and the above equation trivially holds.

To recover the large-$q$ SYK result, using \eqref{eq:an-main}-\eqref{eq:bn-main}, \eqref{asymab} and \eqref{ag}, we identify the following \cite{ Bhattacharjee:2022lzy}
\begin{align}
    \eta = \frac{2}{q}\,,~~~\mathcal{J}^2 = \gamma^2 (1-u^2)\,~~~2 \gamma u = \tilde{\lambda}\,,
\end{align}
in the $O(1/q)$ expansion. This implies that the leading order in $q$, the Krylov complexity, and its variance in the SYK model vary as $\propto 1/q$  with the expected growth;  see \eqref{cc} and \eqref{sykv0} with $\gamma = \alpha = \mathcal{J}$),  already obtained before.

%\begin{figure}[t]
%   \centering
%\includegraphics[width = 8.3cm, height=5.2cm]{Kcomp.pdf}
%\caption{The behavior of Krylov complexity \eqref{kcplot} for different dissipation strengths. The black dashed line indicates the behavior of $\sim e^{2t}$ (for $u=0$) of a closed system. We choose $\eta = 1.5$. The figure is taken from \cite{Bhattacharjee:2022lzy}.} \label{fig:kcomp}
%\end{figure}

It is interesting to seek a physical interpretation of the plateau structure and the dissipative timescale in Fig.\,\ref{fig:kcomp}, which appears to hold for generic all-to-all systems. This is intuitive from the operator growth perspective. Since the dissipation strength is linear in the operator size, the dissipation acts stronger as the operator grows. 
The scrambling rate balances the dissipative strength at a timescale logarithmic in the dissipative strength, leading to the observed plateau. In an infinite system, scrambling persists indefinitely, preventing any halt in operator growth. The Markovian approximation (weak dissipation limit) suggests that dissipation alone is insufficient to reduce operator size in such cases. However, in systems of finite size, the size of operators can diminish when dissipation is weak, particularly at later times.

We also provide an intuitive understanding of the observed plateau from the perspective of quantum measurement.  A notable parallel is drawn between Eq. \eqref{ms2} and the Lindblad equation, Eq. \eqref{st4}, where the jump operators undertake a measurement-like role. Essentially, the environment conducts a continuous measurement with an indeterminate outcome.
%- unlike the typical deterministic measurement. 
Since the quantum measurement is a non-unitary operation, it steers the system away from its unitary trajectory. Hence, given that the measurement rate is analogous to the dissipation strength, an increase in dissipation thwarts the exponential growth typically seen in unitary evolution. Hence, the observed plateau solely results from the measurement process of the unknown environment. Similar findings for the Lyapunov exponent in the dissipative SYK model have been reported \cite{Liu:2024stj, Garcia-Garcia:2024tbd}.

At low temperatures and near equilibrium, the dissipative Sachdev-Ye-Kitaev (SYK) model, equipped with linear jump operators \eqref{jumpop}, adheres to a set of equations of motion reminiscent of two coupled non-Hermitian SYK models connected by a Keldysh wormhole \cite{GarciaKeldysh1}. This type of wormhole is distinct from the Euclidean or traversable wormholes typically discussed in the context of AdS/CFT correspondence \cite{Gaowormhole2017, Maldacena:2018lmt, PhysRevD.103.046014}. The dissipative parameter $\mu$ mimics the coupling between two sides of SYK and possibly relates to the length of the wormhole \cite{Bhattacharjee:2022lzy}. The interplay between scrambling and decoherence leads to a plateau, and dissipation intensifies the coupling across the wormhole's bifurcations. A deeper exploration of these phenomena, particularly the dual geometry of \eqref{jumpop} and the generic random Lindbladian \eqref{jpp0} presents an exciting avenue for future research.

%\subsection{Dissipative chaos bound}

\subsection{Pole structure of autocorrelation and spectral density}

In Section \ref{sec:moment1}, we observed that the pole structure of the autocorrelation function has a direct correlation with its growth dynamics. In particular, assuming smooth behavior of the Lanczos coefficients, the pole nearest to the origin plays a crucial role in dictating the expansion of the Lanczos coefficients and, consequently, the Krylov complexity. Given this relationship, it becomes pertinent to investigate the alterations in the pole structure under the influence of dissipation. This is especially relevant when considering the diagonal coefficients of the Lindbladian, denoted by ${a_n}$. We consider the following hypothetical autocorrelation function \cite{Bhattacharjee:2023uwx}
\begin{align}
    \mathcal{C}(\mu,t) = \frac{\sqrt{\alpha^2+\mu^2}}{\alpha}\,\text{sech}\left( t \sqrt{\alpha^2 + \mu^2} + \sinh^{-1}(\mu/\alpha) \right)\,, \label{poleautodiss}
\end{align}
where $\alpha$ is the parameter governing the growth of $b_n$, while $\mu$ represents the dissipation factor. It is evident that in the absence of dissipation, Eq. \eqref{poleautodiss} simplifies to $\mathcal{C}(0,t)= \text{sech}(\alpha t)$, aligning with the closed system scenario in \eqref{bb}. The nearest pole to the origin of the autocorrelation function is situated along the imaginary $t$ axis (Fig.\,\ref{fig:autofig}). Given that the autocorrelation function is normalized to unity at $t=0$, we can employ the recursive algorithm \eqref{mombn} to obtain
\begin{align}
\begin{split}
        a_n = i\mu (2n+1) \sim i \mu \chi n\,, ~ n\geq 0\,,~~
    b_n = \alpha n \,, ~ n\geq 1\,. \label{abd}
\end{split}
\end{align}

\begin{figure}[t]
   \centering
\includegraphics[width=0.72\textwidth]{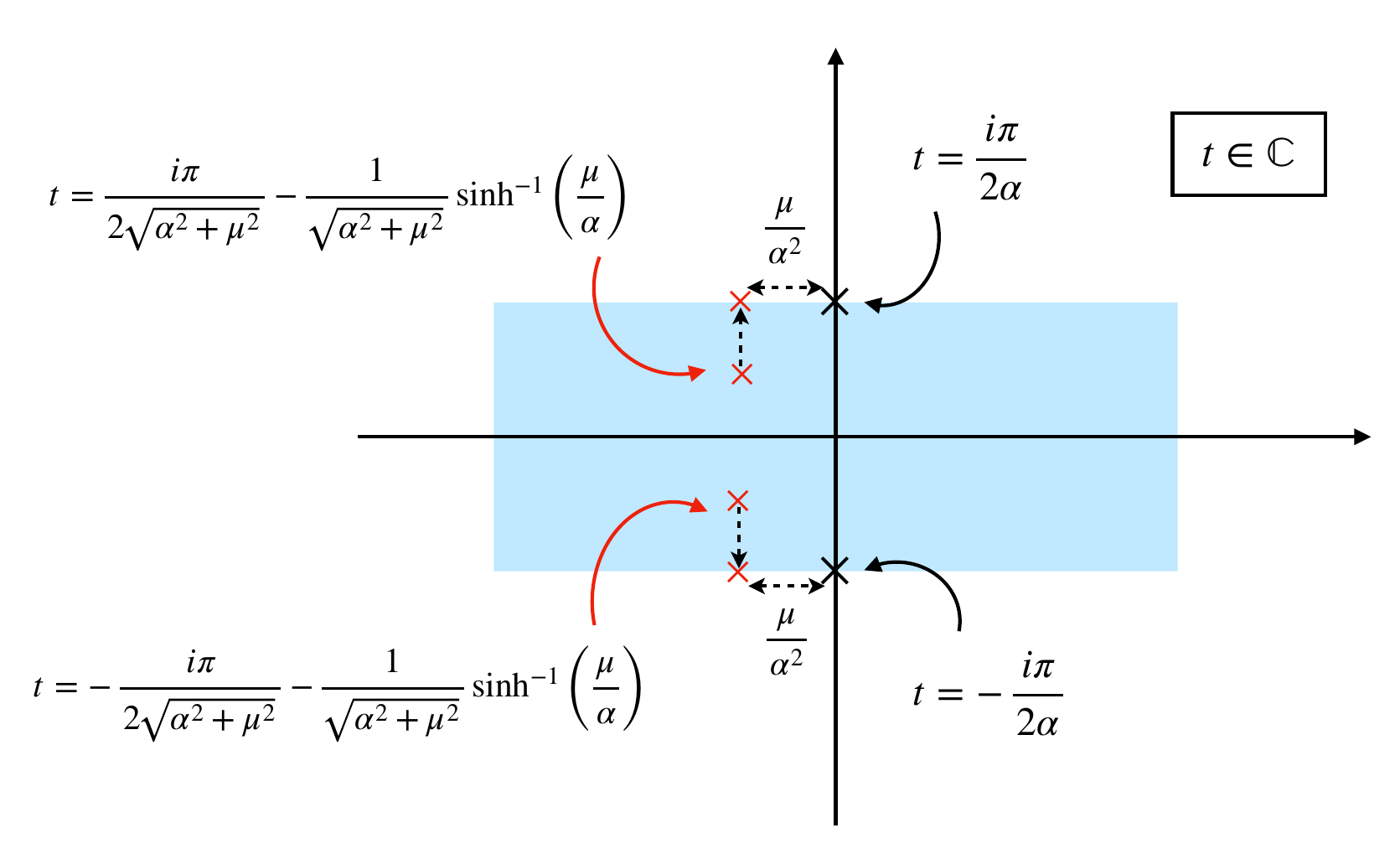}
\caption{The pole structure of the autocorrelation function \eqref{poleautodiss} showing the universal growth \eqref{asymab}. The red crossed marks indicate the pole in generic dissipation strength, which shifts to the negative real axis (considering $\mu \geq 0$) in the weak dissipation limit. In such a limit, the blue-shaded region shows the analytic region of the autocorrelation function. The vertical distance (along the $y$ axis) of the pole from the origin governs the growth of $b_n$ while the horizontal distance (along the $x$ axis) governs the growth of $a_n$ in this limit. The smooth behavior of the Lanczos coefficients is assumed. The diagram is taken from \cite{Bhattacharjee:2023uwx}.} \label{fig:poleshifted}
\end{figure}

The above coefficients can also be obtained by applying the Toda chain method in Sec.\,\ref{todadiscussion} upon the replacement $t \rightarrow - i \tau$, and setting $\tau_0 = 0$. These expressions represent the asymptotic limits of the Lanczos coefficients, as hypothesized in \eqref{asymab} in the weak dissipation limit. However, it is important to note that the autocorrelation function \eqref{poleautodiss} is not the sole function yielding these coefficients; any function $f(\mu, t)$ satisfying $f(0,t) = f(\mu,0) = 0$ 
%and suitably chosen $f(\mu,t)$ 
will also suffice. The poles of the proposed autocorrelation function \eqref{poleautodiss} are described by \cite{Bhattacharjee:2023uwx}
\begin{align}
    t_{\pm} = \pm \frac{i \pi}{2\sqrt{ \alpha^2 +\mu^2}} - \frac{1}{ \sqrt{ \alpha^2 +\mu^2}}  \sinh^{-1}\left(\frac{\mu}{\alpha}\right)\,. \label{autopolediss}
\end{align}
Remarkably, these findings are valid for any value of $\mu$, not just small ones, as the derivation of \eqref{abd} did not rely on a small $\mu$ approximation. Nonetheless, for the sake of continuity with the discussions on Markovian dissipation in Sec.\,\ref{marchao}, we consider $\mu$ to be small, leading to the poles \cite{Bhattacharjee:2023uwx}
\begin{align}
     t_{\pm} = \pm \frac{i \pi}{2 \alpha} - \frac{\mu}{ \alpha^2} + O(\mu^2)\,. \label{mmo}
\end{align}
The pole structure, as depicted in Fig.\,\ref{fig:poleshifted}, reveals a discernible shift of the closest poles along the negative real axis when weak dissipation is present, while their distance from the imaginary axis (assuming $\mu \geq 0$) mirrors that of the closed system. Thus, the distance to the imaginary axis dictates the growth of $b_n$, and the lateral shift indicates the growth of $a_n$. In cases of more generic dissipation, not necessarily small, the poles seem to affect the growth of $b_n$ through a diagonal shift (a combined shift in both $x$ and $y$ axis). However, the autocorrelation function considered here is a simplified model meant to illustrate the general hypothesis in \eqref{asymab}. Calculating an exact autocorrelation function under generic dissipation would require delving into non-Markovian dynamics \cite{breuer_2016_NonMarkov, Rivas_2014_NonMarkov, Li_2021_NonMarkov,deVega2017}, which falls beyond the scope of the present discussion.

\begin{figure}[t]
   \centering
\includegraphics[width=0.5\textwidth]{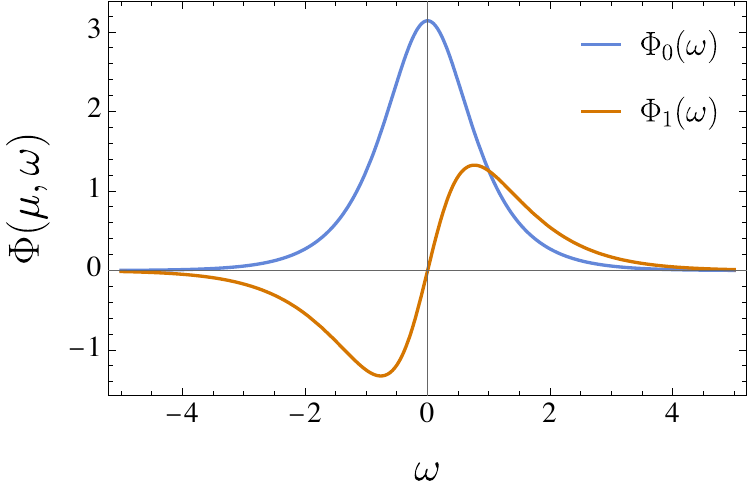}
\caption{Behavior of the leading term and the subleading term of the spectral function in weak dissipation regime. The leading term shows an exponential decay $\exp(-\#|\omega|)$ while the subleading term exhibits product-exponential decay of the form $\omega \exp(-\#|\omega|)$ in the high-frequency regime, according to \eqref{ledo1}-\eqref{ledo2}. We assume the smooth behavior of the Lanczos coefficients and take $\alpha =1$ for the figure. The figure is adapted from \cite{Bhattacharjee:2023uwx}.} \label{fig:specdiss}
\end{figure}

The pole structure of the autocorrelation function influences the decay profile of the spectral function. In the presence of dissipation, the generalized expression for the spectral function becomes
\begin{align}
    \Phi(\mu,\omega) = \int_{-\infty}^{\infty} dt \;  e^{- i \omega t} \,\mathcal{C}(\mu,t) \,,
\end{align}
where $\mathcal{C}(\mu,t)$ is the autocorrelation function. Utilizing the specific autocorrelation function \eqref{poleautodiss} under consideration,  the spectral function is found to be  \cite{Bhattacharjee:2023uwx}
\begin{align}
  \Phi(\mu, \omega) &=    \frac{\pi}{\alpha} \text{sech}\left(\frac{\pi  \omega }{2 \sqrt{\alpha^2+\mu^2}}\right) e^{\frac{i \omega }{\sqrt{\alpha^2+\mu^2}} \sinh^{-1}\left(\frac{\mu }{\alpha }\right)} \,,\label{mufreqBB}
\end{align}
which holds for generic $\mu$, beyond the small dissipation approximation.  In the absence of dissipation ($\mu = 0$), this equation reduces to the closed system result. However, when considering weak dissipation and expanding to the first order in $\mu$, one obtains
\begin{align}
    \Phi(\mu, \omega)\Big|_{\mu \rightarrow 0} = \Phi_0(\omega) + i \mu  \, \Phi_1(\omega) + O(\mu^2)\,, \label{ldn}
\end{align}
where $\Phi_0(\omega)$ represents the closed-system spectral function for $\mu = 0$, and $\Phi_1(\omega)$ corresponds to the first-order correction in $\mu$. These terms are explicitly expressed as \cite{Bhattacharjee:2023uwx}
\begin{align}
    \Phi_0(\omega) &= \frac{\pi}{\alpha} \sech\left(\frac{\pi \omega}{2 \alpha}\right) \sim \frac{\pi}{\alpha}e^{-\pi|\omega|/(2\alpha)}\,, \label{ledo1}\\
    \Phi_1(\omega) &= \frac{\pi \omega}{\alpha^3} \sech\left(\frac{\pi \omega}{2 \alpha}\right) \sim \frac{\pi \omega}{\alpha^3} e^{-\pi|\omega|/(2\alpha)} \,, \label{ledo2}
\end{align}
with the latter equations indicating the high-frequency behavior. Interestingly, while the leading term exhibits an exponential decay, the subleading term decays as a product-exponential function. Despite this, the overall decay remains exponential even within the weakly dissipative regime, as illustrated in Fig.\,\ref{fig:specdiss}. In all cases, the smooth behavior of the Lanczos coefficients is assumed. A parallel approach to studying the spectral function in a large $q$ SYK model has been presented in \cite{NSSrivatsa:2023qlh}.

\section{Krylov complexity in Quantum Field Theories and Holography} \label{QFTholo}

\subsection{Krylov complexity in Quantum Field Theories} \label{QFT}

In this section, we briefly discuss the Krylov space method in quantum field theory, focusing on the particular case of conformal field theories (CFT) \cite{Dymarsky:2021bjq} and simple free and holographic models \cite{Dymarsky:2021bjq,Avdoshkin:2022xuw}. Conformal field theories are special due to their scale-invariant properties and conjecturally dual to the gravitational theory via the AdS/CFT correspondence. The integrable and the chaotic properties of such CFTs are under active investigations \cite{Avdoshkin:2022xuw, Camargo:2022rnt, Kundu:2023hbk, Vasli:2023syq, Khetrapal2023, ArpanBMSKrylov, Malvimat:2024vhr, Iizuka:2024ldf, Chattopadhyay:2024pdj}.

These questions can be tackled using Krylov subspace methods. Because of the infinite number of degrees of freedom, generating an orthonormal basis directly may not be illuminating. However, we can resort to an alternative approach, namely, the moment method and the Toda chain technique, as discussed in Sec.\,\ref{todadiscussion}. This is possible because the autocorrelation function can be computed in many theories exactly due to conformal symmetry. A UV or an IR cutoff may be required in certain cases, or a theory should be compactified on a thermal circle \cite{Avdoshkin:2022xuw, Camargo:2022rnt}. The starting point is the finite-temperature Wightmann two-point autocorrelation function, defined with the help of the  ``Wightmann'' symmetric inner product 
\begin{align}
    \langle  \mathcal{O}  (t)\, \mathcal{O} \rangle_{\beta}^{(\mathrm{W})} = \frac{\mathrm{Tr}(e^{-\beta H/2} \, \mathcal{O}  (t) \, e^{-\beta H/2}\,   \mathcal{O})}{\mathrm{Tr}\,e^{-\beta H}} = \frac{\mathrm{Tr}(e^{-(\frac{\beta}{2} - it) H} \,  \mathcal{O} \,e^{-(\frac{\beta}{2} + it) H} \,  \mathcal{O})}{\mathrm{Tr}\,e^{-\beta H}}\,, \label{wight1}
\end{align}
where we have used the time-evolved operator $\mathcal{O}(t) = e^{i H t} \,  \mathcal{O}\,e^{-i H t}$ and $\beta$ is the inverse temperature. Denoting the density matrix by $\rho = e^{-\beta H}/\mathrm{Tr}(e^{-\beta H})$, the above \eqref{wight1} can be recast into 
\begin{align}
     \langle \mathcal{O}(t)\,  \mathcal{O} \rangle_{\beta}^{(\mathrm{W})} =  \mathrm{Tr}\big(\rho\, e^{i H(t - i\beta/2)}   \mathcal{O} \,e^{-i H(t -  i\beta/2)} \,  \mathcal{O}\big) = \mathrm{Tr}\big(\rho\,   \mathcal{O} (t - i\beta/2)  \,  \mathcal{O}\big) = \langle  \mathcal{O}  (t - i\beta/2)\, \mathcal{O} \rangle_{\beta}^{\mathrm{th}} \,,
\end{align}
where the thermal two-point function at inverse temperature $\beta$ is defined as
\begin{align}
    \langle  \mathcal{O} (t)\,  \mathcal{O} \rangle_{\beta}^{\mathrm{th}} =  \mathrm{Tr}\big(\rho \,  \mathcal{O}  (t) \,  \mathcal{O}\big)\,. \label{thermal}
\end{align}
In the Euclidean time $\tau = i t$, the Wightmann two-point function is given by \cite{Dymarsky:2021bjq}
\begin{align}
    \mathcal{C}(\tau) := \langle \mathcal{O}  (\tau)\,  \mathcal{O} \rangle_{\beta}^{(\mathrm{W})}  := \langle  \mathcal{O}  (-i(\tau + \beta/2))\, \mathcal{O} \rangle_{\beta}^{\mathrm{th}} \,.
\end{align}
This is a universal relation between the Wightmann and thermal two-point functions. Given the thermal function, it can always be converted into the Wightmann function using the above relation. Unless explicitly mentioned, we will always consider the Wightmann inner product as our definition of the autocorrelation function.

Let us consider the example of 2$d$ CFT in $\mathbb{R}^2$. The autocorrelation function is given by the Wightmann inner product of the form \cite{Dymarsky:2021bjq}
\begin{align}
    \mathcal{C}(\tau) = \sec(\pi \tau/\beta)^{2 \Delta}\,, \label{autocft1}
\end{align}
where $\Delta$ is the operator scaling dimension and $\beta$ is the inverse temperature. This function has poles on the real axis of the Euclidean time at $\tau=\pm\beta/2$. This is a universal behavior in any field theory, which comes from the singularity when two local operators collide ${\cal C}(\tau)\propto |\tau\mp \beta/2|^{-2\Delta}$, when $\tau\rightarrow \pm \beta/2$. The order of the pole singularity $\Delta$ is the conformal dimension of $\cal O$.
The two-point function ${\cal C}(\tau)$ is related to the more standard ${\cal C}(t)$ discussed, e.g.,~in the context of large $q$ SYK model above by the Wick rotation.

The Toda equations when $\cal C$ is given by \eqref{autocft1} can be solved explicitly,  this is one of the simple cases solved through the ansatz \eqref{TodaAnsatz}  \cite{Dymarsky:2021bjq},
\begin{align}
    \uptau_n (\tau) = \frac{G(2+n) G(1+n + 2\Delta)}{G(2 \Delta) \Gamma(2\Delta)^{n+1}}   (\pi/\beta)^{n (n+1)} \sec (\pi  \tau/\beta)^{(n+2\Delta) (n+1)}\,,
\end{align}
where $G(n)$ is the Barnes Gamma function defined by  $G(n) := \prod_{k=2}^{n-2} k!$ and satisfying the property $G(n+1) = G(n) \Gamma(n)$. Using the Toda chain technique with $\tau_0 = 0$, for which Eq. \eqref{autocft1} equals unity, the Lanczos coefficients read \cite{Dymarsky:2021bjq}
\begin{align}
\begin{split}
       a_n &= 0 \,,~~~ n\geq 0\,, \\
  b_n &= \frac{\pi}{\beta} \sqrt{n(n-1+2 \Delta)} \,, ~~~ n \geq 1\,.  \label{anbncft}
\end{split}
\end{align}
It is natural to check the above results using the moment method in the special case of $\Delta=1$. For this, we compute the power spectrum. Putting $\tau = i t$, we find $\mathcal{C}(t) = \sech(\pi t/\beta)^2$, and thus \cite{Camargo:2022rnt}
\begin{align}
    \Phi (\omega) = \int_{-\infty}^{\infty} d t \,e^{- i\omega t} \, \sech^2(\alpha t) = \frac{\beta^2 \omega }{\pi  \sinh (\beta  \omega/2)} \,. \label{mom2}
\end{align}
Since the autocorrelation is even in $t$, the odd moments vanish. The even moments are given by
\begin{align}
    m_{2n} = \frac{1}{2\pi} \int_{-\infty}^{\infty} d \omega\, \omega^{2n} \,\Phi (\omega) = \frac{2}{\pi^2 \beta^{2n}} \left(4^{n+1}-1\right) \zeta (2 n+2) \Gamma (2 n+2)\,,
\end{align}
where $\zeta (z)$ is the Riemann-zeta function. Applying the moment method, we recover the same Lanczos coefficients in \eqref{anbncft} with $\Delta = 1$. The linear growth persists indefinitely, with the slope dictated by  $\alpha = \pi/\beta$. In this case the Krylov complexity growth rate is $\lambda_K := 2\alpha = 2\pi/\beta = \lambda_{\mathrm{MSS}}$, saturating the Maldacena-Shenker-Stanford (MSS) bound \cite{Maldacena2016}.

The unexpected linear growth of the Lanczos coefficients in free CFT puts the validity of the universal operator growth hypothesis into question. However, linear growth arises due to the infinite UV cutoff in quantum field theory. To overcome this, we must put the field theory on a lattice such that the lattice spacing acts as a regulator of the theory. Alternatively, we can introduce a UV regulator into the frequency space, which can be either \emph{hard} or \emph{smooth}. Let us consider the hard UV regulator $\Lambda$, such that the moments are given by \cite{Camargo:2022rnt}
\begin{align}
    m_{2n} = \frac{1}{2\pi} \int_{-\Lambda}^{\Lambda} d \omega\, \omega^{2n} \,\Phi (\omega)\,. \label{momreg}
\end{align}
This regulates the integration range from $(-\infty, \infty)$ to $[-\Lambda, \Lambda]$. Evaluating the expression \eqref{momreg} is a tiresome task. However, in the asymptotic limit for large $n$, the integral is dominated by the frequency $|\omega| = \Lambda$, and the ratio of the consecutive moments is controlled by \cite{Camargo:2022rnt}
\begin{align}
    \lim_{n \rightarrow \infty} \frac{m_{2n+2}}{m_{2n}} \sim \frac{\omega^{2n+2}}{\omega^{2n}}\bigg|_{|\omega| = \Lambda} = \Lambda^2\,. \label{momratio1}
\end{align}
This behavior fundamentally affects the growth of the Lanczos coefficients. In other words, the Lanczos coefficients cease to grow indefinitely and approach a constant. The saturation value $b_{\mathrm{sat}}$ is controlled by the ratio of the moments \cite{Camargo:2022rnt}
\begin{align}
    \lim_{n \rightarrow \infty} \frac{m_{2n+2}}{m_{2n}} = 4 b_{\mathrm{sat}}^2\,. \label{momratio2}
\end{align}
Combining \eqref{momratio1} and \eqref{momratio2}, we obtain the saturation value \cite{Camargo:2022rnt}
\begin{align}
    b_{\mathrm{sat}} \sim \Lambda/2\,,
\end{align}
which is linearly proportional to the UV regulator. For a finite $\Lambda$, the indefinite growth of Lanczos coefficients ceases and reaches a plateau. Hence, one needs to look at the Lanczos coefficients beyond the UV cutoff to determine the chaotic nature of quantum field theories. For a soft regulator, see \cite{Camargo:2022rnt}.

On the other hand, the IR scale has a different effect. This includes considering the massive theory, where the bare mass in the field theory behaves as an IR cutoff, or the theory is placed in a compact manifold. In such cases, the Lanczos coefficients split into two smooth branches,  even and odd.
In the case of massive theory two branches grow linearly, with the same slope $\alpha =\pi T$, but different intercepts, this behavior is called ``persistent staggering'' \cite{Avdoshkin:2022xuw, Camargo:2022rnt}.

The case of compact manifolds is more complex. In this case even and odd branches of $b_n$ grow linearly, but with different slopes \cite{Avdoshkin:2022xuw}
\begin{align}
\label{twoslopes}
    b_n= \left\{
    \begin{array}{cr}
         \alpha_e n+\gamma_e+o(1) & \,\,n\,\,\, {\rm is\, even}\,, \\
         \alpha_o n+\gamma_o+o(1) & n\,\,\, {\rm is\, odd}\,,  
    \end{array}
    \right.
\end{align}
both different from $\pi T$, demonstrating behavior going beyond the universality of the original operator growth hypothesis of \cite{parker2019}.

Using the integral over Dyck paths formalism a particular combination of the coefficients $\alpha_e,\alpha_o$ can be related to the singularity of ${\cal C}(\tau)$,  located at $\tau=\pm \beta/2$, and the combination of $\gamma_e,\gamma_o$ can be related to $\Delta$ \cite{Avdoshkin:2022xuw}.

\subsection{Krylov complexity in Holography} \label{Secholo}

One of the interesting settings to study Krylov complexity is provided by holography. 
Holographic theories in the semiclassical gravity regime, with the bulk geometry being the black hole in $AdS$, exhibit exponential growth of OTOC. This makes holography a natural playground to study the relation between the exponential growth of Krylov complexity and OTOC.

Furthermore, by providing a non-perturbative definition of quantum gravity in the bulk via boundary QFT, holographic theories serve as fertile ground to study the relation between gravity and quantum complexity. That led to a number of influential conjectures geometrizing quantum complexity in the bulk \cite{Brown16prd,Jefferson:2017sdb,Brown2018,belin2022does}. This makes holography a natural starting point to investigate the relation between Kyrlov complexity and its more established QFT analogs.

Formulating Krylov complexity in holography is analogous to the field theory case. The starting point is thermal Wightman-ordered two-point function \eqref{wight1}, i.e.,~with the operators placed on opposite points on the thermal circle, and for convenience analytically continued to Euclidean time, thus removing the question of time ordering. Computing the Lanczos coefficients and Krylov complexity is then completely analogous to the previous section, as is expected, since the holographic duality is, in the end, one of the descriptions of field theory. 

There are only a handful of holographic examples in which the Krylov complexity has been evaluated so far, some of them numerical and rely on the semiclassical bulk approximation, as in \cite{Dymarsky:2021bjq}. There are also calculations in thermal AdS${}_3$ and BPZ black hole
dual to 2d CFTs at small and high temperature \cite{Avdoshkin:2022xuw}. These calculations are in line with other QFT results and we already referred to them while discussing field theory or Krylov complexity at finite temperature in the preceding sections. The main takeaway is that holography helps solidify support for the extension of the Maldacena-Stanford-Shenker bound on OTOC growth \eqref{boundTextended}.
Another point worth mentioning is that holographic theories, at least in a certain case of thermal AdS${}_3$ exhibit the two-slopes behavior \eqref{twoslopes}, which goes beyond the universality, originally outlined in \cite{parker2019}. This case is notable also because the Krylov complexity is trapped to IR values: an initial growth of $K(t)$ stops at an early time independent of the UV cutoff, after which $K(t)$ oscillates. This means the asymptotic or time-averaged value of Krylov complexity is independent of the value of UV cutoff, which marks a sharp distinction with the behavior of computational or holographic complexities that are explicitly UV cutoff-dependent \cite{Avdoshkin:2022xuw}.

Discussion of Krylov complexity in holography would not be complete without mentioning possible bulk manifestations of $K(t)$. It is essentially an established lore in holography that in the limit of classical gravity, all physically meaningful quantities in boundary field theory should have a clear geometric interpretation in the bulk. One remarkable example is the Ryu-Takayanagi (RT) prescription, which calculates entanglement in field theory \cite{RT}. 
Asking the same question for Krylov complexity is suggestive, especially because the thermal two-point function, from which $K(t)$ can be mathematically derived, admits a full geometric description in the so-called geodesic approximation. Yet this question does not seem to have a simple answer for standard $d\geq 2$ holography, but we note that an interesting geometric proposal was formulated recently for the Jackiw–Teitelboim (JT) gravity 
\cite{Rabinovici2023DSSYK}.

We briefly elaborate on the implications of this proposal. The bulk Hilbert space in the perturbative regime of Lanczos spectra i.e., $n \sim o(1)$ is constructed using the tridiagonal form of the transfer matrix in the double-scaled SYK (DSSYK) model. The ensemble average partition function of the DSSYK Hamiltonian is given by
\begin{align}
    \langle Z(\beta + it)\rangle|_{\beta =0} = \langle \mathrm{Tr}(e^{-i H t}) \rangle = \langle 0|e^{-\beta H}|0\rangle,
\end{align}
where $T$ is the tridiagonal transfer matrix and $|0 \rangle$ is the zero-chord state, an effective infinite-temperature TFD state in the averaged theory \cite{Rabinovici2023DSSYK}. The Lanczos coefficients can be computed via the moment method of the partition function, yielding $\mathsf{b}_n = \sqrt{\frac{1-q^n}{1-q}}$. Here, $q \in (0,1]$ is a parameter of the DSSYK model encoding the ratio of interacting fermions to the total number of fermions \cite{Berkooz:2018jqr}. It also controls the DOS, transitioning from a semicircle distribution to a normal distribution in two extreme limits, analogous to spin glasses \cite{Erdos2014, Berkooz:2018qkz}.

This result was first obtained in Refs.\,\cite{Rabinovici2023DSSYK, Nanunpub} in different limits, such as the triple-scaled limit where $q \rightarrow 1$, focusing near the ground state of the Hamiltonian. This is also the limit where the gravitational theory emerges. It was shown that Krylov space is a consistent formulation to describe the bulk Hilbert space. In the perturbative regime, the Krylov state complexity has a dual description of the length (size) of the wormhole \cite{Rabinovici2023DSSYK, Lin:2022rbf}. However, this description breaks down in the non-perturbative regime of Lanczos coefficients, known as the Lanczos descent.

Since the boundary theory must have a finite-dimensional Hilbert space, this requires finding a finite-dimensional Hamiltonian with full non-perturbative Lanczos spectra, whose perturbative description matches the previous result of Ref.\,\cite{Rabinovici2023DSSYK}. This was recently achieved in Ref.\,\cite{Nandy:2024zcd, Balasubramanian:2024lqk} by obtaining fully non-perturbative Lanczos coefficients from the RMT description of the DSSYK model, where the perturbative regime indeed agrees with \cite{Rabinovici2023DSSYK}. The finite-dimensional Hilbert space immediately predicts a saturation of complexity while the Krylov basis provides a physical status of the auxiliary chord space. The conclusion offers a novel interpretation of Krylov (state) complexity as a fully non-perturbative description of wormhole size or the volume of the Einstein-Rosen bridge in gravitational theory, suggesting promising avenues for future research.

\section{Krylov complexity and Integrability} \label{secIntSys}

\begin{figure}
\centering
\includegraphics[width=0.95\linewidth]{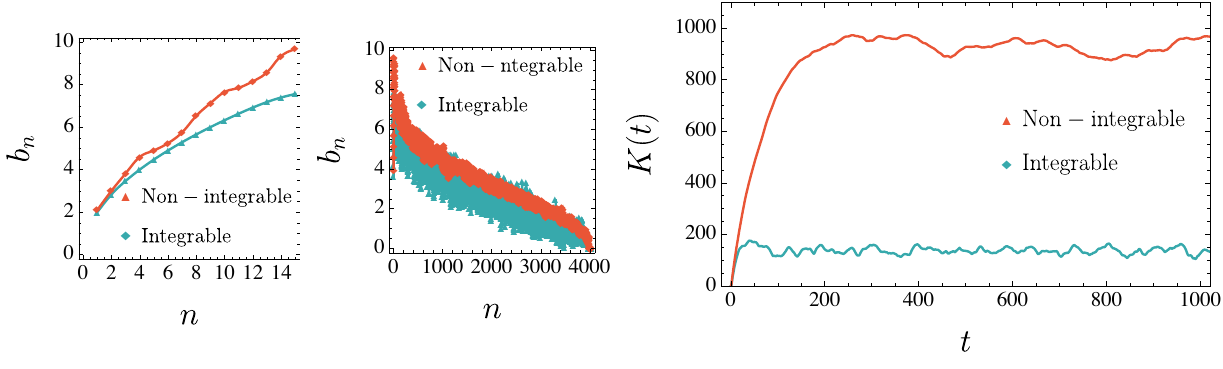}
\caption{Left and middle figure show the initial growth and the full Lanczos spectrum for the mixed-field Ising model (MFIM) $H_{\mathrm{MFIM}} = - \sum_{j=1}^{N-1}  \sigma^{z}_{j} \sigma^{z}_{j+1} - g \sum_{j=1}^{N} \sigma^{x}_j - h \sum_{j=1}^{N} \sigma^{z}_{j}$ in the integrable and the non-integrable limit. The parameter choices are
$(g,h) = (1,0)$ (integrable) and $(g,h) = (-1.05,0.5)$ (non-integrable) \cite{Banuls11, Roberts2015}. We chose the initial operator $\sigma_1^z$ with periodic boundary conditions, with system size $N =6$. The initial Lanczos coefficients exhibit sublinear growth in the integrable regime and linear growth in the non-integrable regime. In both regimes the coefficients terminate at $n = D_K = 4033$, saturating the bound \eqref{Kbound}. However, the large $n$ Lanczos coefficients show greater fluctuation in the integrable regime compared to the non-integrable limit. This is incorporated in the log variance, namely $\Delta b_n^{\mathrm{integrable}} \gg \Delta b_n^{\mathrm{non-integrable}}$. This figure uses the moving average of order $6$ in both regimes to reduce the noise, which does not impact the behavior. The right figure shows the behavior of the Krylov complexity in both regimes. The complexity saturates at a higher value in non-integrable regimes compared to the integrable regime due to the ``Krylov localization'' \cite{Rabinovici:2022beu, Rabinovici:2021qqt}. The figures are adapted from \cite{Bhattacharya:2023zqt} with different system parameters.}
\label{IsingKcomp}
\end{figure}

Recent research on Krylov complexity has revealed its potential as a discerning tool for differentiating between integrable and non-integrable systems, an idea that goes back to Parker et al. \cite{parker2019}. This distinction has been demonstrated through concrete examples within XXZ spin chains \cite{Rabinovici:2022beu, Rabinovici:2021qqt, Bhattacharya:2023zqt}, Ising spin chains \cite{Bhattacharya:2022gbz, Bhattacharya:2023zqt, Espanyol23, Aranyanonlocal}, Bose-Hubbard model \cite{BhattacharyyaBoseHubbard2023}, and Floquet systems \cite{AditiFloq, Yates2022floq, Nizami23, Yeh:2023fek}, encompassing both their integrable and non-integrable variants. The term ``non-integrable'' is used here to denote systems where integrability is disrupted by adding a specific term that can either strongly \cite{AvivaSantosQchaosXXZ, Pandey20, SantosXXZ} or weakly break the integrability \cite{szasz2021weak, Fitos23, Olexei23}. Figure \ref{IsingKcomp} (top row) illustrates the initial and entire Lanczos spectrum for the mixed-field Ising model (MFIM) in both integrable and non-integrable limits, with the initial operator $\sigma_1^z$. Due to the instability of the coefficients, the Full orthogonalization (FO) method \cite{Rabinovici:2020ryf} is performed. To suppress the noise, the moving average of order $6$ in both the limits has been performed. The Lanczos coefficients exhibit sublinear growth in the integrable case and linear growth in the non-integrable scenario. The Lanczos spectrum terminates at $n = D_K$, constrained by the finite size of the system. Intriguingly, the Krylov dimension $D_K$ remains the same across both scenarios. In particular, the Krylov dimension bound \eqref{Kbound} is saturated in both cases, depending on the chosen initial operator.

Further, the Lanczos spectrum in the integrable limits exhibits greater disorder compared to its chaotic analogs. To quantify the disorder within the Lanczos coefficients, one can define the logarithmic variance of the Lanczos sequence as follows
\cite{Rabinovici:2022beu, Rabinovici:2021qqt}, 
\begin{align}
    \Delta b_n = \mathrm{Var}\big(\log(b_n/b_{n+1})\big)\,. \label{distbn}
\end{align}
This measure effectively captures the degree of disorder among the Lanczos coefficients, which was analyzed in quantum billiards \cite{Hashimoto23}, the Bose-Hubbard system \cite{BhattacharyyaBoseHubbard2023}, the SYK model \cite{Menzler:2024atb}, and random matrices \cite{Scialchi:2023bmw}. In scenarios where integrability is strongly broken, integrable systems exhibit a higher degree of disorder in the Lanczos spectrum than their non-integrable counterparts. This disorder correlates with an auxiliary off-diagonal Anderson hopping model, leading to a phenomenon ascribed as ``Krylov localization'' \cite{Rabinovici:2021qqt}.

The level of disorder present in the Lanczos sequence influences the late-time saturation value of Krylov complexity.  This saturation value is reduced in the integrable phase, yet it increases as the system transitions towards the chaotic phase, as depicted in the bottom row of Fig.\,\ref{IsingKcomp}. This observation was initially made in the XXZ chain with an integrability-breaking term \cite{Rabinovici:2022beu, Rabinovici:2021qqt} and was subsequently identified in other models such as the Ising spin chain \cite{Bhattacharya:2022gbz, Bhattacharya:2023zqt, Espanyol23, Madhok23} and the Bose-Hubbard model \cite{BhattacharyyaBoseHubbard2023}. Nonetheless, the saturation value remains below the threshold of $D_K/2$, a benchmark typically met in genuinely chaotic systems such as the SYK$_4$ model \cite{Rabinovici:2020ryf}. However, this pattern is not universally applicable, as the XXZ chain and the Ising chain do not always conform to this behavior \cite{Rabinovici:2022beu, Rabinovici:2021qqt, Bhattacharya:2023zqt}. Correspondingly, the saturation value of Krylov complexity is much larger in the non-integrable limit, yet lower than $D_K/2$ (see Fig.\,\ref{IsingKcomp}).

\begin{figure}
\hspace*{-0.5 cm}
\includegraphics[width=1.02\linewidth]{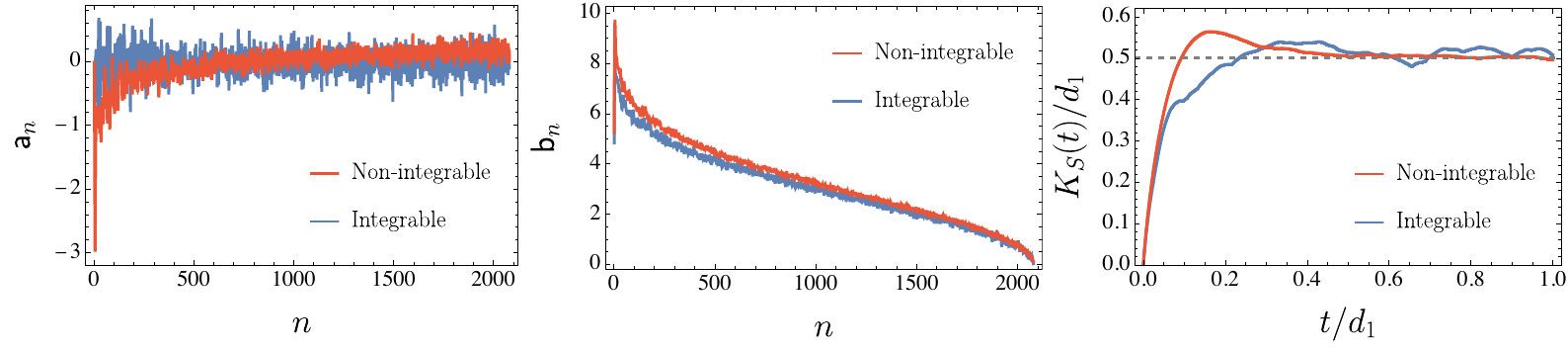}
\caption{(Left and middle) The Lanczos spectrum and (right) the Krylov state (spread) complexity for the MFIM model: $H_{\mathrm{MFIM}} = - \sum_{j=1}^{N-1}  \sigma^{z}_{j} \sigma^{z}_{j+1} - g \sum_{j=1}^{N} \sigma^{x}_j - h \sum_{j=1}^{N} \sigma^{z}_{j}$ in the integrable and the non-integrable limit. The parameter choices are
the same as Fig.\,\ref{IsingKcomp}, except we chose the open boundary condition. We chose the parity $P = +1$ block, with Krylov dimension $d_1 = 2080$. The initial state is chosen as the infinite-temperature TFD state. The figures are adapted from \cite{Camargo:2024deu}.}
\label{IsingKcompspread}
\end{figure}

Analogously, we compute the Krylov state (spread) complexity in the MFIM model for both integrable and non-integrable regimes (see Fig.\,\ref{IsingKcompspread}). Given that the Hamiltonian splits into two parity sectors with unequal dimensions, we selected the parity sector with parity $P = +1$. In the non-integrable limit, eigenvalue statistics follow the Wigner-Dyson distribution corresponding to the GOE, whereas the integrable limit exhibits Poissonian statistics. We chose the infinite-dimensional TFD state as the initial state. The Lanczos coefficients and the complexity are illustrated in Fig.\,\ref{IsingKcompspread}. In the non-integrable case, we observe a peak followed by saturation, which is absent in the integrable case. Notably, the saturation is identical for both integrable and chaotic cases, depending solely on the dimension of the Hilbert space.

\section{Applications of Krylov complexity}
\label{secKryapp}

\subsection{Applications to Quantum Control}
\label{secQControl}

Adiabatic driving offers a powerful scheme for quantum state preparation in quantum science and technology. Adiabatic strategies provide the rationale for adiabatic quantum computation and quantum annealing \cite{Lidar18}. Their implementation is hindered by the presence of noise, uncontrolled sources of errors, and the coupling to the surrounding environment. It is thus desirable to find alternative nonadiabatic driving schemes without the requirement for slow driving.  
This is the scope of {\it shortcuts to adiabaticity}  \cite{Chen10,Torrontegui11,GueryOdelin19}. Among the techniques used for their engineering, counterdiabatic driving (CD) \cite{Demirplak03,Demirplak05,Demirplak08}, also known as transitionless quantum driving \cite{Berry09}, stands alone as a universal strategy.
Its original formulation focuses on driven quantum systems evolving unitarily.
Consider an uncontrolled reference Hamiltonian with a point-like spectrum and spectral decomposition $H_0(\lambda)=\sum_{n=1}^dE_n(\lambda)|n(\lambda)\rangle\langle n(\lambda)|$, that is modulated by a single time-dependent parameter $\lambda(t)$ for simplicity. In the  limit of slow driving, the time-evolution of an initial eigenstate $|n(\lambda_0)\rangle$  follows the adiabatic trajectory
$|\psi_n(t)\rangle =\exp[i\upphi_n(t)]|n(\lambda(t))\rangle$\,, 
where the phase factor is the sum of the dynamical phase and the geometric phase
 \begin{equation}
 \upphi_n(t)=-i\int_0^tds E_n(\lambda(s))
+i\int_{\lambda_0}^{\lambda_t}d\lambda \langle n
 |\partial_\lambda n\rangle\,.
\end{equation}
The adiabatic trajectory $|\psi_n(t)\rangle$ with respect to $H_0$ is the exact solution of the time-dependent Schr\"odinger equation $i\partial_t|\psi(t)\rangle=H|\psi(t)\rangle$ when the dynamics is generated by a different Hamiltonian $H$. 
The latter can be written as the sum $H=H_0+H_{\rm CD}$ of the uncontrolled Hamiltonian and the CD term $H_{\rm CD}=\dot{\lambda}A(\lambda)$,  where
\begin{eqnarray}
A(\lambda)=
i\sum_{n=1}^d\left[|\partial_\lambda n\rangle\langle n|-\langle n|\partial_\lambda n\rangle |n\rangle\langle n|\right] \,,
\end{eqnarray}
is also known as the adiabatic gauge potential.
CD in many-body systems generally involves nonlocal multiple-body interactions \cite{delcampo12,Takahashi13,Saberi14,Damski14}. This has motivated the development of schemes to approximate the CD auxiliary Hamiltonians by variational methods \cite{Opatrny14,Saberi14,Sels17} or otherwise \cite{Takahashi13,Mukherjee16}. While CD schemes for many-body systems are hard to implement in analog quantum devices, they are amenable to digital quantum schemes \cite{Fauseweh24,delcampo12,Saberi14}. Harnessing the advantage of CD to steer the dynamics with the flexible implementation of digital schemes is the basis of digitized counterdiabatic quantum algorithms \cite{Chandarana22,
Hegade21,Hegade21factorization,Hegade22,Hegade22DCQO,Chandarana23}. Even in this context,  the truncated CD controls are desirable and derived as leading orders in various series expansions. 
The integral representation of the CD term
\cite{Hastings04,Hastings05,Hastings2010Locality}
\begin{align}
 A(\lambda) = -\frac{1}{2}
 \lim_{\eta\to 0}
 \int_{-\infty}^\infty ds\,{\rm sgn}\,(s)e^{-\eta|s|} e^{iH_0(\lambda)s}\partial_\lambda H_0(\lambda)e^{-iH_0(\lambda)s}\,,
 \label{AGPint}
 \end{align}
 motivates the nested commutator expansion
 \cite{Claeys19,Pandey20}
 \begin{equation}
 A(\lambda)=i\sum_k \alpha_k(\lambda)
 \mathcal{L}_\lambda^{2k-1}\partial_\lambda H(\lambda)\,, \label{var}
 \end{equation}
 where the coefficients $\alpha_k$ are often determined by a variational approach.
As an alternative, the Krylov expansion of the CD term has been presented in \cite{Takahashi24, Bhattacharjee23, Lim:2024jtf}. Choosing $ \mathcal{O} =\partial_\lambda H(\lambda)$ and using the Krylov expansion of $ \mathcal{O} (s)=  \sum_{n=0}^{D_K-1} i^n\varphi_n(s)  \mathcal{O}_n$ in Eq. (\ref{AGPint}) yields 
\begin{eqnarray}
 A(\lambda)=i b_0 \sum_{k=1}^{d_A} \alpha_k(\lambda)\,\mathcal{O}_{2k-1}\,.
\end{eqnarray}
Here, $b_0^2=(\partial_\lambda H,\partial_\lambda H)$ while $d_A=D_K/2,(D_K-1)/2$ for even and odd Krylov dimension $D_K$, respectively.
The expansion coefficients are then fixed in terms of the Lanczos coefficient, 
circumventing the need for their approximate determination through a variational approach. Specifically, for even $D_K$, they are set by the iterative relations
\begin{eqnarray}
\alpha_1 = -\frac{1}{b_1}, \quad \alpha_{k+1}= -\frac{b_{2k}}{b_{2k+1}}\alpha_{k}\,,
\end{eqnarray}
while they can be found as a solution to a linear matrix equation for off $D_K$ \cite{Takahashi24}.

Knowledge of the Krylov expansion makes it possible to relate the features of the CD term and $A(\lambda)$ with the properties of the system through the operator growth hypothesis and the analysis of the Lanczos coefficients.
The norm of the CD term is used to quantify the cost of CD protocols \cite{Demirplak08}. It is further related to fidelity susceptibility and the quantum geometric tensor \cite{delcampo12,Funo17} and using the Krylov expansion, one finds \cite{Takahashi24}
\begin{eqnarray}
(A,A)=b_0^2\sum_{k=1}^{d_A}\alpha_k^2\,.
\end{eqnarray}

The expansion of the CD term in Krylov space is likely to prove useful in other applications. 
In the conventional approach, the CD term enforces parallel transport in the instantaneous eigenbasis of the uncontrolled Hamiltonian. 
%In other contexts, alternative choices are preferable.  
By contrast, generalizations of the CD to open quantum systems involve parallel transport of the generalized eigenstates of the Liouvillian \cite{Vacanti14} or the instantaneous eigenstates of the reduced density matrix of the system, also known as natural orbitals \cite{Alipour20}.  
The rationale behind CD can also be applied to parameter estimation in quantum metrology \cite{Giovannetti11}. In this context, optimal strategies involve parallel transport along the operator given by the parametric derivative of the generator of evolution \cite{Pang17,Yang22}.  The Krylov expansion of the CD term can also be utilized in numerical methods involving parallel transport along a family of quantum states, e.g.,  of matrix product states in tensor network algorithms \cite{Kim2023variational,Keever2023adiabatic}.

\subsection{Applications to Quantum Computing}
\label{secQComp}

While Krylov subspace methods have a long tradition in conventional computing, their implementation for many-body systems in quantum computers is not straightforward. They are computationally costly as the dimension of the Krylov basis scales exponentially with the system size. 
In addition, quantum computers based on the circuit model are naturally suited to implement unitaries rather than powers of the generator of evolution \cite{Manenti23}.
Recent progress has advanced the application of Krylov subspace methods to quantum computers by circumventing these challenges.
 
Variants of Krylov subspace methods have been put forward, replacing powers of the Hamiltonian with unitaries. Such an approach is suited for approximated real-time evolution as well as imaginary-time evolution. The latter provides a natural scheme for the preparation of ground states and thermal states \cite{Motta20,Yeter-Aydeniz20}. This has given rise to a family of hybrid quantum-classical algorithms \cite{Bharti21}.
An alternative approach relies on replacing the need for Hamiltonian powers with combined unitary evolutions \cite{Seki21}.
Additional advances have focused on the exact construction of the Krylov basis in a quantum computer without relying on the simulation of real or imaginary time evolution \cite{Kirby23}.  This approach has the advantage of reducing exponential classical cost, being achievable in polynomial time and memory.

These recent efforts focus on the Hamiltonian as the generator of time evolution. The use in quantum computers of Krylov subspace methods for open quantum systems governed by Lindbladians and other generators is an enticing prospect that may be facilitated by progress in quantum simulation of open quantum systems \cite{Muller12,Ciccarello22}.

\section{Open problems} \label{secOP}

In what follows, we mention some open problems regarding the formalism of the Krylov subspace method for quantum dynamics, leaving aside applications that are expected to be many and broadly spread out. 
At the time of writing, most Krylov subspace methods in quantum dynamics remain restricted to time-independent generators.
Recent efforts have focused on extensions to time-dependent systems that can be described by a Floquet operator, using a Krylov expansion involving its powers \cite{AditiFloq, Yates2022floq, Nizami23, Yeh:2023fek, ZhengFloquetKrylov} and a unitary quantum circuit with Trotterized evolution \cite{suchsland2023krylov}. The use of the Floquet operator for the construction of the Krylov basis can be leveraged to describe the dynamics with arbitrary (nonperiodic) time-dependent Hamiltonians.  The dynamics is described as a diffusion problem in the Krylov lattice, but one with nearest-neighbor hopping that is inhomogeneous and time-dependent \cite{Takahashi25}.  
As an alternative approach, one may consider approximating the dynamics under a time-dependent generator by a step-wise sequence with constant generators. In addition, an arbitrary unitary evolution can be described using the Magnus operator, making the case for a Krylov expansion using its powers \cite{Blanes09}.

Beyond unitary dynamics, the progress reviewed in Sec.\,\ref{secOpen} has focused on  Markovian quantum systems with no memory, described by the Lindblad master equation  \cite{lindblad_generators_1976,gorini_completely_1976,rivas_open_2012}. The extension to general non-Markovian evolutions constitutes an interesting prospect \cite{rivas_open_2012,breuer_2016_NonMarkov,deVega2017}.
It is known that any time-continuous evolution described by a density matrix can be associated with a master equation of a generalized Lindblad form, with time-dependent rates and Lindblad operators \cite{Alipour20,GarciaPintos22}, making the extension of the Krylov basis construction in such setting desirable.  
Beyond the time-continuous case,  stochastic evolutions are essential in the treatment of fluctuating Hamiltonians \cite{vanKampen11, budini_quantum_2001},  the quantum-jump approach associated with the stochastic unraveling of master equations \cite{Plenio98,carmichael2007statistical}, and the theory of continuous quantum measurements \cite{Jacobs06,jordan2024quantum, Wiseman_Milburn_2009, Jacobs_2014}.  
In any such generalization, it would be interesting to investigate whether relations resembling those in the time-independent case hold between the set of Lanczos coefficients, the correlation function for operators, and the survival probability for quantum states. 

Regarding the notion of quantum state, we have presented the use of Krylov subspace methods for time-dependent operators in Sec.\,\ref{secObservable}, pure states in Sec.\,\ref{secStates}, and mixed density matrices in Sec.\,\ref{secDensity}. Quantum evolution can be discussed in many other representations. Among the phase-space quasiprobability distributions, the most celebrated is that introduced by Wigner \cite{Wigner32,Hillery84}, to which Krylov subspace methods have recently been applied \cite{basu2024complexity}. Other phase space distributions such as the $P$ and $Q$ distributions are frequent in quantum foundations \cite{Zachos05},  many-body physics \cite{Polkovnikov10}, and quantum optics \cite{MandelWolf1995,Barnett02}.  

Another open problem focuses on understanding the role of symmetry in relation to the dynamics in Krylov space, including the behavior of the Lanczos coefficients and the growth of Krylov complexity. 
For time-independent Hamiltonians, the complexity algebra leads to the identification of different classes of evolutions, involving not only the Liouvillian but also the anti-Liouvillian and the Krylov complexity operator, as discussed in Sec.\,\ref{secKdyn}. 
By contrast, the traditional symmetry classifications in quantum physics focus primarily on the generator of evolution. 
Dyson's three-fold way classification led to the introduction of the Gaussian and circular ensembles distinguished by the Dyson index $\upbeta_D$ \cite{Dyson1962}. 
Altland and Zirnbauer enriched this classification, including time-reversal, particle-hole, and chiral symmetries \cite{Atland1997}. This classification has been generalized to non-Hermitian matrices describing Hamiltonians as well as Lindbladians \cite{Kawabata2019,GarciaGarcia2022,Kawabata2023,Lucas2023PRX}.
It remains to be seen whether such symmetry classes imprint a clear signature on the dynamics in Krylov space.  
Further, the integrable structure of the Krylov dynamics in isolated systems has been established in terms of the Toda flow  \cite{dymarsky2020a}. One may wonder whether instances of integrable dynamics in Krylov space can be identified under more general evolutions, e.g., in non-Hermitian and open systems. And if such instances exist, do they have any connections with other notions of integrability? In particular, notions of integrability in open systems have been introduced by mapping the vectorized generator of evolution to non-Hermitian integrable models that are Bethe-ansatz solvable and satisfy the Yang-Baxter equation.

The development of such extensions and the analysis of their usefulness in applications remain to be explored and offer a tantalizing prospect for further studies.

%%%%%%%%%%%%%%%%%%%%%%%%%%%  CONCLUSION  %%%%%%%%%%%%%%%%%%%%%%%%%%%%%%%%%%
\section{Conclusion} \label{secConc}
In this review, we have provided a comprehensive account of the use of Krylov subspace to characterize the evolution of quantum systems. While the underlying tools are established entities within linear algebra—primarily utilized to execute efficient tridiagonalization of matrices for eigenvalue extraction—their significance has burgeoned within the realm of physics. Thanks to recent progress via the operator growth hypothesis, these methods provide a framework for the fundamental characterization of many-body quantum systems, their time evolution, the mechanisms underpinning thermalization, and quantum chaos, and their complexity.

Our discourse methodically elucidates the construction of the Krylov space, employing the Lanczos algorithm for unitary evolution. The discussion extends to encompass both pure and mixed quantum states, furnishing a comprehensive formulation for each. Progressing further, we delve into a suite of generalizations that incorporate the Arnoldi iteration and the bi-Lanczos algorithm, addressing the challenges posed by non-unitary evolution in open quantum systems. As an adjunct to the Lanczos technique, the moment method—a viable alternative construction from the two-point autocorrelation function is introduced. These analytical frameworks are pivotal in the ongoing quest to unravel quantum chaotic attributes within the contexts of quantum field theory and holography via AdS/CFT correspondence.

Significant strides have been made analytically, particularly in relation to coherent states. These advancements have shed light on the geometric essence of quantum systems and established constraints on their fundamental characteristics, such as the quantum speed limits. Despite their conceptual simplicity, these tools wield the capacity to grapple with an array of complex systems. Examples include Random Matrix Theory (RMT), spin chains, and the Sachdev-Ye-Kitaev (SYK) models - the latter sharing a close connection to gravitational theories. Through these paradigms, we glean insights that inform our understanding of the fundamental bound of quantum chaos and its generalization. Such understanding is instrumental in unraveling the intricacies of quantum integrability and quantum control problems. Throughout this text, we intersperse analytical and numerical examples to reinforce the underlying theoretical framework.

To date, the study of Krylov subspace methods for quantum dynamics has been mostly confined to theoretical physics, applied mathematics, and computer science. As progress is made in combining Krylov methods with quantum algorithms, their implementation in quantum devices seems feasible in quantum computers. The development of experimental methods for probing Krylov complexity measures is in a nascent stage. However, the groundwork laid by existing methods—such as those employed to simulate the spectral form factor and the out-of-time-ordered correlator (OTOC) in digital quantum simulators—provides a promising foundation \cite{vermersch_probing_2019, green_experimentalOTOC_2022, joshi_probing_2022}. Nonetheless, this endeavor presents a formidable challenge that is essential to foster progress harnessing the development of quantum processing units towards computational supremacy in quantum computation and quantum simulation of nonequilibrium phenomena \cite{Arute2019,Zhong2021,Madsen2022,Ebadi2022,Kim2023,Shaw2024,king2024computational}.

%%%%%%%%%%%%%%%%%%%%%%%%%%%  ACKNOWLEDGEMENTS  %%%%%%%%%%%%%%%%%%%%%%%%%%%%%%%%%%
\section*{Acknowledgments}
We are thankful to Budhaditya Bhattacharjee, Hugo A. Camargo, Xiangyu Cao, Pawe\l{} Caputa, Nicoletta Carabba, Aurelia Chenu, Pieter W. Claeys,  
Íñigo L. Egusquiza, Niklas H\"ornedal, Norihiro Iizuka, Victor Jahnke, Norman Margolus, Javier Molina-Vilaplana, Masahiro Nozaki, Tanay Pathak, Tomaž Prosen, Zdenek Strakos, Shinsei Ryu, Lucas S\'a, Lea F. Santos, Aninda Sinha, Julian Sonner, Ruth Shir, Kazutaka Takahashi, Jing Yang, Zhuo-Yu Xian, and Zhenyu Xu for insightful discussions. We acknowledge financial support from the Luxembourg National Research Fund (project  No. grant\,17132054 and 16434093, and Attract Grant No.\,15382998). One of these projects has received funding from the QUANTERA II Joint Programme with cofunding from the European Union's Horizon Europe research and innovation programme. The work of P.N. is supported by the Japan Society for the Promotion of Science (JSPS) Grant-in-Aid for Transformative Research Areas (A) ``Extreme Universe'' No.\,21H05190. A.D. acknowledges support by the NSF under grant 2310426. For the purpose of open access, the authors have applied a Creative Commons Attribution 4.0 International (CC BY 4.0) license to any Author Accepted Manuscript version arising from this submission.

%%%%%%%%%%%%%%%%%%%%%%%%% LIST OF SYMBOLS %%%%%%%%%%%%%%%%%%%%%%%%%%%%%
    
\section*{List of Symbols and Acronyms}
\begin{table}[H]
    \centering
    \resizebox{!}{.48\textheight}{
    \begin{tabular}{ll}
%        $a^*$ & Complex conjugate of $a$\\ 
        $\mathscr H$ & Hilbert space \\ 
        $\id$ & Identity operator \\
        $T$ & Temperature \\ 
        $\beta$ & Inverse temperature \\ 
        $\upbeta_D$ & Dyson beta index \\
        $Z(\beta)$ & Partition function at inverse temperature $\beta$ \\ 
        $d$ & Hilbert space dimension \\ 
        $D_K$ & Krylov dimension of operators\\ 
        $t ~\mathrm{or} ~\tau$ & Lorentzian or Euclidean time \\ 
        $\mathcal{C}(t)$ & Autocorrelation function of time $t$\\ 
        $m_n$ & Liouvillian moments of the autocorrelation function\\
        $\Phi(\omega)$ & Spectral function of frequency $\omega$\\
        $\uptau_n (\tau)$ & Toda function of Euclidean time $\tau$\\
        $H$ & Hamiltonian \\ 
        $E_\ell,\;  \ket{\ell}$ & Eigenvalues, eigenvectors of $ H$ \\ 
        $\mathcal L$ & Liouvillian superoperator \\ 
        $\mathcal L_o$ & Lindbladian superoperator \\ 
        $\mathcal M$ & Anti-Liouvillian \\ 
        $\omega_{nm}$ & Eigenvalues of the Liouvillian $\mathcal L$ \\ 
        $|\omega_{nm})$ & Eigenvectors of Liouvillian $\mathcal L$ \\ 
        $ \mathcal{O}(t)$ & Observable in Heisenberg picture \\ 
        $|\mathcal{O}(t))$ & Vectorized observables\\ 
        $| \mathcal{O}_n)$ & Krylov basis (orthonormal version) for observables\\ 
        $|\mathsf{O}_n)$ & Krylov basis (monic version) for observables\\
        $a_n, b_n$ & (Bi)-Lanczos coefficients for the (Lindbladian) Liouvillian\\ 
        $\varphi_n(t)$ & Amplitudes in Krylov basis for observables \\ 
        $\mathcal{K}$ & Krylov complexity superoperator\\ 
        $K(t)$ & Krylov complexity \\
        $\Delta K(t)^2$ & Krylov variance \\
        $\ket{\Psi (t)}$ & pure quantum state \\ 
        $P (t)$ & Purity of a quantum state \\ 
        $S(t)$ & Survival amplitude \\ 
        $\mu_n$ & Hamiltonian moments of the survival amplitude \\ 
        $\ket{K_n}$ & Krylov basis element for pure quantum states \\ 
        $\mathsf{a}_n, \mathsf{b}_n$ & Lanczos coefficients for the Hamiltonian\\ 
        $\uppsi_n (t)$ & amplitudes in Krylov basis for pure states \\ 
        $\mathcal{K}_S$ & Krylov spread complexity operator \\ 
        $K_S(t)$ & Krylov spread complexity of states \\ 
        $ \rho(t)$ & Density matrix \\ 
        $|\rho(t))$ & Vectorized density matrix\\ 
        $|\rho_n)$ & Krylov basis for density operators\\ 
        $\phi_n(t)$ & amplitudes in Krylov basis for density matrices \\
        $L_k$ & Jump operators\\
        $h_{n,m}$ & Arnoldi coefficients\\
        $|p_n\rrangle$ & Bi-Lanczos vectors\\ 
        $\langle \cdots \rangle_{\beta}^{\mathrm{W}}$ & Wightman inner product \\
        $\langle \cdots \rangle_{\beta}^{\mathrm{th}}$ & Thermal inner product \\
        vec & vectorization\\
        RMT & Random Matrix Theory\\
        SFF & Spectral Form Factor\\
        OTOC & Out-of-Time-Order Correlator\\
        SYK & Sachdev-Ye-Kitaev\\
        MSS & Maldacena-Shenker-Stanford\\
        TFD & Thermofield double\\
        QFT (CFT) & Quantum (Conformal) field theory\\
        AdS & Anti-de Sitter\\
        OQS & Open Quantum System(s)\\
         CD & Counterdiabatic driving \\
    \end{tabular}}
    \label{tab:my_label}
\end{table}

%%%%%%%%%%%%%%%%%%%%%%%%% A BRIEF LIST OF REFERENCES OF PARTICULAR INTEREST %%%%%%%%%%%%%%%%%%%%%%%%%%%%%
    
\section*{A Brief List of References of primary topics}
\begin{table}[H]
    \centering
    \resizebox{!}{.22\textheight}{
    \begin{tabular}{ll}
        Proposal of the ``universal operator growth hypothesis'' and Krylov complexity  & \cite{parker2019} \\ 
        Proposal of Krylov state (spread) complexity & \cite{Balasubramanian:2023kwd} \\
        Subtleties regarding the hypothesis, saddle-dominated scrambling & \cite{Dymarsky:2021bjq, Bhattacharjee:2022vlt, huh2023} \\
        Krylov entropy and higher-order cumulants & \cite{barbon2019, Caputa:2021ori, Bhattacharjee:2022ave} \\ 
        Details of numerical algorithms & \cite{viswanath1994recursion, parker2019, Rabinovici:2020ryf} \\ 
        Krylov complexity in SYK and related models & \cite{parker2019, Jian:2020qpp, Rabinovici:2020ryf, Bhattacharjee:2022ave, HeSYK2022} \\
        Coherent states, complexity algebra, and quantum speed limits & \cite{caputa2021, Hornedal2022, Hornedal2023, Patramanis:2023cwz, Haque:2022ncl} \\ 
        Integrability and localization & \cite{Cao:2020zls, Rabinovici:2022beu, Trigueros:2021rwj} \\ 
        Relation with entanglement and modular Hamiltonian & \cite{Patramanis:2021lkx, CaputamodularH24}\\ 
        Krylov complexity in density matrices & \cite{Caputa2024} \\ 
       Krylov complexity in QFT and CFT & \cite{Dymarsky:2021bjq, Camargo:2022rnt, Avdoshkin:2022xuw, Caputa:2021ori, Kundu:2023hbk, Vasli:2023syq} \\ 
        Krylov complexity in quantum mechanics and billiards & \cite{Hashimoto23, Camargo:2023eev}\\
        Toda chain formalism & \cite{dymarsky2020a}\\
        Application to open quantum systems, Arnoldi and bi-Lanczos algorithms & \cite{Bhattacharya:2022gbz, Liu:2022god, Bhattacharjee:2022lzy, Bhattacharya:2023zqt, Bhattacharjee:2023uwx, NSSrivatsa:2023qlh, Carolan:2024wov}\\
        Driven, Floquet systems and unitary circuits & \cite{AditiFloq, Yates2022floq, Nizami23, Yeh:2023fek, suchsland2023krylov, ZhengFloquetKrylov,Takahashi25} \\ 
        Quantum spin chains & \cite{Rabinovici:2022beu, Rabinovici:2021qqt, Bhattacharya:2023zqt, Bhattacharya:2022gbz, Bhattacharya:2023zqt, bento2023krylov, Espanyol23, Aranyanonlocal} \\ 
        Connection to Nielsen and circuit complexity & \cite{Craps:2023ivc, Lv:2023jbv} \\ 
        Relation to quantum scars & \cite{Bhattacharjee:2022qjw, Nandy:2023brt} \\ 
        Relation with holography & \cite{Jian:2020qpp, Kar2022, Rabinovici2023DSSYK, Nandy:2024zcd, Balasubramanian:2024lqk} \\ 
        Quantum control, shortcuts to adiabaticity and gauge potentials & \cite{Takahashi24, Bhattacharjee23}\\ 
        Phase space and Wigner negativity & \cite{basu2024complexity} \\ 
        Application to RMT & \cite{Balasubramanian:2022dnj, Balasubramanian:2023kwd, Erdmenger2023, Nandy:2024zcd} 
    \end{tabular}}
    \label{tab:my_label}
\end{table}

%%%%%%%%%%%%%%%%%%%%%%%%%%%  BIBLIOGRAPHY  %%%%%%%%%%%%%%%%%%%%%%%%%%%%%%%%%%%%%%
%\normalem
\bibliographystyle{elsarticle-num} 
\bibliography{QDKS}

\end{document}